\documentclass[microtype]{gtpbook} %_pb for Amazon paperback and Google
\pdfoutput=1
\usepackage{pinlabel,adjustbox,labelfig,fancyhdr,makeidx}
\makeindex

\title[A new paradigm]{A new paradigm for the universe}

    %% cosmology
  %% black holes
  %% gal and stell struc
  %% gal struc and dyn
  %% geometrodyn
\renewcommand{\chaptermark}[1]{\markboth{Chapter \thechapter\qua #1}{}}
\renewcommand{\sectionmark}[1]{\markright{\thesection #1}}
\def\sh#1{\section{#1}}%\addcontentsline{toc}{subsection}{#1}

\def\ssh#1{\subsection*{#1}\addcontentsline{toc}{subsection}{#1}}
\def\astar{${\rm A}^*$}
\def\K{\mathcal{M}}
\def\AC{\Theta}
\def\MCF{\Phi}
\def\ID{inertial drag}
\def\id{inertial drag}
\def\sun{{\rm sun}}
\def\wrt{with respect to}
\def\gc{globular cluster}
\def\Gc{Globular cluster}
\def\rdot{\kern 2pt\dot {\kern -1 pt r}}
\def\rdd{\text{\em\"r}}
\def\thetadot{\dot\theta}

\def\re{\mathbb{R}}
\makeop{inert}
\makeop{IF}
\def\dtwo{d^{\mkern 2mu 2}\mkern -1mu}
\def\thinm{\mkern 2mu}
\def\ss{spherically-symmetric}
\def\BH{black hole}

\def\AM{angular momentum}
\makeop{NM}
\def\lfrac#1#2{#1/#2}
\def\strutt{\vrule width 0pt height 20pt}
\newtheorem*{theorem}{Theorem}

\makeop{SO}
\makeop{deS}
\makeop{Ric}
\makeop{div}
\makeop{Exp}
\makeop{Isom}

\def\H{\mathbb{H}}
\def\P{\mathbb{P}}
\def\xx{\langle x,x\rangle}
\makeop{Cont}
\makeop{const}
\def\H{\mathbb{H}}

\newtheorem{prop}{Proposition}

\def\th{\,^{\rm th}}
\def\d{\partial}
\def\ind#1{\index{#1}#1}

\begin{document}

\vglue 3in
\thispagestyle{empty}

\cl{\huge\bf A new paradigm for the universe}

\vglue 2in

\cl{\sc\huge Colin Rourke}

\np

\thispagestyle{empty}

\index{version}\index{publication details}

\leavevmode $\phantom{99}$

\vspace{2in}

\iffalse
This is Google version 4 of ``A new paradigm for the universe''\nl

\bigskip

This version is also available from Amazon in Kindle format and as a
paperback with publication data:

ISBN: 9781973129868\nl
Imprint: Independently published\nl

\bigskip

An earlier version is available on the arXiv as {\tt astro-ph/0311033}
and on the author's web page

\bigskip

\fi
\iffalse

This is Kindle version 5 of ``A new paradigm for the universe''\nl

It is available from Amazon in Kindle format and as a paperback with
publication data:
\medskip

ISBN: 9781973129868\nl
Imprint: Independently published\nl

\bigskip

This version is also available from Google Play

\bigskip

An earlier version is available on the arXiv as {\tt astro-ph/0311033}
and on the author's web page

\bigskip

\fi
%\iffalse
This is arXiv version 5 of ``A new paradigm for the universe''

{\tt arxiv.org/abs/astro-ph/0311033v5}

also available from the author's web page
\bigskip\bigskip

The most up-to-date version is available from Google
Play and from Amazon in Kindle format and as a paperback with
publication data:
\bigskip

ISBN: 9781973129868\nl
Imprint: Independently published\nl
%\fi

\np
\pagenumbering{roman}

\vglue -1in

{\def\thesection{}\section[Foreward]{\protect{\hglue 2in \normalsize
      Foreword}}}
 
{\small\leftskip 25pt\rightskip 25pt\parskip 4pt plus 2pt

  \markboth{Foreward}{}
  \index{foreward}

The universe is a wonderful, amazing place.  We know a lot about it
through a series of fantastic observations using ground and space
based telescopes and other equipment, including recently, equipment to
detect gravitational waves.  There is a consensus model (the
``standard model'') for the way it works which starts with the ``Big
Bang'' and expands \ldots\ BUT there are several unsatisfactory
features to this model and the purpose of this book is to present a
complete new model which fits all the observations and does not share
these unsatisfactory features.

The main unifying idea of this book is that the principal objects in
the universe form a spectrum unified by the presence of a massive or
hypermassive black hole.  These objects are variously called quasars,
active galaxies and spiral galaxies.  The key to understanding their
dynamics is angular momentum and the key tool, and main innovative
idea of this work, is a proper formulation of ``Mach's principle''
within Einstein's theory of General Relativity (EGR) using Sciama's
ideas.\index{quasar-galaxy spectrum}

In essence, what is provided here is a totally new paradigm for the
universe.  In this paradigm, there is no big bang, and the universe is
many orders of magnitude older than current estimates for its age.
Indeed there is no natural limit for its age.  The new model for the
underlying space-time of the universe is based on a relativistic
analogue of the sphere, known as \ind{de Sitter space}.  This is a
highly symmetrical space which makes the model fully Copernican in
both space and time, in other words the universe looks much the same
in the large at any place or time.  By contrast the current standard
model of mainstream cosmology is Copernican only in space and not in
time.  This means that the view of the universe expounded here is
similar to the steady state theory proposed and defended by Fred Hoyle
(and others) in the last century, but it is not the same as their
theory, which proposed an unnatural continuous creation hypothesis;
like the big bang, this hypothesis breaks commonly accepted
conservation laws.  \index{Hoyle, Fred}%

It is worth mentioning that, by contrast with many attempts to find a
model for the universe with no big bang, this book does not propose
any new physics.  It fits squarely within EGR.  But it is necessary to
make a new hypothesis for the inertial dragging effect of rotation in
order to formulate the version of Mach's principle needed (Sciama's
principle) within the framework of EGR.  This formulation solves one
of the main philosophical objections to Mach's principle namely the
causal problems that a naive formulation runs into.
\index{Einstein!general relativity}

\index{EGR|see{Einstein general relativity}}
\index{GRB|see{gamma ray bursts}}
\index{CMB|see{cosmic microwave background}}
\index{PCP|see{perfect Copernican principle}}
\index{coherency postulate|see{Weyl}}

\vglue 7pt plus 2pt

{\bf AMS classification}\quad\makeatletter
\href{http://www.ams.org/mathscinet/search/mscdoc.html?code=\ifx\@primclass\relax\@secclass\else\@primclass\ifx\@secclass\relax\else,(\@secclass)\fi\fi}{{\ifx\@primclass\relax\@secclass\else\@primclass\ifx\@secclass\relax\else;
    \@secclass\fi\fi}}\makeatother\np}

%\pagenumbering{roman}
%\thispagestyle{fancyplain}

\vglue -.5in

{\def\thesection{}\section[Preface]{\hglue 2in \normalsize Preface}}

\markboth{Preface}{}\index{preface}

{\small\leftskip 25pt\rightskip 25pt\parskip 4pt plus 2pt

I started with the intention of writing a book intelligible to a
general reader with a scientific interest.  Some of the material that
I wrote in this endeavour is included as \fullref{app:beginners}
``Introduction to relativity''.  Readers who have little previous
knowledge, or wish to have their basic knowledge refreshed, should
read this appendix before the main text.

However I quickly realised that the main material of the book is far
too technical to treat at an elementary level in a book of modest
proportions and I have not tried to avoid technicalities in the main
body of the text.  But I have tried to make the introductory parts of
the book and of each chapter accessible to a general reader and I hope
that a reader who has only a little technical knowledge will be able
to find sufficient material to read to understand the main ideas
presented here.

Several parts of this book are based on joint work with Robert MacKay
and I thank him for allowing me to use this material and also for a
thoughtful critical read.  I would also like to thank Rosemberg Toala
Enriques for the use of the material in the draft three author paper
\cite{BHQR} on quasars.  Special thanks are due to Robert MacKay, Ian
Stewart and Rob Kirby for unfailing
\index{MacKay, Robert}\index{Stewart, Ian}\index{Kirby, Rob}%
\index{Toala Enriques, Rosemberg}support through the discouraging
process of attempting to publish this work in serious scientific
journals.  It seems that self-publishing is the only vehicle open to
an author who challenges the received orthodoxy.

\vglue 0.5in
Colin Rourke\nl
October 2017\nl
\nl
Mathematics Institute, University of Warwick, Coventry CV4 7AL, UK\nl
\href{mailto:cpr@msp.warwick.ac.uk}{\tt cpr@msp.warwick.ac.uk}
\qquad \href{http://msp.warwick.ac.uk/~cpr}{\tt
    http://msp.warwick.ac.uk/\char'176cpr}

\np}

{\def\thesection{}\section[About the author]{\protect{\hglue 1.85in
      \normalsize About the author}}}
 
{\small\leftskip 25pt\rightskip 25pt\parskip 4pt plus 2pt

\cl{\includegraphics[width=1.5in]{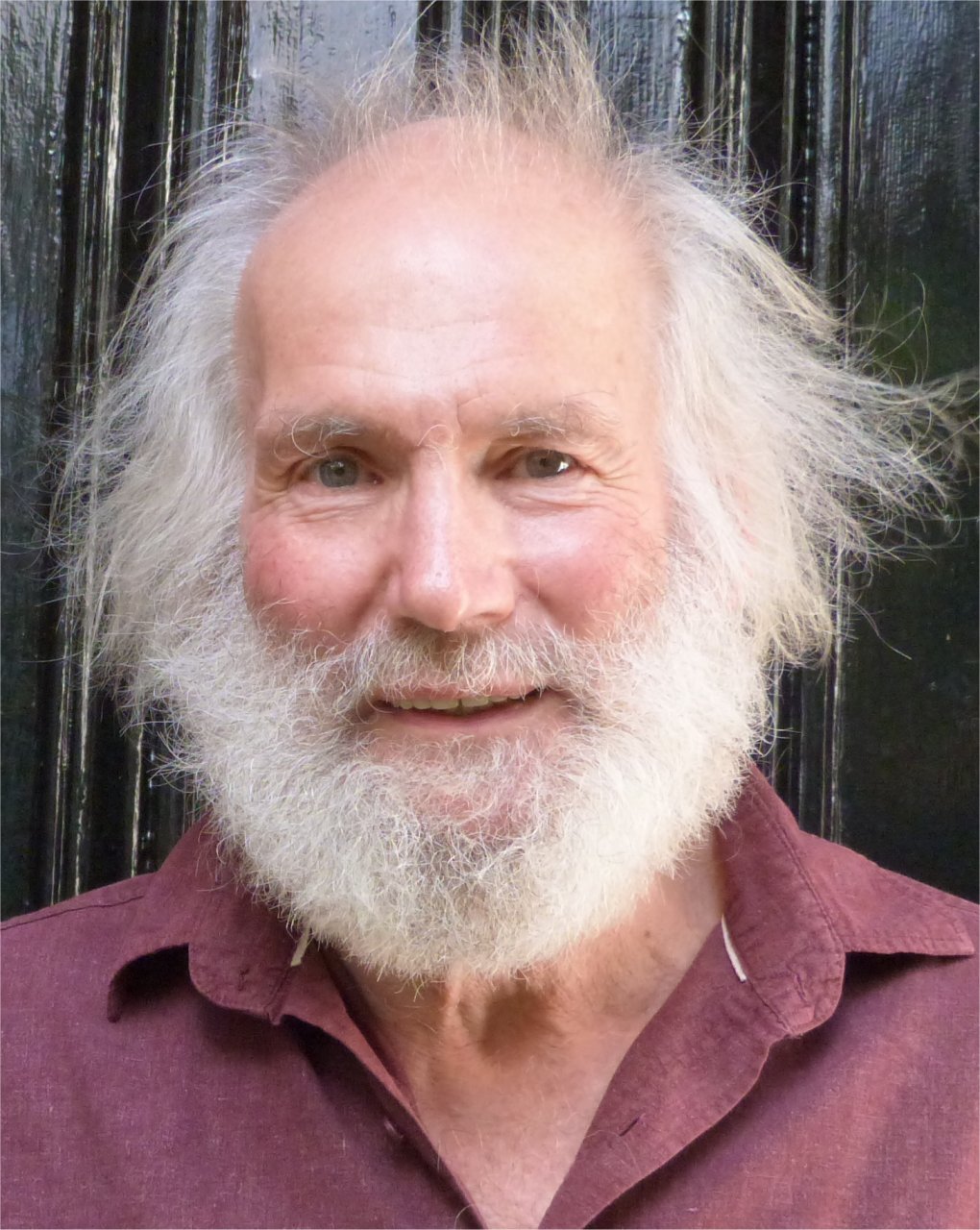}}

\index{Rourke, Colin}

Colin Rourke is a professor emeritus of mathematics at the University
of Warwick, and has also taught at the Princeton Institute for
Advanced Study, Queen Mary College London, the University of Wisconsin
at Madison, and the Open University, where he masterminded rewriting
the pure mathematics course; he has recently retired from lecturing
after completing a half-century (of 50 years of lecturing).  He has
written papers in high-dimensional PL topology, low-dimensional
topology, combinatorial group theory and differential topology.

In 1996, dissatisfied with the rapidly rising fees charged by the
major publishers of mathematical research journals, Colin decided to
start his own journal, and was ably assisted by Rob Kirby, John Jones
and Brian Sanderson.  That journal became Geometry \& Topology.  Under
Colin's leadership, GT has become a leading journal in its field while
remaining one of the least expensive per page.  GT was joined in 1998
by a proceedings and monographs series, Geometry \& Topology
Monographs, and in 2000 by a sister journal, Algebraic \& Geometric
Topology.  Colin wrote the software and fully managed these
publications until around 2005 when he cofounded Mathematical Sciences
Publishers (with Rob Kirby) to take over the running.  MSP has now
grown to become a formidable force in academic publishing.  With his
wife Daphne, he runs a smallholding in Northamptonshire, where they
farm Hebridean sheep and Angus cattle.\index{MSP}
\index{Geometry \& Topology}\index{Jones, John}%
\index{Sanderson, Brian}\index{Kirby, Rob}

In 2000 he started taking an interest in cosmology and published his
first substantial foray on the ArXiv preprint server in 2003.  For the
past ten years he has collaborated with Robert MacKay also of Warwick
University with papers on redshift, gamma ray bursts and natural
observer fields.  He now feels that he has mastered the basis of a
completely new paradigm for the universe without either dark matter or
a big bang.  This new paradigm is presented in this
book.\index{MacKay, Robert}

\medskip\index{Sheldon, Pip}
\rightline{\tiny Photo: Pip Sheldon\hspace{25pt}}

\newpage}

\thispagestyle{empty}

{\small\leftskip 25pt\rightskip 25pt \vglue -1in
\lhead[\fancyplain{}{\small\bfseries\thepage}]%
      {\fancyplain{}{\small\bfseries Contents}}
\rhead[\fancyplain{}{\small\bfseries Contents}]%
      {\fancyplain{}{\small\bfseries\thepage}}
\parskip 0pt plus 1pt\tableofcontents
\addcontentsline{toc}{subsection}{Table of contents}}

\np
\thispagestyle{empty}
\np

\mainmatter\pagenumbering{arabic}
\renewcommand{\sectionmark}[1]{\markright{\thesection\qua #1}}
\lhead[\fancyplain{}{\small\bfseries\thepage}]%
      {\fancyplain{}{\small\bfseries\rightmark}}
\rhead[\fancyplain{}{\small\bfseries\leftmark}]%
      {\fancyplain{}{\small\bfseries\thepage}}

\chapter{Introduction}\label{sec:intro}

Cosmology -- the study of the universe in the large -- is a topic
which arouses a great deal of public interest, with serious articles
both in the scientific press and in major newspapers, with many of the
theories and concepts (eg the ``big bang'' and ``black holes'')
discussed, often in depth.  The observations that support these
discussions use sophisticated and expensive machinery both on the
earth (eg the LIGO equipment used to detect gravitational waves) and
in orbit around the earth (eg the Hubble telescope).  There is a
consensus for the theoretical framework supporting and interpreting
these observations, which is known as the ``\ind{standard model}''.  It
starts with the big bang and expands from there to fill the entire
visible universe.  The image presented is of a complete and full
theory that (with a few minor unsolved problems) explains all the
observations that we have.\index{LIGO}\index{Hubble
  telescope}\index{big bang theory}

\index{big bang theory!problems|(}

This image is (like many images) completely false.  There are huge
problems with the standard model, some of which will be discussed
shortly.  It is the aim of this book to present an alternative model
to the standard one, which avoids the most glaring of these problems.
The model is complete in outline with several topics covered in full
detail, including the dynamics of galaxies and the nature of quasars.

\sh{Two philosophical problems}\label{sec:twopp}

There are two major philosophical problems with the standard model.
Firstly it is based directly on Einstein's theory of ``General
Relativity'' (hereinafter referred to as EGR) which does not satisfy
``Mach's principle'' in general.  This principle centres around the
philosophically compelling idea that the concept of acceleration or
rotation must be connected to the main distant mass of matter in the
universe.  Rotation is the simplest to think about.  An observer can
tell that he is rotating without leaving his closed windowless
spaceship, because there are forces that he experiences (for example
\ind{Coriolis force}) that he does not experience if he is not rotating.
But what possible difference is there between him rotating, and him
being still with the universe rotating around him?  The conclusion is
that the forces he experiences are due to some mysterious effect of
the rotation of the universe around him.  These considerations have
passed into general circulation as ``Mach's principle'' which is
usually summarised as stating that the local concept of inertial frame
(a frame in which there is no acceleration or rotation) is correlated
with the distribution and motion of all the matter in the universe.
Any theory which aims to accurately describe the universe must have
such a property in some form.  EGR does not.\index{Mach!principle}

The second problem is the (again philosophically compelling) principle
known as the Copernican (or cosmological) principle.  This is the
principle that no particular location should be special.  There is no
centre to the universe: no ``fingers of God''.  This should also be
true of time as well as space, there should be no special times: we
should not live in a special time any more than we live in a special
place.  This space-time non-speciality principle will be called the
``Perfect Copernican Principle'' or PCP (which can also be read
as the ``Perfect Cosmological Principle'').%
\index{perfect Copernican principle}

Obviously the current consensus model does not satisfy this principle
because it starts with a very special event, the big bang, taking
place at a very special time.

\sh{Dark matter}\index{dark matter|(}

There are two other major problems.  The first (the so-called ``dark
matter'' problem) is recognised as a major problem, whilst the second
(the ``Arp problem'') is not recognised but ought to be.  The dark
matter problem arises from the observed rotation curves for galaxies.
Typically the curve (of tangential velocity $v$ against distance from the
centre $r$) comprises two approximately straight lines with a short
transition region.  The first line passes through the origin, in other
words rotation near the centre has constant angular velocity
(plate-like rotation); the second is horizontal, in other words the
tangential velocity is asymptotically constant, see
\fullref{fig:modbeg} (right).  Furthermore, observations show that the
horizontal straight line section of the rotation curve extends far
outside the limits of the main visible parts of galaxies and the
actual velocity is constant within less than an order of magnitude
over all galaxies observed (typically between 100 and 300km/s) see
\fullref{fig:rotsSR}.

Galactic rotation curves are so characteristic (and simple to
describe) that there must be some strong structural reason for them.
They are very far indeed from the curve obtained with a standard
Keplerian model of rotation under any reasonable mass distribution.
In a Keplerian model if $v$ is asymptotically constant then the mass
inside radius $r$ is asymptotically equal to a constant times $r$ and
tends to infinity with $r$.

Nevertheless, in spite of the huge mass needed, a Keplerian model is
exactly what is assumed in the consensus model.  To square the circle,
current theory hypothesises the existence of a huge amount of matter.
Since this matter is not observed, it is called called ``dark''.  It
needs to be distributed in precisely the right way to make Keplerian
rotation fit the rotation curve.  This is extremely implausible for
several reasons which will be discussed in \fullref{sec:rot_curve}.
There is also a companion problem for the dynamics of spiral galaxies.
The standard model has no satisfactory model for galactic dynamics
which explains the persistent spiral structure widely observed.  The
new model presented here solves the dark matter problem and gives a
full model for galactic dynamics.  It does this by building a limited
version of Mach's principle into EGR (and this also deals with the
first philosophical problem).  In essence all that is needed is to use
a suitable relativistic model instead of a Keplerian one.%
\index{dark matter|)}

In passing, it is worth mentioning that there is another ``problem''
often mentioned, namely ``\ind{dark energy}''.  The author's view is that
there is no problem here.  Dark energy is just a name for the term in
the field equations that provides global curvature.  It is not a
problem any more than the curvature of the earth's surface is a
problem: it is just part of the description of the universe!

\sh{The Arp problem}\index{Arp, Halton!problem}

Halton Arp was a talented observer who provided key observations
supporting Hubble's theory of the expanding universe.  He also
observed many examples of quasars with intrinsic (gravitational)
redshift against the current mainstream dictat that all redshift must
be cosmological.  A particularly striking example is reproduced in
\fullref{fig:NGC7603}.  It is commonsense that the alignments seen in
this configuration of galaxies and quasars are not due to chance and
there are many similar such in Arp and other's observations
\cite{Arp,Getal}.  Objects 2 and 3 (quasars) have cosmological
redshift around $z=0.030$ (for the filament) and the remainder must be
instrinsic (presumably gravitational).  As often happens when a
consensus view is challenged by direct evidence, the evidence is
ignored and the challenger discredited.  Arp was sidelined by the
mainstream cosmological community and denied observation time on the
big telescopes.  One of the main aims of this book is to rehabilitate
Arp's reputation (unfortunately posthumous), and if it succeeds in
this aim it will have been worth writing.  As will be seen later, it
is in fact quite easy to correct the standard model for quasars to
allow for instrinsic redshift, indeed, once again, the correction is to
use a simple piece of relativistic geometry.%
\index{Arp, Halton}\index{big bang theory!problems|)}

\sh{A little history}\label{sec:hist}
\index{cosmology!short history|(}

It is worth giving a quick history of cosmology to explain how the
current theory has become stuck in a groove of self-justifying
explanations which deny the reality of evidence that contradicts the
consensus view.  One problem has already been hinted at.  Despite
being nominally based on General Relativity, the theory tends to opt
for non-relativistic models (such as the Keplerian model for galactic
rotation) and avoid the flexibility offered by fully relativistic
models.  Another example is the universal time assumed in the standard
model in contrast to the relativistic nature of time (depending on the
observer) in general relativity.  Without a universal time, the
observations of expansion do not imply that the universe must have
started with a big bang.

Cosmology arose directly from Einstein's theory of general relativity.
This theory is so beautiful and well-formed that it rapidly became
accepted as the basic geometry of the universe.  Einstein himself
started model building very early in the development of the theory.
He became obsessed with the problem of finding a static solution to
his equations and introduced his ``\ind{cosmological constant}'' in order to
do this.  Incidentally this constant is the source of ``\ind{dark energy}''
mentioned earlier.  He built models based on the simplest manifolds
for space (flat space or the 3--sphere) with a universal time
parameter.  These models, which should by now have been discarded as
overly simple, have stuck and the current standard model also uses very
simple manifolds with a universal time parameter.  It is the over
simplicity of the models used which leads to the fallacy that observed
expansion implies global expansion which in turn implies the big bang.

This happened despite the model proposed early on by Willem de Sitter
who collaborated with Einstein.  This is the analogue in space-time of
a sphere in ordinary space and has a similarly complete set of
symmetries: there is a symmetry that carries any point to any other
and indeed one that carries any time-like geodesic to any other and
hence this model has the PCP.  It fits Einstein's equations (with
cosmological constant) and there is no big bang.  Unfortunately for
cosmology, de Sitter space was abandoned as a model for the universe
very early in its development.  De Sitter himself used a very
unnatural metric which obscured the symmetry and an influential
mathematician, Hermann Weyl, proposed a widely accepted ``coherency
postulate'' \cite{We} which supposes that no astronomical object can
suddenly appear into view or has so appeared at a finite time in the
past.  All must have been always visible.  This cuts de Sitter space
down to just the expanding frame based on the home geodesic (the one
on which our galaxy is travelling) and loses the PCP.  It makes the
point at time $-\infty$ on the home geodesic a special point: the
universe seems to have this point as an origin, see
\fullref{fig:Mosch} (left).  In this figure the blue lines are
geodesics along which objects permitted by Weyl's hypothesis are
travelling, with the central one being the home geodesic.  The black
lines are transverse (flat) space slices.  Thus the hypothesis leads
the way to the current ``standard models'' which start with a Big Bang
at a finite time in the past.\index{standard model}%
\index{Weyl!coherency postulate}\index{Weyl}%
\index{perfect Copernican principle}

Another early model with the PCP was the Bondi--Gold--Hoyle
\ind{steady state theory} (SST): this hypothesises that matter is
created by empty space at exactly the correct rate to compensate for
the observed Hubble expansion -- a hypothesis also briefly toyed with
by Einstein (see \cite{ESS}) as a possible alternative to (or
explanation of) the \ind{cosmological constant}, as a means of attaining a
static solution to the field equations.  Hoyle was an energetic
proponent of this theory against the Big Bang theory (a sarcastic name
that he invented) and his writings on the subject are well worth
revisiting.  Eventually he gave way because of the evidence of
historic change in the composition of the universe from quasar
observations -- evidence that was in fact badly misinterpreted, see
\fullref{sec:killam} and \fullref{sec:quasars}.  Hoyle's SST could
still be correct and in deference to his enthusiasm, a model with the
PCP will be called a ``Hoyle Universe''.  The model outlined in this
book, based on de Sitter space, is indeed a Hoyle Universe, but it
does not have the unnatural continuous creation hypothesis of the SST,
which, like the Big Bang breaks commonly accepted conservation laws.%
\index{cosmology!short history|)}

For more (or perhaps better) history the reader is directed to
\cite[Chapter 1]{Longair}.

\sh{The quasar--galaxy spectrum}\index{quasar-galaxy spectrum}

A major theme of this work is that there is a spectrum of related
phenomena.  This is the quasar--galaxy spectrum.  The unifying element
is the presence of a massive (or hypermassive) \ind{black hole}.  The
position of a quasar or galaxy on the spectrum is determined entirely
by the size of this associated black hole, which varies from $10^6$
solar masses (sm), or less, for a small quasar such as Sagittarius
\astar, \index{Sgr A*} through $10^9$ to $10^{11}$ sm for a so-called
active galaxy and up to $10^{14}$ sm, for a full size mature spiral
galaxy. An aside here: the phrase ``so-called'' for active galaxies is
used because one of the main theses of this work is that all galaxies
are highly active and that, for spiral galaxies, this activity
manifests itself in the very spiral structure that characterises them.%
\index{galaxy}\index{galaxy!active}\index{galaxy!spiral}%

These phenomena are systematically misunderstood in the consensus
theory.  At the smaller end, the quasar end, there is an observed
redshift which can be very large (up to $z=8$ or more---much more as
will be seen later) and, for reasons which will be explained shortly,
the current mainstream view is that this redshift is entirely
cosmological (due to the expansion of the universe).  This implies
that these objects are very distant, extremely massive, created just
after the big bang and have a truly phenomenal power output, which is
very hard to explain.  One of the major tasks of this work is to
explain how this view has arisen and how it can be changed to the view
that, by contrast, quasars are typically small, nearby objects with a
modest power output easily modelled by a simple spherical accretion
mechanism.\index{quasar}\index{quasar!redshift}

The key to this misunderstanding and to the correct model for quasar
energy production is angular momentum.  Very early in the study of
quasars it was decided that the behaviour of angular momentum gives a
compelling reasons for believing that quasar redshift is cosmological.
Quasars are typically believed to be based around a very dense object,
probably a black hole, and their energy production is believed to be
due to \ind{accretion} from the surrounding medium.  Particles fall into the
gravitational well of the central mass and the gravitational energy is
released by interaction between different infalling particles.  Now
given a small but very heavy object, a particle approaching with a
small tangential velocity will have its tangential velocity magnified
by conservation of angular momentum and there will be a radius of
closest approach.  It is very unlikely to actually fall into the
central gravitational well.

The same thing happens for the full flow of infalling matter from the
surrounding medium which will typically have a nonzero angular
momentum around the black hole.  This gives an obstruction to
accretion which was found not long after quasars were discovered, for
\index{Michel}example Michel \cite[Section 4, p 158]{michel} (1976)
states:

\begin{quote}\em
\ldots One must, however, somehow transfer away most of the angular
momentum that the infalling gas had relative to the centre of mass.
It seems physically plausible that the effect of such angular momentum
would be to choke down the inflow rates.  For example, even when
magnetic torques are included \ldots one finds that the `infall'
solutions terminate at finite distances from the origin in analogy
with the minimum approach distance of a single particle trajectory
having non-zero initial angular momentum. \ldots
\end{quote}

These considerations have led to the subject being dominated by the
theory of ``accretion discs''.  The idea is that, since infalling
matter cannot flow smoothly into the central black hole, it must
typically settle into a rotating structure of some kind, which is
called (whatever its actual shape) an accretion disc.  Then
interaction between infalling matter and this structure allows energy
to be produced.\index{accretion!disc}

A consequence of this is that redshift, which is frequently observed
in quasar radiation, is generally believed to be cosmological and not
gravitational (or intrinsic).  Indeed if the observed radiation comes
from an accretion disc and the redshift is caused by the gravitational
field of the nearby black hole, then because the disc varies in its
distance from the black hole over its extent, the spectral lines
observed would be wide (a phenomenon known as ``redshift gradient'')
\index{redshift!gradient}\index{redshift!intrinsic}\index{redshift!gravitational}
and not the narrow lines that are observed.

This in turn implies that the universe has varied in its constitution
over the observable past.  As remarked above, a quasar with a large
cosmological redshift must be a massive object with a huge energy
output.  But there are no observations of huge sources of energy close
to us like these (supposed) near the big bang.  This provides strong
supporting evidence for the \ind{big bang theory}, which entails a
continuous change in the constitution of the universe.  A steady state
model cannot contain a big bang.

It was considerations like these that caused Fred Hoyle
\index{Hoyle, Fred} to abandon his
continuous creation model which is fully Copernican in both space and
time, ie with no observable global change over time.
\index{Copernican principle}

\sh{Killing the angular momentum obstruction}\label{sec:killam}
\index{angular momentum!obstruction}
\index{angular momentum!obstruction!killing}

One of the main theses of this book is that the angular momentum
obstruction to accretion \emph{can be killed by the black hole itself}
and this implies that quasars can be relatively small, nearby objects
and the universe could be Copernican in time as well as space.  Thus
Hoyle's model could still be correct (though it is not the model
proposed here).

The key to killing this angular momentum obstruction is to work in a
relativistic framework and not the Newtonian framework implicitly
assumed in the above discussion.  A relativistic effect---the dragging
of \ind{inertial frame}s, abbreviated to
``\ind{inertial drag}''---allows the black hole to compensate for the
angular momentum of the infalling gas$/$plasma stream and for an
energy production model to be established with radiation coming from a
thin spherical region (the \ind{Eddington sphere}) which can be very
close to the event horizon of the black hole and subject to an
arbitrarily high gravitational redshift, with the cosmological
redshift small in comparison.  Because the production sphere is thin,
there is little redshift gradient.  \index{inertial drag}

As mentioned above there is some very strong evidence in the
observations of Arp and others \cite{Arp,Getal} that quasars do in
fact possess intrinsic, ie gravitational, redshift.  This and the
angular momentum considerations just mentioned led Arp to propose some
fantasy physics explanations for this redshift.  The explanation
proposed in this work uses only well-accepted (and definitely not
fantasy) physics and is fully consistent with Arp's
observations.\fnote{The one new hypothesis that is made in this work,
  the inertial drag force, see below, does not play any role in this
  explanation.}

\index{redshift!intrinsic}\index{Arp, Halton}

\sh{Embedding Mach's principle in EGR}
\index{Mach!principle}

Alongside inertial drag, the other main ingredient for the new
paradigm presented in this book is ``{Mach's principle}'' outlined
above.  It is necessary to explain how to embed the version of Mach's
principle that is needed for this work into EGR (Einstein's General
Relativity).  There are obvious causal problems in a naive statement:
how exactly does distant matter communicate with local matter to
determine the local inertial frame?  and does the influence happen
instantaneously or travel at the speed of light?  These problems will
be avoided by restricting to a limited version of the principle due to
Sciama \cite{Sciama} which is quantitative rather than philosophical
and which is referred to as Sciama's principle.  The discussion is
further simplified by concentrating on rotation at the expense of
general non-inertial motions.  EGR deals well with acceleration, so
this makes sense for the purposes of embedding the principle within
EGR.\index{Sciama}\index{Sciama!principle}

A new hypothesis is needed for the dragging effect of a rotating body
on the inertial frames near it.  The precise behaviour that is needed
is not a consequence of Einstein's equations and the hypothesis
amounts to assuming that a rotating mass has a non-zero effect on the
stress-energy tensor near it -- in other words stops the space near it
being a true vacuum.  This gives a natural way to understand how
inertial drag propagates: the disturbance to the local vacuum is akin
to a gravity wave and propagates at the speed of light.  Furthermore
reading back from the rest of the universe, the local background
inertial frame is created by the rest of the universe by a similar
propagation effect from all the rest of the matter (a brief aside
here: this makes sense only if the sum is finite -- or quasi-finite --
this will be explained in the next chapter).  \index{inertial
  drag!propagation}

It is worth briefly comparing the new inertial drag hypothesis made in
this book with the dark matter hypothesis made in current mainstream
cosmology.  At first sight they may appear to be similar.  Both
correct the rotation curve for galaxies.  But the dark matter
hypothesis amounts to assuming the existence of inert matter, which
has no effect other than gravitational attraction, and cannot
otherwise be detected.  The inertial drag hypothesis on the other hand
amounts to assuming a new effect of a rotating body on the field
outside it.  It embodies a limited version of Mach's principle which,
as has been seen, is philosophically compelling, and must be embodied
in any theory that seeks to accurately describe reality.  Thus, unlike
the dark matter hypothesis, the \id\ hypothesis is a necessary part of
a complete theory.  Furthermore, the \id\ hypothesis also underlies a
good model for the dynamics of spiral galaxies, whereas the dark
matter hypothesis leaves this problem unsolved.  More detail on this
point will be given later in the book (\fullref{subsec:Add}).

\sh{Outline of the rest of the book}

Mach's principle is discussed in \fullref{sec:Sciama}, after which
\fullref{sec:rot_curve} derives the inertial drag effect, that allows
quasars to cancel out the angular momentum obstruction to accretion,
and fuels the dynamics of galaxies.  In this chapter it is applied to
model the rotation curve for galaxies without needing ``dark matter''.

Next in \fullref{sec:quasars} the subject of quasars is taken up in
earnest.  Here it is explained how \ID\ allows \BH s to absorb the
\AM\ in infalling gas$/$plasma and to grow by accretion.  The
spherical accretion model \index{accretion!spherical}that this allows
is joint work with Rosemberg Toala Enriques and Robert MacKay
\cite{BHQR}.  This work is still in draft form, but nevertheless the
model fits observations extremely well, including those of Arp
\cite{Arp}, and also explains the apparently paradoxical results of
\ind{Hawkins} \cite{H}.  This section contains a first description of
the pivotal quasar--galaxy spectrum.  Technical details from
\cite{BHQR} are deferred to \fullref{app:3author}.

After this the second main task of the book is tackled in
\fullref{sec:spiral_struc}, namely to provide a model for the spiral
structure of full-size galaxies, such as the Milky Way, which lie at
the other end of the quasar--galaxy spectrum.  The nature of these
objects is also much misunderstood by mainstream cosmology.  Spiral
galaxies all contain a central hypermassive black hole (of mass
$10^{11}$ sm or more), which controls the dynamic by the same inertial
drag effect that allows accretion in quasars, and which is surrounded
by an accretion structure responsible for generating the visible
spiral arms.  Another aside here: there is a special misunderstanding
with the Milky Way, where Sgr\astar\ with a mass of only
$4.3\times10^6$ sm, far too light to have any dynamic effect on the
galaxy, is believed to be the central black hole.  This
misunderstanding will be cleared up at a later stage.%
\index{accretion!structure}\index{Sgr A*}

Between quasars and spiral galaxies lie ``active'' galaxies for which
accretion structures have been directly observed.  This is the only
part of the quasar--galaxy spectrum which is more-or-less correctly
understood by mainstream cosmology.  There will be a lot more to say
about the whole quasar--galaxy spectrum later in this work.
\index{quasar-galaxy spectrum}

As mentioned above, there is, inside a full size spiral galaxy, an
accretion structure, called ``\ind{the generator}'', which is
responsible for generating the spiral arms.  This is described in
\fullref{sec:spiral_struc} where a full model for the resulting spiral
structure is derived.  The generator feeds the roots of the spiral
arms with a pure light element mixture (H and He with a trace of Li).
This is the same mixture of elements that is hypothesised to have been
created in the big bang just before the time of the last scattering
surface from the cooling of a hot plasma of quarks, and the process is
similar.  The residue of these streams, not condensed into stars,
escapes the galaxy and feeds the intergalactic medium and this
explains the observed proportion of these elements in the universe
(which is one of the so-called ``pillars of the big bang theory'').%
\index{light elements}

\fullref{sec:obs} and \fullref{sec:cosm} cover observations and
consequences for cosmology.  Included here are a comprehensive
rebuttal of the big bang theory and explanations for redshift and the
Cosmic Microwave Background (CMB), which are the other two pillars.
There is also an explanation for Gamma Ray Bursts (GRB).  Technical
details for several of the topics are again deferred to appendices.

\np\thispagestyle{empty}

\chapter{Sciama's principle}\label{sec:Sciama}
\index{Sciama!principle|(}
\index{Mach!principle}

This chapter is concerned with a discussion of Mach's principle and
the restricted version that is needed for the dynamical applications
(to quasars and spiral galaxies) in the rest of the book.  The final
form of the principle (the Weak Sciama Principle) hypothesises an
inertial dragging effect from a rotating body which drops off
asymptotically with $k/r$ where $k$ is a constant and $r$ is distance
from the centre.  A reader who is happy to accept this principle can
omit this chapter without loss.  The precise assumption is repeated
near the beginning of the next chapter.

\sh{Inertial frames and Mach's principle}

In any dynamical theory there are certain privileged frames of
reference in which the laws of Newtonian physics hold to first order.
These frames are variously called ``inertial frames'' or ``rest
frames''.  They are characterised by a lack of forces correlated with
acceleration or rotation.  In Newtonian physics there is a universal
inertial frame referred to as ``\ind{absolute space}'' and in
Minkowski space the standard coordinates provide an inertial frame at
the origin.  Then Lorentz transformations carry this frame to an
inertial frame at any other point, providing inertial frames for
special relativity.  General relativity is built on Minkowski space
which in turn provides inertial frames for this theory, see
\fullref{sec:genrel}.  Berkeley \cite{Berkeley} and Mach \cite{Mach}
criticised Newton's assumption of absolute space.  Berkeley suggested
that the local rest frame could be defined by distant ``fixed'' stars.
Mach's book \cite[Ch II.VI (p 271 ff)]{Mach} contains a devastating
critique of Newton's assumptions and is well worth reading.  It was
extremely influential and Einstein acknowledged a debt to his ideas.
Mach's basic point is that one should never assume anything that is
not directly connected to observations of some kind and in particular
the concept of the local inertial frame must be defined in terms of
(theoretically) observable quantities.  Some detail from Mach is given
in \fullref{sh:Mach} below.\index{inertial frame}\index{rest
  frame}\index{Mach}\index{Berkeley}

The basic property of inertial frames is that they are only defined up
to uniform linear motion.  Given any inertial frame, a frame which is
in uniform linear motion with respect to the given frame is also an
inertial frame.  Thus ``the'' inertial frame at a point $P$ in fact
means an equivalence class of frames, two frames in the class being
in mutual uniform linear motion \wrt\ each other.  (For this reason,
calling them ``rest'' frames is highly misleading and this terminology
will not be used again.)\index{uniform motion}

Mach's ideas have passed into general circulation as ``Mach's
principle'' which is usually summarised as stating that the local
concept of inertial frame is correlated with the distribution and
motion of all the matter in the universe.  However there are many
other ways of interpreting the principle and there is a huge
literature on the subject.  At its weakest, the principle is
interpreted as merely stating that all phenomena must have their
origin in some material source (see eg \cite{SWG}), and it has even
been interpeted as an assumption about the nature of the big bang (Tod
\cite{Tod}).

For the purposes of this book, a statement is needed which is more
precise than these but not so wide ranging.  What is needed is a local
version which applies to rotation of inertial frames and which is
quantified precisely.

\sh{Sciama's principle}\label{subsec:Sp}

The version that is used is close to the version in Sciama's thesis
\cite{Sciama}.  Sciama makes a bold attempt to base a full theory of
dynamics on Mach's principle.  His idea is that the inertial frame at
any point $P$ in the universe is determined by the inertial frames at
every other point $Q$.  The contribution from $Q$ is nonzero only if
there is a mass $m_Q$ at $Q$ and then the contribution is (a)
proportional to this mass and (b) inversely proportional to the
distance $r_{Q}$ between $P$ and $Q$.  In other words the contribution
is $$m_Q\thinm \IF_Q / r_Q,$$ where $\IF_Q$ means the inertial frame
at $Q$.  The idea is that this should be summed over ``all the matter
in the universe''.\index{Sciama!thesis}

To make sense of this sum it is necessary to make a number of
assumptions.  Firstly, in order to add up contributions, it is
necessary to work in a linear framework and the simplest way to do
this is to work with a perturbation of flat (Minkowski) space, which
is exactly what Sciama does.  The underlying Minkowski space provides
``standard'' reference frames at each point and the motion of any
frame can be measured with respect to this standard, and also provides
a space in which to measure the distance $r_Q$ used in the summation.

Working within a perturbation of Minkowski space limits the theory to
weak fields, but it suffices for most of this work.  When working near
the massive centre of a galaxy, use can be made instead of a
perturbation of any \ss\ metric, eg the Schwarzschild metric, which
allows stronger fields.

Secondly, in order for the summation to converge, the ``universe''
needs either to be finite or to be ``quasi-finite'' in the sense that
only a finite part contributes to the sum.  More detail on this point
is given below.

Finally, it is necessary to keep $r_Q$ from getting too small or else
the contribution of $m_Q$ will be far too large.  This can be done
either by ignoring masses which are close to $P$, since the factor
$1/r$ implies that the sum is dominated by distant matter, see the
discussion below, or, if there is a significant and very massive body
(eg the black hole at the centre of a galaxy) nearby, then the sum can
be normalised as explained below.

To formulate the principle quantitatively use the notation $\NM_P$ for
the the non-uniform motion of the inertial frame at $P$ and ditto $Q$,
in other words its acceleration and/or rotation measured \wrt\ to the
local reference frame, then the inertial frame at $P$ is given by the
reference frame plus $\NM_P$ and the principle states that
\begin{equation}\label{eq:SP}\tag*{\textbf{Sciama's principle}}
\hspace*{-55pt}\NM_P = K\sum_Q \frac{m_Q}{r_{Q}} (\NM_Q).
\end{equation}
This statement is digested from Sciama's introduction and the precise
formulation in terms of the field \cite[Equation (1), page
  37]{Sciama}.  It is called \emph{Sciama's principle} in order to
distinguish it from Mach's principle. Here $K$ is a normalising
factor which will be dicussed further below.

Notice that this principle is completely symmetric.  The effect of
$Q$'s motion on the inertial frame (IF) at $P$ is exactly similar to
the effect of $P$'s on the frame at $Q$.  And note that the effect is
\emph{coherent} in the sense that an acceleration or rotation of the
frame at $Q$ causes an acceleration or rotation of the frame at $P$
with the \emph{same} direction or sense.  Sciama describes this
symmetry eloquently in his introduction, for example: \emph{``\dots
  the statement that the Earth is rotating and the rest of the
  universe is at rest should lead to the same dynamical consequences
  as the statement that the universe is rotating and the Earth is at
  rest, \dots''}

Also notice that using $\NM_P$ in the summation implies that the
inertial effect of matter in uniform linear motion is ignored.  This
is correct for small masses or for larger masses sufficiently distant
that gravitational induction effects can be ignored.

With a caveat that this needs needs to be treated with care in special
cases, this will be adopted as a working hypothesis which fits the
intuitive idea of inertial effects:

\textbf{Working hypothesis}\qua
\text{\sl Uniform linear motion has no inertial effect.}
\index{uniform motion!inertial effect}
\index{uniform motion!inertial effect!working hypothesis}
\index{working hypothesis|see{uniform motion}}

Sciama is clear that his principle is incompatible with Einstein's
General Relativity (EGR) and is attempting to create an alternative
theory.  Later it will be seen precisely how the principle is
incompatible with EGR and it will be explained how to modify EGR to
include the principle for rotation (by interpeting the principle as
adding a stress field that causes the \id\ and radiates from the
rotating mass).

Sciama starts to derive a full gravitational theory from this
principle.  He specialises to a ``field'' (a vector field) defined on
Minkowski space and as he makes clear this is an interim approach
which will need improvement is a subsequent promised sequel paper
(which in fact was never written).  In order for the summation to
converge, the ``universe'' needs either to be finite or to be
``quasi-finite'' in the sense that only a finite part contributes to
the sum.  More detail on this point is given in the next paragraph.

Sciama discusses three cases in detail:

(a)\qua The effect of distant matter on the local IF.  

The factor $1/r$ is chosen to make distant matter dominate.  In order
to get a finite sum, Sciama assumes standard Hubble expansion and then
it is natural to limit the summation to the visible universe (in other
words to ignore parts that are regressing faster than $c$).  It is
worth remarking in passing, that it is not necessary to assume the
existence of a big bang (BB) to satisfy this quasi-finite hypothesis.
There are models for the universe with redshift fitting observations
but with no BB (cf \fullref{sec:red}), the simplest of which is the
expanding part of de Sitter space; there is also the (now largely
ignored) \ind{continuous creation} model of Hoyle et\,al \cite{HBN}.  The
effect of distant matter needs to be normalised to unity.  For
example, if the whole universe is rotating about $P$ with angular
velocity $\omega$, then this should induce a rotation of $\omega$ in
the IF at $P$, in other words the situation should be exactly the same
as if all were at rest.  Similarly for acceleration.  Thus
\begin{equation}\label{eq:Norm}
K\sum_Q \frac{m_Q}{r_{Q}} =1
\end{equation}
where the sum is taken over all accessible matter $Q$ (ie within
the visible universe).  One way to arrange this is to assume that
$K=1$ and 
\begin{equation}\label{eq:Unnorm}
\sum_Q \frac{m_Q}{r_{Q}} =1
\end{equation}
This makes perfect sense provided that $r_Q$ is never small (if $r_Q$
is allowed to tend to zero, the contribution from $m_Q$ goes to
infinity, which is absurd) and this is effectively what Sciama does.
A more sensible way is to normalise by setting
\begin{equation}\label{eq:Knorm}
K=
1/\sum_Q \frac{m_Q}{r_{Q}}
\end{equation}
which compensates for large local masses and this is what will be done
when the principle is applied near the large central mass of a galaxy.

Equation (\ref{eq:Unnorm}) implies a fundamental relation between the
various gravitational and cosmological constants which Sciama derives
\index{fundamental relation!cosmological constants} as \cite[Equation
  (7)]{Sciama}.  He points out that this is, within reasonable limits,
in accord with observations.  Misner, Thorne and Wheeler (MTW)
\cite[below 21.160]{MTW} make exactly the same point using more modern
observations\fnote{There are about $10^{11}$ galaxies in the visible
  universe of weight about $0.03$ ($10^{11}$ solar masses) at
  distances varying up to $10^{10}$, where natural units are used
  ($G$ (Newton's gravitatonal constant) $=c=1$ and everything is in
  measured in years).}.  \index{Misner, Thorne and Wheeler!fundamental
  relation}This provides a preliminary justification for the key
factor $1/r$.  A better justification comes with the simple model
Sciama describes, where the field naturally decays like $1/r$.  His
model however is too simplistic (as he readily acknowledges) and in
fact coincides with one of the standard approximations to EGR, namely
``\ind{gravitomagnetism}''.  Shortly, there will be other cogent
reasons for the factor $1/r$.

For the other two cases he uses the model. 

(b)\qua A locally isolated mass.  Here Sciama finds Newtonian
attraction to first order (and in fact it is always attraction).

(c)\qua A locally rotating frame.  Here he finds the usual Newtonian
story (Coriolis forces etc).

\sh{An excerpt from Mach's critique}\label{sh:Mach}
\index{Mach!excerpt from critique}

\begin{figure}[ht!]
\labellist
\small\hair 3pt
 \pinlabel $d$ [r] <0pt,0pt> at 194 100
 \pinlabel $P$ [r] <0pt,0pt> at 193 3
 \pinlabel $x$ [b] <0pt,0pt> at 280 192
 \pinlabel $r$ [tl] <0pt,0pt> at 292 95
 \pinlabel $Q$ [b] <0pt,2pt> at 391 192
 \pinlabel $u$ [b] <3pt,0pt> at 440 200
\endlabellist
\centering
\includegraphics[width=.4\hsize]{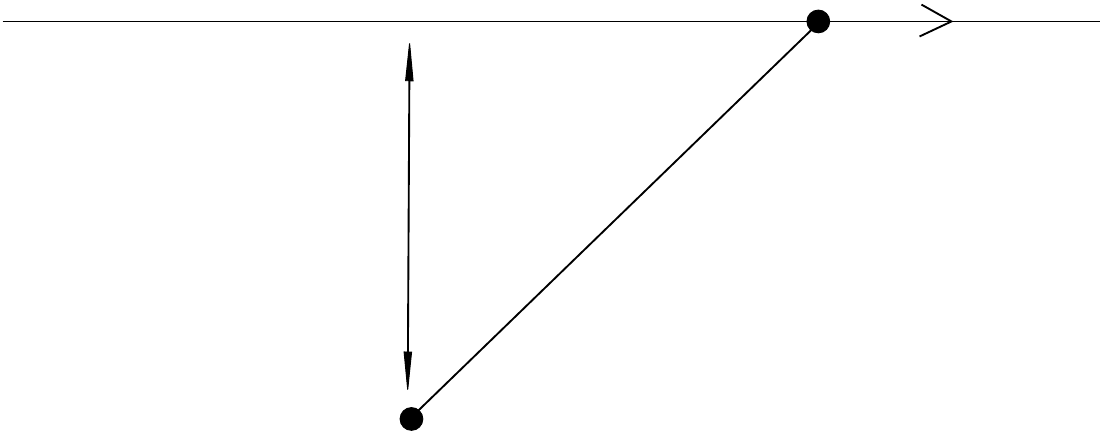}
\caption{Proof of Mach's formula for apparent acceleration of bodies
  in uniform relative motion. Here $\lfrac{dx}{dt} = u$ (const) and
  $\lfrac{dr}{dt}=v$.  Differentiating $r^2=\dtwo+x^2$ twice,
  gives $r\,\lfrac{\dtwo r}{dt^2}+v^2=u^2$ and hence
  $\lfrac{\dtwo r}{dt^2}=(\lfrac1r)\,(u^2-v^2)$ with $|v|=|(\lfrac
  xr)\,u| < |u|$.}
\label{fig:mach}\index{Mach!apparent acceleration}
\end{figure}
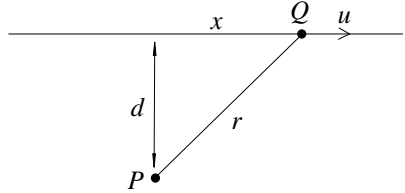

There is a passage in Mach's critique which can be used to provide
further support for the factor $1/r$.  In Ch II.VI.7 (page 286) of
\cite{Mach} he points out that if two bodies move uniformly in
ordinary 3--space then each sees the other as having non-zero
acceleration along the common line of sight.  Uniform motion does not
\emph{appear} uniform.  Indeed if $r$ is the distance between the
bodies, then 
\begin{equation}\label{eq:mach1}
\frac{\dtwo r}{dt^2} = \frac1r(u^2-v^2)
\end{equation} 
where $u$ is the absolute value of the relative velocity and
$v=dr/dt$.  This is readily proved from Pythagoras' Theorem, see
\fullref{fig:mach}.  And notice that $|v|<|u|$.
Thus to describe even uniform motion in terms of observation is quite
complicated.  On the next page he gives a formula for the mean
acceleration of a body $P$ with respect to a system of other masses
(weighted by their mass) namely 
\begin{equation}\label{eq:mach2}
\sum m_Q \frac{\dtwo r_Q}{dt^2}\Big/\sum m_Q
\end{equation}
where $m_Q$ is at distance $r_Q$ from $P$.  The notation (but not the
formula) has been changed in order to show the connection of Mach's
analysis with Sciama's principle.  Because now, if it is assumed that
all bodies move uniformly with bounded mutual velocity and if
(\ref{eq:mach1}) is substituted in (\ref{eq:mach2}), the following
formula for the acceleration of $P$ in terms of the other masses is
found $$\sum \frac{m_Q}{r_Q}b_Q$$ where $b_Q=(u_Q^2-v_Q^2)/\sum m_Q$
are all bounded by $1/\sum m_Q$ times the square of the bound for the
mutual velocities.  This is very close to Sciama's principle (ignoring
rotation).  To see the connection, drop the assumption of an absolute
space where all this was supposed to take place.  Keep only the
observations.  This equation can be interpreted as specifying the
``absolute'' acceleration of $P$ (and hence the IF at $P$) in terms of
data at $Q$ and if these data are labelled ``inertial effect'' then
this would obtain precisely Sciama's principle.

It is important to remark that this discussion is not intended to
suggest that uniform motion has an inertial effect; a small mass
moving uniformly has negligible inertial effect, though a large mass
has some effect due to inductive effects from its gravitational field.
What is intended is that the formula that it is sensible to use to
estimate the local inertial frame is likely to include a factor $1/r$
since apparent acceleration due to uniform motion does indeed include
such a factor.  The discussion is intended to support the contention
that inertial effects drop off like $1/r$.

\sh{Rotation}
\index{rotation!inertial effect}

Non-inertial motions are combinations of acceleration and rotation
(and inertial motions).  Now EGR deals well with acceleration.  This
is in some sense its major application, but as will be seen, it does
not deal well with rotation.  So for the purposes of embedding
Sciama's principle in EGR, it makes sense to concentrate on
rotation.  

Sciama's principle applied to rotation says that rotation of a mass
$m_Q$ at $Q$ contributes $K\thinm m_Q\thinm \omega_Q/r_Q$ to the
rotation of the IF at $P$ where $\omega_Q$ is the angular velocity of
$m_Q$.

It is important to notice that it is the angular \emph{velocity} of
$m_Q$ which contributes to the sum and not the angular \emph{momentum}
of $m_Q$ about $P$.  This behaviour (and a further final argument
supporting the key factor $1/r$) can be deduced from a simple
dimensional agument.  There is a highly relevant passage in MTW
\cite{MTW} discussing precession of the Foucault pendulum which is
worth quoting extensively.  It starts on page 547 para 3 with the
margin note \emph{The dragging of the inertial frame}.  It has been
edited very slightly to make the notation fit with the present
discussion and to suppress mention of conventional units.  In this
book \emph{natural units}, with $G=c=1$ and everything measured in
years are used for most of the calculations; here $G$ is Newton's
gravitational constant and not Einstein's tensor which is also
commonly denoted $G$.

\begin{quote}\small\em 
Enlarge the question.  By the democratic principle that equal masses
are created equal, the mass of the earth must come into the
bookkeeping of the Foucault pendulum.  Its plane of rotation must be
dragged around with a slight angular velocity, $\omega_{\rm drag}$,
relative to the so-called ``fixed stars.''  How much is $\omega_{\rm
  drag}$?  And how much would $\omega_{\rm drag}$ be if the pendulum
were surrounded by a rapidly spinning spherical shell of mass $m_{\rm
  shell}$ and radius $r_{\rm shell}$ turning at angular velocity
$\omega_{\rm shell}$?
\index{Misner, Thorne and Wheeler!democratic principle}
\index{Misner, Thorne and Wheeler!inertial drag}

Einstein's theory says that inertia is a manifestation of the geometry
of space-time.  It also says that geometry is affected by the presence 
of matter to an extent proportional to the factor $G/c^2$ \emph{(ie 1 in
natural units)}.  Simple dimensional considerations leave no room
except to say that the rate of drag is proportional to an expression
of the form
\begin{equation}\tag{21.155}\label{eq:mtw}
\omega_{\rm drag} = k\frac{m_{\rm shell}}{r_{\rm shell}}\omega_{\rm shell}.
\end{equation}
Here $k$ is a factor to be found only by detailed calculation. \dots
\end{quote}

Details of the dimensional argument used here will be given later.
The authors continue by discussing the results of Lense and Thirring
where $k$ is calculated to be $4/3$ assuming a specific approximation
which is in fact identical to the Sciama field. There will be more to
say about this shortly.

At this point it is worth making an observation.  The Sciama field can
be seen as a first approximation to a full-blown theory of dynamics
based on Mach's principle.  Since it coincides with \ind{gravitomagnetism},
which is a first approximation to EGR, it follows that no local
observations, where the fields are weak (for example the
\ind{Gravity Probe B} experiment \cite{GPB}) can distinguish
between EGR and a theory of
dynamics based on Mach's principle.  One of the main theses of this
book is that there is however strong experimental evidence in favour
of the latter from observations of galaxies.

\sh{The weak Sciama principle}\label{subsec:weakSp}

Continuing the discussion of the MTW quotation and equation
(\ref{eq:mtw}), their ``democratic principle'' is close to Sciama's
principle, at least in its universality, referring as it does to all
(accessible) matter in the universe.  The equation itself is precisely
the principle for the contribution of the mass $m_{\rm shell}$.  And
notice that it is implied that the dragging effect of the earth should
be coherent with the earth's rotation.  This point is so obvious that
it may easily be overlooked and is only mentioned because shortly a
model will be examined where the dragging is not always coherent.  To
see the connection with Sciama's principle for many distinct rotating
masses, consider the following thought experiments.  Replace the shell
by a ring of matter at distance $r=r_{\rm shell}$.  Nothing changes
qualitatively.  The constant $k$ reflects the precise geometry of the
setup and may change.  Now imagine that the ring is a necklace of $n$
beads all of the same mass $m$.  By the democratic principle, each has
the same effect $\omega'_{\rm drag}= \omega_{\rm drag}/n$ and, if $P$
is the centre of the ring and $Q$ one of the beads, then $Q$
contributes $k\thinm m\thinm \omega/r$ to the inertial frame at $P$
where $\omega$ is the angular velocity of $Q$ moving around $P$.  But
the local motion of $Q$ is exactly the same as a (uniform) linear
motion of velocity $\omega r$ along the tangent together with a
rotation on the spot of $\omega$.  Using the working hypothesis, the
linear motion has no inertial effect and the formula for the drag is
now exactly Sciama's principle in this case, namely:

\textbf{Weak Sciama Principle}\qua {\sl A mass $m$ at distance $r$ from $P$
  rotating with angular velocity $\omega$ contributes a rotation of
  $k\thinm m\thinm \omega/r$ to the inertial frame at $P$ where $k$ is
  constant.}

This \emph{\ind{weak Sciama principle}} is the statement that is needed for
the dynamics of galaxies.  The constant $k$ is a normalising factor
which needs to be set in context.  When the principle is used in the
next chapter (equation \ref{eq:nett}), this will be made precise.

Incidentally it can now be seen why the working hypothesis implies
that angular momentum is the wrong measure of the inertial effect of
one mass on another.  A uniform linear motion has no inertial effect,
but, adding a linear motion to $Q$ may well have a strong effect on
its angular momentum about $P$.  Conversely rotation need not
correlate with angular momentum: If $Q$ is in fact a point mass, then
rotation of $Q$ with angular velocity $\omega$ has no angular momentum
about $P$ whereas motion in a circle around $P$ with the same angular
velocity does have angular momentum.

The weak Sciama principle is not Machian in even the weakest version
(that all effects are due to observable source).  It makes no attempt
to completely specify the IF at $P$ in terms of all the matter in the
universe and indeed it leaves open the possibility that the IF at $P$
may be affected by unknown events (perhaps they are outside the
visible horizon --- cf MacKay--Rourke \cite{GRB} and
\fullref{app:GRB}).  But the advantage of a local statement of this
type is that it avoids the causality problems implicit in any global
statement and it is open to direct verification using local
observations.  One of the main theses of this book is that it is
indeed strongly supported by observations of galaxies and in
particular their characteristic rotation curves.

\sh{The Lense--Thirring effect}
\index{Lense-Thirring effect}
Like the full Sciama principle, the weak principle only makes sense in
an approximation to Minkowski space and this is exactly how it will be
used (the formulation is given near the start of the next chapter).
Early work of Lense and Thirring \cite{LT} mentioned above, calculated
the \id\ due to a heavy rotating body assuming a specific
approximation to EGR.  To be precise they calculated the \id\ due to a
rotating spherical shell for points nearby.  As seen above, this
effect is roughly in accord with Sciama's principle for points inside
the shell, but as will be seen shortly, it is hopelessly wrong
outside.

The approximation they used is the same as that used by Sciama and is
known as \ind{gravitomagnetism}.  The equations correspond formally to
Maxwell's equations and the effect can be understood by thinking of
electromagnetism.  Motion of matter corresponds to electrical current
and a circular motion induces a linear magnetic effect.  The dragging
effect corresponds to magnetic lines of force with the induced
rotation having the line as axis with rotation around the line in the
positive sense.  Thus a rotating body behaves like a magnet and causes
\id\ which is coherent near the poles but anti-coherent to the side
where the magnetic lines run back between the poles.

This has some very counter-intuitive consequences.

(a)\qua  Uniform linear motion has rotational inertial effects.

(b)\qua  A rotating body drags some frames nearby in the opposite
direction to the rotation causing the drag.

(c)\qua In general the direction of drag is unrelated to the rotation
which induces it.

Effect (b) was picked up by \ind{Rindler} \cite{Rindler} and correctly
labelled ``anti-Machian''.  However his conclusion that Mach's
principle needs to be treated with care ``one simply cannot trust
Mach!'' is bizarre.  The philosophical reasons for Mach's principle
are compelling and it must be incorporated in any theory that
describes reality.  It is the Lense--Thirring effect that must be
wrong.  In any case, it is not necessary to appeal to Mach's principle
to see that \id\ should be coherent.  As will be seen in a couple of
lines, a simple thought experiment using general principles of
symmetry and continuity will establish this fact.

\sh{Central rotation}

Perhaps the Lense--Thirring effect is wrong because of the
approximation used, so now turn to theories without approximation,
including EGR.  Consider a dynamical theory, which may not be EGR, but
which is metrically based and which specialises to special relativity
locally in same way that EGR does, with a similar equivalence
principle.  Here is a simple thought experiment which shows that, in
any such theory, frame dragging due to a central rotating body exists
and is coherent.

\begin{figure}[ht!]
\labellist
\small\hair 2pt
 \pinlabel {inner body} at 143 143
 \pinlabel {outer body} at 139 26
 \pinlabel {``infinity''} [bl] at 243 242
\endlabellist
\centering
\includegraphics[width=.5\hsize]{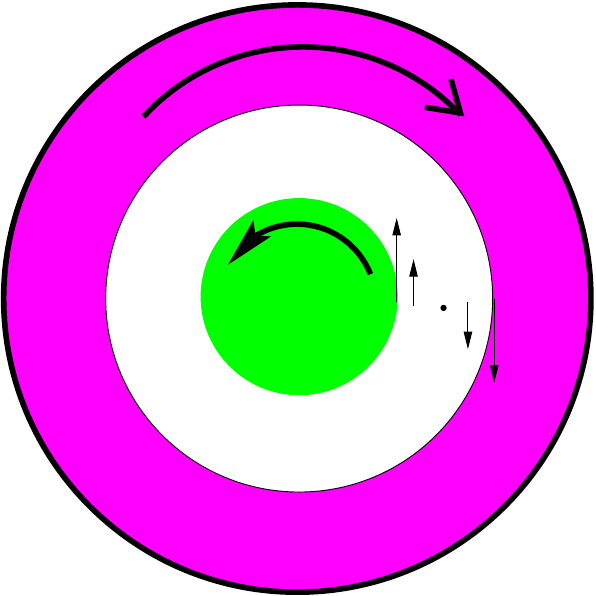}
\caption{Inertial drag between two heavy bodies}
\label{fig:drag}
\end{figure}

Imagine that the universe is a 3--sphere (spatially) and that it is
filled with two very heavy bodies (both 3--balls) with a comparatively
small (vacuum) gap between them.  Suppose that these bodies are in
relative rotation.  Then by symmetry, frames half way between the
bodies will rotate at the average speed and by continuity the
\id\ will move towards rotation with each of the bodies as one moves
away from the centre.  Diagramatically the situation is pictured in
\fullref{fig:drag}.  Note that in the figure the bodies are
represented as nested.  To get the correct view think of the outer
circle labelled ``infinity'' as the diametrically opposite point to
the centre of the inner body.  To make sense of \id\ here, assume
that the space between the two bodies has a flat background metric,
but do not assume anything about the space inside the bodies.

Now shrink the inner body to be the central rotating body and imagine
the outer body to be the rest of the heavy universe.  It is
unreasonable to suppose that the qualitative description of inertial
drag changes during the shrinking process and therefore, in the final
metric, the central body will induce coherent \id.

Now appeal to the same dimensional considerations as used in the
passage quoted from MTW (above) to deduce the weak Sciama principle.
Let the rotating body be labelled $Q$ and have mass $m$ and angular
velocity $\omega$.  For simplicity, consider a point $P$ on the
equatorial plane of the rotating body at distance $r$ from the centre.
It is a commonsense assumption that the dragging effect at $P$ is
proportional to $m\thinm\omega$ and, being a pure rotation of the
local inertial frame, has dimension $1/T$ where $T$ means ``time''.
(Notice that there is no sensible meaning to the centre of rotation
for this effect.  Two rotations which have the same angular velocity
but different centres differ by a uniform linear motion and inertial
frames are only defined up to uniform linear motion.)  Now in
relativity time, mass and distance all have the same dimension.  Thus
$m\thinm\omega$ is dimensionless and the only sensible formula for the
induced \id\ is $k\thinm m\thinm\omega/r$, possibly normalised (note
in passing that normalising constants such as $k$ are dimensionless
and do not affect this argument).

Go further with this thought experiment.  Assume now that the universe
is $\re^3$ spatially with the heavy inner rotating body at the origin.
And imagine that the outer body is the outside of a sphere of radius
$R$ say and is in fact at rest.  Now let $R$ tend to infinity and, as
it does so, control the mass of the outer body to keep its inertial
effect near the inner body constant.  In the limit, the outer body is
replaced by an asymptotically flat metric near infinity and then,
outside the inner body, is a metric which is stationary (the whole
construction was stationary) axially-symmetric and asymptotically flat
at infinity.  Assume now that the theory being considered is in fact
EGR.  Then this metric must coincide with the Kerr metric which is
well-known to be the unique metric satisfying Einstein's equations for
a vacuum with these properties.  Equally well-known, the inertial drag
effects of the Kerr metric drop off like $1/r^3$.  Thus this metric,
constructed using the thought experiment, continuity and dimensional
arguments, does not satisfy Einstein's equations; or rather, to be
very precise, if it is assumed that \emph{the space constructed is a
  vacuum outside the hypothesised masses}, then the metric does not
satisfy Einstein's equations.

One may wonder why the dimensional argument does not equally apply to
the Kerr metric.  This is because the \ind{Kerr metric} has a well-defined
angular momentum but no well-defined angular velocity.  Thus the
\id\ effect of the Kerr metric must be proportional to angular
momentum NOT to mass times angular velocity.  But angular momentum has
dimension $T^2$ and to get a drag effect of dimension $1/T$ a formula
of the type $k\thinm A/r^3$ is needed where $A$ is angular momentum.

It is worth at this point recapping why angular momentum is the wrong
measure for inertial effects.  This is a simple consequence of the
working hypothesis that linear motion has no inertial effect.  Angular
momentum can be altered by adding a linear motion.  Angular velocity
cannot be so altered.

\sh{Adding Sciama's principle to EGR}\label{subsec:Add}

At this point the story seems to have run into an impasse.  Assuming
the universe obeys standard relativity (EGR) then the version of
Mach's principle that is needed does not hold.  Inertial drag drops
off as $1/r^3$ in EGR and not $1/r$.

There are two sensible ways out of this impasse.

(1)\qua The \emph{revolutionary} approach is to abandon EGR and build
a new theory which satisfies Sciama's principle.

(2)\qua The \emph{conservative} approach is to continue to use EGR but
add a hypothesis within EGR that implies Sciama's principle.  As seen
above, this is impossible assuming the space between bodies is a
vacuum, so this approach entails hypothesing that \emph{space near a
  rotating body is not a vacuum} and the thought experiment conducted
above is impossible because the space between the rotating bodies is
not a vacuum.
\index{conservative approach}\index{revolutionary approach}

This book adopts the conservative approach.  Apart from avoiding the
non-trivial problem of finding a theory to replace EGR, this approach
has one great technical advantage: it provides a mechanism for Mach's
principle (at least as it applies to rotation) which does not run into
causal problems. 

The hypothesis added to EGR is that any rotating body disturbs the
local space-time by dragging inertial frames near it coherently by an
amount proportional to the rotating mass times its angular velocity,
with the influence dropping off asymptotically with $k/r$ where $r$ is
distance from the centre of gravity of the rotating mass and $k$ is
constant.  The precise formula is given in the next chapter, where
there is also an interpretation in terms of the metric.

In a vacuum, EGR does not have this inertial drag effect.  The Kerr
metric which is the only rotationally symmetric vacuum metric flat at
infinity and valid in EGR has a drag effect dropping off much faster
than this (asymptotically with $k/r^3$).  So the hypothesis amounts to
assuming that a rotating mass has a non-zero effect on the
stress-energy tensor near it -- in other words stops the space near it
being a true vacuum.  It also gives a natural way to understand how
inertial drag propagates: the disturbance to the local vacuum is akin
to a gravity wave and propagates at the speed of light.  Furthermore
reading back from the rest of the universe, the local background
inertial frame is created by the rest of the universe by a similar
propagation effect from all the rest of the matter.  Thus the
hypothesis gives a natural causal framework for Mach's principle.  An
example of this causal framework working in practice would be the case
where a rotating body undergoes a sudden change (eg breaking up) which
changes the inertial drag field that it causes.  This makes a
disturbance in the local space-time (a sort of gravity wave) which
propagates at the speed of light with no causal problems.
\index{inertial drag!propagation}

Another consequence is that a rotating body interacts directly with
surrounding matter and indeed energy can be extracted in a similar way
to the \ind{Penrose effect} which extracts energy from the Kerr metric.
This implies that the rotation will eventually radiate away.  This is
an extremely small effect for ordinary rotating bodies and only becomes
significant for rotating black holes where the energy radiating away
fuels the surrounding dynamic as will be seen in the next few
chapters.  The effect of this can be seen graphically in the spiral
structure of full-size galaxies, eg the so called Whirlpool galaxy,
\fullref{fig:more_gals}, left.
\index{galaxy!Whirlpool}

Superficially the change to vacuum that the new hypothesis entails may
seem like an alternative formulation of ``dark matter'' but it is in
fact quite different.  The dark matter hypothesis amounts to assuming
the existence of inert matter, which has no effect other than
gravitational attraction, and cannot otherwise be detected.  It is an
\emph{incident} hypothesis in the sense that it contains nothing more
than what is needed to correct the rotation curve; it is a ``fudge
factor'', designed to correct a shortfall.  The inertial drag
hypothesis on the other hand amounts to assuming a new effect of a
rotating body on the field outside it.  It is justified by Mach's
principle which, as has been seen, is philosophically compelling and
must be embodied in any theory that seeks to accurately describe
reality.  Thus it is a \emph{necessary} hypothesis in the sense that
it needs to made, independently of the rotation curve, in order to
encode the necessary Mach principle.  The \id\ field that is assumed
to exist can be detected directly by its effect on inertial frames so
it has an existence independent of the rotation curve that it
serendipitously also predicts.

In EGR a rotating mass does in fact have an effect on the field
outside the body, but this is confined to the skew-symmetric part of
the field (the \emph{Weyl} tensor, or \emph{trace-free} part of the
curvature).  So the new hypothesis implies that a rotating body also
affects the other part, the \emph{Ricci} curvature.  Einstein's
equations for a vacuum are equivalent to the vanishing of the Ricci
curvature (see \fullref{sec:Eequns}).  Thus, if the Ricci curvature is
nonzero, then the field is not an Einstein vacuum.
\index{curvature!Weyl tensor} \index{curvature!trace-free}

\sh{Sciama's principle and black holes}
\index{black hole!effective radius}

Applying Sciama's principle to black holes entails assuming that a
black hole has a well-defined angular velocity as well as a
well-defined angular momentum.  Equivalently a black hole has an
effective radius, $r_{\rm eff}$, related to angular momentum $\Omega$ and
angular velocity $\omega$ by 
\begin{equation}
\Omega = M \omega r_{\rm eff}^2.
\label{eq:AmAv}
\end{equation}
For a black hole the fiction is that the actual radius is zero (total
gravitational collapse) and hence angular velocity is not determined.
So this assumption is equivalent to replacing conventional theory by
the more sensible assumption that, in the collapse to a black hole,
matter reaches a small but non-zero size.

\sh{Coda}

Sciama's initiative, to base a dynamical theory on Mach's principle as
formulated in Sciama's principle, has never been followed up and this
approach to dynamics remains dormant.  One of the aims of this book is
to reawaken this approach.  Sciama did return to the topic of Mach's
principle in \cite{SWG}.  However this paper abandons Sciama's
principle and formulates Mach's principle in one of its weakest forms,
namely that all phenomena have their origin in some material source or
boundary condition.  Moreover the theory exposited in \cite{SWG} is
EGR which as has been seen is incompatible with even the weak Sciama
principle.

\index{Sciama!principle|)}

\chapter{The rotation curve}\label{sec:rot_curve}
\index{rotation curve|(}
\index{galaxy!rotation curve}

The \emph{rotation curve} of a galaxy with an equatorial plane (for
example a spiral galaxy has its spiral arms lying roughly in such a
plane) is the plot of tangential velocity against distance from the
centre for a particle (star or similar) moving in the equatorial
plane.  In practice it is not possible to observe one star, but rather
the general motion of all stars (or other radiating matter) in the
equatorial plane.  This makes the observed nature of rotation curves
all the more striking.  Typically the curve (of tangential velocity
against distance from the centre) comprises two approximately straight
lines with a short transition region.  The first line passes through
the origin, in other words rotation near the centre has constant
angular velocity (plate-like rotation); the second is horizontal, in
other words the tangential velocity is asymptotically constant, see
\fullref{fig:modbeg} (right) below.  Furthermore, observations show
that the horizontal straight line section of the rotation curve
extends far outside the limits of the main visible parts of galaxies
and the actual velocity is constant within less than an order of
magnitude over all galaxies observed (typically between 100 and
300km/s) see \fullref{fig:rotsSR}.

Galactic rotation curves are so characteristic (and simple to
describe) that there must be some strong structural reason for them.
They are very far indeed from the curve obtained with a standard
Keplerian model of rotation under any reasonable mass distribution.
In a Keplerian model, suppose that the mass within a radius $r$ of the
centre is $M(r)$ then equating centrifugal force with gravitational
attraction gives
$$\frac{v^2}r=\frac{G\,M(r)}{r^2}$$ where $v$ is tangential velocity
and $G$ is Newton's gravitational constant (taken to be 1 in natural
units).  Thus if $v$ is asymptotically constant then $M(r)$ is
asymptotically equal to a constant times $r$ and tends to infinity
with $r$.

Nevertheless, in spite of the huge mass needed, a Keplerian model is
exactly what is assumed in current cosmological theory.  To square the
circle, current theory hypothesises the existence of a huge amount of
matter.  Since this matter is not observed, it is called called
``dark''.  It needs to be distributed in precisely the right way to
make Keplerian rotation fit the rotation curve.  This is extremely
implausible for several reasons.  Firstly it has just bben seen that the
quantity of \ind{dark matter} required is huge and tends to infinity
with the radius of fit, which as mentioned above appears to be
unbounded.  Secondly it is unreasonable to suppose that exactly the
right distribution of dark matter happened (by condensation) for every
galaxy and thirdly, the final arrangement with most of the matter on
the outside is dynamically unstable.  For stability in a rotating
system (such as the solar system or Saturn's discs) there must be a
strong central mass to hold it together.  Failing this the system will
tend to condense into smaller systems.  Finally despite the best
efforts expended in the search, nor hair nor hide of dark matter has
been found to date.\index{angular momentum!problem}

This chapter presents a solution to these problems using a quite
different point of view.  The suggestion made here is that the centre
of a typical galaxy contains a huge rotating body (probably a black
hole) and that the \id\ effects coming from this rotating mass are
responsible for the observed rotation curves.  

There is strong evidence that the masses of galaxies exceed the mass
of the visible parts by some orders of magnitude.  This goes back to
Zwicky 1933 \cite{Zwicky} \index{Zwicky}who used the virial theorem to estimate the
mass of galaxies in the \ind{Coma Berenices cluster} and discovered that the
mass exceeds luminosity mass by a factor of about $10^2$.  In current
cosmological theory, this missing matter is identified with the
invisible ``dark matter'' needed to make Keplerian motion fit the
rotation curve.  In the solution presented here, this extra matter is
concentrated in the heavy rotating centre which controls the dynamics
by \id\ effects.

Assume that there is a standard background space (Minkowski or
Schwarzschild space) and use an approximation to this background.
Sciama's principle as discussed in \fullref{sec:Sciama} implies that
the central rotating mass creates an \id\ field dropping off like
$k/r$, which causes inertial frames to rotate \wrt\ the background.
With this assumption, it is not hard to solve the equations to find
the tangential velocity in an equatorial orbit as a function of $r$
(distance from the centre), and every equatorial
orbit has the salient feature of observed rotation curves, namely a
horizontal asymptote.  This asymptote is \emph{the same for all
equatorial orbits} and hence any average over many orbits will also
have this asymptote and this explains the observed rotation curve.

This provides strong evidence for the (weak) Sciama principle with
\id\ drop off asymptotically at $k/r$ as promised at the end of
\fullref{subsec:weakSp}.

\sh{The \ind{weak Sciama principle}}

Sciama's principle (\fullref{subsec:Sp}) implies that the rotation of
the local inertial frame (IF) is the sum
$$\sum_Q \frac{m_Q}{r_{Q}} \omega_Q$$ where the sum is taken over all
(accessible) masses $m_Q$ in the universe where $m_Q$ is at distance
$r_Q$ and rotating with angular velocity $\omega_Q$ and the sum is
suitably normalised .

For the purposes of this work, only the weak version is needed:
(\fullref{subsec:weakSp}).

\textbf{Weak Sciama Principle (WSP)}\qua {\sl A mass $M$ at distance
  $r$ from $P$ rotating with angular velocity $\omega$ contributes a
  rotation of $k\thinm M\thinm \omega/r$ to the inertial frame at $P$
  where $k$ is constant.}

In the main application $M$ will be the (heavy) centre of a galaxy,
but the analysis applies to any axially-symmetric rotating body which
does not need to be assumed to be heavy.

To fix notation, consider a central mass $M$ at the origin in
$3$--space which is rotating in the right-hand sense about the
$z$--axis (ie counter-clockwise when viewed from above) with angular
velocity $\omega_0$.  Assume a flat background space-time, away from
$M$, with sufficient fixed masses at large distances to establish a
non-rotating IF near the origin, if the effect of $M$ is ignored.  Let
$P$ be a point in the equatorial plane (the $(x,y)$--plane) at
distance $r$ from the origin.  The rotation of the inertial frame at
$P$ is given by adding the contribution from $M$ to the contribution
from the distant masses.  Because $P$ is near a large mass, it makes
sense to normalise the sum as in equation \ref{eq:Knorm}.  This is
equivalent to using a weighted sum, in other words the inertial frame
at $P$ is rotating coherently with the rotation of $M$ by the average
of $\omega_0$ weighted $kM/r$ and zero (for the distant fixed masses)
weighted $C$ say.  Further normalise the weighting so that $C=1$
(which is the same as replacing $k/C$ by $k$) which leaves just one
constant $k$ to be determined by experiment or theory.  The nett
effect is a rotation of
\begin{equation}\label{eq:nett}
\frac{(kM/r)\times\omega_0 +1\times 0}{(kM/r) + 1} =
\frac{A}{r+K}\ \ \text{where}\ \ K=kM\ \ \text{and} \ \ A=K\omega_0.
\end{equation}

{\bf Note}\qua If the full Sciama principle is assumed and that
$\sum_Q m_Q/r_Q=1$ (equation \ref{eq:Unnorm}), which as was seen has
some observational evidence to support it, then $C$ and $k$ are both 1
and $K=M$ and $A=M\omega_0$.  However the choice of $k=1$ is not
relevant to the arguments presented in this or subsequent chapters.
Nothing that is proved depends on knowing the exact relationship
between $K$ and $M$.

\sh{The dynamical effect of the \id\ field} \label{sec:ideffects}
\index{inertial drag!dynamic effect}

The key to the rotation curve is to understand the way in which the
\id\ field affects the dynamics of particles moving near the origin.
For simplicity work in the equatorial plane.  Assume that the IF at
$P$ (at distance $r$ from the origin) is rotating \wrt\ the background
with angular velocity $\omega(r)$ counter-clockwise.  When computing
rotation curves, the formula for $\omega(r)$ just found
(\ref{eq:nett}) will be used but for the present discussion it is just
as easy to assume a general function.  The IF at $P$ can be identified
with the background space, but it is important to remember that it is
rotating.  As remarked in \fullref{subsec:weakSp} there is no sensible
meaning to the centre of rotation for an inertial frame.  Two
rotations which have the same angular velocity but different centres
differ by a uniform linear motion and inertial frames are only defined
up to uniform linear motion.  Thus it can be assumed for simplicity
that all the rotations have centre at the origin.  Then the IFs can be
pictured as layered transparent sheets, each comprising the same
point-set but with each one rotating with a different angular velocity
about the origin.  Each sheet corresponds to a particuar value of $r$.
It is necessary to be very clear about the nature of motion in one of
these frames.  A particle moving with a frame (ie one stationary in
that frame) has no \emph{inertial velocity} and its velocity is called
\emph{rotational}.  In general if a particle has velocity $\mathbf{v}$
(measured in the background space) then
$$
\mathbf{v}=\mathbf{v}_{\rm rot}+\mathbf{v}_{\rm inert}
$$ where its \emph{\ind{rotational velocity}} $\mathbf{v}_{\rm rot}$ is the
velocity due to rotation of the local inertial frame and
$\mathbf{v}_{\rm inert}$ is its \emph{\ind{inertial velocity}} which is the
same as its velocity measured \emph{in} the local inertial frame.
Note that $\mathbf{v}_{\rm rot} = r\omega(r)$ directed along the
tangent.\index{velocity!inertial}\index{velocity!rotational}

\begin{figure}[ht!]
\centering
\includegraphics[width=2.5in]{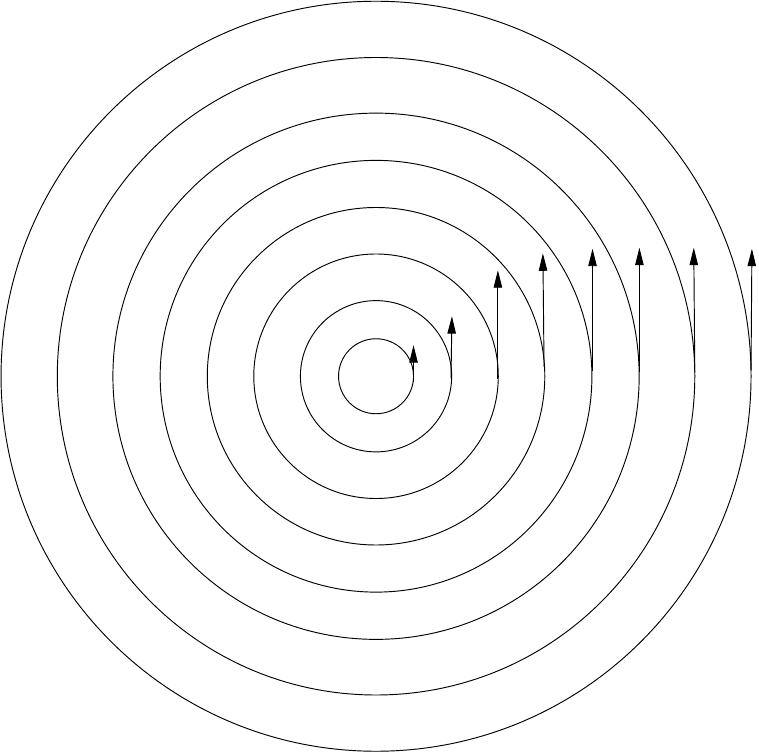}
\caption{Rotational velocities in the \id\ field near a rotating body}
\label{fig:IFs}
\end{figure}

The reader might find \fullref{fig:IFs} helpful at this point.  

Inertial velocity correlates with the usual Newtonian concepts of
\ind{centrifugal force} and conservation of angular momentum.%
\index{angular momentum!conservation}

As a particle moves in the equatorial plane it moves between the
sheets so that a rotation about the origin which is rotational in one
sheet becomes partly inertial in a nearby sheet.  For definiteness,
suppose that $\omega(r)$ is a decreasing function of $r$ and consider
a particle moving away from the origin and at the same time rotating
counter-clockwise about the origin.  The particle will appear to be
being rotated by the sheet that it is in and this causes a tangential
acceleration.  This acceleration is called the \emph{\ind{slingshot effect}}
because of the analogy with the familiar effect of releasing an object
swinging on a string.  But at the same time the particle is moving to
a sheet where the rotation due to \id\ is decreased and hence part of
the tangential velocity becomes inertial and is affected by
conservation of angular momentum which tends to decrease the angular
velocity.  These two effects balance each other out in the limit and
this explains the flat asymptotic behaviour.  Below this is proved
analytically, but first, here is a metrical interpretation of the
hypothesised inertial drag effect being used.

\sh{A metrical interpretation of \id}
\index{inertial drag!metrical interpretation}

Define an \emph{\id\ metric} \index{inertial drag!metric}by adding a variable rotation factor to a
\ss\ metric.  The primary metrics of interest are obtained from the
flat (Minkowski) metric and the Schwarzschild metric, but the proof of
the rotation curve applies to any metric of this type.  The
\id\ metric based on the Schwarzschild metric is likely to be close to
the metric that will eventually be chosen if the conservative approach
(cf \fullref{subsec:Add}) is generally adopted and serves to motivate
the search for this metric.

Furthermore, as will be seen in the next chapter, a model for quasars
based on the \ind{Schwarzschild metric} successfully explains a good deal of
the observations of these strange objects and this strongly suggests
that this metric is a real reflection of reality at least in
particular cases.

The most general \ss\ metric can be written in the form:
\begin{equation}\label{eq:metric-gen}
ds^2 = -B\,dt^2 + A\,dr^2 + r^2\,d\Omega^2
\end{equation}
where $A$ and $B$ are positive functions of $r$ and $t$ on a suitable
domain.  Here $t$ is time, $r$ is ``distance from the centre'' (but
see the note below) and $d\Omega^2$, the standard metric on the unit
2--sphere $S^2$, is an abbreviation for $d\theta^2 + \sin^2\theta \,
d\phi^2$.  Orient the 2--sphere so that the $z$--axis passes through
it at the north pole where $\theta=\pi/2$.  The $(x,y)$--plane (pasing
through the origin and perpendicular to the $z$--axis) is the
\emph{equatorial plane} where $(r,\phi)$ are polar coordinates.  The
Schwarzschild--de Sitter metric is the case
$$B=\frac1A=1-\frac{\Lambda r^2}3 - \frac{2M}r$$ with $\Lambda$ and
$M$ constants.  By Birkhoff's theorem (cf \fullref{sec:Birk}) this is
the only case where the metric satisfies Einstein's vacuum equations
with cosmological constant in some region.  In this case the metric is
necessarily static in this region.  The special cases $M=\Lambda=0$
and $\Lambda=0$ give the Minkowski and Schwarzschild metrics
respectively.\index{Birkhoff's theorem}

{\bf Note}\qua It is important to observe that $r$ is a coordinate
\emph{which is not precisely the same as distance in the metric}.  It
is chosen so that the sphere of symmetry at coordinate $r$ has area
$4\pi r^2$.  Distance measured in the metric along a radius near this
sphere is not the same as change in the coordinate $r$ (this only
happens if $A$ takes the value 1 near the point under consideration).

The \emph{\id\ metric} is formed by adding a variable rotation about
the $z$--axis.  This is done by replacing $\phi$ by $\phi-\omega t$.
The metric is no longer diagonal
\begin{equation}\label{eq:metric-mod}
ds^2 = (-B+\rho^2\omega^2)\,dt^2 + A\,dr^2 + r^2\,d\Omega^2 -
2\rho^2\omega^2\,d\phi\,dt
\end{equation}
where $\rho=r\sin\theta$. 

If $\omega$ is constant this is the same metric viewed through
rotating glasses, but the whole point is to allow $\omega$ to vary.
Starting with the Schwarzschild--de Sitter metric and making this
substitution with variable $\omega$, gives a metric which no longer
satisfies Einstein's vacuum equations: indeed the change made is the
\emph{metrical embodiment of the hypothesised \id\ field}. It is not
hard to see that the inertial frame at a point rotates about a line
parallel to the $z$--axis with angular velocity the value of $\omega$
at that point.  This is clear if $\omega$ is constant and in general,
provided $\omega$ is continuous, it follows from the locality of
inertial frames.  So to fit with \id\ as formulated in \eqref{eq:nett}
it is necessary to set $\omega=\lfrac{A}{(r+K)}$ (at least in the
$(x,y)$--plane).  However it is easy to work with a general function
$\omega$ and specialise when needed.  The orbits of particles moving
on geodesics in the equatorial plane will now be investigatied and,
provided $\omega$ decreases like $A/r$ as $r\to\infty$, the orbits
will be found to fit observed rotation curves.

\begin{figure}[ht!]
\labellist
\small\hair 2pt
 \pinlabel $\phi$ [b] <2pt,0pt> at 103 202
 \pinlabel $t$ [l] <2pt,0pt> at 233 108
\endlabellist
\centering
\includegraphics[width=2in]{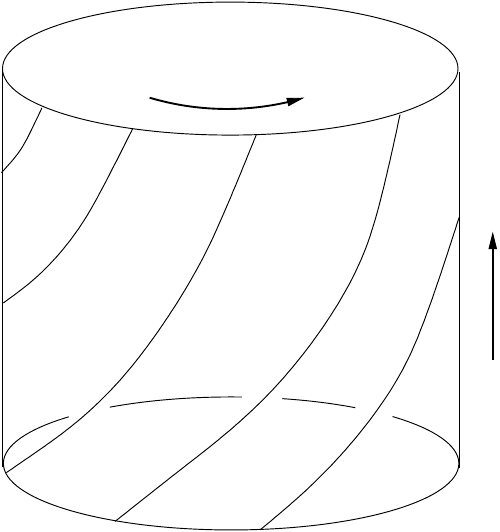}
\caption{Geodesics on the cylinder $r=z=$const in the \id\ metric}
\label{fig:metric}
\end{figure}

The reader might find \fullref{fig:metric} helpful for visualising
geodesics in the \id\ metric and understanding the \id\ effects.
It shows geodesics on a typical cylinder $r=z=$const.

\sh{Conservation of angular momentum}\index{angular momentum!conservation}

The discussion continues with the analytic derivation of the rotation
curve (the relation between $v$ and $r$) for the orbit of a particle
in the equatorial plane moving with total velocity $\mathbf v$, which
has tangential component $v$ (perpendicular to the line through the
origin).  Recall from the discussion in \fullref{sec:ideffects} above
that there are two opposing effects at work: the slingshot effect,
which tends to increase $v$ with $r$ and conservation of angular
momentum which tends to decrease it.  These two effects are calculated
together. The proof works in any \id\ metric, where the particle moves
along a geodesic.  (The special case of flat Minkowski space with
\id\ effects was motivated in \fullref{sec:ideffects}.)

The derivation starts with a proof of conservation of angular
momentum, which is a property of any system with a central force (or
space-time geometry which simulates a central force).  It is not
restricted to Newtonian physics.  The proof is adapted from Newton's
proof of the equal area law for planetary orbits (which law is exactly
the same as conservation of angular momemtum).  For the time being
ignore $\omega$ (or set it equal to zero).
\index{angular momentum!conservation!Newton's proof}
\begin{figure}[ht!]
\labellist
\small\hair 2pt
 \pinlabel $A$ [r] <0pt,0pt> at 19 233
 \pinlabel $P'$ [br] <0pt,0pt> at 36 275
 \pinlabel $B$ [l] <0pt,0pt> at 105 275
 \pinlabel $P$ [l] <0pt,0pt> at 104 202
 \pinlabel $O$ [t] <0pt,0pt> at 35 0
 \pinlabel $u'$ [b] <0pt,0pt> at 67 275
 \pinlabel $u$ [t] <0pt,-1pt> at 56 219
 \pinlabel $r'$ [r] <1pt,0pt> at 33 121
 \pinlabel $r$ [l] <1pt,0pt> at 71 111
 %\pinlabel $b$ [l] <0pt,0pt> at 104 244
 %\pinlabel $a$ [r] <1pt,2pt> at 25 252
\pinlabel {$\bf u$} [lb] <0pt,0pt> at 65 242
\endlabellist
\centering
\includegraphics[width=1.5in]{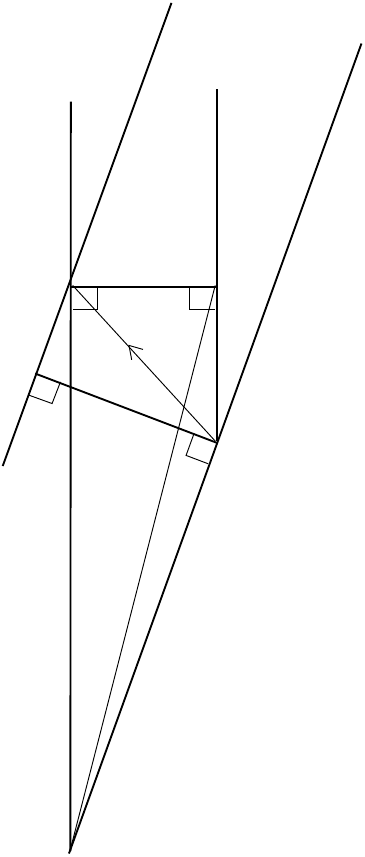}
\caption{Proof of conservation of angular momemtum}
\label{fig:CAM}
\end{figure}
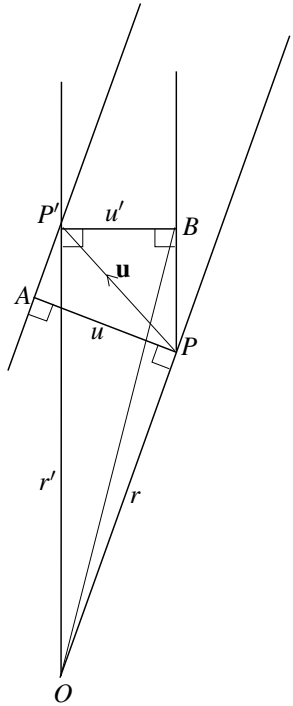

The idea is to replace the central force by a series of central
impulses at equally spaced (small) intervals of time.  Consider
\fullref{fig:CAM}.  At a particular time the particle (of small unit
mass) is at $P$ and has just received a central impulse resulting in
velocity ${\bf u}$.  Its tangential velocity at $P$ is $u=|AP|$.  One
small interval of time later the particle is at $P'$ and receives
another central impulse (along the line $OP'$) which does not change
its tangential velocity $u'=|P'B|$.  But the triangle $OPP'$ can be
regarded as having base $r= |OP|$ and height $u$ or base $r'=|OP'|$
and height $u'$ hence
\begin{equation}\label{eq:CAM}
  ur=u'r'
\end{equation}
in other words the angular momentum at $P$ is the same as that at
$P'$.

To obtain the result for an arbitrary continuous central force,
take the limit of a sequence of central impulses.  Note the proof does
not use any property of the central force other than that it acts
towards the centre.  Nor does it assume that $r$ represents a genuine
distance in the metric under consideration.  All that is needed is
that Euclidean geometry correctly describes the relationship between
$r$ and distances perpendicular to radii near $P$and $P'$ which is
precisely how $r$ was chosen.

\sh{The fundamental relation}
\index{fundamental relation!rotation curve}

Now reinstate $\omega$.  Note that ``force'' in the model is a
property of local space-time geometry.  In the case that $\omega$ is
constant, the inertial frame (rotating with $\omega$) is the same as
the unrotated case and in this frame the force is central.  Therefore
by locality it is central in the general case in the inertial frame.
Therefore the proof just given makes sense in the inertial frame at
$P'$, in other words rotating with angular velocity $\omega'=
\omega(P')$, though, as will be seen, in the limit the same result is
obtained if it is assumed that the frame is rotating with angular
velocity $\omega(P)$.  To find the required relationship between $v$
and $r$ write $v$ for the full tangential velocity at $P$ and $v'$ at
$P'$.  Since the frame is rotating at $\omega'$, $v=u+\omega'r$ and
$v'=u'+\omega'r'$.  Write $v'= v+\delta v$, $u'=u+\delta u$, $r'=
r+\delta r$ and $\omega'= \omega+\delta\omega$.

Since $ur=u'r'$ (equation \ref{eq:CAM}), substituting for $u',v'$ and
simplifying gives
\begin{equation}\label{eq:udr+rdu}
u\,\delta r+r\,\delta u =0\,.
\end{equation}
But 
$$\delta u=u'-u=v'-\omega'r'-(v-\omega'r)=v'-v - \omega'(r'-r) =
\delta v- \omega'\,\delta r$$ 
and substituting for $u,\delta u$ in \eqref{eq:udr+rdu} gives
$$(v-\omega'r)\,\delta r + r (\delta v- \omega'\,\delta r) =0$$ which
gives $$r\,\delta v=2r\omega'\,\delta r - v.$$ It is now possible to
replace $\omega'$ by $\omega$ to first order (as forecast) and going
to the limit yields the \emph{fundamental relation} between $v$ and
$r$:
\begin{equation}\label{eq:fund}
\fboxrule1pt\fboxsep5pt\fbox{$\displaystyle\frac{dv}{dr}=2\omega-\frac vr$}
\end{equation}
The fundamental relation can be understand intuitively as follows.
The slingshot effect intuitively produces an acceleration $dv/dr =
\omega$.  On the other hand $v_{\inert}=v-\omega r$ is the
``inertial'' tangential velocity (corrected for rotation of the local
inertial frame) and therefore conservation of angular momentum
produces a deceleration in $v$ of $v_{\inert}/r$ or an acceleration
$dv/dr = \omega- v/r$.  Adding the two effects gives the relation.

\sh{Solving to find rotation curves}

Given $\omega$ as a function of $r$, \eqref{eq:fund} can be solved to
give $v$ as a function of $r$.  Rewrite it as
$$r\,\frac{dv}{dr}+v=2\omega r\,.$$
The LHS is $d/dr\,(rv)$ and the general solution is
\begin{equation}\label{eq:gensol}
v=\frac1r\Bigl(\int2\omega r\,dr + \text{const}\Bigr).
\end{equation}
It is now clear that any prescribed differentiable rotation curve can
be obtained by making a suitable choice of continuous $\omega$.
\index{rotation curve!general solution}

Of interest here are solutions which, like observed rotation curves,
are asymptotically constant and inspecting \eqref{eq:gensol} this
happens precisely when $\int2\omega r\,dr$ is asymptotically equal to
$Cr$ for some $C$ and this happens precisely when $2\omega$ is
asymptotically equal to $C/r$.  This proves the following result.

\begin{theorem}The equatorial geodesics in the \id\ metric
  \eqref{eq:metric-mod} have tangential velocity asymptotically equal
  to constant $C$ if and only if $\omega$ is asymptotically equal
  to $A/r$ where $C=2A$.
\end{theorem}

\sh{The basic model}
\index{rotation curve!basic model}

Now specialise to the case $\omega=\lfrac{A}{(r+K)}$ which gives the
value of \id\ formulated in \eqref{eq:nett}.  The constant $C$ in the
theorem is no longer needed and it is reused.

From \eqref{eq:gensol} \def\strutt{\vrule width 0pt height 20pt}
\begin{align}
v&=\frac1r\left(\int\frac{2Ar}{r+K}\,dr +
C\right)=\frac{2A}r\left(\int1-\frac{K}{r+K}\,dr\right) + \frac Cr\notag\\
 &= 2A -\frac {2AK}r \log\left(\frac rK +1\right) + \frac Cr\strutt\label{eq:v}
\end{align}
where $C$ is a constant depending on initial conditions.  For a
particle ejected from the centre with $v= r\omega_0$ for $r$ small,
$C=0$, and for general initial conditions there is a contribution
$C/r$ to $v$ which does not affect the behaviour for large $r$.  For
the solution with $C=0$ there are two asymptotes.  For $r$ small,
$v\approx r\omega_0$ and the curve is roughly a straight line through
the origin.  And for $r$ large the curve approaches the horizontal
line $v=2A$.  A rough graph is given in \fullref{fig:modbeg} (left)
where $K=A=1$.  The similarity with a typical rotation curve,
\fullref{fig:modbeg} (right), is obvious.  Note that no attempt has
been made here to use meaningful units on the left.  See
\fullref{fig:rotsmod} below for curves from the model using sensible
units.

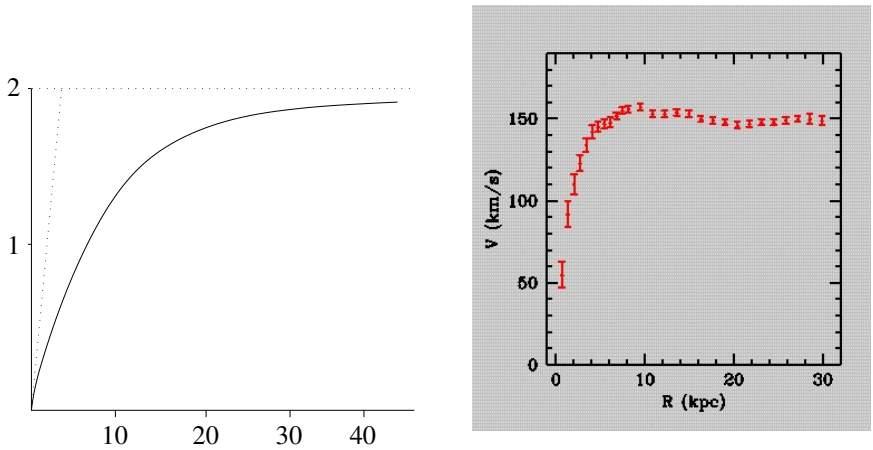
\begin{figure}[ht!]
\labellist\small
\pinlabel 2 [r] at 14 457
\pinlabel 1 [r] at 14 247
\pinlabel 10 [t] at 130 17
\pinlabel 20 [t] at 252 17
\pinlabel 30 [t] at 364 17
\pinlabel 40 [t] at 463 17
\endlabellist
\cl{\includegraphics[width=.4\hsize]{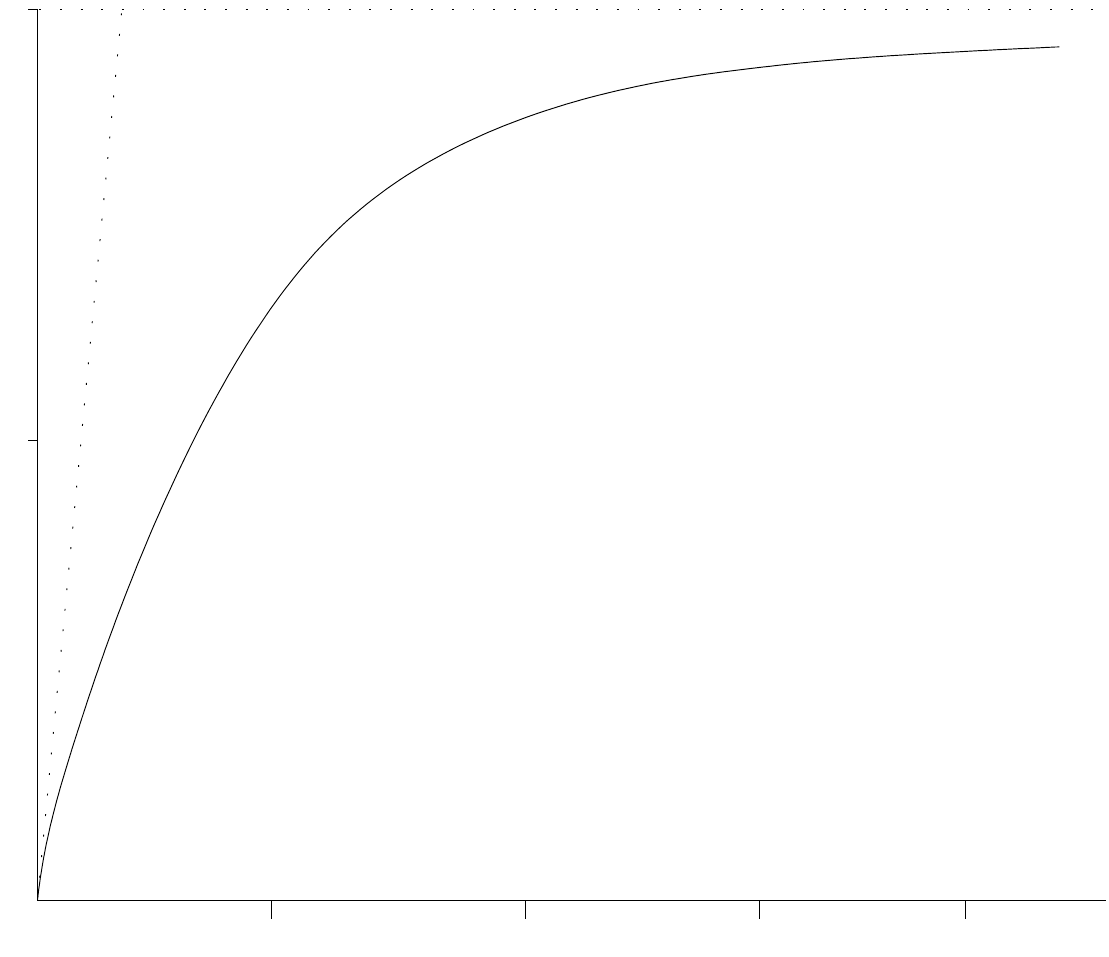}\qquad\includegraphics[width=.4\hsize]{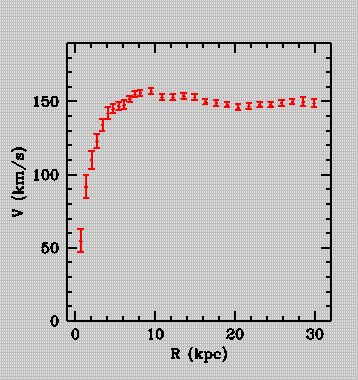}} 
\caption{The rotation curve from the model (left) and for the galaxy
  NGC3198 (right) taken from Begeman \cite{Beg}}
\label{fig:modbeg}\index{galaxy!NGC3198}
\end{figure}

There are other shapes for rotation curves; see \cite{SR} for a
survey.  All agree on the characteristic horizontal straight line.
\fullref{fig:rotsSR} is reproduced from \cite{SR} and gives a good
selection of rotation curves superimposed.  In \fullref{fig:rotsmod}
is a selection of rotation curves again superimposed, sketched using
Mathematica\fnote{The notebook {\tt Rots.nb} used to draw this figure
  can be collected from \cite{Nb} and the values of the parameters
  used read off.} and the model given here.  The different curves
correspond to choices of $A,K$ and $C$.  The similarity is again
obvious.  The units used differ.  In the model given here natural
units are used so that a velocity of .001 is 300km/s and a distance of
45,000 is 15Kpc approx.
\begin{figure}[ht!]
\cl{\includegraphics[height =1.8in]{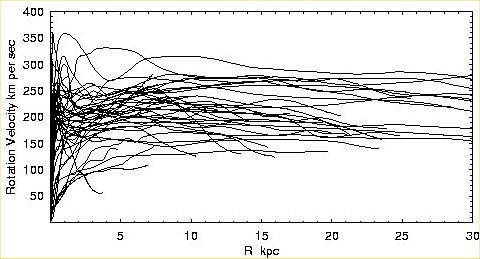}}
\caption{A collection of rotation curves  from \cite{SR}}
\label{fig:rotsSR}
\index{rotation curve!observed}
\end{figure}

\begin{figure}[ht!]
\cl{\includegraphics[width=.7\hsize]{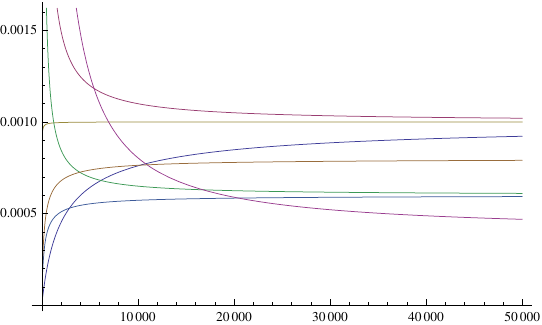}}
\caption{A selection of rotation curves from the model}
\label{fig:rotsmod}
\index{rotation curve!from model}
\end{figure}

It is worth commenting that the observed rotation curve for a galaxy
is not the same as the rotation curve for one particle, which is what
has been modelled here.  When observing a galaxy, many particles are
observed at once and what is seen is a rotation curve made from
several different rotation curves for particles, which may be close
but not identical.  So it is expected that the observed rotation
curves have variations from the modelled rotation curve for one
particle, which is exactly what is seen in Figures \ref{fig:modbeg}
(right) and \ref{fig:rotsSR}. 

The next chapter is devoted to the other main application of \id,
namely quasars.  Then in \fullref{sec:spiral_struc} the analysis given
here will be extended to find equations for orbits in general (not
just for the tangential velocity) and, using a hypothesised central
generator, the spiral arm structure will be modelled as well.  The
basic idea is that the central mass accretes a belt of matter which
develops instability and explodes feeding the roots of the arms.
Stars are formed by condension in the arms and move outwards as they
develop.  Thus a typical star is on a long outward orbit and the
rotation curve observed for stars in a spiral arm is formed of many
such similar orbits.  But this full picture is not necessary to
explain the observed rotation curves, since the tangential velocity
for all orbits has the same horizontal asymptote.
\fullref{sec:spiral_struc} is more specific about the size of the
central mass in a galaxy.  These vary from $10^{9}$ to $10^{14}$ solar
masses with the range $10^{9}$ to $10^{11}$ corresponding to so-called
``active galaxies'' and the range $10^{11}$ to $10^{14}$ to full-size
spiral galaxies.  The central masses for the curves in
\fullref{fig:rotsmod} vary from $3\times10^{11}$ to $10^{14}$ solar
masses and it is useful to know that a mass of 1 in natural units is
$3\times10^{11}$ solar masses.\index{galaxy!central mass}

\sh{Postscript}

As remarked earlier, the effect described in this chapter is
independent of mass.  However for rotating bodies of small mass the
effect is unobservably small.  For example the sun has $K\approx$ 3km,
assuming $K=M$, and $\omega_{\rm Sun}=2\pi/25$ days.  Thus the
asymptotic tangential velocity $2A=2K\omega_{\rm Sun}$ is 6km per 4
days or .06 km per hour.
\index{rotation curve|)}

\chapter{Quasars}\label{sec:quasars}
\index{quasar|(}

quasar\nl
kwe\i za:, kwe\i sa:,\nl
\emph{noun} Astronomy\nl
noun: quasar; plural noun: quasars

{\leftskip 0.5in

    a massive and extremely remote celestial object, emitting
    exceptionally large amounts of energy, which typically has a
    starlike image in a telescope. It has been suggested that quasars
    contain massive black holes and may represent a stage in the
    evolution of some galaxies.

}

Origin\nl
1960s: contraction of quasi-stellar.

\rightline{\emph{Google dictionary definition (October 2017)}}
\index{quasar!Google definition}

\begin{figure}[ht!]
\cl{\includegraphics[width=.4\hsize]{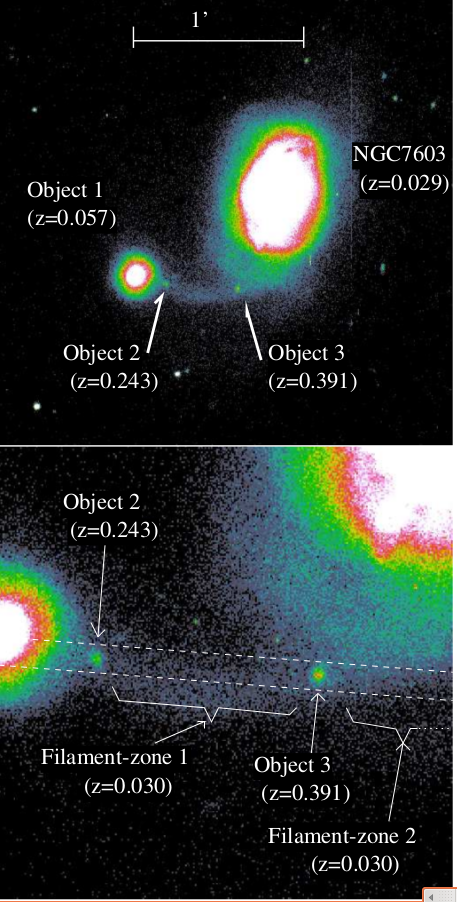}}
\caption{NGC 7603 and the surrounding field. R-filter, taken on the
  2.5 m Nordic Optical Telescope (La Palma, Spain). Reproduction of
  Figure 1 of \cite{LCG}}\index{galaxy!NGC7603}
\label{fig:NGC7603}
\end{figure}

\bigskip
Quasars were first observed in the 1960's.  Through a telescope they
appear to be stars but they exhibit strange features not shared by
ordinary stars.  They have spectra which often appear to be hugely
redshifted and they vary irregularly with time scales that range from
hours to months.  Early in the study of quasars a heated controversy
raged about these huge redshifts.\index{redshift!controversy} Are they
cosmological due to the expansion of the universe? or are they
gravitational due to the near presence of a massive object (eg a black
hole)?  The cosmological explanation implies that quasars with large
redshifts are extremely distant objects with truly phenomenal power
outputs which are very hard to explain.  By constrast the
gravitational explanation allows the possibility that they are modest
size objects, not too distant and with easily modelled power outputs.
As can be seen from the Google definition, the cosmological
explanation is the currently accepted one.  This is despite some
incontrovertible evidence in the form of observations of Halton Arp
\index{Arp, Halton}
and others \cite{Arp,Getal} that quasars are often closely associated
with galaxies with the redshift for the quasars significantly higher
than that for the associated galaxies, a striking example of which is
reproduced in \fullref{fig:NGC7603}.

This example contains two \ind{Arp quasars} (objects 2 and 3) strongly
associated with a galaxy (and companion) both of lower redshift.
Lopez Corredoira and Gutierrez \cite{LCG} report $z = 0.0295$ and $B =
14.04$ mag for the main galaxy, NGC 7603 and comment: ``A fact that
attracted attention is its proximity to NGC 7603B (Object 1
hereafter), a spiral galaxy with higher redshift $z =0.0569$, moreover
a filament can be observed connecting both galaxies. They also found
two objects superimposed on the filament with redshifts
$0.394\pm0.002$ and $0.245\pm 0.002$ for the objects closest to and
farthest from NGC 7603, Objects 3 and 2, respectively.
$B$--magnitudes corrected for extinction (due to the filament) are
respectively $21.1\pm1.1$ and $22.1\pm 1.1$.''

It is commonsense that the alignments seen in \fullref{fig:NGC7603}
are not due to chance and there are many similar such in Arp and
other's observations \cite{Arp,Getal}.  Objects 2 and 3 have
cosmological redshift around $z=0.030$ (for the filament) and the
remainder must be instrinsic (presumably gravitational).  As often
happens when a consensus view is challenged by direct evidence, the
evidence is ignored and the challenger discredited.  Arp was sidelined
by the mainstream cosmological community and denied observation time
on the big telescopes.  If this book can serve to rehabilitate Arp's
reputation (unfortunately posthumous) it will have been worth writing.

The purpose of this chapter is to explain how the same \id\ phenomenon
used in the last chapter to model rotation curves can be used to
restore a sensible explanation for these observations and to establish
a simple model for quasars with modest power output that explains all
the observations.

\sh{Angular momentum and \id}\label{subsec:locking_effect}

As explained in the introduction, the strongest argument supporting
the current consensus view (that redshift in quasars is cosmological)
comes from a consideration of angular momentum.  Assume that a quasar
contains a large central mass (presumed to be a black hole) and that
its energy production is due to \ind{accretion} from the surrounding
medium.  Particles fall into the gravitational well of the central
mass and the gravitational energy is released by interaction between
different infalling particles.  Now given a small but very heavy
object, a particle approaching with a small tangential velocity will
have its tangential velocity magnified by conservation of angular
momentum and there will be a radius of closest approach.  It is very
unlikely to actually fall into the central gravitational well, and the
same thing happens for the full flow of infalling matter from the
surrounding medium, which will typically have a nonzero angular
momentum around the black hole.  (See the Michel quote in
\fullref{sec:intro}.)  This gives an obstruction to accretion, which
was found not long after quasars were discovered, and led to the
subject being dominated by the theory of accretion discs.

But now assume that the central mass is rotating and that the
infalling particle is in the equatorial plane.  Take \id\ effects into
consideration.  From \eqref{eq:gensol} the angular momentum per unit
mass of the particle can be read as $vr=\int2\omega r\,dr +
\text{const}$.  This is the \emph{apparent} angular momentum (as
calculated by a distant viewer).  To find the \emph{true} angular
momentum (ie as measured in the local inertial frame) 
replace $\mathbf{v}$ by $\mathbf{v}_{\rm inert}$, which entails
subtracting $\omega r$ from $v$, so the true angular momentum (per
unit mass) is
\begin{equation}\label{eq:ang-mom}
\fboxrule1pt\fboxsep5pt\fbox{$\displaystyle vr-\omega r^2=\int2\omega
  r\,dr - \omega r^2 + \text{const}$}
\end{equation}
By suitable choice of the integration constant, there are solutions
with low angular velocity (either true or apparent) for $r$ small and
significant angular velocity for larger $r$ and it follows that the
effect of the inertial drag is that \emph{the rotating body can absorb
  angular momentum}.  And notice that this holds for almost any
dynamical theory, in particular general relativity.  (It also works
with almost any nontrivial function $\omega$.)

Now if angular momentum can be nullified by central rotation, then it
does not force the existence of an accretion disc and a simple
\ss\ model for accretion can be used.  Here is another description
of the effect being used here which gives further information.  The
formula for $\omega$ (equation \ref{eq:nett}) implies that IFs near
the origin rotate at roughly the same rate, in other words all fit
with a plate-like rotation.  If the speed of this rotation is the same
as the effective rotation of the infalling matter then the latter
rotation will be ``rotational'' (due to the rotation of the IF) and
not ``inertial''.  Thus conservation of angular momentum (which acts
only on inertial velocity) will not change it and the inflow will be
radial in the local inertial frame.  Moreover there is a feedback
effect working in favour of this.  If the incoming matter has excess
angular momentum, then it will tend to contribute to the central
rotation which therefore changes to increase the inertial drag effect
until the two balance again.  Conversely, if there is a shortfall, the
black hole will slow down.  In other words, once locked on the ambient
conditions that allow the black hole to accrete, there is a mechanism
for maintaining that state.
\index{inertial drag!locking effect}

At this point recall that given a black hole with angular velocity
$\omega$ and angular momentum $\Omega$ then $\Omega = M \omega r_{\rm
  eff}^2$ (\ref{eq:AmAv}) where $r_{\rm eff}$ is the effective
radius of the black hole, assumed to be small but not zero.  Thus a
very small change in angular momentum corresponds to a reasonable
change in the angular velocity.  This makes the locking effect
described above more responsive and effective.

If the incoming matter has angular momentum about a different axis
than the rotation axis for the black hole, then a similar feedback
effect will cause the black hole rotation axis to change into
alignment with axis for the incoming matter.

The conclusion is that the angular momentum obstruction for accretion
can effectively be ignored and a spherically symmetric accretion model
used.  \index{accretion!spherical}A suitable model based on the
Schwarzschild metric is studied in joint work with Robert MacKay and
Rosemberg Toala Enriques \cite{BHQR}.  In this model gravitational
redshift can take arbitrary values\index{redshift!gravitational}
\begin{equation}\label{eq:red-prev}
z = 1.27\times 10^7\, \K^{-1}\, n^{-1}\, T^{1.5}\, [1/(2X)]
\end{equation}
where $\K$ is the black hole mass in solar masses, $n$ is density of
the ambient gas$/$plasma in number of particles per cubic metre
(assumed to be Hydrogen atoms or protons) and
$T$ is temperature in degrees Kelvin.  $X$ is an absorption factor
which can be taken to be $1/2$ (ie ingore the factor in square
brackets).  The full technical details of the model (hereafter called
the ``three-author model'') are given in \fullref{app:3author}, where
equation (\ref{eq:red-prev}) is proved.  Notice one important point
about this equation.  The mass of the \BH\ $P$ appear inverted so that
(other parameters being equal) \emph{redshift decreases with
  \BH\ mass}.  This was observed directly by Arp and caused him to
invent some fantasy physics to explain it because he was not aware of
the considerations discussed here.
\index{redshift!decrease with mass}

Also in the appendix are many worked examples including NGC7603 and
associated objects.  One particular quasar is worth mentioning here
because of the (in fact false, as will be seen) importance that it has
for the Milky Way, namely Sagittarius \astar.\index{Sgr A*}
\index{Sagittarius A*|see{Sgr A*}}This quasar is regarded
as problematic by the mainstream quasar community because its level of
radiation is 8 orders of magnitude below the Eddington limit.  It is
suggested here that this is due to a very high redshift, $z=10^4$,
which causes the power to be attenuated by the square of this, namely
$10^8$.  Full details and supporting evidence from the luminosity
graph can be found at the end of \fullref{sec:data}.

This chapter finishes with an outline of this three-author model and
discussion of evidence and previous work on quasars.  Note that this
chapter and the related appendix use MKS units and not the natural
units used in other chapters.

\sh{Outline of the three-author model}
\index{quasar!three-author model}

As remarked earlier this model for black hole radiation (aka quasar
radiation) is \ss, fully relativistic and based on the Schwarzschild
metric.  There are fully relativistic Schwarzschild black hole models
to be found in the literature, for example the models of Flammang,
Thorne and Zytkow \cite{FTZ} quoted by Meier \cite[page 490]{meier}.
But the significance of these models, and in particular their
redshift, has been ignored, presumably because of the angular momentum
obstruction discussed above.  Thus the excellent fit with observations
that is found has been overlooked.

In the three-author model, black holes radiate by converting the
gravitational energy of incoming matter into radiation and, since only
a fraction of the available energy is radiated back out, they accrete
mass and grow over time.  There will be a good deal more to say about
this growth in later parts of the book.  It is highly suggestive of a
life-form.

The basic set-up considered is a \BH\ floating in a gas of Hydrogen
atoms (the \emph{medium}), which might be partially ionised (ie form a
plasma), with the radiation coming from accretion energy.  Matter
falls into the \BH\ and is accelerated.  Interaction of particles near
the \BH\ changes the ``kinetic energy'' (KE) of the incoming particles
into thermal energy of the medium and increases the degree of
ionisation.  The thermal energy is partially radiant and causes the
perceived \BH\ radiation.

Kinetic energy is not a relativistic concept as it depends on a
particular choice of inertial frame in which to measure it.  It is for
this reason that it  has been placed in inverted commas.  Nevertheless,
it is a very useful intuitive concept for understanding the process
being described here.

The following simple considerations suggest that most of the KE of the
infalling matter is converted into heat and available to be radiated
outwards.  A typical particle is very unlikely to have purely radial
velocity.  A small tangential velocity corresponds to a specific
angular momentum.  As the particle approaches the \BH, conservation of
angular momentum causes the tangential velocity to increase.  Thus the
KE increase due to gravitational acceleration goes largely into energy
of tangential motion.  Different particles are likely to have
different directions of tangential motion and the resulting mel\'ee of
particles all moving on roughly tangential orbits with varying
directions is the main vehicle for interchange of KE into heat and
hence radiation.  Very little energy remains in the radial motion, to
be absorbed by the \BH\ as particles finally fall into it.  Thus the
overall radial motion of particles is slow. In terms of the models of
\cite{FTZ}, the ``breeze solutions'' for radial flow \cite[Figure
  12.2, page 489]{meier} are being used.  Far away from the \BH, where
density is close to ambient density, and therefore low, this process
converts angular momentum into radial motion with little loss of
energy and serves to allow the plasma to settle into the inner
regions, where the density is higher and the particle interactions
generate heat and radiation.

\sh{Three important spheres}\label{sec:3spheres}

For simplicity of exposition now assume that the medium is a Hydrogen
plasma and the heavy particles are therefore protons.  This is true in
the higher temperature parts of the model, for example once the
Eddington sphere is reached, see below.  But there is no material
difference if the medium is in fact a partially ionised Hydrogen gas.

Observations of quasars often show the presence of other atomic
material in the radiation zone so that this simplifying assumption may
need revision at a later stage.

There are three important spheres.  The outermost sphere is the
\emph{\ind{Bondi sphere}} of radius $B=2GMm_H/3kT$ defined by equating
the average velocity of protons in the medium with the escape velocity
at radius $B$.  Here $M$ is the black hole mass, $G$ is the
gravitational constant, $k$ is Boltzmann's constant, $T$ is
temperature and $m_H$ is the mass of a proton.

The significance of the Bondi sphere is that protons in the medium are
trapped (on average) inside this sphere because they have KE too small
to escape the gravitational field of the \BH.  The mass of matter per
unit time trapped in this way is called the \emph{accretion rate} $A$
and can be calculated as
\begin{equation}
A = 2 B^2 n \sqrt{2\pi kTm_H} 
\label{eq:A-prev}
\end{equation}
where $n$ is the density of the medium (number of protons per unit
volume).\index{accretion!rate}

Details for these calculations are given in \fullref{sec:Bondi}. 

Proceeding inwards, the next important sphere is the \emph{\ind{Eddington
  sphere}} of radius $R$ which is defined by equating outward radiation
pressure on the protons in the medium with inward gravitational
attraction from the \BH.  More precisely, the outward radiation
pressure acts on the electrons in the medium which in turn pull the
protons by electrical forces.  This is the same consideration as used
to define the Eddington limit for stars and this is why the same name
has been used.  At the Eddington sphere the gravitational pull on an
incoming proton is balanced by the outwards radiation pressure
(mediated by electrons) and, assuming the radiation pressure is just a
little bigger, the acceleration of the incoming proton is replaced by
deceleration and the KE of infall is absorbed by the medium and
available to feed the radiation.  It is a definite hypothesis that
there is an Eddington sphere, but the final model that is constructed
using this hypothesis does fit facts pretty well, and this justifies
it.

It is helpful to think of the Eddington sphere as a transition barrier
akin to the photosphere of a star.  Indeed the Eddington radius $R$ is
also the radius at which photons get trapped in the medium and for
this reason is also known as the trapping radius.  This can be seen by
thinking of the forces that define it the other way round.  The
incoming matter flow exerts a force on the outward radiation and when
these two are in balance, the outward radiation is stopped and photons
are trapped.

Thus at the Eddington sphere two things are happening: the
infalling protons are stopped and their KE released into the general
pool of thermal energy and the outward flow of radiation is also
stopped.  Thus radiation from the \BH\ is generated by activity in the
close neighbourhood of the Eddington sphere and this is the place
where redshift of the outward radiation due to the gravitational pull
of the \BH\ arises.

The region outside the Eddington sphere is optically thin whilst the
region inside is optically thick.  The radiation that is emitted comes
from a narrow band near the Eddington sphere and which is all at
roughly the same distance from the central black hole.  This allows
the radiation to exhibit a consistent redshift.
\index{optically thick region}
\index{optically thin region}

Precise formulae that determine the Eddington radius in terms of the
other parameters are given in \fullref{sec:KE}.

The final sphere is the familiar Schwarzschild sphere or event horizon
of radius $S = 2GM/c^2$ where $M$ is the \BH\ mass.  

The region between the Schwarzschild and Eddington spheres is called
the \emph{\ind{active region}} and the region between the Bondi sphere and
the Eddington sphere, the \emph{\ind{outer region}}.  A simplifying
assumption is made that nearly all the KE that powers the \BH\ is
released in the active region.  This means that any KE turned into
heat by particle interaction in the outer region is ignored.  This is
justified by the fact that this region has low density, close to the
ambient density, so that most particle interactions are between
particles sufficiently far apart to conserve kinetic energy.  It is
useful to think of this region as a ``settling region'' where angular
momentum is converted into radial motion, allowing the plasma to
settle towards the active region.  See also the discussion below
equation (\ref{eq:PreTrap}) and in \fullref{sec:conc}.

One other simplifying assumption is made: it is assumed that there is
no significant increase in temperature near the Bondi sphere due to
the \BH\ radiation, ie $T$ is the ambient temperature.

\sh{Previous work on quasars and gravitational redshift}

This chapter finishes with a review of the historical reasons for
abandoning the idea that quasars might have significant instrinsic
(gravitational) redshift and why they do not apply to the model.

The principal reason (angular momentum) has already been fully
explained.  There are four main further reasons:

\medskip\goodbreak
(1)\qua Redshift gradient (see the discussion in \cite{fifty-years} on
pages 3--4)

If redshift is due to a local mass affecting the region where
radiation is generated, then the gravitational gradient from approach
to the mass would spread out the redshift and result in very wide
emission lines.  This effect is called ``redshift gradient''.
\index{redshift!gradient}

In the model, although the energy production takes place throughout
the active region, the emitted radiation is generated only at (or
near) the \ind{Eddington sphere} which is all at the same distance from the
central mass and subject to the same redshift.  Thus the model has the
observed property that emission lines are moderately narrow.

\medskip\goodbreak
(2)\qua Forbidden lines (cf Greenstein--Schmidt \cite{G-S})

Many examples of \BH\ radiation show so-called \ind{forbidden lines}, which
can only be produced by gas or plasma at a fairly low density.  The
assumption that \emph{all} the radiation is produced by a low density
region leads to an implausibly large and heavy mass (see \cite[page 1,
  para 2]{G-S}). 

In the three-author model, the region directly adjacent to the
Eddington sphere is at roughly ambient density which is, in all
examples that are examined in the appendix, low enough to support
forbidden lines (more details on this will be given in
\fullref{sec:calc}).  A narrow shell of low density near the Eddington
sphere is excited by the radiation produced at the sphere and produces
radiation in turn.  It is here that forbidden transitions take place
and result in the observed forbidden lines.

\medskip\goodbreak (3)\qua Mass and variability problems (cf
Greenstein--Schmidt \cite{G-S}, Hoyle--Fowler \cite{H-F})
 
The mass problem is a rider on the forbidden line problem but also
applies to attempts at models for gravitational redshift without
significant redshift gradient.  As remarked above, assuming that all
the radiation is produced by a low density region leads to an
implausibly large and heavy mass.  The same thing happens if one tries
to produce a region with sufficient local gravitational field to
provide a base for the radiation production, without redshift
gradient, as for example in Hoyle and Fowler \cite{H-F}.  This problem
is compounded by the fact that quasars typically vary with time scales
from days to years.  For variability over a short timescale, a small
production region is needed (significantly smaller than the distance
that light travels in one period).

It is worth remarking in passing that this problem is unresolved by
the current assumption that all quasar redshift is cosmological.  This
implies that quasars are huge and very distant so that special (and
unnatural) mechanisms are invoked to explain variability.

In the three-author model, the size of the radiation producing region
is small enough.  The \BH\ sizes that fit observations are in the
range $10^3$ to $10^8$ solar masses.  For quasars with significant
intrinsic redshift, the radius of the Eddington sphere has the same
order of magnitude as the Schwarzschild radius, and for $10^8$ solar
masses this is $3 \times 10^{11}$ metres or $10^3$ light seconds or
about 20 light minutes.  Thus the natural mechanism for variability,
namely orbiting clouds or more solid bodies causing periodic changes
in observed luminosity, fits the facts perfectly.
\index{quasar!variability!natural explanation}

It is also worth observing here that there is a quite remarkable paper
of M\,R\,S \ind{Hawkins} \cite{H}, which proves an apparently paradoxical
result, namely that a certain sample of quasars exhibits redshift
without time dilation.  The paradox arises from the fact that redshift
and time dilation are identical in general relativity.  Indeed they
are identical in any theory based on space-time geometry.  What
Hawkins actually finds is a sample of quasars with varying redshift
for which the macroscopic variation in light intensity does not
correlate with the redshift.  The resolution of the paradox is that
the mechanism that produces the redshift and the mechanism which
causes the variability are not subject to the same gravitational
field.  This is precisely how the model works.  The redshift is caused
by the central \BH\ and the variability is caused by orbiting clouds
etc, much further out, and in a region of lower redshift.  For more
detail on the Hawkins paper and its meaning see \fullref{sec:hawk}.  Properly
understood, the paper proves conclusively that quasars typically have
intrinsic redshift.

\medskip\goodbreak
(4)\qua Statistical surveys

Stockton \cite{S} is widely cited as a proof that quasar redshift is
cosmological.  He takes a carefully selected sample of quasars and
searches for nearby galaxies within a small angular distance and at
close redshift.  Out of a chosen sample of 27 quasars, he finds a
total of 8 which have nearby galaxies with close redshifts.  He
assumes that all these quasars have significant intrinsic redshifts
and are therefore not actually near their associated galaxies.  He
then calculates the probability of one of these coincidences occurring
by chance at about 1/30, and concludes that the probability of this
number of coincidences all occurring by chance is about 1.5 in a
million.\index{Stockton survey}

The conclusion he draws is that all quasar redshift is cosmological.

The fallacy is obvious from this summary.  It may well be that many of
the quasars in the survey do not have significant intrinsic redshift
and therefore some of these coincidences are not chance events.  As
can be seen from equation (\ref{eq:red-prev}) the three-author model
allows the gravitational redshift of a quasar to vary from near zero
to as large as you please.  Roughly speaking, redshift is small
(orders of magnitude smaller than 1) if the mass is big or the medium
is dense and cold.  Conversely, with a small mass and a hot thin
medium, the redshift can be several orders of magnitude greater than
1.  There is a natural progression for a quasar, as it accretes mass
and grows heavier, to start with a very high gravitational redshift
and gradually evolve towards a very low one.  Without a sensible
population model for quasars, it is difficult to comment on the number
of coincidences that Stockton finds, but it is highly plausible that
heavy quasars (with low gravitational redshift and central masses of
say $10^7$ to $10^9$ solar masses) gravitate towards galactic clusters
and therefore have nearby galaxies at a similar cosmological redshift.
This would provide a natural framework for the Stockton survey within
the model.

Stockton does discuss the possibility that quasars may have both small
and large intrinsic redshifts (see \cite[page 753, right]{S}), but the
discussion is marred by assuming that the two classes must be
unrelated objects.  The three-author model has a natural progression
between the two classes.

There is a more modern survey by Tang and Zhang \cite{T-Z} which also
claims to prove that all quasar redshift is cosmological.  But
examining the paper carefully, what is actually proved is that some
particular models for quasar birth and subsequent movement are
incompatible with observations.  To comment properly on this paper a
good population model for quasars would again be needed.  But it is
worth briefly mentioning that at least one of their models (ejection
at $8\times 10^7$ m/s from active galaxies with a lifespan of $10^8$
years) does fit facts fairly well, see \cite[figure 1, page 5]{T-Z}.
The ejection velocity is implausibly large, but the lifespan could
easily be 50 times larger allowing for a plausible ejection velocity
of say $10^7$ m/s and a better fit with the data.

Finally, there is another interesting argument given by Wright
\cite{LAF} ``proving'' that quasar redshift is all cosmological from
details of the spectra.  This is the \ind{Lyman-alpha-forest} argument.  The
observations he cites give useful information about the outer region.
This, and the fallacy in the argument, will be discussed near the end
of \fullref{app:3author} in \fullref{sec:conc}.

The story continues in \fullref{app:3author} where full technical
details of the three-author model and the fit with data can be found.

But to finish this chapter here are some comments on quasar growth.
It has been seen that quasars grow by accretion and lose their
intrinsic redshift (as observed by Arp, but explained using
non-standard physics).  If, as Arp suggests, they are ejected from
mature galaxies, then there is a natural way to think of them as young
galaxies.  As they grow and gain mass, they will take on more and more
features of active galaxies and finally develop into mature spiral
galaxies (discussed in the next chapter).  As a highly speculative
example, the grouping of four objects (two galaxies and two quasars)
seen in \fullref{fig:NGC7603} could be a ``family'' group: two adults
and two children.  Indeed the quasar--galaxy spectrum has all the
appearances of forming the dominant lifeform for the universe.  This
topic is taken up again in sections \ref{sec:gen} and \ref{sec:lords}.
\index{quasar|)}\index{quasar-galaxy spectrum}

\np\thispagestyle{empty}

\chapter{Spiral structure}\label{sec:spiral_struc}

\index{galaxy!spiral structure|(}
The chapter combines ideas from the last two chapters to give a
complete description of the dynamics of spiral galaxies.

\sh{Introduction}\label{sec:Sp_intro}

\begin{figure}[ht!]
\cl{\includegraphics[width=.43\hsize]{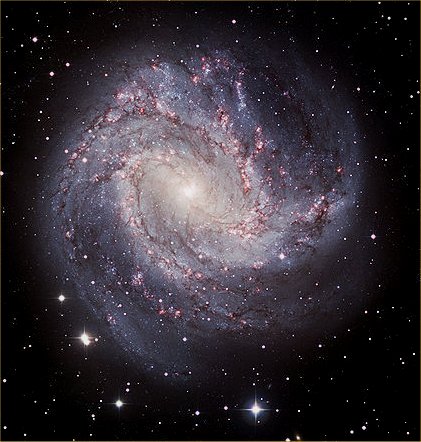}\qquad\includegraphics[width=.45\hsize]{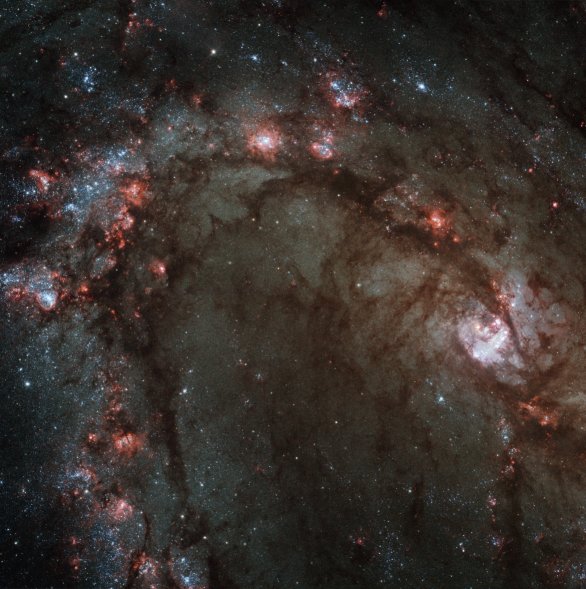}}
\caption{M83 Southern Pinwheel.  Left: image from European Southern
  Observatory \cite{ESO}.  Right: close-up from Hubble site \cite{Hubble}}\label{fig:M83}
\end{figure}

Spiral galaxies (Figures \ref{fig:M83}, \ref{fig:galaxies},
\ref{fig:more_gals}) are surely the most beautiful objects in the
universe and it comes as a shock to find that there is no proper
theory for their structure in current cosmology.

\begin{figure}[ht!]
\cl{\includegraphics[height =1.4in]{figs/rotation}\qquad\qquad\includegraphics[height =1.4in]{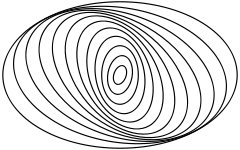}}
\caption{Left: the rotation curve for the galaxy NGC3198 reproduced
  from \cite{Ber} (sourced from Begeman \cite{Beg}).  Right: the
  standing wave theory, reproduced from
  wikipedia}\label{fig:rot-stand}
\end{figure}

\begin{figure}[ht!]
\cl{\includegraphics[height=1.62in]{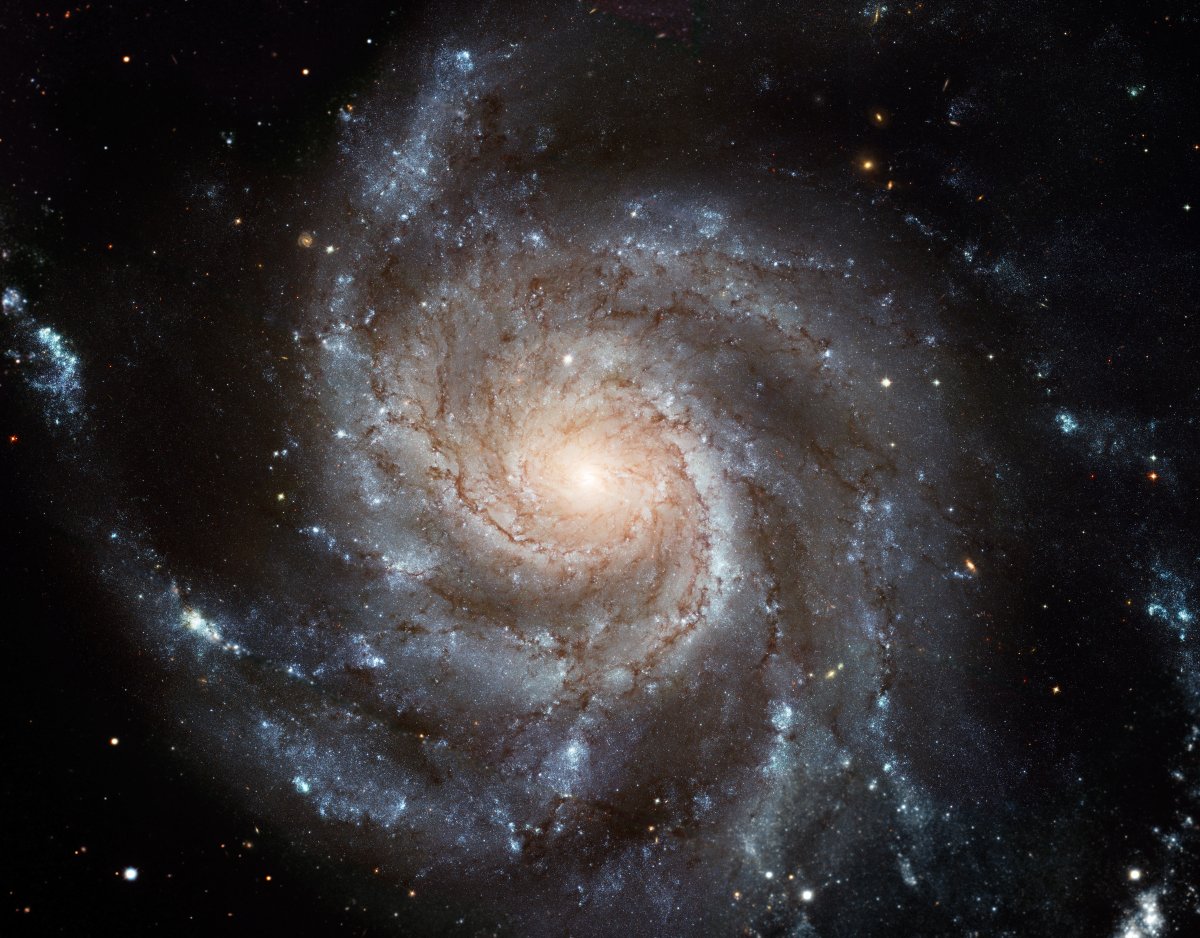}\qquad\includegraphics[height=1.62in]{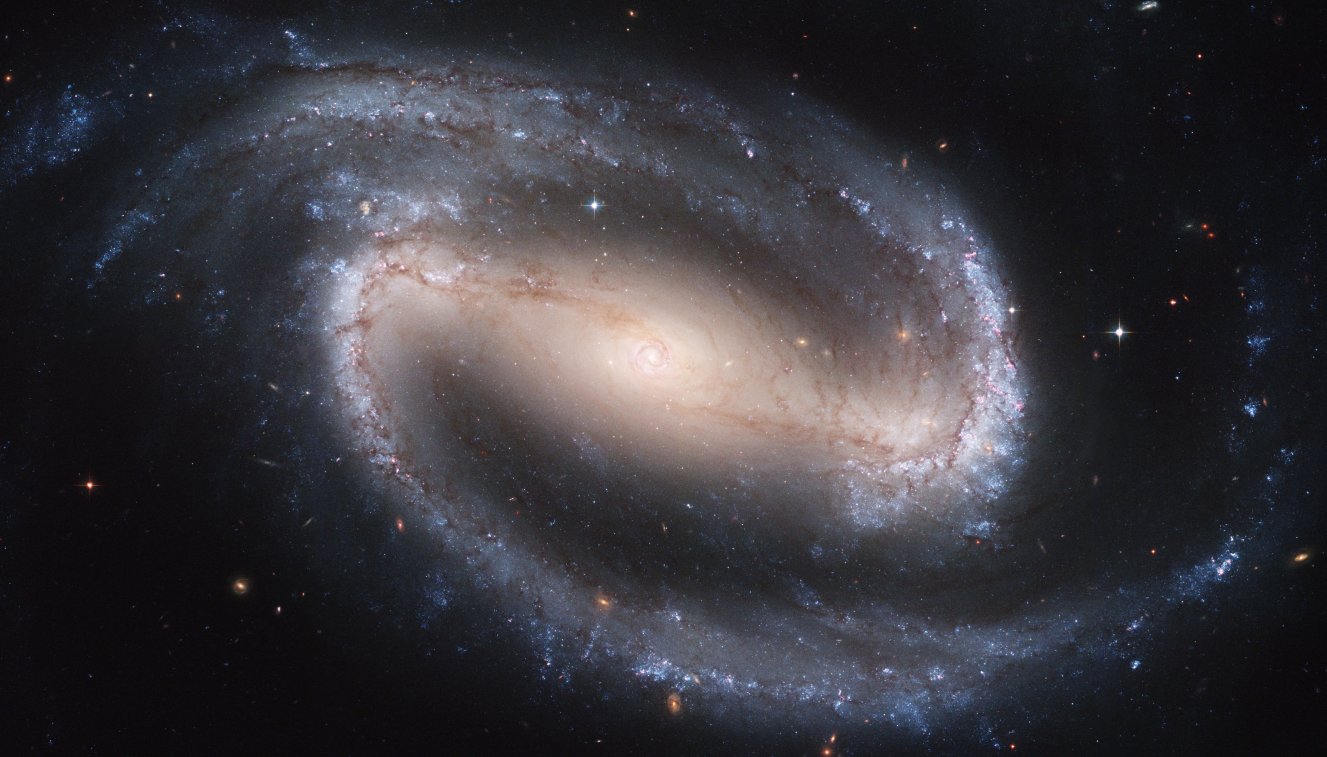}}
\caption{M101 (left) and NGC1300 (right): images from the Hubble site
\cite{Hubble}} \label{fig:galaxies}
\end{figure}

The main problem arises from the assumption that stars move on roughly
circular orbits.  In order for a spiral structure to be maintained
stably over several revolutions, with all stars moving on circular
orbits, it is necessary for tangential velocity to be roughly
proportional to distance from the centre which means that the rotation
curve is far from the one observed, \fullref{fig:rot-stand} (left).
This problem is known as the ``\ind{winding dilemma}''.  In order to
solve this problem conventional cosmology proposes that the spiral
arms are not real but virtual.  It proposes that they are in fact
``standing waves'' or ``density waves'', \fullref{fig:rot-stand}
(right).  Although this theory gives plausible spirals, the nature of
the arms in real galaxies, \fullref{fig:galaxies} (right) or
\fullref{fig:more_gals} (left) does not fit it at all.  Real arms are
composed of a spiral curve of intense star producing regions and
associated high luminosity short-life stars, see for example the close
up of M83 from the Hubble site \fullref{fig:M83} right.  There is no
trace of the orbits that are supposed to form the density wave outside
the actual spiral arm. 
\index{galaxy!density wave theory}% 
\index{galaxy!standing wave theory} %
To be a little fairer to the \ind{standing wave theory}, there is a
rider to the theory which suggests that a shock-wave effect causes
short-life stars to appear as the standing wave moves.  Indeed it is
clear from any galactic picture that the main luminosity of typical
spiral arms comes from high luminosity short-life stars, but the
short-life stars produced by a shock wave would last long enough to
blur the arms and the pictures are quite clear: no such blurring
occurs.  Moreover, this model begs the question of where the
continuous supply of pre-stellar material comes from to support this
creation process.

In this chapter a quite different solution to this problem is
proposed.  The idea is, instead of assuming that stars move on roughly
circular orbits, to assume that they move outwards along the arms as
they rotate around the centre.  Thus the familiar spiral structure is
like the visible spiral structure in a \ind{Catherine wheel}, the arms being
maintained by stars moving along them, and there is no need for any
special pleading to explain the observed structure.  Moreover, the
motion can be modelled and the stable spiral structure demonstrated .
The model has one crucial feature in common with the standing wave
theory in its shock-wave version: the visible spiral structure does
consist mainly of short-life stars and star-producing regions and this
common feature is crucial for accurate modelling because it allows for
elapsed time along the arms of about $10^8$ years which fits the
models constructed here.  More details on this are given in
\fullref{sec:math}.  The problem of supply of pre-stellar material is
solved in the model by continuous replenishment from the centre of the
galaxy.

Two assumptions are needed, the first of which was anticipated in
\fullref{sec:rot_curve} to explain the observed rotation curve, namely
that the centre of a normal spiral galaxy such as the Milky Way
contains a hypermassive black hole, of mass $10^{11}$ solar masses or
more.  The second assumption is that this black hole is ringed by an
accretion torus of a very precise type, which is called the
\emph{\ind{generator}} or \emph{\ind{belt}} and which is responsible
for generating the streams of material which feed and maintain the
spiral arms.  The belt is an example of the accretion structures
hypothesised for the nuclei of ``active'' galaxies (for which central
super-massive black holes, of $10^{8}$ to $10^{10}$ solar masses, have
been directly observed) and used to explain their observed radiation.
``Active'' has been placed in quotation marks because of one of the
main theses of this book, that all galaxies are active: the activity
of a spiral galaxy is responsible for its spiral structure; indeed
there is no distinction between active galaxies and ``normal'' spiral
galaxies.  The real difference is that the central black hole in a
normal galaxy is masked from view by the matter which is trapped in
accretion structures near it, the most prominent of which is the
central bulge.  The spectrum of black hole based objects will be
discussed further in \fullref{sec:obs}.\index{galaxy!bulge}

The assumption of a hypermassive central black hole in a spiral galaxy
directly contradicts current beliefs of the nature of Sagittarius
\astar\ and this problem together with other observational matters
will be dealt with in the next chapter (\fullref{sec:astar}).  Very
briefly, Sgr\astar\ \index{Sgr A*} and the stars in close orbit around
it form an old globular cluster near the end of its life with most of
the matter condensed into the central black hole.  It is not at the
centre of the galaxy but merely roughly on line to the centre and it
is about half-way from the sun to the real galactic centre which is
invisible to us.

As remarked above, in the new model for galactic dynamics proposed
here, young stars in a galaxy are moving outwards as well as around
the centre.  This general outward movement has not been observed,
although there are some old observations of Oort, Kerr and Westerhout
\cite{OKW} which show outward movement in gas clouds but which are
generally misinterpreted (see \fullref{sec:21}).  Indeed, early
observations of Lindblad, using Shapley's maps of globular cluster,
suggested that stars in the neighbourhood of the sun move on circular orbits (see
\cite[page 16]{BM}) and this has created an id\'ee fixe that all stars
in galaxies move on roughly circular orbits with any contrary
observations explained away on an ad hoc basis.  In the model proposed
here, motion of stars is far from Keplerian, being strongly controlled
by inertial drag effects from the (rotating) centre.  The result is
that the outward progress takes a very long time---commensurate with
the lifetime of a star---and hence the outward velocity, far out from
the centre where the Sun lies, is rather smaller than (about one tenth
of) the observed rotational velocity.  Thus the new model is
consistent with the Lindblad observations.  For more detail here, see
the analytic models constructed in \fullref{sec:math}.

The general picture which emerges is of a structure stable over an
extremely long timescale (at least $10^{12}$ years) with stars born
and aging on their outward journey from the centre and returning to
the centre to be recycled with new matter to form new solar systems.
The tentative suggestion is that galaxies have a natural lifetime of
\index{galaxy!natural lifetime}
perhaps $10^{16}$ years with the universe considerably older than
this.  The consequences of these suggestions for cosmology as a whole
will be discussed in the next chapter; here note that the theory of
galactic dynamics presented in this chapter does not depend on this
timescale.  Indeed it could at a pinch be consistent with the current
standard model for the universe as a whole starting with the big bang.
But the author's opinion is that the \ind{big bang theory} is a serious
mistake.  For more detail here see \fullref{sec:bb} .

\index{galaxy!NCG1300}
\index{galaxy!M101}

\sh{The \ind{generator}}\label{sec:gen}

The full dynamics of spiral galaxies will be developed in the next and
following sections, but first here is an outline of the proposed
generator for the spiral arms.  The story is part of the main story of
the book, namely the quasar--galaxy spectrum, which will be taken up
again in \fullref{sec:spec} and following sections.  The keys to
understanding the generator are the familiar ones: angular momentum
and inertial drag.  Previous chapters have covered the start of the
story: \id\ effects allow a black hole (aka quasar) to cancel out the
angular momentum obstruction to accretion and feed on the surrounding
medium and hence grow in size.  As it grows the mass increases and
its intrinsic (aka gravitational) redshift decreases.  For a very
small quasar such as Sgr\astar\ of mass $10^{6.6}$ solar masses, the
intrinsic redshift can be very large (in this case a figure of
$z=10^4$ is indicated by observations), but for a larger quasar (of
size say $10^8$ solar masses) $z=0.05$ is more typical, but there is a
huge variation, see the tables at the end of \fullref{sec:data}.

\begin{figure}[ht!]
\cl{\includegraphics[height=0.3\hsize]{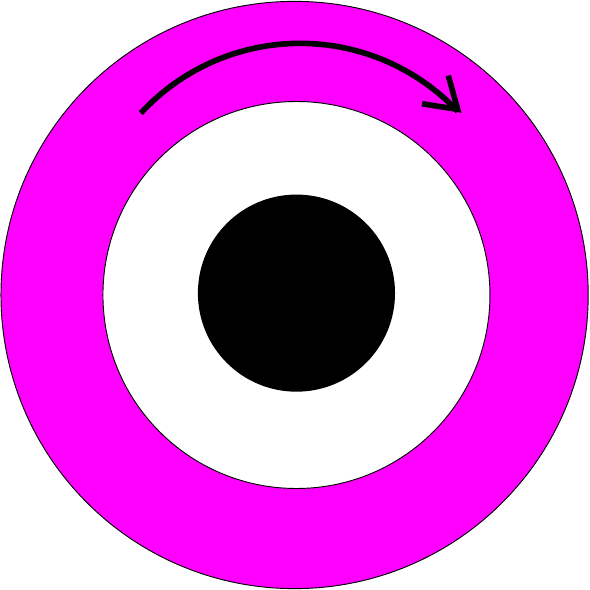}\qquad\includegraphics[height=.3\hsize]{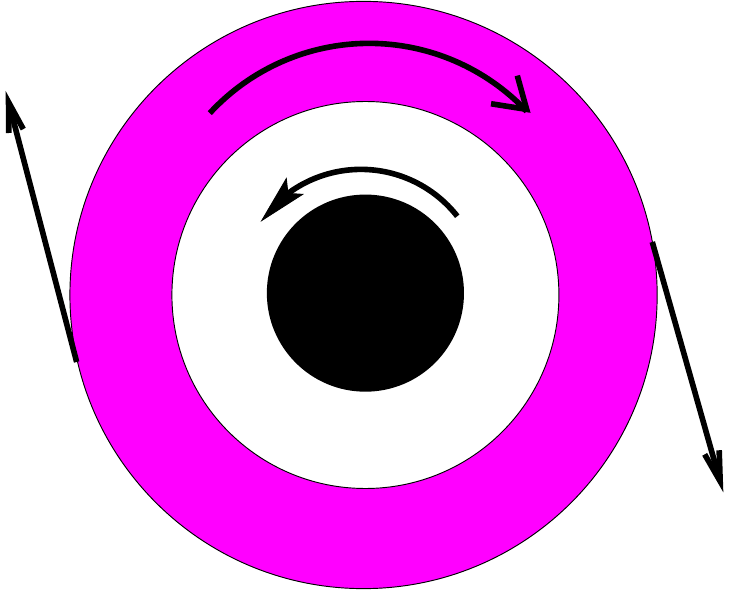}}
\caption{Left: the belt rotates clockwise.  Right: ejected matter causes the
  whole system to rotate anti-clockwise.}\label{fig:belt}
\end{figure}

As the mass grows and the instrinsic redshift decreases, the simple
spherical accretion model, described in \fullref{sec:quasars} and
\fullref{app:3author}, breaks down because the accretion rate is too
great for smooth accretion to take place.  The outer settling region
\index{quasar!spherical accretion!breakdown}%
develops instabilities; there is evidence for this starting in the so
called ``Lyman-alpha forest'' (see the end of \fullref{sec:data}).
The flow accumulates near the Eddington sphere choking the inflow.  A
rotating toroidal accretion structure (the belt), similar to the
conventional theory, forms \fullref{fig:belt} (left).

Notice that as the material in orbit around the black hole grows in
mass and extent, it increasingly masks the central black hole and the
virial theorem typically used to estimate this mass becomes less
useful.  This point in the spectrum is where ``active'' galaxies (of
mass in the range $10^8$--$10^{10}$ solar masses) start.
\index{black hole!masking}

The quasar starts to produce explosive outflow (jets) and to morph
into an active galaxy.  The angular momentum locking effect, described
in \fullref{subsec:locking_effect}, is now no
longer stable because the jets carry away angular momentum and cause
the whole system to rotate in the \emph{opposite direction} to the
rotation of the belt.  So the central black hole is rotating to the
left (say) and a surrounding belt is rotating to the right
\fullref{fig:belt} (right).  (Note that throughout this chapter
anticlockwise or positive (positive value of $\thetadot$) is used in
all illustrations for the main rotation of the central body; the belt
has negative (clockwise) rotation.)  This implies that the angular
momentum in the inertial frames is augmented by the inertial drag
effects described earlier and the effective energy in the belt
similarly augmented.  Thus energy is being fed into the belt structure
directly from the black hole itself.  There are also two other sources
of energy for the belt: accretion energy as for any quasar and energy
(heat) caused by interference due to \id\ between layers: a sort of
``friction'' effect.  With all this energy going into the belt, it
becomes extremely hot and a plasma of quarks forms nearest the centre,
condensing into a normal plasma of ionised H and He nuclei, with a
trace of Li, further out.  Conditions here are similar
\index{belt!energy build-up}\index{belt!big bang conditions}%
to those hypothesised to have occurred just after the big bang and the
resulting mix of elements is the same.  The energy results in
explosions causing the jets mentioned already.  As mass increases the
jets become massive and permanently established and manifest
themselves as the familiar spiral arms of the galaxy as explained
below.  \index{belt!jets}\index{generator!jets}

At the same time there is a build-up of matter trapped near the
central black hole, visible as the familiar bulge, which totally masks
the black hole, and the fiction that the central black hole of the
Milky Way could be only $10^{6.6}$ solar masses is not obviously wrong
(though it is completely incompatible with the dynamics presented in
this chapter).\index{black hole!masking}

\begin{figure}[ht!]
\cl{\includegraphics[width=0.5\hsize]{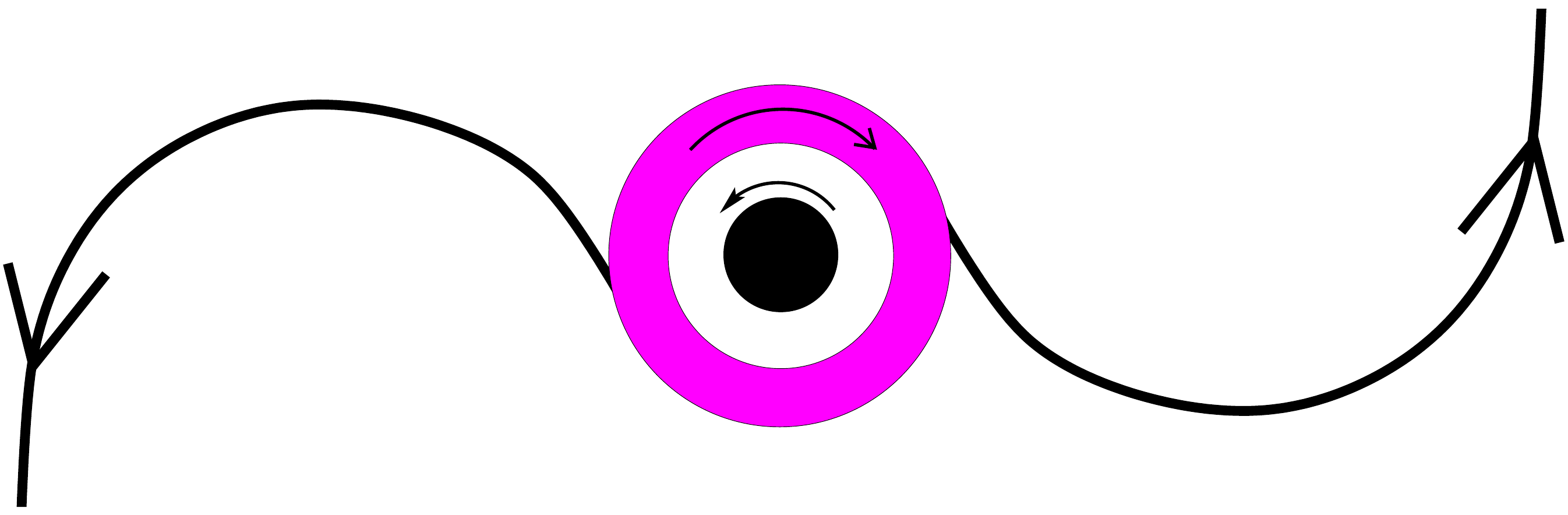}}
\caption{Inertial drag carries ejected matter anticlockwise and a
  balance is reached.  Arms form.}\label{fig:arms}
\end{figure}

Once the central black hole starts rotating, the \id\ effects
calculated in Sections \ref{sec:ideffects} and \ref{sec:dyn} come into
play and, further out, matter ejected from the centre starts to rotate
{\em with} the hole and against the rotation of the belt.  Matter is
lost from the outer regions and, if ejected from the centre fairly
slowly so that the \id\ effect dominates, carries away angular
momentum of the opposite sign, \fullref{fig:arms}.
\index{galaxy!stable rotation}
\index{galaxy!stability of rotation velocity}
There is a stable situation in which the loss of angular momentum in
both directions is in balance: highly energetic particles ejected from
the belt are not strongly affected by \id\ effects and carry away
clockwise angular momentum; less energetic particles are affected and
carry away anticlockwise angular momentum.  This balancing effect is
why there is strong stability in the limiting tangential velocity, in
other words why rotation velocity is roughly constant over all
galaxies observed, \fullref{fig:rotsSR}.  Note that \fullref{fig:arms}
shows orbits {\em not\/} arms.  It should be compared with
\fullref{fig:loose} left.  Inertial drag causes the roots of the arms
to appear to precess clockwise and the snapshot of orbits that is seen
has the familiar spiral form (as demonstrated in the Mathematica drawn
figures in \fullref{sec:math}).

Energy is lost from the black hole because of the matter ejected from
the belt, but energy is recovered by matter falling into the active
region near the black hole so that the whole structure is stable over
an immense timescale.  In the next chapter compatibility of this model
with the timescale of the big bang is discussed.  This global loss of
energy implies that there is no further growth in general, indeed
there is now steady loss due to radiation energy and matter lost to
the system, to balance accretion.

The spiral arms form as follows.  The explosions from the belt
mentioned above are the mechanism which feeds the spiral arms.  These
do not occur in random places: most normal galaxies have a pronounced
bilateral symmetry with two main opposing arms (eg Figures
\ref{fig:M83}, \ref{fig:galaxies} and \ref{fig:more_gals}).  There is
no intrinsic reason for this to happen, but it is a stable situation.
Once two arms have formed, then the gravitational pull of these arms
will form bulges at the roots of the arms and encourage explosions
there to feed the arms.  The bilateral symmetry arises because the
bulges are tidal bulges, caused by the pull of the nearby spriral
arms, and tidal bulges always have bilateral symmetry.  This tendency
to bilateral structure is weak and looking at a gallery of galaxies
many examples where it fails to form or where other weak arms have
formed as well as the two main arms can be found.
\index{galaxy!bilateral symmetry}

Notice that ejection from the belt is generally in the direction of
the belt rotation, which is opposite to the direction of the black
hole rotation and this as will be explained later is why the roots of
the arms generally have a noticeable offset (see the galaxy examples
referred to above).  To be precise, there is a constant $C$ in the
model which sets the tangential velocity at the root of the arms, and
with this set negative, the arms are offset, cf \fullref{sec:math}.
\index{galaxy!roots offset}

How the structure fits with detailed observations of our galaxy, the
Milky Way, and other nearby galaxies is explained in the next chapter,
and more detail on the composition of the arms and of the
corresponding stellar population distribution is given.  Here it only
needs to be noted that the arms are fed by a stream of gas$/$plasma
(comprising H, He and a trace of Li) ejected from the belt which
condenses into stars which then form the visible spiral arms.

\sh{The full dynamic}
\index{galaxy!full dynamic}

The construction of the model that will explain spiral structure
starts here.  The first step is to extend the analysis of
\fullref{sec:rot_curve} to obtain a full model for orbits in the
galactic plane and not just a formula for the rotation curve of such
an orbit.  The analysis applies to any rotating mass, but the results
are only significant for truly enormous masses such as the
hypothesised central mass in a galaxy.

Equation \ref{eq:v} (in \fullref{sec:rot_curve}) gave a formula for
the tangential velocity in an orbit.  What is needed is a formula for
the radial velocity (again in terms of $r$) and these two will
describe the full dynamic in the equatorial plane, which can then be
used to plot orbits.

\index{galaxy!radial forces}
Intuitively there are two radial ``forces'' on a particle: a
\ind{centripetal force} because of the attraction of the massive centre and
a \ind{centrifugal force} caused by rotation in excess of that due to
inertial drag.  Thus a formula for radial acceleration of the
following form is expected
\begin{equation}\label{eq:rdd}
\rdd = \frac{v_{\inert}^2}r - F(r)
\end{equation}
where $v_{\inert}=v-\omega r$ and $F(r)$ is the effective central
``force'' at radius $r$, per unit mass.  The same notation as in
\fullref{sec:rot_curve} is used here and in particular
$\omega=\omega(r)$ is the \id\ at radius $r$.

This will be proved in a similar way to the proof of conservation of
angular momentum given in \fullref{sec:rot_curve}, using a geometrical
argument which is valid in the \id\ metric.

%\sh{Radial acceleration}

\begin{figure}[ht!]
\labellist
\small\hair 2pt
 \pinlabel $A$ [r] <0pt,0pt> at 19 233
 \pinlabel $P'$ [br] <0pt,0pt> at 36 275
 \pinlabel $B$ [l] <0pt,0pt> at 105 275
 \pinlabel $P$ [l] <0pt,0pt> at 104 202
 \pinlabel $O$ [t] <0pt,0pt> at 35 0
 \pinlabel $u'$ [b] <0pt,0pt> at 67 275
 \pinlabel $u$ [t] <0pt,-1pt> at 56 219
 \pinlabel $r'$ [r] <1pt,0pt> at 33 121
 \pinlabel $r$ [l] <1pt,0pt> at 71 111
 \pinlabel $b$ [l] <0pt,0pt> at 104 244
 \pinlabel $a$ [r] <1pt,2pt> at 25 252
\pinlabel {$\bf u$} [lb] <0pt,0pt> at 65 242
\endlabellist
\centering
\includegraphics[width=1.5in]{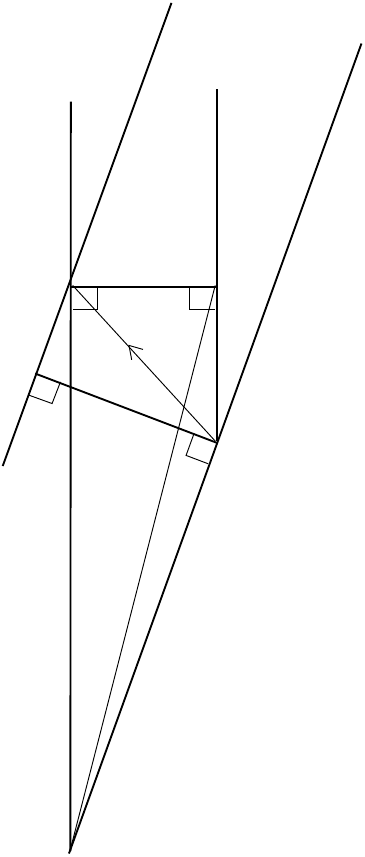}
\caption{Diagram for radial acceleration}
\label{fig:CAMa}
\end{figure}

The idea is the same, namely to replace the central force by a series
of central impulses at equally spaced small intervals $\delta t$ of
time and then take the limit as $\delta t\to0$.  Start by setting
$\omega$ equal to zero.  Consider \fullref{fig:CAMa} (a copy of
\fullref{fig:CAM} with extra labels).  Recall that the motion of a
particle (of small unit mass) in the equatorial plane is being
considered and that, at a particular time, it is at $P$ and has just
received a central impulse resulting in velocity ${\bf u}$.  $a=
|AP'|$ is the outward velocity (ie $\rdot$) at $P$ (after the central
impulse) and $b=|PB|$ is the outward velocity at $P'$ before the
central impulse.  The effect of the central impulse is to subtract
$F(r')\,\delta t$.  Therefore if $a'$ denotes the value of $\rdot$ at
$P'$ then $a'=b-F(r')\,\delta t$ or
\begin{equation}\label{eq:deltaa}
a-b=-F(r')\,\delta t-\delta a
\end{equation}
where $a'=a+\delta a$.  But by Pythagoras $a^2 + u^2 = ||{\bf u}||^2 =
b^2+(u')^2$ and hence
$$
(a-b)(a+b)=\delta u\,(u+u')
$$
where $\delta u = u'-u$ as before. Then substituting for $a-b$ from
\eqref{eq:deltaa} gives
\begin{equation}\label{eq:a+b}
(a+b)(-F(r')\,\delta t-\delta a)=(u+u')\,\delta u.
\end{equation}
But recall from equation (\ref{eq:CAM}) that $ur=u'r'$ which implies
\begin{equation}\label{eq:udr+rdubis}
u\,\delta r+r\,\delta u =0
\end{equation}
to first order where $\delta r = r'-r$ as before.  Now multiply 
\eqref{eq:a+b} by $r$, reverse sign and substitute for $r\,\delta u$
from \eqref{eq:udr+rdubis} to obtain:
\begin{equation}\label{eq:nearly}
r\,(a+b)(\delta a+F(r')\,\delta t)=u(u+u')\,\delta r
\end{equation}
But to first order $a+b = 2 \delta r/\delta t$ (recall that $a$ is
$\rdot$), $F(r')=F(r)$ and $u+u'=2u$.  Thus \eqref{eq:nearly}
simplifies to
$$
\frac{\delta a}{\delta t} +F(r) =  \frac{u^2}r.
$$
In the limit $\lfrac{\delta a}{\delta t}$ becomes
$da/dt=d\!\rdot/dt=\rdd$, which proves
\begin{equation}\label{eq:u-version}
 \rdd =  \frac{u^2}r-F(r).
\end{equation}
Now reinstate $\omega$.  Exactly as in the previous proof, by locality
the proof just given makes sense in the inertial frame at $P$ in other
words rotating with angular velocity $\omega= \omega(P)$.  But $u =
v-\omega r = v_{\inert}$ and \eqref{eq:rdd} is proved.

\sh{Computing radial velocity}\label{sec:dyn}
\index{galaxy!radial velocity}

Now specialise to the case $\omega=\lfrac{A}{(r+K)}$ (equation
\ref{eq:nett}) which was the formula for \id\ coming from the Weak
Sciama Principle.  Here $A=K\omega_0$ and $K=kM$, where $M,\omega_0$
are the mass and angular velocity of the central mass, and $k$ is a
weighting constant which can be taken to be 1 for purposes of
exposition.  The following formula for $v$ (equation
\ref{eq:v}) was found:
\def\strutt{\vrule width 0pt height 20pt}
\begin{align}
v&=\frac1r\left(\int\frac{2Ar}{r+K}\,dr +
C\right)=\frac{2A}r\left(\int1-\frac{K}{r+K}\,dr\right) + \frac Cr\notag\\
 &= 2A -\frac {2AK}r \log\left(\frac rK +1\right) + \frac Cr\strutt\label{eq:v-recap}
\end{align}
where $C$ is a constant which can be read from the tangential velocity
for small $r$.  This implies:
\begin{equation}\label{eq:vinert}
v_{\inert}=2A -\frac {2AK}r \log\left(\frac rK +1\right) + \frac Cr\strutt - \frac{Ar}{K+r}
\end{equation}
Moreover for the purposes of investigation assume that $F(r)$ is the
inverse square law $F(r)=M/r^2$.  This is correct for the \id\
metric based on Minkowski space (with Newtonian physics to first
order) and is a good approximation for Schwarzschild and
Schwarzschild--de Sitter provided $r$ is not small.  Thus:
$$\rdd = \frac{v_{\inert}^2}r - \frac M{r^2} = \frac1r\Bigg[2A -\frac {2AK}r \log\left(\frac rK +1\right) + \frac Cr\strutt - \frac{Ar}{K+r}\Bigg]^2-\frac M{r^2}$$
Multiplying by $\rdot$ and integrating wrt $t$ (using a computer
integration package) gives
\begin{align}\label{eq:energy}
\tfrac12 \rdot^2 = \int \rdd dr =& - \frac{C^2}{2r^2} + \frac{M - 2AC}r +
\frac{A^2K}{K + r} + A^2\log(K + r) \\
&+   \frac{2AK(C + 2Ar)\log(1 + r/K)- (2 AK\log(1 + r/K))^2}{r^2} + E\notag
\end{align}
where $E$ is another constant determined by the overall energy of the
orbit.  From this equation $\rdot$ can be read off (in terms of $r$).
Moreover since there is a formula for $v$, there is also a formula for
$\dot\theta = v/r$ (where notation has been changed to use the usual
polar coordinates $(r,\theta)$ in the equatorial plane instead of
$(r,\phi)$ as used in \fullref{sec:rot_curve}).  From this it is
possible express $\theta$ and $t$ in terms of $r$ as integrals.  These
integrals are not easy to express in terms of elementary functions but
Mathematica is happy to integrate them numerically and this can be
used to plot the orbits of particles ejected from the centre.  Now use
the hypothesis of \fullref{sec:gen} that the centre of a normal galaxy
contains a belt structure, which emits jets of gas$/$plasma, which
condense into stars.  The orbits of these stars can be modelled and a
``snapshot'' of all the orbits taken at an instant of time, in other
words a picture of the galaxy can be given.  Excellent models for the
observed spiral structure of normal spiral galaxies are found.  This
in done \fullref{sec:math}.

\begin{figure}[ht!]
\cl{\includegraphics[height=2in]{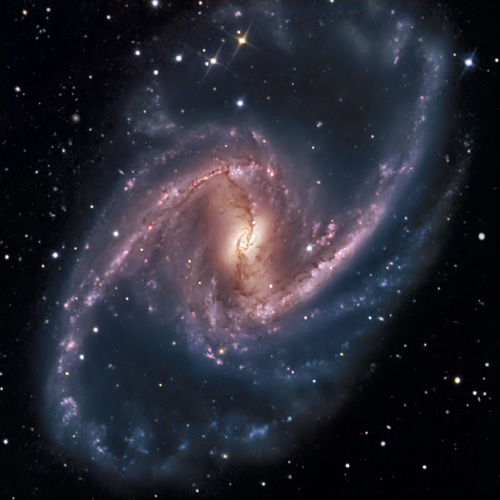}\qquad\includegraphics[height=2in]{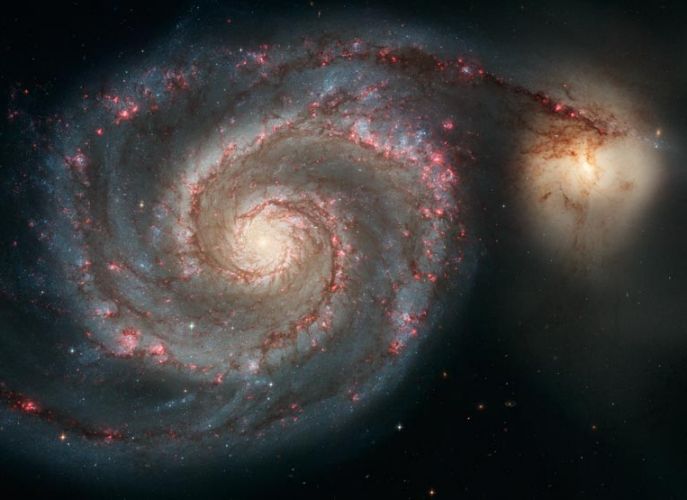}}
\caption{NGC1365 and M51 images from NASA and Hubble site resp}
\label{fig:more_gals}\end{figure}
\index{galaxy!NCG1365}
\index{galaxy!M51}

\sh{Simplified equations}
\index{simplified equations}

There is a very convenient simplification for the equations given in
the last section, which helps to explain how \id\ controls the
dynamic.  For most of an orbit in a galaxy $r\gg K$, since $r$ varies
up to $10^5$ for the main visible disc whilst $K\approx M\approx .1$.
This makes the fraction $\lfrac A{(K+r)}$ close to $\lfrac Ar$ and the
formulae for $v$ and $v_{\inert}$ reduce to $2A+C/r$ and $A+C/r$
respectively and then $$\rdd = \frac{A^2}r + \frac{AC-M}{r^2} +
\frac{C^2}{r^3}.$$ There are good reasons for setting $C<0$ (see
\fullref{sec:gen}) so that the $AC/r^2$ term acts to increase the
gravitational pull.  But the positive terms $A^2/r$ and $C^2/r^3$
offset the central gravitational pull (the first for large $r$ and the
second for small $r$) and this allows long slow outward orbits which
fill out the spiral arms.

\sh{Mathematica generated pictures}\label{sec:math}
\index{mathematica!spiral structure}

Below is the basic Mathematica notebook which generates galaxy
pictures from the dynamics found in \fullref{sec:dyn} above.  The
notation is as close as possible to the notation used before.  {\tt
  A}, {\tt K}, {\tt r} and {\tt v} are $A,K,r$ and $v$ resp.  $E$ and
$C$ have been replaced by {\tt EE} and {\tt CC} because {\tt E} and
{\tt C} are reserved symbols in Mathematica.  $M$ has been replaced
by three constants {\tt Mcent}, {\tt Mdisc} and {\tt Mball}.  This is
to allow an investigation of the effect of significant non-central
mass on the dynamic.  {\tt Mcent} acts exactly as $M$ above whilst
{\tt Mdisc} and {\tt Mball} act as masses of a uniform disc or ball of
radius {\tt rmax}.  Setting {\tt Mball = Mdisc = 0} reduces to the
case of just central mass considered above.  {\tt inert} is
$v_{\inert}$ and the other variables should be obvious from their
names.  

\index{mathematica!main notebook}
\begin{verbatim}
A = 0.0005; Mcent = .03; EE = -.00000345; CC = -10; 
B = .00000015; Mball = 0; Mdisc = 0; K :=  Mcent;
rmin = 5000; rmax = 50000; iterate = 1000; step = (rmax - 
    rmin)/(iterate - 1); 
v := 2*A - 2*K*A*Log[1 + r/K]/r + CC/r; inert := v - A*r/(K + r);
Plot[{inert, v}, {r, rmin, rmax}, AxesOrigin --> {0, 0}]
rdoubledot := inert^2/r - Mcent/r^2 - Mdisc/rmax^2 - Mball*r/rmax^3;
Plot[{rdoubledot}, {r, rmin, rmax}, AxesOrigin --> {0, 0}]
energy := -CC^2/(2*r^2) + (Mcent - 2*A*CC)/r - Mdisc*r/rmax^2 + 
   Mball*r^2/(2*rmax^3) + A^2*K/(K + r) + A^2*Log[K + r] + 
   2 A*K (CC + 2*A*r) Log[1 + r/K]/(r^2) 
   - (2 A*K*Log[1 + r/K]/r)^2 + EE;
Plot[{energy}, {r, rmin, rmax}, AxesOrigin --> {0, 0}]
rdot := Sqrt[2*energy];
Plot[{rdot}, {r, rmin, rmax}, AxesOrigin --> {0, 0}]
ivalue := rmin + (i - 1)*step;
thetadot := v/r;
dthetabydr := thetadot/rdot ;
dtbydr := 1/rdot; 
thetavalues = 
 Table[NIntegrate[dthetabydr, {r, rmin, ivalue}], {i, iterate}]
tvalues = Table[NIntegrate[dtbydr, {r, rmin, ivalue}], {i, iterate}]
ListPolarPlot[{ Table[{thetavalues[[i]] - B*tvalues[[i]], ivalue},
       {i, iterate}] , 
    Table[{thetavalues[[i]] - B*tvalues[[i]] + Pi, ivalue}, 
       {i, iterate}] }]
\end{verbatim}

\index{galaxy!precession of roots}
The program uses equations \ref{eq:energy} and \ref{eq:v-recap} to
express $dt/dr$ and $d\theta/dr = \thetadot/\rdot=v/(r\rdot)$ in terms
of $r$ and then integrates numerically with respect to $r$ in steps of
size {\tt step}.  It then plots the resulting values of $\theta$ in
the $(r,\theta)$ plane.  If {\tt step} is small this gives a good
approximation to the orbit of a particle.  To plot the spiral arms, it
is necessary to allow the roots of the arms to precess.
There is a new constant $B$ which is the (apparent) rate of
precession.  Suppose that the roots are at radius $r_0$, where \id\ is
approximately $A/r_0$, then the inertial frame at that radius is
rotating with respect to the background Minkowski metric with angular
velocity approximately $A/r_0$, and, if the roots are stationary in
that frame, they {\it appear} to be precessing with this angular
velocity.  $B$ adds a linear term to $\theta$ to realise this.  With
$B$ set to 0, the program sketches orbits.  With $B$ set nonzero the
program sketches a snapshot of the spiral arms at a particular time.
There might be some other effect causing precession and $B$ can be
adjusted to fit any such effect.  In any case, it is necessary to
guess $r_0$ in order to set $B=A/r_0$.  Since the program starts at
$r=${\tt rmin}, a good first guess for $B$ is $A/${\tt rmin}.

\begin{figure}[ht!]
\cl{\includegraphics[width=.6\hsize]{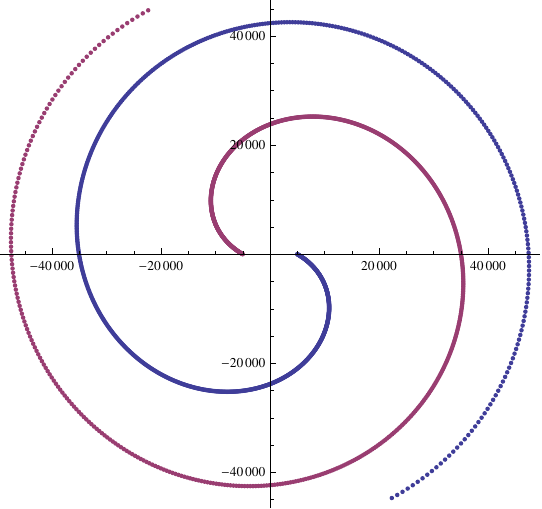}}
\caption{Output from the program as printed}
\label{fig:basic}
\end{figure}

\begin{figure}[ht!]
\cl{\includegraphics[width=.45\hsize]{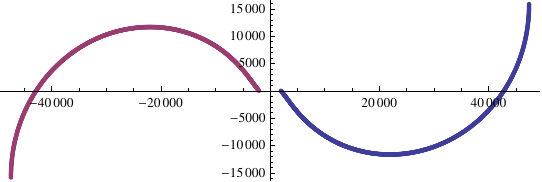}\qquad\includegraphics[width=.45\hsize]{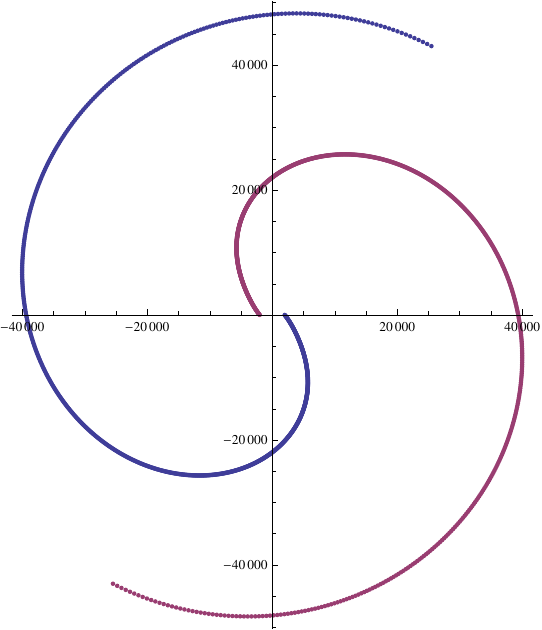}}
\caption{Left: orbits.  Right: loose spiral}
\label{fig:loose}
\end{figure}

\begin{figure}[ht!]
\cl{\includegraphics[width=.45\hsize]{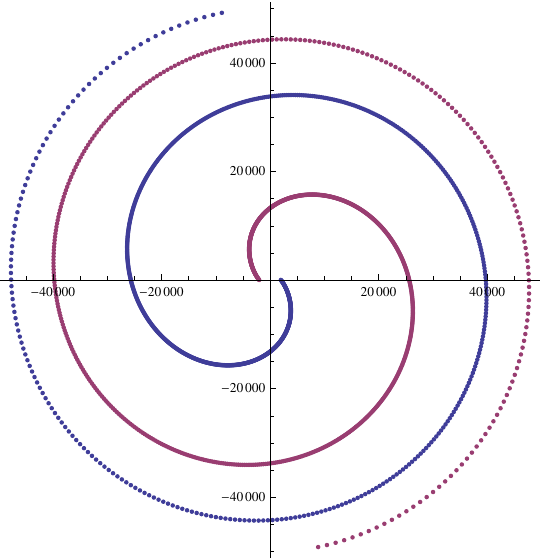}
\qquad\includegraphics[width=.45\hsize]{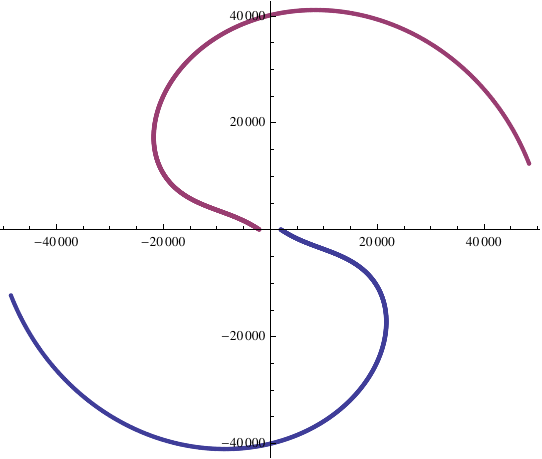}}
\caption{Left: tighter spiral.  Right: bar.}
\label{fig:tighter}
\index{galaxy!interactive program!bar}
\end{figure}

\index{galaxy!interactive program} The program is intended for
interactive use and the reader is recommended to investigate the
output.  Copies of the notebook with the settings used here can be
collected from \cite{Nb}.  Note that these are all the same program;
just the pre-set settings vary.  Details of these settings are given
in the descriptions which follow.  For \fullref{fig:basic} use {\tt
  Basic.nb} and for Figures \ref{fig:loose} (left and right) and
\ref{fig:tighter} (left) use {\tt Full.nb}.  For \fullref{fig:tighter}
(right) use {\tt Bar\_galaxy.nb}.  Here are some hints on using it.
As remarked earlier, the sketches are discrete plots obtained by
repeated numerical integration.  The number of plot points is set by
{\tt iterate}.  Start investigating with {\tt iterate = 100} which
executes fairly quickly and then set {\tt iterate = 1000} for good
quality output.  The plots are calculated in terms of $r$ {\em not\/}
time.  The time values can be read from the {\tt tvalues} table which
is printed as part of the output.  $r$ varies in equal steps from {\tt
  rmin} to {\tt rmax} which need to be preset.  You can't run to the
natural limit for $r$ (when $\dot r =0$) but have to stop before this
happens.

Set {\tt A} to fit the desired asymptotic tangential velocity $2A$.
For example to get $2A \approx 300$km/s set {\tt A = 0.0005}.  Set
{\tt Mcent} to the desired central mass.  For example $10^{11}$ and
$10^{12}$ solar masses are $M=.03$ and $.3$ respectively.  Leave $K$
set to equal {\tt Mcent} unless you want to experiment with large
values (which will increase the \id\ effect for a fixed mass).  Start
with $B$ set to $A/${\tt rmin} and adjust to get the desired spiral
pitch.  The integration constants $C$ and $E$ affect the picture
mostly near the middle and outside resp.  There are theoretical
reasons for setting $C$ to be negative because of the nature of the
spiral arm generator (see \fullref{sec:gen}) and with $C$ set
negative, the roots of the spiral arms are offset in a way seen in
many galaxy examples.  $E$ is a key setting as it determines the
energy of orbits and hence the overall size of the galaxy.  To get the
most realistic pictures you need $\dot r$ to go to almost to zero at
the maximum for $r$ which you get by fine tuning $E$.  To help with
this tuning, the program plots the graphs of $v, v_{\inert}, \rdd$,
energy and $\rdot$ so that you can adjust to get $\rdot$ and $\rdd$
near zero at {\tt rmax}.

Here now are some plots of orbits and galaxy arms obtained from this
program.  These should be compared with the images of real galaxies
that are reproduced in (Figures \ref{fig:M83}, \ref{fig:galaxies} and
\ref{fig:more_gals}).

\fullref{fig:basic} is the output from the program as printed above.
$M$ has been set to $10^{11}$ solar masses (all central) with
tangential velocity asymptotic to 300km/s, {\tt rmin} has been set to
5,000, $B$ to $A/${\tt rmin} and {\tt rmax} to 50,000 light years
(corresponding to a visible diameter of 100,000 light years).  Time
elapsed along the visible arms is $5.5\times 10^7$ years.  The nature
of the visible spiral arms will be discussed carefully in
\fullref{sec:obs}.  Here merely note that the visible arms correspond
to strong star-producing regions and bright short life stars, which
burn out or explode in $10^5$ to $10^7$ years.  Thus a total time
elapsed of $5.5\times 10^7$ years allows several generations of stars
to be formed and to create the heavy elements necessary for planets
such as the earth to be formed.

Figures \ref{fig:loose} (left and right) and \ref{fig:tighter} (left)
have the same settings with only $B$ varied.  The settings are similar
to \fullref{fig:basic} but with a small realistic contribution to the
mass coming from {\tt Mdisc} and {\tt Mball} which are both set to
0.01 (1/3 of the central mass).  {\tt rmin} has been reduced to 2000
to get nearer to the centre.  Elapsed time for all three is the same
and again is $5.5\times 10^7$ years.  $B=0$ for \fullref{fig:loose}
left, so these are actual orbits and $B$ has been set to $10^{-7}$ and
$2\times 10^{-7}$ resp for the other two to give a loose and a tighter
spiral.  Finally in \fullref{fig:tighter} right $M$ has been reduced
to 0.01 ($10^{10.5}$ solar masses) and the settings chosen ($C=-5$ and
$B=5\times10^{-8}$) to give a realistic bar galaxy.  Elapsed time here
is $10^8$ years.
Two classic examples of bar galaxies are NGC1300,
\fullref{fig:galaxies} (right), and NGC1365, \fullref{fig:more_gals}
(left).  Both of these appear to contain two rather different
structures: spiral arms and a superimposed dusty bar.  The model given
in \fullref{fig:tighter} (right) models an amalgam of these so there
is a need for a better model for bar galaxies, and this is discussed
in the next two subsections.

%See also Ian Stewart \cite[Chapter
%  12]{INS} for other ideas for modelling bar galaxies.

%\index{Stewart, Ian}
\index{galaxy!bar}
\index{galaxy!spiral structure|)}
\index{galaxy!bar!structure|(} \index{galaxy!bulge}

\sh{The bulge}\label{sec:bulge}

In order to model bar galaxies more accurately it is necessary to
consider another feature of galaxies which has only been mentioned in
passing up to now, namely the central bulge.  This is a chaotic
collection of stars and other material lacking the dynamic coherence
of the spiral arms.  Most stars are on fairly tight orbits around the
central black hole and there is a very large range of star types
observed.  There is a predominance of poulation II (cool red stars)
and this accounts for the red colour of the bulge seen in many galaxy
photos.

In terms of the accretion model constructed in \fullref{sec:quasars}
and \fullref{app:3author} the bulge is analogous to the settling
region where incoming matter loses its kinetic energy (KE) by
interaction and settles towards the central black hole.  The KE of
incoming matter keeps up the energy levels and it is also fed from the
central black hole in the same way as the generator.  So there is
analogous activity with small jets (not so organised as for the main
jets that create spiral arms) and new star streams.  The general
appearance is of a cloud of stars which is usually spherical.  But if
affected by nearby strong gravitational fields it can take other
shapes and this is precisely what happens in a bar galaxy.

\begin{figure}[ht!]\index{galaxy!NGC4394}
  \hglue .7in\includegraphics[height=2.2in]{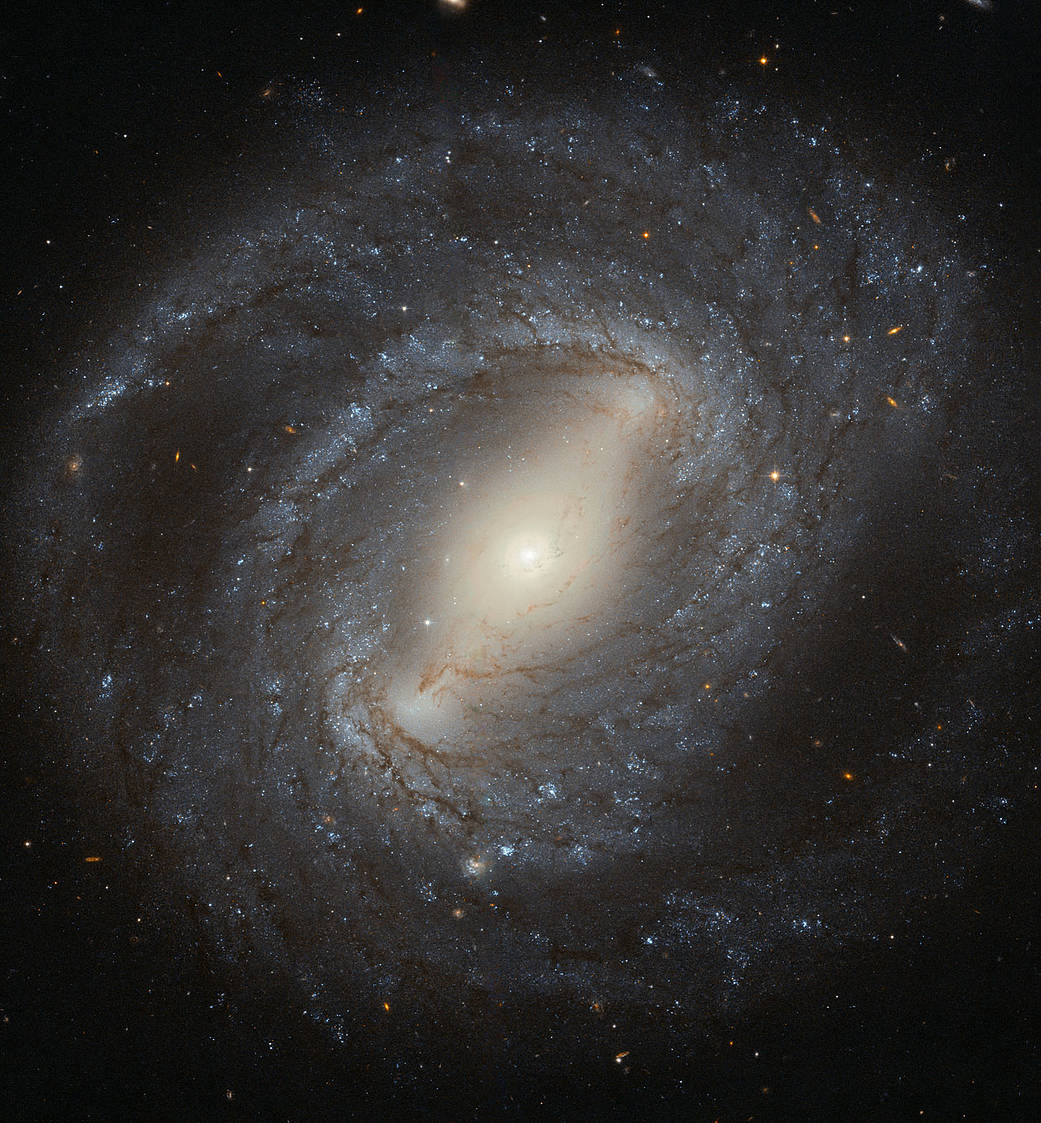}\qquad\vbox{\includegraphics[height=1.09in]{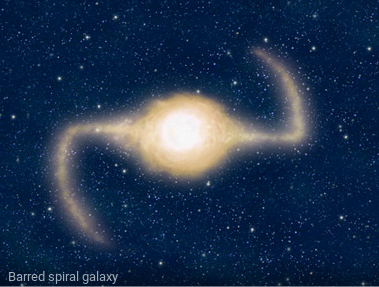}\nl\includegraphics[height=1.09in]{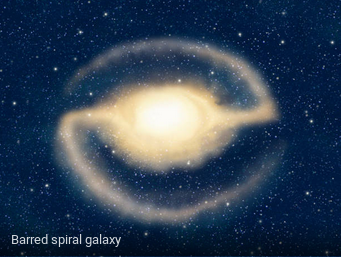}}
\caption{NGC4394 (left) and archetypal bar galaxies (right)}
\label{fig:arch_bars}\end{figure}

\sh{Bar galaxies}\label{sec:bar}

\begin{figure}[ht!]
\cl{\includegraphics[height=2in]{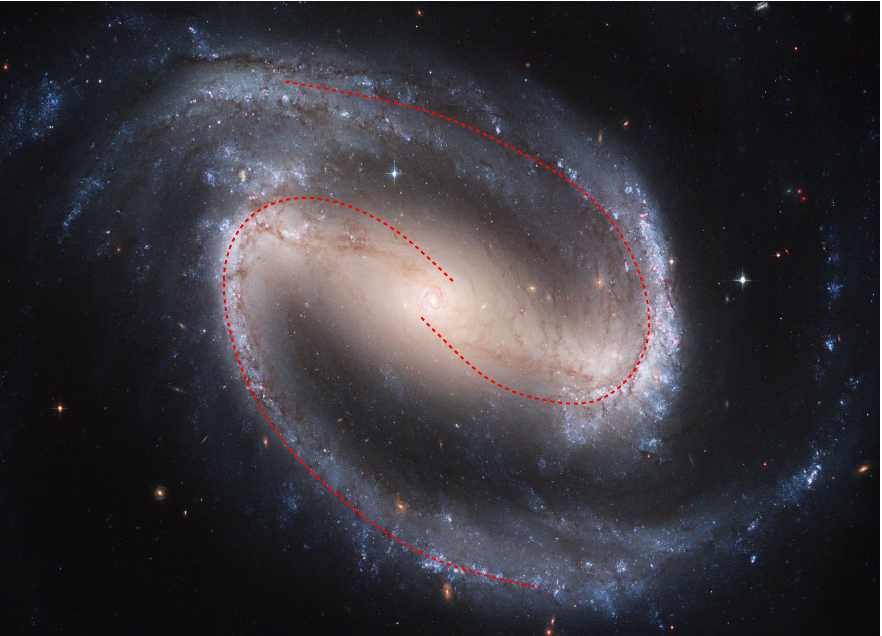}\qquad\includegraphics[height=2in]{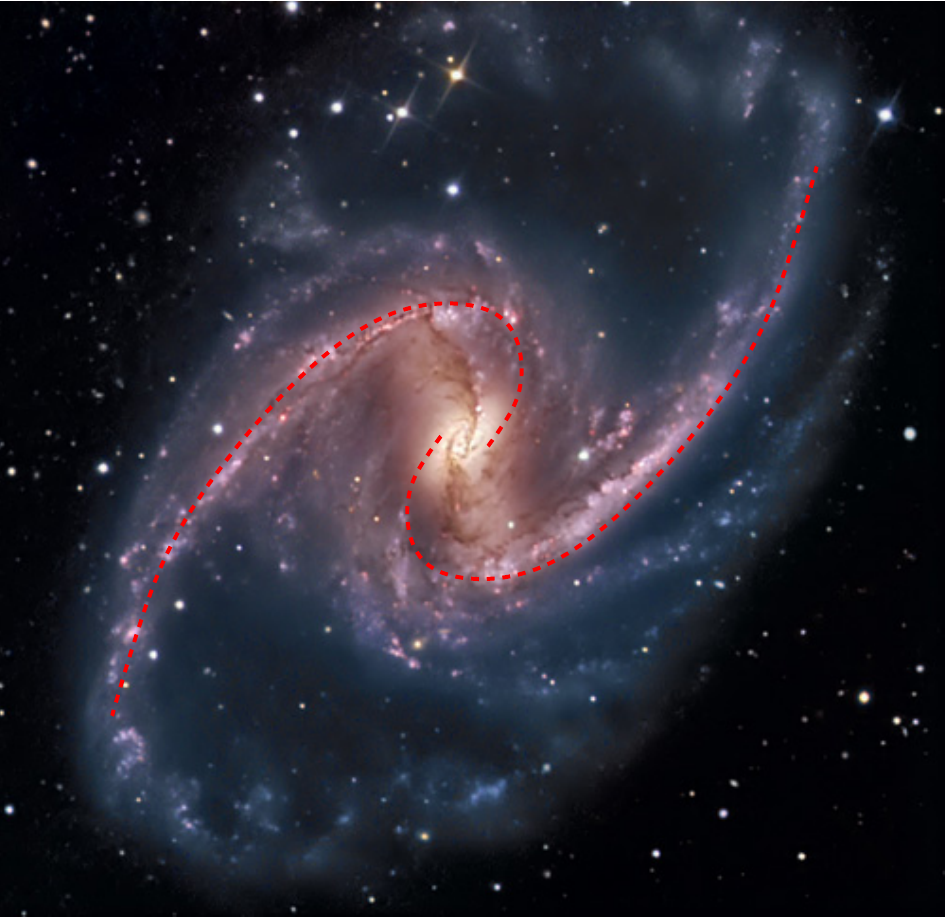}}
\caption{NGC1300 (left) and NGC1365 (right) with spiral arms indicated}
\label{fig:bar_dots}\end{figure}

Turning now to bar galaxies, here is a description from the Hubble
site: ``NGC 4394 is the archetypal barred spiral galaxy, with bright
spiral arms emerging from the ends of a bar that cuts through the
galaxy's central bulge'', \fullref{fig:arch_bars} (left). The two
pictures on the right of the figure are standard sketches of barred
spirals based on this archetype but not on any real galaxy.  Real
galaxies never look like either of these!  NGC 4394 itself has an
extensive but rather chaotic spiral structure extending right into the
central region marked by spiral lanes of dust. The arms are not well
defined but what can be seen very clearly is that they do NOT emerge
from the ends of the bar.  Where the spiral arms emerge in a barred
galaxy can be seen more clearly in the two classic bar galaxies
pictured above, namely NGC1300, \fullref{fig:galaxies} (right), and
NGC1365, \fullref{fig:more_gals} (left).  In both cases the arms can
be traced back to the centre as indicated in \fullref{fig:bar_dots}.

Not all of the indicated arms are visible.  For NGC1365 they are
more-or-less visible all the way to the centre but for NGC1300, only a
short section can be seen inside the radius of the bar.  There are
three possible explanations for this.  Firstly the bar itself occludes
the arms and secondly the nature of the arms (clouds of pre-stellar
material condensing into violent star-producing regions) makes it
likely that the arms are invisible near the roots; this is analogous
to the way a gas flame burns: the stream of plasma needs to condense
into clumps to form stars and ignite.  The third possibility is that
the generator has stopped emitting material for a while.  Looking at a
collection of galaxy photos it is possible to find clear examples
where spiral arms stop and start, presumably because the generator
runs out of ``fuel'' and needs to accrete some more.

The bar itself is not part of the spiral structure but a distorted
central bulge, the distortion being due to the nearby massive streams
of matter that are feeding the main spiral arms.  In fact the bar and
the arms are dynamically disjoint: orbits in the bar are local but
elongated by the distorion from the pull of the arms; orbits in the
arms are (like all arm orbits) long slow outward spirals lasting
$10^8$ years or more.

\index{galaxy!bar!structure|)}
\np\thispagestyle{empty}

\chapter{Observations}\label{sec:obs}
\index{Milky Way|(}

Previous chapters have established a new model for spiral galaxies
based on Sciama's principle.  This has provided a satisfactory
explanation for observed rotation curves without using ``dark
matter'', and accurately modelled the spiral structure.  The salient
features of this new model are (1) a central rotating mass (presumably
a black hole) of $10^{11}$ to $10^{14}$ solar masses which controls
the dynamic via \id\ effects and (2) a counter-rotating belt structure
similar to the accretion disc around the (rather smaller) black holes
in so-called ``active'' galaxies which feeds the roots of the spiral
arms with pure H--He ions (with a trace of Li).  The whole galaxy has
a cyclical structure with matter ejected from the centre, condensing
into stellar systems, moving outwards along the arms, burning out and
falling back into the centre to be recycled.  Thus the outward flow of
gas mixes with dust and debris outside the belt and, as it flows
outwards, condenses into violent star-producing regions illuminated by
novae and supernova explosions.  These regions sythesise the heavier
elements needed for planets such as the earth to support chemically
based life-forms.  Solar systems containing such planets are formed
further out along the arms.
\index{galaxy!cyclical structure}

This chapter considers detailed observations from our own galaxy which
support this new model.  More information on the entire structure will
emerge in the process.

Topics covered are: early 21cm observations, \fullref{sec:21}, stellar
populations, \fullref{sec:pop}, the nature of Sgr\astar,
\fullref{sec:astar}, the position of the Sun, \fullref{sec:sun},
globular clusters, \fullref{sec:gcs}, and, as a related appendix
(\fullref{app:lsv}), an extended discussion of local stellar
velocities.  In particular, the new model provides natural
explanations for the non-existence of hypothetical type III stars and
for the peculiarities of the velocity ellipsoid for local stellar
velocities.

\sh{\ind{21cm emission observations}}\label{sec:21}

The first comment is that the outward flow of gas along the arms of
the Milky Way was clearly observed by Oort, Kerr and Westerhout
\cite{OKW} in 1958 and in subsequent surveys.  The correct
interpretation was made at the time but was later changed to attribute
these observations to a hypothetical bar structure (for which there is
little other evidence) see Binney and Merrifield \cite[pages
  17--18]{BM}.  The idea is that the bar provides a massively
asymmetrical central gravitating mass which allows for highly
non-circular orbits, some parts of which fit the observed gas flows.
This explanation is implausible for the same reason that the dark
matter explanation of the rotation curve is implausible.  In both
cases a rotating dynamical system is proposed which is supposed to be
stable but does not have a dominating central mass to provide
stability.

It is worth remarking that a satisfactory model for bar galaxies is
provided in Sections \ref{sec:bulge} and \ref{sec:bar} and that in
this model the bar itself is part of the central bulge structure, in
which orbits are generally chaotic.  It is also worth remarking that
similar gas flows have been observed in other galaxies.

\sh{Stellar populations}\label{sec:pop}
\index{stellar populations}

In the model for galaxies proposed here, stars are formed by
condensation in the outward flowing streams of gas coming from the
central belt structure, loosely called ``the generator''.  The outer
layer of the belt is a plasma of H and He ions with traces of Li and
other particles.  Conditions here are similar to those hypothesised to
have occurred just after the big bang and the consequent mix of light
elements is the same.  As modelled in \fullref{sec:spiral_struc}, the
outward flowing gas streams form into the familiar spiral arm
structure.  It is in these arms that stars condense.  The residue of
the gas streams, not condensed into stars, escapes the galaxy and
feeds the intergalactic medium and this explains the observed
proportion of light elements in the universe (which is one of the
so-called ``pillars of the big bang theory'').

Near the roots of the arms, this condensation creates the observed
violent star-producing regions with novae and supernovae.  Here
heavier elements are synthesised in abundance and moving outwards
along the arms, the composition in the background gas stream alters to
include dust and debris from this synthesisation and stars condensing
further out have higher metalicity\fnote{Metal is used here, with the
  misuse common in astronomy, to mean all elements heavier than He.}.
Thus for stars in the neighbourhood of the sun, fairly far out from
the centre along an arm, there is a natural inverse correlation
between the age of a star and its metalicity.  Later a good estimate
for the distance of the Sun from the centre of the galaxy will be
found.

This is usually described in terms of ``stellar populations'':
population II stars are older stars with low metalicity formed near
the roots of the arm in which the sun lies whilst population I stars
are younger stars formed further out, after enough population II stars
have exploded as supernovae to provide the higher metalicty in these
stars.  The Sun is a population I star.

Under the big bang hypothesis, there should be a third population
(\ind{population III stars}) formed immediately after the big bang from pure
H--He with zero metalicity.  These stars have never been detected.  In
the model proposed in this book, they would have to be formed at the
very roots of the arms.  But, because of the cyclical nature of the
model, outside the belt the galaxy is heavily polluted with dust and
debris of various kinds coming from stellar systems falling back into
the centre to be recycled.  Thus the pure stream of H--He is quickly
contaminated with traces of metals.  Therefore stars formed even very
near the roots will be contaminated with metals and be population II
stars.  Thus the model naturally explains the different stellar
populations and why there are no population III stars observed.
Notice that the difference between population I and population II
stars is not their age, but where they are formed in the arms.  Stars
formed near the centre will be older by the time they reach the
neighbourhood of the sun than stars formed further out.  Thus for
stars near the sun, there is a inverse correlation between metalicity
and age, as is observed.

The model for star creation suggested here implies that stars like the
sun formed away from the centre will typically be surrounded by
planets condensed from the heavier lumps of debris in the vicinity.
This suggests that solar systems like ours are the norm rather than
the exception.  This prediction has in fact already been verified by
many recent observations.

\begin{figure}[ht!]
\includegraphics[width=\hsize]{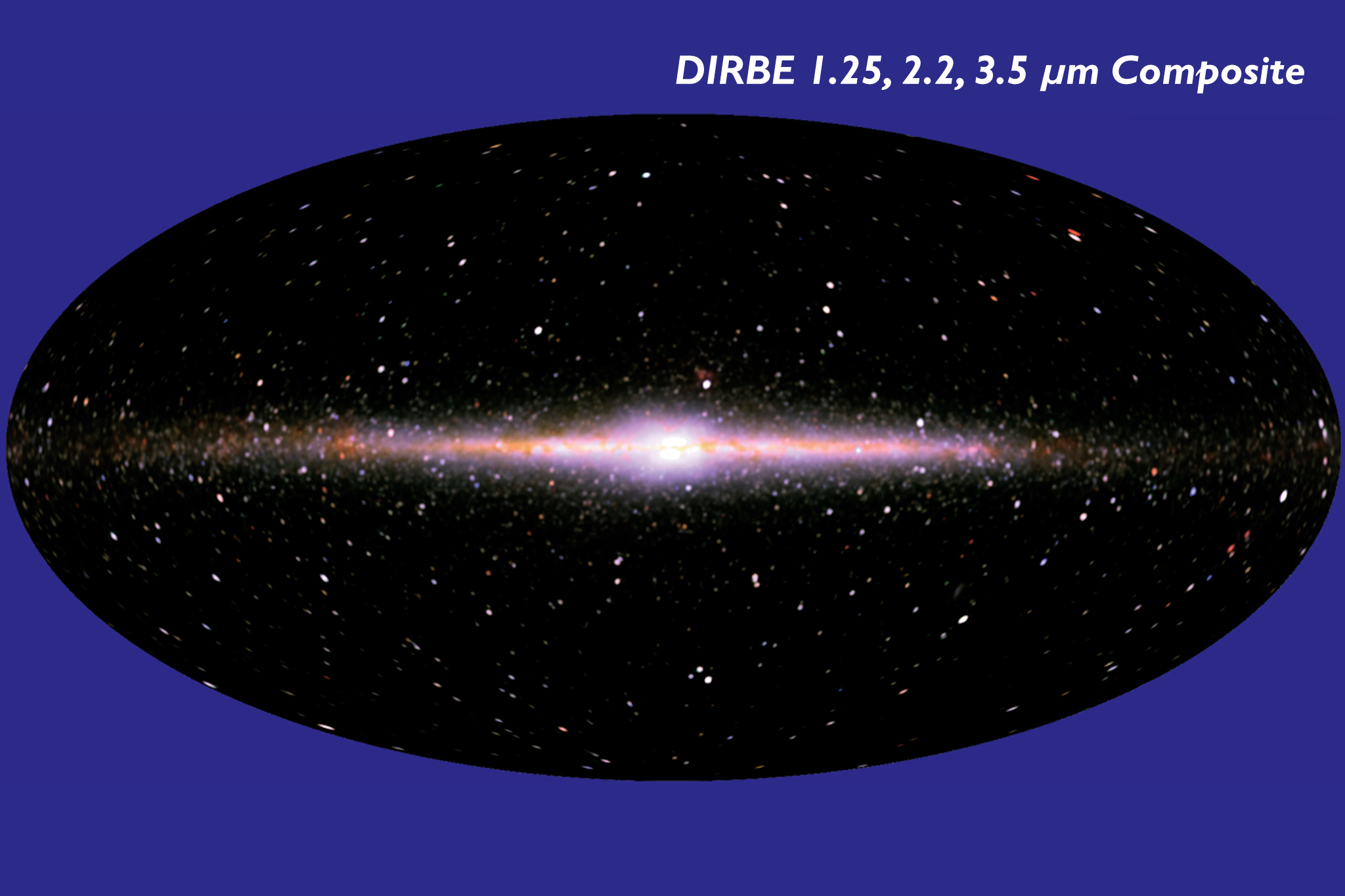}
\caption{Composite image of our galaxy from the COBE
  satellite}\label{fig:COBE}\index{COBE image}
\end{figure}

\sh{Sagittarius \astar}\label{sec:astar}
\index{Sgr A*|(}

There is a strong radio source at Sgr\astar\ which has been the
subject of many observations.  Observations of the proper motion of
Sgr\astar\ using a very long baseline interferometer, due to Reid et
al \cite{Reid}, suggested that the observed motion could be ascribed
to the orbital motion of the Sun and that Sgr\astar\ might in fact be
at rest.  Further the orbits of stars near Sgr\astar\ have been
carefully monitored over a period of twenty years or so.  These
observations establish that this object is massive (about
$4.3\times10^6$ solar masses, presumably a black hole) and at a
distance from the Sun of about 8.3kpc.  For a good overview see
Gillesen et al \cite{Gill}.

The suggestion by Reid et al that Sgr\astar\ might be at rest with
respect to the galaxy as a whole has led to the belief that it is in
fact at the centre of the Milky Way and this has become now an
accepted ``fact'' with Gillesen et al for example describing
Sgr\astar\ as ``the Massive Black Hole in the Galactic Center''.
\index{Sgr A*!Reid}\index{Sgr A*!Gillesen}
However it is not nearly massive enough to drive the dynamic of a
full-size spiral galaxy, and therefore this conclusion directly
contradicts one of the main hypotheses of this book.  Thus it is
necessary to advance another explanation for these observations.

\index{globular cluster!containing Sgr A*}
\Gc s have total mass varying up to around $10^7$ solar masses and
central black holes have been detected in many clusters.  Moreover
there is a well-established theory for mass concentration and black
hole formation in clusters, see \cite{wiki:gc}.  Indeed this is a
natural phenomenon as clusters age.  Stars will burn out and collapse
and mass concentration will cause a group of collapsed stars to
coalesce into a single black hole.  The group of stars orbiting
Sgr\astar, together with Sgr\astar\ itself have all the
characteristics of a globular cluster near the end of its life with
most of the mass coalesced into the central black hole and the
remaining stars in orbit around the centre.  

\begin{figure}[t!]
\includegraphics[width=\hsize]{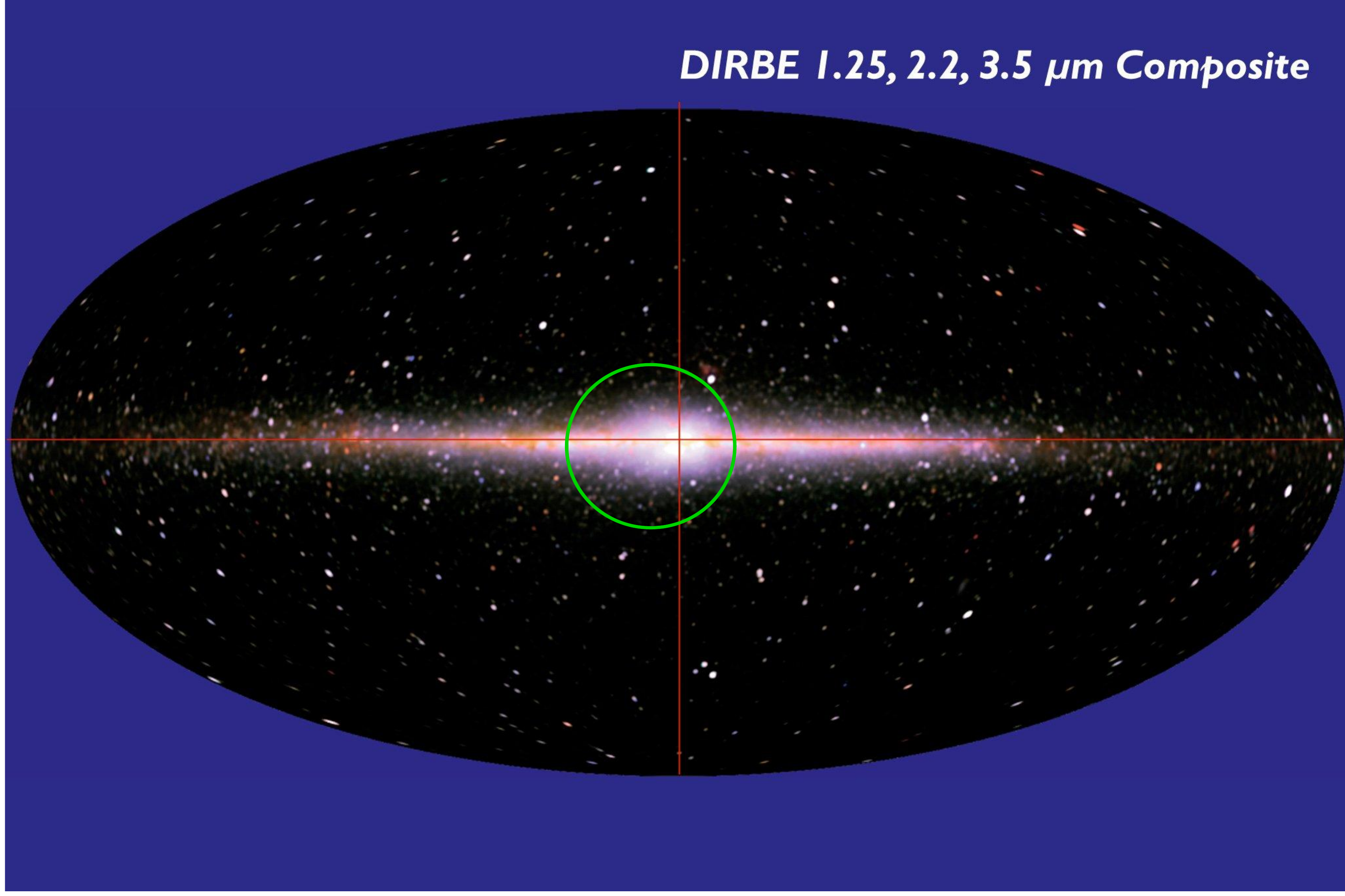}
\caption{Image of our galaxy from the COBE
  satellite with axes and bulge highlighted}\label{fig:COBE2}
\end{figure}

At this point the truly wonderful image of our galaxy, the Milky Way,
\fullref{fig:COBE} obtained from data collected by the COBE satellite
\cite{COBEimage}\index{COBE image}
should be considered.  This image provides clear
evidence that Sgr\astar\ is not at the centre of the galaxy.  The
image uses Mollweide projection, which preserves area and central
symmetry.  Because of the conviction that Sgr\astar\ is at the centre,
this has been located dead centre in the image.  The horizontal scale
is galactic longitude covering the full $360^\circ$ and it is linear.
If Sgr\astar\ was truly at the centre of the galaxy then this image
would be symmetrical about both the central vertical and horizontal
axes.  It is clearly not.  The bulge peaks rather to the left of
centre and the main disc (seen edge-on) is also displaced to the left.
Not quite so obvious, but also clearly visible, is vertical asymmetry,
with the disc displaced slightly downwards from the central horizontal
line.  To help these asymmetries to be seen, the image has been
reproduced in \fullref{fig:COBE2}
with the Mollweide axes superimposed in red and with a green circle
centred on the widest part of the bulge.  The bulge itself is not
symmetric with a curious right hand smaller bulge superimposed.  Below
an explanation for this is suggested, and the smaller bulge has been
ignored in placing the green circle.  But even if it is not ignored
and the two bulges are averaged, then it is still displaced to the
left.  (There is also a blown-up image of the centre of
\fullref{fig:COBE} on the same site \cite{COBEbulge} with the
asymmetry of the bulge and the vertical displacement both very clearly
visible.)  Because of the non-circular nature of spiral arms there is
no reason to expect the main disc to appear symmetrical.  But it is
pretty symmetrical albeit displaced to the left.  However vertical
symmetry and symmetry in the central bulge is expected.  The asymmetry
corresponds to a displacement of Sgr\astar\ by $1^\circ$ upwards from
the true centre of the galaxy and between $3^\circ$ and $5^\circ$ to
the right (depending on whether the secondary bulge is ignored or the
two are averaged).

So Sgr\astar\ is not at the centre of the galaxy.  Is there any reason
to suppose that it is at rest?  This assumption has become
self-fulfilling with other velocities measured against it.  If this
assumption is dropped, there is no direct evidence to reinstate it.
It would be necessary to measure average velocities for the galaxy as
a whole, compensating for redshift due to a heavy centre (not
Sgr\astar) if any.  There were early estimates using globular clusters
(due to Shapley, see \cite[page 8ff]{BM}), but no apparent recent
global data to replace this assumption.

Now consider the bright region near Sgr\astar\ which is associated
with the smaller right-hand bulge.  Here is a suggestion for what this
might be, which also leads to a suggestion to explain the stellar
composition of the cluster.  It is suggested that the image here is
looking straight down part of the arm coming out of this side of the
central bulge and seeing an amalgam of strong star-producing regions
which accounts for the brightness of the radio image.  This implies
that the region containing Sgr\astar\ is full of pre-stellar material
(dust and light elements) out of which stellar systems are condensing.
The cluster has moved into this region, which accounts for the strange
stellar composition --- predominantly Wolf-Rayet and Type O with a
sprinkling of young stars (the so-called \ind{paradox of youth}).  The young
stars could result from the capture of clouds of pre-stellar material
which have condensed into stars.  The Wolf-Rayet and Type O stars are
heavy old stars consistent with extreme age for the cluster as a
whole.
\index{Sgr A*|)}

\sh{Where is the Sun?}\label{sec:sun}
\index{Milky Way!location of sun}

Since Sgr\astar\ is not the centre of the galaxy, there is no direct
way to measure the distance of the Sun from the centre.  There is
however a good deal of indirect evidence which places it at 17kpc
($5\times 10^4$ in natural units) or more from the centre.  It is
necessary to consider what is actually seen when looking at a spiral
galaxy.  There are several images reproduced in
\fullref{sec:spiral_struc} to look at (Figures \ref{fig:M83},
\ref{fig:galaxies} and \ref{fig:more_gals}).  In all cases it is clear
that the visible spiral arms are characterised by intense star
producing regions populated by massive short life stars and that a
region of smaller older stars such as our immediate neighbourhood
would very probably appear quite dark from a distance.  So it is
expected to be some way outside the main visible disc (which is
typically about $10^5$ in diam).

There is also the timescale to consider.  The visible arms mostly
comprise massive short life stars which burn out or explode in $10^5$
to $10^7$ years.  This fits well with the models constructed in
\fullref{sec:spiral_struc} where matter takes from $10^7$ to $10^8$ to cover the
length of the arms from centre.  This gives time for several
generations of stars to be formed and to create the heavy elements for
population I stars (like the Sun) to contain (not to mention the
earth).  The Sun is about $5\times 10^9$ years old and
probably formed about half way along one of the arms of the galaxy.
By now it must have moved beyond the visible arms.  It is worth
commenting that the spirals found in the models constructed in
\fullref{sec:spiral_struc} have very shallow
pitch near the outside (where both $\rdot$ and $\rdd$ are small) and
the outward movement slows down very considerably there.  This means
that Sun may be just a short way outside the visible arms,
more-or-less on the edge of the visible disc at about $5\times 10^4$
out from the centre.

Finally there is conclusive evidence again from the COBE satellite
image, \fullref{fig:COBE}.  The visible arms clearly lie to one side
of the Sun.  They thin down to almost nothing for about half (or a
little more) of the full circle represented by the centre line on the
diagram.  This puts the Sun right on the edge the main disc, or just
outside, at again about $5\times10^4$ from the centre.

Incidentally the estimates found here agree closely with those made by
Harlow Shapley \index{Shapley, Harlow} in around 1918 based on
distances to globular clusters (see \cite[page 8ff]{BM}).  These were
later revised downwards and it is tentatively suggested that there may
have been a systematic error in these revisions.

\sh{\Gc s}\label{sec:gcs}
\index{globular cluster}

\Gc s comprise mostly population II stars.  So they are formed very
close to the central part of the galaxy.  It is suggested that the
instability in the central region, fed directly by energy from the
black hole, occasionally throws a huge flare of gas (the usual H--He
mixture) in a direction other than in the galactic plane.  This could
happen as a short-life ``storm'' structure.  An analogy would be a
cyclone forming in the earth's atmosphere.  Such a flare could
condense to form a tight cluster of population II stars: a globular
cluster in fact.

There are about 200 \gc s in a galaxy and they have lifetimes of
$10^{10}$ years or more so, to maintain the population, there need be
only one new cluster formed every $10^7$--$10^8$ years.  Thus this model
makes it possible that the constitution of a galaxy might be
more-or-less constant over a timescale several orders of magnitude
greater then current estimates.  In the next chapter these ideas are
pursued and their consequences for global cosmology discussed.

\sh{Local stellar velocities}
\index{local stellar velocities}

There has been a huge effort expended mapping the velocities of stars
in the neighbourhood of the sun.  There are some paradoxical
properties of these excellent observations.  In particular, the
symmetries in velocity variations that would be expected from the
current dynamical model of the galaxy (with stars moving in circular
orbits) are not observed.  The ``velocity ellipsoid'' which expresses
this variation does not have the line from the Sun to the galactic
centre as a principal axis, as would be expected from symmetry; the
deviation of these two directions is called ``\ind{vertex deviation}''.
Further, vertex deviation varies systematically with stellar age.  The
dynamical model proposed in this book has no such symmetry and these
paradoxical aspects disappear.  Further vertex deviation and its
correlation with age have very natural explanations.
\index{velocity!ellipsoid}

The discussion is fairly technical and has been postponed to an
appendix (\fullref{app:lsv}).
\index{Milky Way|)}

\chapter{Cosmology}\label{sec:cosm}

\index{big bang theory|(}
This chapter discusses cosmological consequences of the model for
galactic dynamics constructed in the earlier chapters and starts by
considering the big bang theory.  No abstract theory in physics has
ever captured the general imagination in the way that this theory has.
Even the fine detail has passed into everyday usage.  Here for example
is an excerpt from a review from \emph{The Guardian}:

\begin{quote}\small

{\em 
The modern hunger to accord food spiritual ``meaning'' seems a
relatively recent development: it is refreshing to note the absence of
such inflated claims, for example, in the much-loved 1931 American
cookbook The Joy of Cooking, by Irma Rombauer. {\rm[Description of
    low-key rhetoric in this book omitted]}

\index{big bang theory!inflation}
Yet since then foodist rhetoric has, like the early universe,
experienced a period of rapid \ind{inflation}. The foodist movement is
desperate to claim other cultural domains as inherent virtues of food
itself, so as not ever to have to stop thinking about stuffing its
face. Food becomes not only spiritual nourishment but art, sex,
ecology, history, fashion and ethics\dots}

Extracted from: The Guardian 29 Sep 2012, Review section, Steven Poole
``Get stuffed''

\end{quote}

Given such universal appreciation of the fine detail of the theory, it
seems churlish to prove that it is wrong.  But unfortunately this is
the case.

\index{big bang theory!pillars}After dismissing the big bang theory,
the three so-called pillars of the theory: the distribution of light
elements, the cosmic microwave background (CMB) and redshift are
discussed.  As has been mentioned before, the observed distribution of
light elements is accounted for using the proposed central generator
for spiral arms in \fullref{sec:spiral_struc}, see \fullref{sec:dist}.
The other two pillars (redshift, \fullref{sec:red} and the CMB,
\fullref{sec:CMB}) use the new model of the universe proposed in this
book, see \fullref{sec:deS}.  Other topics discussed are gamma ray
bursts, \fullref{sec:deS} and \fullref{sec:revis}, the origin of life, \fullref{sec:life} and
an extended discussion of the quasar--galaxy spectrum,
\fullref{sec:spec}.\index{big bang theory!pillars}

\sh{The big bang?}\label{sec:bb}

It has been observed several times that the model of galaxies that is
proposed could be stable over a huge timescale (perhaps $10^{16}$
years or more).  There is a natural cycle with matter ejected from the
centre condensing into star populations with metalicity increasing
with distance from the centre.  Stars move out along the visible
spiral arms and burn out before gravitating back towards the centre to
be recycled.  The contrary hypothesis, that the galaxy is only just
older than the oldest known stars (or not quite as old as the oldest
known \gc s -- see below) is just about possible, but is not credible
in the light of galactic observations.  There is a continued vigour to
the star producing regions visible in all galaxies, which suggests a
steady renewal of material from the centre and a long-term steady
state.
\index{galaxy!long-term steady state}

Furthermore there are now several pieces of direct evidence that the
big bang hypothesis is wrong.  Globular clusters have just been
mentioned.  Although this is quite an old piece of evidence it is
nevertheless completely solid; widely ignored, it effectively subverts
the big bang theory.

Stellar evolution theory is very well-established, having an excellent
fit with a huge body of observations; evolution for globular clusters
is based firmly on stellar evolution.  It is as solid a theory as any
theory in physics.  There are globular clusters in \emph{this galaxy}
which are 15 billion years old or more.  This means that the galaxy
itself must have been around for a good while longer than that.  The
big bang happened 13.7 billion years ago.  There is a rather amusing
chapter in Binney and Merrifield \cite[Chapter 6]{BM} about
this.\fnote{There are some recent attempts to square this circle (see
  for example \cite{wiki:cap}) but they feel like fudges to the
  author.}
\index{globular cluster!15 billion years old}

To add to this, there are several recent observations of what should
be features of the early universe, showing for example galaxies in the
formative stages, which stubbornly refuse to show anything other than
normal galaxies that might be seen nearby.  The first clue that
something was very much amiss, was provided by the space-based Hubble
telescope.  In 2003 the Hubble telescope was pointed at a dark part of
sky, where it is possible to see back to near the big bang, and left
running for a long time.  The resulting ``\ind{Hubble ultra-deep field}''
\index{HUDF|see{Hubble ultra-deep field}}
(HUDF) is published on the web \cite{HUDF}.  It contains a wealth of
information, some of which is so important for the arguments in this
book, that \fullref{app:HUDF} has been devoted to its properties.
Here the point being made is that there are clear full-size galaxies
in this image which are so small (and therefore remote) that they are
far too close to the big bang to have developed.  A typical example is
the \ind{very distant spiral galaxy} (VDSG).

\begin{figure}[ht!]
\cl{\includegraphics[height=1.5in]{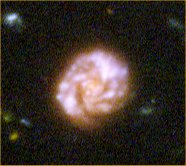}\qquad\qquad\raise 0.27in\hbox{\includegraphics[height=.9in]{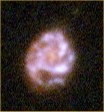}}}
\caption{The very distant spiral galaxy (left) and a possible smaller
  example (right)}\label{fig:VDSG}
\end{figure}

\medskip
To follow arguments about the HUDF here and in \fullref{app:HUDF}, the
reader is recommended to download a copy of the highest resolution
jpeg of the HUDF as instructed in the bibliography at \cite{HUDF}.  To
help find a particular galaxy or image, intrinsic coordinates are
given from the bottom left, where the height and width are 1 unit and
coordinates are taken mod 1 (so that a negative number is a coordinate
from the right or top).  The VDSG is at $(.40,.26)$.  A snippet of the
field with this galaxy in it is reproduced as \fullref{fig:VDSG}
(left).

\medskip
\medskip
The image is somewhat distorted and this is a characteristic feature
of the HUDF which will be explored in \fullref{app:HUDF}.  In brief,
there is a background noise of gravitational waves which causes
optical distortion.  But assuming that this galaxy is what it appears
to be, a full-size spiral of say $10^5$ light years diameter, then by
measuring the image and knowing that the HUDF has a linear size of 2.4
arc minutes, it can be calculated that this galaxy is $11\times 10^9$
light years away, and was fully formed a mere $2.7\times 10^9$ years
after the big bang.  This is far far too early for a full-size galaxy
to have formed under standard theory.  There is another even smaller
example at $(.17,.36)$ \fullref{fig:VDSG} (right).  This image is
probably too distorted to definitely label as a full-size spiral, but
if it is assumed that this too is a galaxy of $10^5$ light years
diameter, then it is $13\times 10^9$ light years away and formed
within 700 hundred thousand years after the big bang!
\medskip

The discovery of distant far-too-large spirals near the big bang has
been confirmed several times using different observations.  Here for
example is a news item from Physics.org \cite{Phys-org} about Abel 383
a gravitational lensing image from the Hubble site \cite{Hubble}:
\newpage
\begin{quote}\small

\cl{April 12 2011: First galaxies were born much earlier than expected}
\medskip
{\em 
The giant cluster of elliptical galaxies in the centre of this image
contains so much dark matter mass that its gravity bends light. This
means that for very distant galaxies in the background, the cluster's
gravitational field acts as a sort of magnifying glass, bending and
concentrating the distant object's light towards Hubble. These
gravitational lenses are one tool astronomers can use to extend
Hubble's vision beyond what it would normally be capable of
observing. Using Abell 383, a team of astronomers have identified and
studied a galaxy so far away we see it as it was less than a billion
years after the Big Bang. Viewing this galaxy through the
gravitational lens meant that the scientists were able to discern many
intriguing features that would otherwise have remained hidden,
including that its stars were unexpectedly old for a galaxy this close
in time to the beginning of the Universe. This has profound
implications for our understanding of how and when the first galaxies
formed, and how the diffuse fog of neutral hydrogen that filled the
early Universe was cleared.}

Credit: NASA, ESA, J Richard (CRAL) and
J-P Kneib (LAM)\nl
Acknowledgement: Marc Postman (STScI)
\end{quote}

And another from Nature (1 April 2009) \cite{Nature}:

\begin{quote}\small

\cl{News: Early galaxies surprise with size}

\cl{\em Astronomers revise galaxy-formation models with the
  discovery}\cl{\em that early galaxies could have grown fat---fast.}

\cl{Eric Hand}

{\em Slurping up cold streams of star fuel, some of the Universe's first
galaxies got fat quickly, new observations suggest. The findings could
overturn existing models for the formation and evolution of galaxies
that predict their slow and steady growth through mergers.

Researchers using the Subaru telescope in Hawaii have identified five
distant galaxy clusters that formed five billion years after the Big
Bang. They calculated the mass of the biggest galaxy in each of the
clusters and found, to their surprise, that the ancient galaxies were
roughly as big as the biggest galaxies in equivalent clusters in
today's Universe.

The ancient galaxies should have been much smaller, at only a fifth of
today's mass, based on galaxy-formation models that predict slow,
protracted growth. ``That was the reason for the surprise -- that it
disagrees so radically with what the predictions told us we should be
seeing,'' says Chris Collins of Liverpool John Moores University in
Birkenhead, UK. Collins and his colleagues publish the work today in
Nature \cite{Collins}.}

\end{quote}

The quote has been curtailed.  The rest is about patching up the
theory.  It is necessary to be blunt about all these observations.
They show that the big bang theory is \emph{wrong}.  Of course,
because so much has been invested in the theory, no-one has admitted
that it is wrong and indeed a strong fiction is being maintained that
it is being corrected.  This is not going well.  For example here is
an excerpt from the abstract for a cutting-edge seminar given at
Warwick on 22 May 2013:

\begin{quote}\small\em
Once considered the simplest class of galaxy to model and explain,
the assembly history of early type galaxies still presents many
puzzles.  Spectroscopic observations show that the most massive
examples completed their star formation earlier than that in their
less massive counterparts, in apparent contradiction to popularly-held
hierarchical models.
\end{quote}

What is being said is that larger galaxies were formed ealier, which
is obvious if there is no time zero to contend with, but which causes
serious problems when there is a time zero and the galaxy formation is
far too close to it!  The big bang theory has a strong analogy with
the \ind{flat earth theory}.  In terms of this analogy, in these
observations of very distant full-size spiral galaxies, cosmologists
are looking directly at the horizon and watching ships sailing over it
and still insisting that there is nothing beyond it.

So the big bang hypothesis is wrong and alternative explanations are
needed for the evidence that currently supports it.  There are three
so-called ``pillars'' of the big bang theory: redshift, the
distribution of \ind{light elements} in the universe and the
\ind{cosmic microwave background}.  Explanations for all three are
given in the next four sections.
\index{big bang theory|)}

\sh{The distribution of \ind{light elements}}\label{sec:dist}

The explanation for one of the pillars, the distribution of \ind{light
  elements} in the universe, has already been anticipated in outline
in \fullref{sec:pop}.  To recap, recall from
\fullref{sec:spiral_struc} that near the hypermassive black hole in
the centre of a spiral galaxy is an accretion structure called the
``belt'' or ``generator''.  It is extremely hot, being fed energy both
by accretion and by gravitational induction from the black hole, and
this causes a plasma of quarks to form near the black hole.  Moving
outward from the centre the temperature drops until ordinary ionised
matter starts to form.  This is exactly what happens in the standard
big bang model, except that it takes place over space and not time.
This produces the same mix of elements as in the big bang (H and He
and a trace of Li and other particles, with ratio by weight of H and
He roughly 3:1).  There is a level that is equivalent to the last
scattering surface in big bang theory where energy is radiated
outwards.  (An aside: this seems to be analogous to the Eddington
sphere in the quasar model of \fullref{sec:quasars}.)  Again as
explained in \fullref{sec:spiral_struc} the belt also emits the
streams of matter that feed the roots of the spiral arms (with the
same mix of light elements) and the residue of these streams, not
condensed into stars, escapes the galaxy and feeds the intergalactic
medium and this explains the observed proportion of these elements
which is the second of the three pillars.  Notice that the universe as
a whole is also cyclic with galaxies feeding the intergalactic medium
and also being fed by accretion from this medium.

\sh{De Sitter space}\label{sec:deS}
\index{de Sitter space|(}

To explain the other two pillars it is necessary to model the universe
as a whole.  The model is based on de Sitter space, the Hoyle Universe
mentioned in \fullref{sec:hist}.  \fullref{app:deS} is devoted to de
Sitter space, where properties, such as the fact, mentioned in
\fullref{sec:hist}, that all time-like geodesics are equivalent, are
proved, see \fullref{prop:PCP}.  The most elegant description of de
Sitter space (dentoted $\deS$) is that it is the analogue of a sphere
in Minkowski space (the simplest space in which relativity takes
place) in other words the set of points (events) at a fixed distance
from the origin in \emph{Minkowski 5--space} -- ordinary 5--space with
the Minkowski metric:
$$ds^2=-dt^2+dw^2+dx^2+dy^2+dz^2$$
\begin{figure}[ht!]\small
\SetLabels
\R\E(.06*.49) horizons\\
\endSetLabels
%\ShowGrid
\cl{\includegraphics[height=2.05in]{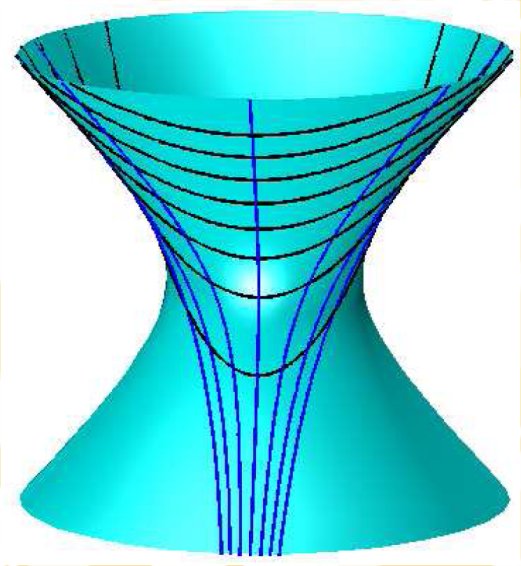}
\qquad\qquad\qquad
\AffixLabels{\includegraphics[height=2in]{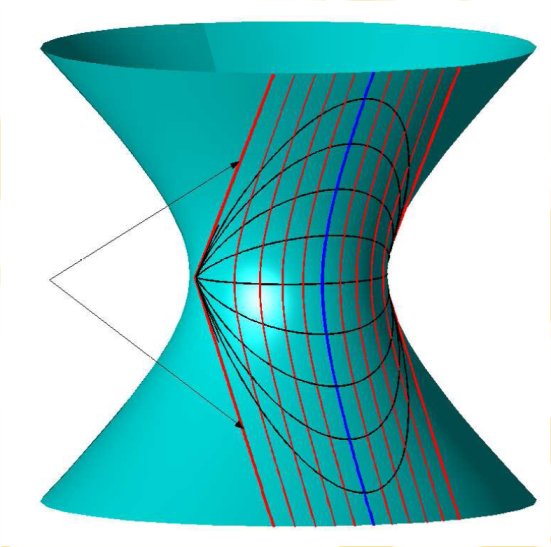}}}
\caption{Figures reproduced from \cite{Mosch} of de Sitter space
  showing the central expanding frame (left) and the de Sitter metric
  (right)}\label{fig:Mosch}
\end{figure}
This enables us to picture $\deS$ as a hyperboloid of revolution as
drawn in \fullref{fig:Mosch} for Minkowski 3--space.  This gives an
accurate represetation of the points (or events) in $\deS$ but a very
misleading idea of the metric.  As in special relativity, there are
motions, \emph{hyperbolic rotations} or \emph{shears} that move points
along rectangular hyperbolas (see \fullref{fig:grid}).  There is
a shear carrying the central horizontal circle into any ellipse given
by intersecting with a plane through the origin, so all of these are
equivalent.  The picture does give a good idea of the linear structure
of $\deS$, geodesics are given by planes through the origin, geodesic
2--spaces by 3--spaces through the origin etc.  The left picture in
\fullref{fig:Mosch} shows the expanding frame based on the central
(home) geodesic (blue) with transverse flat 3--spaces (1--spaces in
the picture).  Note that time and expansion are upwards in the
picture.  The right picture shows the de Sitter metric which has
horizontal geodesics (black ellipses) and transverse time-like flow
lines (orange) only one of which (central blue line) is a geodesic.
Thus although this is a static metric, the time-like flow is highly
unnatural.  This was an unhappy choice of a first metric for the space
and obscured its perfect symmetry (like the 4--sphere it has a 10
dimensional symmetry group).

\begin{figure}[ht!]\small
\cl{\SetLabels
\L(.6*.5) $O$\\
\L\E(.99*.99) future of $O$\\
\L\E(.81*.35) past of $O$\\
\endSetLabels
%\ShowGrid
\AffixLabels{\includegraphics[width = 2.5in]{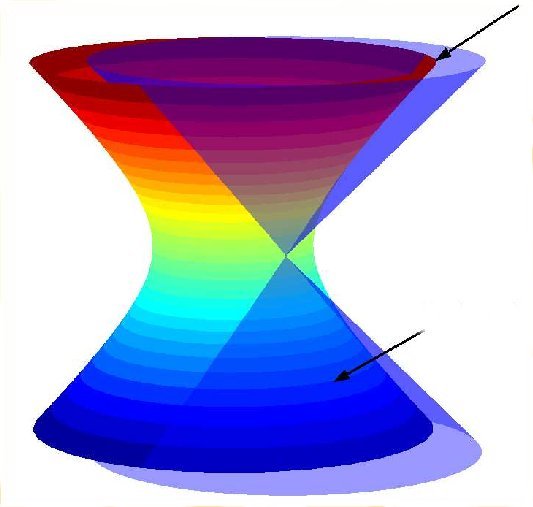}}}
\caption{Light-cones: the light-cone in $\deS$ is the cone on a
  $0$--sphere (two points) in the dimension illustrated, in fact it is
  the cone on a $2$--sphere.  The figure is reproduced from
  \cite{Mosch}.}\label{fig:light}
\end{figure}

Light paths in Minkowski space are straight lines at $45^\circ$ to the
vertical and light paths in de Sitter space are those light paths in
Minkowski space that lie in the hyperboloid, see \fullref{fig:light},
which illustrates a light cone in Minkowski space meeting de Sitter
space in a light cone.  Thinking projectively, a forward light line is
a tangent to the sphere $S_+$ at plus infinity (at the top) and a
backward light line is a tangent to the sphere $S_-$ at minus
infinity.\index{de Sitter space!light cones}

\ssh{Expansion}\index{de Sitter space!expansion}

\begin{figure}[ht!]
\labellist\small\hair 2pt
\pinlabel $A$ [b] <-2pt,0pt> at 224 295
\pinlabel $a$ [bl] <-2pt,-2pt> at 221 266
\pinlabel {horizon for $A$ at $a^*$} <0pt,2pt> [t] at 127 57
\pinlabel $a^*$ [bl] <-6pt,-1pt> at 162 163
\pinlabel $B$ [l] at 60 232
\pinlabel {\tiny $-\infty$} [tr] <5.5pt,0pt> at 46 101
\pinlabel $b$ [br] at 47 116
\pinlabel $S_-$ [br] <1pt,-1pt> at 230 23
\pinlabel $S_+$ [b] <0pt,0pt> at 425 276
\pinlabel {\tiny $\infty$} [b] <0pt,0pt> at 608 196
\pinlabel $a$ [tl] <0pt,0pt> at 607 185
\pinlabel $A$ [l] <0pt,0pt> at 595 69
\pinlabel $B$ [r] <0pt,0pt> at 421 11
\pinlabel $b$ [tl] <0pt,0pt> at 444 35
\pinlabel $b^*$ [tl] <0pt,0pt> at 490 142
\endlabellist
\cl{\includegraphics[width = 0.9\hsize]{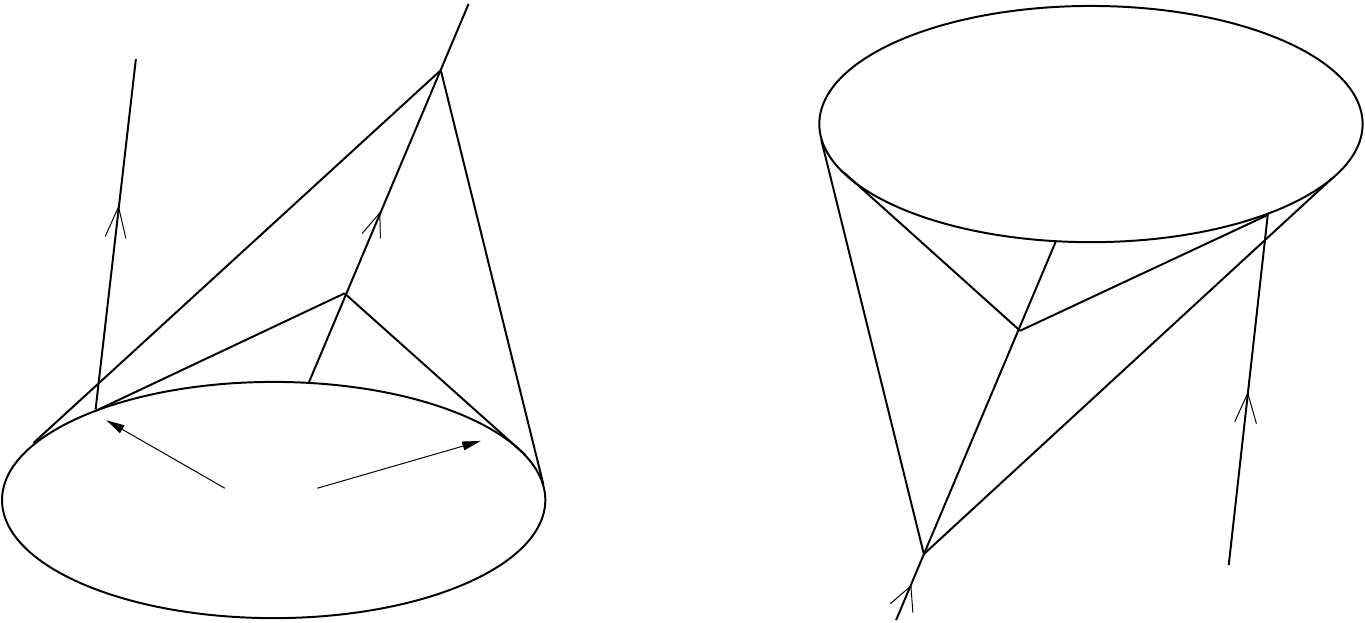}}
\caption{Left: first encounter\qquad\qquad\qquad Right: last contact\qquad} 
\label{fig:pics}
\end{figure}

De Sitter space has a very suggestive property: any two geodesics
which are not essentially the same (ie not both extensions of the same
geodesic interval) always eventually move apart in positive time at an
increasing rate which rapidly tends to exponential separation.  This
is exactly what happens in the current ``standard model'' where \emph{all}
geodesics allowed by Weyl's postulate move apart at all times.  But
the standard model starts with a Big Bang singularity and is very
unsymmetric.  Thus although perfectly symmetrical, and having no Big
Bang singularity, expansion is built into de Sitter space.  There is
a very interesting fact that follows from this: consider two observers
$A$ and $B$ moving on different geodesics.  Since they eventually move
apart faster than the speed of light, communication becomes
impossible.  Indeed there is a definite finite time $b^*$ for observer
$B$ after which a light path from $B$ cannot reach $A$.  But $A$ does
not see this happening: $A$ sees $B$ for ever in $A$'s time.  $B$
appears to be moving away faster and faster getting more and more
redshifted.  The moment $b^*$ when $B$ goes out of contact appears to
$A$ to be at time plus infinity.  This effect is illustrated in
\fullref{fig:pics} on the right.  The left-hand picture shows the dual
effect for geodesics coming into contact and will be described
shortly.\index{Weyl!coherency postulate}
\index{de Sitter space!projective pictures}

The pictures in \fullref{fig:pics} are projective pictures in which
the light spheres, which are at infinity in $\deS$, are represented by
finite spheres, and geodesics are represented by straight lines, with
null geodesics (light lines) tangent to the light spheres.  $S_+$ is
the forward light sphere and $S_-$ the backwards light sphere.  One
can think of these pictures as obtained by projecting from the origin
in Minkowski space onto a hyperplane not through the origin.  In the
pictures a typical light path from $B$ to $A$ leaving $B$ at time $b$
and arriving at $A$ at time $a$ is shown.  In the right-hand picture
the time $b^*$ for $B$ is the time when the light cone for $B$ meets
$S_+$ where $A$ meets it, and after this the future for $B$ does not
contain any points of $A$ and communication ceases.  For $A$ this
happens at time $+\infty$.

\ssh{Contraction}\index{de Sitter space!contraction}

\begin{figure}[ht!]\small
\cl{\includegraphics[height=3in]{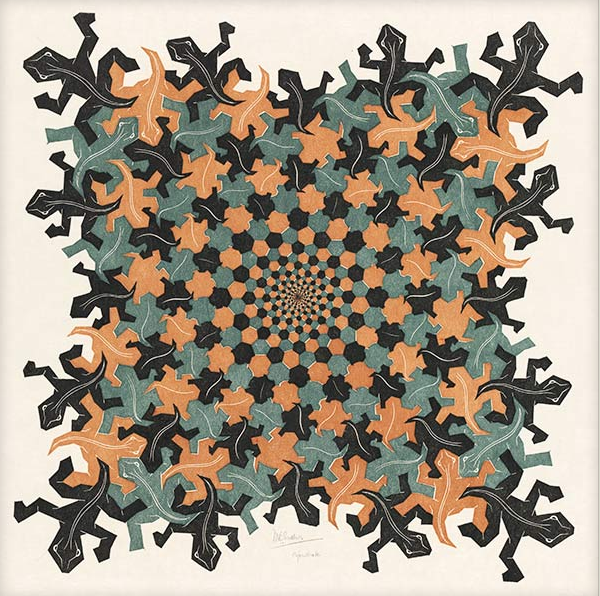}}
  \caption{MC Escher ``Development II'' from \cite{Escher}}
  \label{Escher}
\end{figure}

The perfect symmetry of $\deS$ implies that there is a dual effect for
backward time.  Backwards geodesics also separate exponentially and
there is a contracting frame based on any time-like geodesic
comprising geodesics which converge to the given geodesic in forward
time.  For a picture, take the left picture in \fullref{fig:Mosch}
and turn it top to bottom to get the contracting frame based on the
home geodesic.  Thus contraction is also built into de Sitter space
and the expanding and contracting frames fit together like the
different elements of an Esher
print, see eg \fullref{Escher}.
\index{Escher}

The dual effect to the ``last contact'' shown in \fullref{fig:pics}
(right) is the ``first encounter'' illustrated on the left.  The time
$a^*$ for $A$ is where the backwards light cone for $A$ crosses over
$B$ at time $-\infty$ for $B$.  In a very short period of $A$'s time,
$B$ bursts over $A$'s horizon and its entire history back to $-\infty$
is seen by $A$.  The dual of the infinite redshift at time $\infty$
for $A$ is an infinite blueshift at time $-\infty$ for $B$ and the
energy received by $A$ is a burst of energy starting in the high gamma
ray band and rapidly dropping to ordinary light and radio
waves.\fnote{This can be regarded as an extreme example of
  ``relativistic beaming''.}  This is qualitatively similar to
observed gamma ray bursts (GRBs) and MacKay--Rourke \cite{GRB} (see
also \fullref{app:GRB}) propose this as a possible explanation for
these, and the reader is directed to this paper and appendix for
fuller details and pictures.  The theory fits a lot of the facts but
has one problem.  If $B$ radiates uniformly then the energy received
by $A$ in the burst is infinite and it is necessary to make a
regularity hypothesis to reduce it to the finite bursts observed.
Shortly another way to avoid this problem will be proposed as a
consequence of the model for the cosmic microwave background (CMB).
But first it can be seen that the model already fits redshift
observations.\index{gamma ray bursts}

\sh{Redshift}\label{sec:red}
\index{redshift|(}

Assume that $A$ is our home geodesic and that $B$ is another geodesic
on which a light source (typically another galaxy, also labelled $B$)
is travelling.  Assume that $B$ is not gravitationally bound to our
galaxy, ie not part of the local group.  Typically there is a short
period starting with the gamma ray burst at time $a^*$ when we see the
light from $B$ blueshifted and $B$ appears to be travelling towards
us.  Then the blueshift decreases and gets replaced by redshift: the
galaxy appears to come as close to us as it can and then moves
steadily away.  The key behaviour is (1) the blueshift period is
typically very short (it can be arbitrarily short), (2) the movement
away rapidly tends towards exponential separation (ie an exact Hubble
law) and (3) we see this phase for the entire remaining lifetime of
the universe, which in de Sitter space is for an infinite time.  There
are graphs illustrating this in \fullref{fig:zrho} and \cite[2.5,
  2.6]{GRB}.

Now assume a uniform distribution of light sources radiating to us
from other geodesics.  All but a tiny number appear to be travelling
away from us and to fit closely to an exact Hubble law, because we see
the ones travelling away for an infinite time and the ones coming
towards us for a short time.  Thus almost all (in a measure-theoretic
sense) are close to an exact Hubble law.  It is natural to ignore the
few exceptions as outliers and decide that \emph{all} light sources
are on the Hubble law and \emph{the whole universe is expanding}.
It is an extreme example of ``Observer Selection Bias''.  We ignore the
evidence that the universe is also contracting because it is so slight
in comparison.  Indeed most of the energy from incoming sources (the
contracting bit) arrives in the form of GRBs and, as will be seen
shortly, as the energy that is thermalised and seen by us as the
cosmic microwave background.  Neither of these is thought to be
anything to do with expansion or contraction.  In fact the universe is
in a steady state and the expansion is a grand illusion.%
\index{de Sitter space!expansion!grand illusion}

Unlike the Bondi--Gold--Hoyle SST, the incoming matter that balances
expansion is not created; it does not appear from nowhere.  It is in
the universe all along. It is just that our vision is incomplete.
There is a similar incompleteness for forward time in both de Sitter
space and the standard big bang model.  Although a galaxy appears to
us to be there for ever and just getting further away and more
red-shifted, in fact it goes out of contact with us in a finite time
it its own frame. This view makes it clear that the standard model is
incomplete.  It is only the forward half of the picture.%
\index{steady state theory}\index{observer selection bias}

\textbf{Author's remark}\qua Observer selection bias as described
above was proposed by Robert MacKay during the collaboration leading
to \cite{redshift} as a possible explanation for redshift.  Although
not playing a part in the paper as published, my view now is that it
is the key idea in the correct explanation of redshift.
\index{de Sitter space|)}
\index{redshift|)}

\sh{Cosmic microwave background}\label{sec:CMB}%
\index{cosmic microwave background|(}

So far a model for the universe has been found that fits observed
redshift and (with some reservations) GRBs.  This section covers the
CMB.

The CMB is a highly isotropic thermal radiation field that appears to
emanate from every part of the sky.  In particular it comes from the
apparently dark background where there are no visible stars or
galaxies, hereinafter called the ``dark horizon''.  It is thermal to
better than 1 part in $10^5$ and has a temperature of between 2.725K
and 2.726K.  It is apparently anchored in the Machian rest frame
determined by distant galaxies.  Motion of the earth with respect to
this frame (approx 371km/s towards Leo) can be detected accurately
from dipole anisotropy.  There are small fluctuations in temperature
which, under standard big bang theory, come from quantum fluctuations
in the inflation field hypothesised to have smoothed out the universe
when it was very small.\fnote{The author's opinion is that this part
  of the big bang story is fantasy physics on a par with Arp's
  explanations for redshift reduction with growth in quasars.}

\begin{figure}[ht!]\small
\cl{\includegraphics[width=.9\hsize]{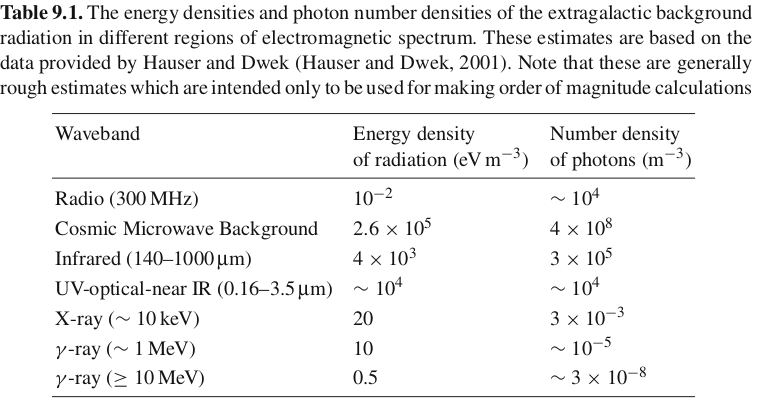}}
  \caption{Table reproduced from \cite{Longair}}
  \label{Longair-table}
\end{figure}

Although typically described as a weak radiation field, the CMB is
about 45 times more energetic than the background starlight field, see
\fullref{Longair-table}.

\ssh{The gravitational fog horizon}

To understand the observed CMB it is necessary to use the fact that
the universe is filled with a ``noise'' of low-level gravitational
waves (arising from the inertial drag fields hypothesised in
\fullref{sec:Sciama}, the relative motions of the massive black holes
at the centres of galaxies, and other sources) which can be observed
in the systematic distortion in distant images seen by the Hubble
space telescope.

This telescope is powerful enough to look back almost to the Big Bang
or, since the Big Bang is not part of this story, to about 1 Hubble
distance, 13.8 billion light years.  In 2003 the Hubble telescope was
pointed at a dark part of sky, where it is possible to see back nearly
one Hubble distance, and left running for a long time.  The resulting
``Hubble ultra-deep field'' (HUDF) is published on the web and
contains a wealth of information.  The most important information for
understanding the CMB is that space-time is not uniform at these
scales.  You can look at the HUDF and see optical distortion due to a
low-level gravitational wave noise signal that fills the universe; see
\fullref{app:HUDF} for details.

This implies that light cannot travel more than a definite finite
distance (about 1 Hubble distance) before it is diverted significantly
from its original direction.  There is an apparent boundary like the
apparent boundary in a fog where light is comprehensively scattered.
So there is a natural horizon where light particles are randomised in
intensity and direction (and, as will be seen shortly, in frequency as
well) by the gravitational fog.  This is what has been called the
``dark horizon''.  The CMB comes from there.

Observations of the CMB carefully correct for all visible galaxies and
other radiation sources, and therefore select this apparent boundary.
Light (as seen by us) travels in typically random ways near the dark
horizon.  It is very unlikely to travel far in any one direction.  It
is, if you like, a random walk.  Thus nearly all the light crossing
the horizon from the far side, and heading towards us, has source
within a region of depth about $1/3$ Hubble distance, $R$ say, behind
the boundary.  Now light passing through a gravitational wave field
exchanges energy with the field.  This is by a process similar to the
Rees-Sciama effect (a special case of the Sachs-Wolfe effect
\cite{wiki:SW}).  The description given in \cite{wiki:SW} can be
expressed in our context as follows.  Think of a random gravitational
wave field as a sequence of gravitational wells and hills which vary
over time.  A photon that goes down a gravity well and then emerges
after the well has become shallower gains energy from the field and
increases its frequency (and vice versa).  And similarly a photon
entering a hill expends energy but does not get all of it back if the
hill becomes smaller before it exits.  Thus light is randomised both
in direction, intensity and frequency by the gravitational wave field
and this is a perfect scenario for thermalisation.

The region $R$ contains about $3\times 10^{10}$ galaxies all
emitting roughly thermally at about 3000K and, after mixing and
thermalisation, the radiation emitted from $R$ in our direction is a
near perfect black body spectrum again of temperature about 3000K.

Now the fog horizon is fixed with respect to us (it is determined by
the distance that we can see clearly) and therefore is part of our
expanding frame, so the radiation is subject to cosmological redshift.
The horizon is about 1 Hubble distance away and the effect of this
redshift (a little over $z=1000$) is to reduce the temperature to the
observed CMB temperature of approximately 2.7K.  This reduction is
exactly the same as in the standard big bang model.

The CMB energy coming in this way from the starlight field near the
fog horizon is not energetic enough (by a factor of about 45, see
\fullref{Longair-table}) and to correct this it is necessary to take
account of the radiation coming from incoming sources in the
contracting frame.  The bulk of this radiation is not subject to
extreme blueshift and contributes, at about the same temperature, to
that which is thermalised by the fog horizon.  This energy boost from
incoming sources accounts for the high level of energy in the CMB
compared to other extraglactic fields.  One can think of this extra
energy as providing a backlight for the horizon.

%[THIS NEEDS PROPER MODELLING]

\rk{Remark}As it is about 1 Hubble distance away, it is suggested that
the dark or fog horizon, where the CMB arises, is called the
\emph{Hubble horizon} in contrast to the infinitely distant \emph{de
  Sitter horizon} used in \fullref{fig:pics}.

\ssh{Horizon effect and dipole anisotropy}

The Hubble horizon is indeed a ``horizon'': a virtual barrier caused
by the behaviour of light.  As such it depends on the observer.  It is
fixed with respect to the observer and therefore the apparent
radiation (the CMB) should be perfectly isotropic (like black body
Hawking radiation at the de Sitter horizon \cite{GH}).  But in the
overview (above) it was mentioned that there is a dipole anisotropy
caused by the motion of the earth.  The explanation for this apparent
contradiction is that the radiation takes a very long time to travel
from the horizon to us.  About one Hubble time in fact.  During this
travel time, the motion of the earth can have (and obviously has)
changed.

\ssh{Quantum fluctuations?}

\begin{figure}[ht!]\centering
\includegraphics[width=.8\hsize]{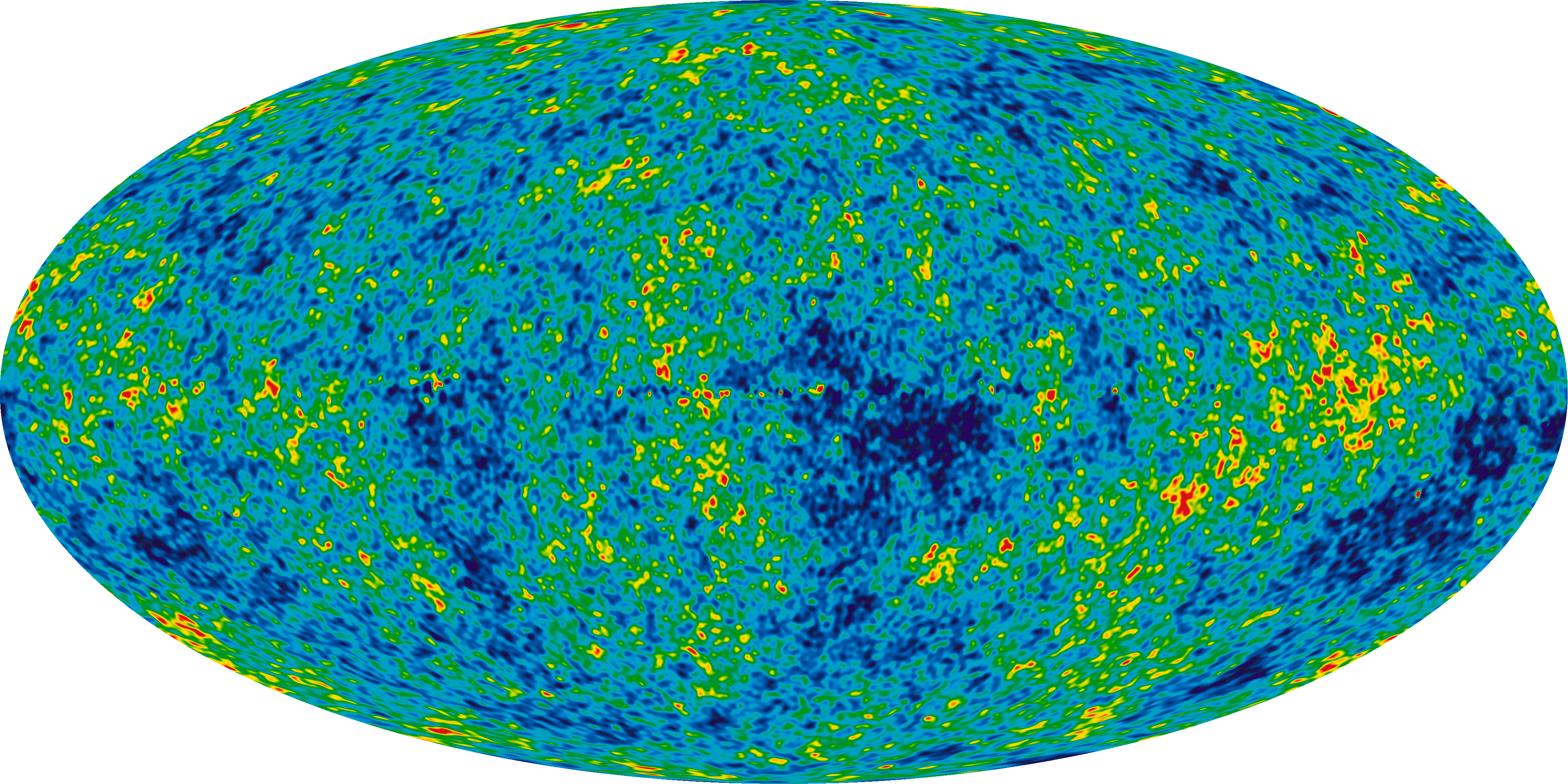}
\caption{CMB temperature map}\label{fig:CMB-temp-map}
\end{figure}

\begin{figure}[ht!]\centering
\includegraphics[width=.5\hsize]{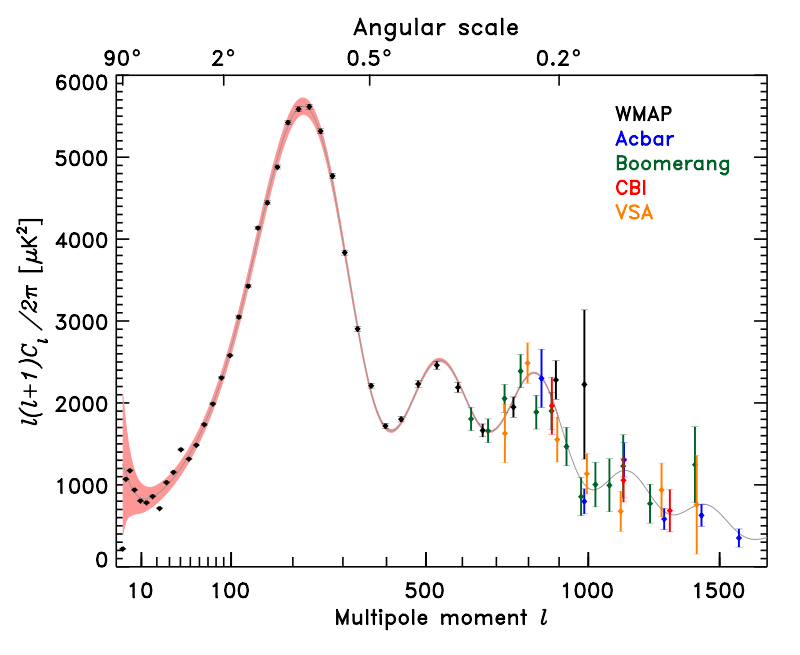}
\caption{CMB Power Spectrum}\label{fig:CMB-Power-Spectrum}
\end{figure}

Now turn to the small fluctuations in power (or equivalently
temperature) observed by many experiments and ascribed to quantum
fluctuations in the inflation field as remarked earlier.  These are
illustrated in the CMB temperature map \cite{map} reproduced in
\fullref{fig:CMB-temp-map}.  This map uses Mollweide projection and is
an amalgam of nine years of data from the WMAP satellite.  It shows
the small variations in temperature of the CMB of the order of
20$\mu$K.  In this map, hotspots are red and yellow and cool areas
green and blue.  

\ssh{The multipole power spectrum}

The power spectrum is usually analysed using spherical harmonics and
this results in the beautiful resonance curve reproduced in
\fullref{fig:CMB-Power-Spectrum}.  Here are some ideas for the origin
of this curve.  Longair has this to say (near the end of section 1.6
on page 22):

\textit{``Equally impressive was the fact that the determination of the
two-point correlation function for galaxies determined from the
large-scale galaxy surveys now overlapped the corresponding angular
scales in the Cosmic Microwave Background Radiation and that these
were in excellent agreement. In the most recent analyses, evidence has
been found for a maximum in the two-point correlation function for
galaxies corresponding to the first peak in the power spectrum of
perturbations in the Cosmic Microwave Background Radiation....''}

The distribution of galaxies as seen by us has a great deal of
regularity.  For example $1^\circ$ is a key angular measurement,
correlated with the likely separation of nearby galaxies and it is
this fact that Longair is referring to in the quotation above.  Thus
the anisotropies in the CMB are likely to be caused by the passage of
the radiation through the galactic field on its way to us from the
Hubble horizon. This is a more natural explanation than the standard
model which has both anisotropies (in the galactic field and the CMB)
caused by the same fantasy physical quantum fluctuations in the
``inflationary field''.  This leaves open the problem of why the
anisotropies exhibit an (aparently acoustic) resonance curve
\fullref{fig:CMB-Power-Spectrum}. There must be something in the large
scale geometry of the universe causing this. But it causes the
distribution of galaxies to have this partial regularity NOT the CMB.
We observe it in the CMB because it is modulated by the galactic field
on its way to our receivers.  Another possibility is that the inertial
drag fields from all the rotating black holes at the centres of
galaxies form a standing wave field that causes the regularity
observed in the power spectrum.

The writer of the wiki page at \cite{wiki:CMB} comments: ``Although
many different processes might produce the general form of a black
body spectrum, no model other than the Big Bang has yet explained the
fluctuations.  As a result, most cosmologists consider the Big Bang
model of the universe to be the best explanation for the CMB.''  The
model proposed here explains these fluctuations.  In order for the big
bang to explain them, a physically implausible hypothesis is needed
(inflation) which breaks nearly all the basic laws of physics;
consequently it is the big bang explanation which lacks credibility.%
\index{cosmic microwave background|)}

\sh{Redshift and GRBs revisited}\label{sec:revis}

Now return to the explanations for redshift and for GRBs given
earlier to see how the CMB explanation has affected them.  Redshift is
quickly dealt with.  Because of the background gravitational wave fog
effect we only see the redshift phase for a \emph{very long time}
(about the current estimate for the age on the universe) and not for
an \emph{infinite} time.  Thus it remains true that nearly all sources
exhibit redshift close to an exact Hubble law and the explanation is
unchanged.

Turning next to GRBs, the infinite blueshift and infinite received
energy are both regularised by the fact that light cannot travel more
than about 1 Hubble distance without being diverted.  The simplest way
to see what happens is to use the duality between expansion and
contraction mentioned in \fullref{sec:deS}.  An outgoing source of
light appears to to be travelling directly away from us and to reach
the Hubble horizon after about 1 Hubble time, and has redshift of
about $z=1000$ when it disappears from view.  Dually an incoming
source pointed directly at us comes bursting over our horizon with a
blueshift of about $z=1000$ (not infinite) and appropriate power.
This is still a very bright burst of energy (a temperature increase of
a factor of $10^3$ and power increase of $10^6$) which fits
observations well.  For GRBs not directed exactly at us, the
enhancements will be rather less.  Indeed the bulk of the incoming
sources will not reach the Hubble horizon and remain beyond it and
contribute to the backlighting effect mentioned earlier.

Here is a quick stab at a calculation.  An incoming source of real
temperature T will cause a GRB starting at up to about 1000T.  This is
quite reasonable: a source of temp about $10^6$K (temp of sun) would
appear to be about $10^9$K which corresponds (using a black body
model) to a wavelength of $3\times10^{-12}$m or $10^{20}$Hz which is
right in the gamma ray band which wiki says is $10^{19}$ Hz and up.

Now not all sources of light that come over the de Sitter horizon are
visible. To be observable they have to come over the Hubble horizon as
well. So there is a (perhaps huge) source of light outside the near
horizon and this feeds this horizon with energy that is re-radiated to
us as the CMB (which has about 45 times the energy of background light
coming from all visible galaxies). The near horizon is ``backlit'' and
it is this backlighting that is observed as the CMB.

\sh{Origin of life}\label{sec:life}

The model for a galaxy has a built-in cyclic nature with solar systems
created by condensation in the arms out of a mixture of the clean gas
stream from the centre and the dust and debris left from stellar
explosions and present as background throughout the galaxy, then
living their lives whilst moving out into the outer dark regions of
the galaxy and finally gravitating back into the centre to be
recycled.

The timescale is huge.  Probably several orders of magnitude greater
than current estimates of the age of the universe.  This is plenty of
time for life to have arisen many times over on suitable planets.
When these planets are destroyed by tidal disruption as they fall into
the centre or by breaking up in collision with other objects, many of
the molecules will survive and become part of the background dust out
of which new planets are made.  Thus in a steady state planets will
start out seeded with molecules (probably in the form of very hardy
viruses) which will help to start life over again.  Indeed standard
selection processes over a galactic timescale will favour lifeforms
which can arise easily from the debris left over from the destruction
of their planetary homes.  This might explain how life arose on earth
rather more quickly than totally random processes can explain.
\index{evolution!galactic timescale}

There is evidence for this in the long-chain hydrocarbon molecules
that are in fact found in meteorites, and in the observations of
\ind{Hoyle and Wickramasinghe} (see for example \cite{HW}).  Indeed
the model proposed here is fully consistent with their ideas on the
present cosmic origin of micro-organisms.

\index{quasar-galaxy spectrum|(}
\sh{The quasar--galaxy spectrum}\label{sec:spec}

This section returns to perhaps the most important consequence of the
ideas presented in this book and probably the best way to understand
them.  The new model for galaxies fits ordinary galaxies into a
spectrum of black hole based phenomena which includes quasars and
``active'' galaxies.  The spectrum is conveniently ordered by the mass
of the central black hole.  As a very rough guide (in solar masses)
these range from $10^{7}$ or less to $10^{14}$ or more as follows:

Quasars: from $10^{7}$ or less to $10^{9}$ 

``Active'' galaxies: $10^{9}$ to $10^{11}$ approximately

``Normal'' galaxies: $10^{11}$ to $10^{14}$ or more

All are highly active.

This spectrum has been discussed several times through the book and in
the remainder of the chapter the main features are recollected.  The
first important point was mentioned very early in the book: quasars
typically exhibit very large intrinsic (gravitational) redshifts as
seen in observations of Halton Arp and others.  Before proceeding it
is worth looking briefly at another proof based on observations of
this fact.

\ssh{The Hawkins paper}\label{sec:hawk}
\index{Hawkins!paper|(}
\index{quasar!Hawkins paper|(}

An independent proof of the existence of gravitational (intrinsic)
redshift in quasars is provided by a paper of Hawkins \cite{H}, which
sets out to prove that quasars show redshift without time dilation (an
impossibility since redshift and time dilation are identical in
relativity and indeed in any metrical space-time theory), but in fact
decisively proves that much of the redshift observed in quasars is
intrinsic.  For full details here, see the paper \cite{HTD} on the
author's web page; what follows is a quick sketch of the arguments.

Hawkins examines a large pool of observations of quasars.  As has been
mentioned before, radiation from quasars typically varies in intensity
periodically over macroscopic time intervals from days to
years.\index{quasar!variability}  He
makes a very careful selection from the pool (some more detail on this
will be given later) and uses some very sophisticated analysis (which
seems sound) to find a collection of quasars for which the macroscopic
intensity variation does not exhibit time dilation correlated
correctly with the observed redshift; indeed for this selection, the
high redshift and the low redshift bins exhibit on average {\em
  exactly the same time dilation}.  For full details, see \cite{H}.

This result is not paradoxical.  What it shows is that for (a large
subset of) this selection of quasars the sources of

(a)\qua the radiation\qquad and\qquad
(b)\qua the time variation

are not in the same place.  To enable discussion let us call these the
{\em generator} and the {\em modulator} respectively.  For the Hawkins
sample, these must be subject to different redshifts, either
cosmological or gravitational or a combination, with the modulator
having lower redshift.
\index{Hawkins!generator}\index{Hawkins!modulator}

There are two possibilities:

(A)\qua The intrinsic redshift arrangement
\index{intrinsic redshift arrangement}

Both are part of the same object (the quasar) and therefore both at
roughly the same distance from us.  This implies that the larger
redshift (affecting the generator) is partly gravitational due to a
nearby mass and that the modulator is further from the large mass and
subject to a lower gravitational redshift.

This arrangement is precisly how the three-author-model described in
\fullref{sec:quasars} and \fullref{app:3author} works.  The generator
is the Eddington sphere dividing the inner optically thick region from
the outer optically thin region.  No direct radiation comes from
inside the Eddington sphere.  The outer region contains strata of gas
or plasma and further out there may be dust or more solid objects, all
of which will typically be trapped in orbit around the central mass.
The radiation from the generator passes through the surrounding layers
on its way to us; the observed variations are due to non uniformity in
these layers, and are naturally periodic with the possibility of
several different periods coming from different layers superimposed.
This is what is observed.  Furthermore there is direct evidence for
these layers in the \ind{Lyman-alpha-forest} that is observed for some high
redshift quasars, see \fullref{sec:conc}.

(B)\qua The microlensing arrangement
\index{Hawkins!microlensing arrangement}

It is clear that the modulator must be on the light path from the
generator to us.  It does not need to be directly associated with the
generator, as in the instrinsic redshift arrangement discussed above,
anywhere on the path will do, provided it lies in a region of lower
cosmological redshift.  One way variations in intensity could arise
would be if the path were subject to variable gravitational lensing
effects or passing through a region of variable density.  Both of
these phenomena are called {\em microlensing}.  There are indeed cases
where this is known to happen (see eg Schild et al \cite{Schild}) and
if this happened to a large proportion of quasars then it would also
explain the Hawkins result.

But is this plausible?  It is not the existence of microlensing that is
in doubt but its pervasiveness.  It would be necessary to assume that
there is a microlensing region happening to lie on the light path from
{\em most quasars to us} and {\em close to us} as well.  This is
highly implausible unless nearly all space acts a microlensing region,
eg if it is filled with suitable gravitational waves.  There is indeed
evidence for a gravitational wave field affecting distant
observations, see \fullref{app:HUDF}, but if this background field
were strong enough to account for observed quasar variation then
everything distant would have similar patterns of variation and no
such variation has been observed for distant galaxies.

The only other way this could work would be if quasars were defined by
the existence of a suitable microlensing region on the path to us.
In other words if quasars were in fact distorted images of distant
galaxies.  But this possibility is again implausible because quasars
have quite different radiation characteristics which could not be
disguised by microlensing.  So although apparently suitable as an
explanation for the Hawkins result, microlensing has to be discarded,
and the only remaining possibility is that a proportion of quasars in
the sample have intrinsic redshift.

A basic question now arises.  For any random sample of objects in the
universe (which for the puposes of this discussion is assumed to be
the standard expanding universe of current cosmology) there should be
a correlation between redshift and time dilation whatever the
mechanism that produces these locally.  This is because the more
distant objects will have both higher redshift, with the addition of
cosmological redshift, and higher time dilation for the same reason.
Hawkins has managed to find a sample which does not have this
property.  Obviously he must have used a non-random selection
criterion at some point.  And indeed he has.  In an attempt to avoid
the effect of another well-known correlation, between magnitude and
redshift in flux limited samples, he has limited his sample to a very
small magnitude range namely between magnitudes $-25.5$ and $-22.5$.
This narrow sample contains high redshift quasars which have low
luminosity and are close to us, and low redshift quasars with high
luminosity which are distant.  The former, being close to us, are
subject to small cosmological time dilation effects and the latter to
large ones.  Thus the redshift--time dilation relation is skewed
against the natural cosmological relation by the presence of these
quasars whose redshift--time dilation is opposite to the natural
relation, and this accounts for the redshifts in the sample not having
the expected correlation with time dilation.
\index{Hawkins!selection}

In passing, it is worth remarking that the well-known correlation
(between magnitude and redshift in flux limited samples) mentioned
above is probably due to observer selection bias.  Most quasars are
probably based around quite small black holes and the nearby ones (ie
the ones with greatest magnitudes) will be the easiest to detect.  The
flux limitation eliminates the nearby ones with low (intrinsic)
redshift and very high magnitude.  Thus in any given flux limited
sample, the higher magnitude quasars are more likely to be the nearby
ones with high (intrinsic) redshift.  \index{Hawkins!paper|)}
\index{quasar!Hawkins paper|)}\index{observer selection bias}

\ssh{Quasars and redshift}

\index{quasar!redshift}
Now return to the main topic, namely the quasar-galaxy spectrum,
starting with quasars.

Quasars have been covered thoroughly in \fullref{sec:quasars} and the
associated \fullref{app:3author}.  Briefly, black holes aka quasars
accrete matter from the surrounding medium and grow in mass.  The key
surface is the \ind{Eddington sphere} which is analogous to the
photosphere of a star.  Inside the Eddington sphere is the
\emph{active} region where radiation is produced by interaction
between infalling particles.  This region is optically thick and only
the boundary (the Eddington sphere) is visible and is where the
radiation that is received comes from.  The Eddington sphere can be
very close to the event horizon and consequently subject to an
arbitrarily high gravitational redshift and this accounts for the
observed high intrinsic redshifts in some quasars.  Because of the
attenuation effect on power output of gravitational redshift [a factor
  $(1+z)^{-2}$], the effective power output from the quasar can be far
lower than the Eddington limit.  Thus small quasars have both high
(insrinsic) redshifts and low luminosity.  One known example here is
Sagittarius \astar\ which has luminosity only $10^{-8}$ of the
Eddington limit and corresponding redshift of $10^4$ (which
incidentally is why this quasar was first detected as a radio source).
But in general quasars of very high redshift are unlikely to be
detected because of their low power output.

As the mass grows with accretion, the distance between the event
horizon and the Eddington sphere increases and gravitational redshift
decreases.  At the same time the central black hole gets increasingly
masked by the accreting matter and more difficult to detect and
measure.  A large black hole tends to accumulate a thick inner region
which masks it from the outside and allows the redshift to be very
small, and conversely a small black hole has only a thin inner region
and a large redshift.  This natural effect explains why the active
nature of normal (spiral) galaxies has not been directly observed.  In
a full size galaxy the central black hole is effectively shielded from
view and the visible matter near the centre (the bulge) is
sufficiently remote from this black hole that the usual way of
estimating the central mass, using the virial theorem, does not yield
any information.

Thus the huge black holes which power the dynamics of spiral galaxies
(see \fullref{sec:spiral_struc}) have not been directly detected and
this is why the false assumption that Sgr\astar\ is the central black
hole for the Milky Way has not been questioned before.

\ssh{Quasars and active galaxies}
\index{quasar!morphing into active galaxy}
  
Moving back down the spectrum to quasars.  The smooth accretion of
matter that happens for small quasars breaks down as the size rises to
about $10^9$ solar masses.  The outer settling region becomes
increasingly chaotic and the smooth accretion of matter into the
central black hole stops.  Matter trapped near the black hole now has
no option but to form a rotating structure (called an accretion disc)
as hypothesised in mainstream quasar theory.  This is the start of the
``active galaxy'' stage for which accretion discs and associated jets
have been directly observed.

The accretion disc continues to grow as the mass of the quasar/galaxy
continues to increase by accretion.

\ssh{Active and spiral galaxies}
\index{galaxy!active morphing into spiral}

Once sufficient matter is trapped in the rotating accretion disc it
begins to collect a significant amount of angular momentum.  Since the
total angular momentum is small (just collected from the pool in the
surrounding medium) the central black hole must rotate the other way
to acheive a balance.  So there is now a rotating black hole with an
orbiting structure, which rotates the other way, and which is referred
to here as the \emph{belt} to emphasise its likely toroidal shape.
For definiteness call the rotation of the inner black hole
``positive'' and that of the belt ``negative''.

Jets produced by the belt will cause negative angular momentum to be
lost to the system and increase the main positive rotation.  There is
now a significant \id\ effect from the rotating black hole which
increases the effective energy in the belt which becomes increasingly
hot.  A stable pair of opposite jets form and feed the roots of the
spiral arms which are now growing.  The belt is now the
\emph{generator} for the spiral structure.  Stars form in the spiral
arms and the whole galaxy radiates into the surrounding space and thus
a limit in size is reached when the radiation balances accretion.
From the model constructed in \fullref{sec:spiral_struc} the limiting
size seems to be around $10^{14}$ solar masses.  

\ssh{The predominant life-form of the universe}\label{sec:lords}
\index{quasar-galaxy spectrum!possible life-form}

It has been seen that quasars, ``active'' galaxies and larger spiral
galaxies are all based around black holes, and that there is a natural
way to suppose that these objects evolve over an extremely long
timescale with points of the spectrum representing different ages of
the same class of objects.  A black hole grows with time by absorbing
matter falling into its gravitational domain and the corresponding
object moves along the spectrum.  Moreover observations of Halton Arp
and others \cite{Arp} suggest that quasar/galaxies have the basic
property of a life-form: reproduction.  Quasars are often closely
connected with parent galaxies and the natural supposition is that
they have been ejected from them, for example, as has been mentioned
earlier, \fullref{fig:NGC7603} shows what could be a family grouping
of two parent galaxies and two offspring quasars.

Arp's observations also show (intrinsic) redshift decreasing with age
which is consistent with the model given in this book where the larger
the central black hole, the smaller the redshift.  (Arp suggests some
outlandish theories to explain this observation, which are quite
unnecessary.)

So a quasar starts life as a comparatively small black hole which
grows heavier with age.  When it reaches the mass of an active galaxy
it starts to throw out small black holes (quasars).  This is the
reproductive stage.  Later it grows into a full size spiral galaxy and
reproduction stops.  Presumably, if it could be recognised, there is a
final senile stage when the black hole disconnects from our space and
the associated galaxy radiates away.

Finally it is worth remarking that nothing whatever is known about the
inner nature of so-called ``black holes''.  There is no such thing in
nature as a singularity; black hole is simply the name given to
another state of matter about which nothing is yet known.  There are
some fascinating observations due to Schild et al \cite{Schild} which
hint at a specific inner structure and which may perhaps shed some
light here. Or perhaps by observing galactic clusters carefully it may
be possible to deduce some of the rules governing this new state of
matter---perhaps to begin to build up a proper physics for black
holes.  One point that needs to be addressed is why galactic centres
are not even more massive.  Black holes can combine to become more
massive.  So perhaps there should have arisen a set of super size
galaxies grazing on ordinary ones etc.  This does not appear to have
happened.  Why?
The reason may be the mechanism described in
\fullref{sec:spiral_struc} which limits size by boiling off excess
matter, or the mechanism may be more elementary.  Black holes over a
certain mass may simply be unstable and spontaneously break up.

\ssh{The lords of the universe}

The main part of the book (before the appendices) finishes with some
wild speculations.  It is natural to think of these black hole based
phenomena as part of \emph{our} universe, but now turn the whole
discussion over and try to see the universe from the point of view of
these, the real inhabitants and creators.  For them this airy space
full of stars and planets must seem just a dream compared with the
solid reality of their being.  From the coincidence observed by Sciama
(equation \ref{eq:Unnorm}) the rough volume of space is a simple
function of the mass of these black holes, as if space is a property
of them.  Further these are the heavy weights which cause the
curvature of space-time that gives the illusion of expansion.  Indeed
they carry nearly all the mass of the universe.
\index{fundamental relation!cosmological constants}

So the natural view from their point of view is that space is a
property of their being.  They create space as we create our dreams.
\index{quasar-galaxy spectrum|)}
\np\thispagestyle{empty}

\appendix
\renewcommand{\chaptermark}[1]{\markboth{Appendix \thechapter\qua #1}{}}

\chapter{Introduction to relativity}\label{app:beginners}

\index{Einstein!special relativity|(}\index{relativity!special|(}
\section*{Special
  relativity}\label{sec:SR}\addcontentsline{toc}{section}{Special
  relativity} 

\section{Causality}\label{subsec:cause}
\index{causality}

{\sl\parskip0pt\obeylines\small
\hfill The Moving Finger writes; and, having writ, 
\hfill Moves on: nor all thy Piety nor Wit, 
\hfill Shall lure it back to cancel half a Line, 
\hfill Nor all thy Tears wash out a Word of it.
\par}
\bigskip

\index{Khayam, Omar}
This is not a book of philosophy nor of poetry.  It is a book of
geometry.  So why has this appendix started with a famous
philosophical poem?  It is because this poem expresses with great
clarity the idea of ``causality'' which is the basis of relativity,
the natural geometry of the universe.  The essence of this famous
quatrain is that the past cannot be altered, cannot be affected by
anything that comes after it.  Nothing that happens in the present
moment can affect any time other than the future.

The most basic concept of relativity is of an \emph{event}.  An event
just means something that has a definite place \emph{where} it happens
and time \emph{when} it happens.  Causality is the relationship
between two events that the first event might affect the second.  The
quatrain says that, for causality to hold, the first event must
precede the other in time.

Here is a contemporary example of causality in action.  On the 11th of
September 2001, in an act of unprecedented evil, two enormous and
immensely strong skyscrapers were intentionally demolished with large
numbers of people trapped inside.  This event has affected almost
every aspect of the current political environment of the whole world.
But this influence has only been felt \emph{after} this event.  At no
time \emph{before} this event was there the slightest foretaste of the
consequent loss of freedom and demonisation of sections of our
communities.  Events only affect the future.  This is what causality
means.

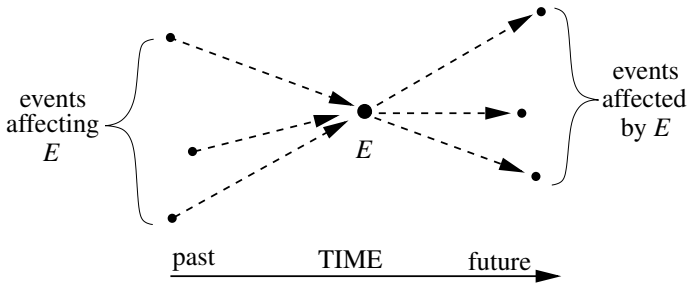
\begin{figure}[ht!]
\labellist
\small\hair 2pt
 \pinlabel events [r] <-5pt,10pt> at 1 160
 \pinlabel affecting [r] <0pt,0pt> at 1 160
 \pinlabel $E$ [r] <-15pt,-10pt> at 1 160
 \pinlabel $E$ [t] <0pt,-10pt> at 266 177
 \pinlabel events  [l] <4pt,10pt> at 504 188
 \pinlabel affected [l] <0pt,0pt> at 504 188
 \pinlabel {by $E$} [l] <7pt,-12pt> at 504 188
 \pinlabel past [b] <0pt,0pt> at 92 8
 \pinlabel TIME [b] <0pt,0pt> at 252 10
 \pinlabel future [b] <-5pt,0pt> at 418 8
\endlabellist
\centering
\includegraphics[width=.5\hsize]{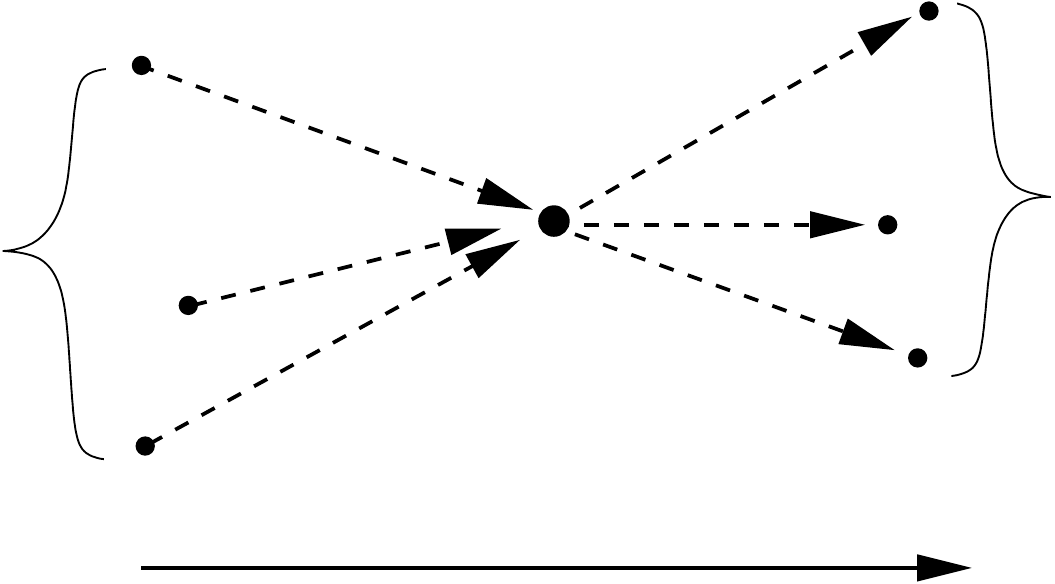}
\caption{Event $E$ is affected by events in the past and affects other
events in the future.}
\label{fig:cause}
\end{figure}

\fullref{fig:cause} is a basic diagram of causality.

Relativists regard time as a dimension of exactly the same quality as
space and they consider the two to be intermingled; they use the
portmanteau word ``\ind{space-time}'' for this intermingling.  A space-time
is a collection of events, ie points in space and time.  Each
event has both a time coordinate (\emph{when} it happens) and a space
coordinate (\emph{where} it happens).  In \fullref{fig:cause} you can
think of the vertical axis (which hasn't been labelled) as space and
then this diagram is a simple example of a ``space-time diagram''.
\index{space-time!diagram}

Another basic concept that will be used repeatedly is that of an
\emph{\ind{observer}}.  An observer just means the idealised path through
space-time of a person.  For each point in time the observer has a
definite position in space, in other words for each time there is an
event, namely that corresponding point in space-time.  The collection
of these points is called the \emph{\ind{world-line}} of the observer.  At
any point on this world-line there is one direction (along the
world-line) which \emph{appears} to be time and the perpendicular
directions \emph{appear} to be space.  But this split into space and
time depends on the world-line.  A different observer will see a
different split.  This is a fundamental point of relativity:

\medskip
\centerline{\textbf{Space and time are relative concepts which depend
    on the observer.}}\index{relativity!of space and time}

\begin{figure}[ht!]
\labellist
\small\hair 5pt
 \pinlabel A [t] <2pt,0pt> at 164 92
 \pinlabel B [b] <0pt,0pt> at 175 30
\hair 2pt
 \pinlabel {world-line of A} [b] <0pt,3pt> at 59 83
 \pinlabel {time for A} [tl] <0pt,0pt> at 195 101
 \pinlabel {world-line of B} [t] <0pt,0pt> at 54 21
 \pinlabel {time for B} [bl] <0pt,2pt> at 212 22
 \pinlabel {space for A} [r] <0pt,0pt> at 158 122
 \pinlabel {space for B} [r] <3pt,0pt> at 168 5
\endlabellist
\centering
\includegraphics[width=.7\hsize]{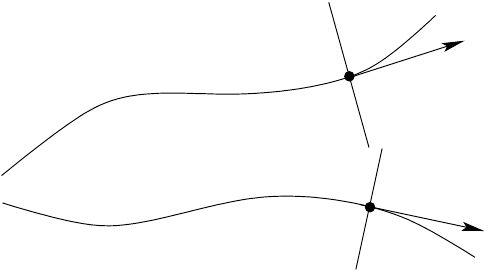}
\caption{Relativity of space and time: A and B are observers}
\label{fig:rel}
\end{figure}
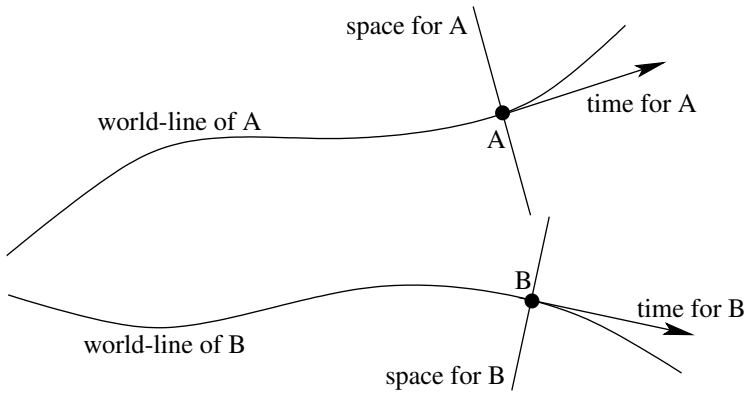

\section{The \ind{speed of light} and \ind{Michelson--Morley}}\label{subsec:MM}

As well as the basic idea of causality, one key fact is needed: no
information or effect of any kind can travel faster than the speed of
light (about $3{\times}10^8$ metres per second).  This limits the
effect of a present event, not just to the future, but to those times
and places in the future that can be reached at a speed up to the
speed of light.  So for two events to be causally related, it must be
possible for a message originating at the first event to reach the
second event at a speed less than or equal to the speed of light.  A
\emph{\ind{light-line}} is the world-line of a photon -- a particle of
light.  In a space-time diagram a light line is a straight line.  For
each point of space-time and for each direction in space, there is a
light line originating at that point going in that direction.

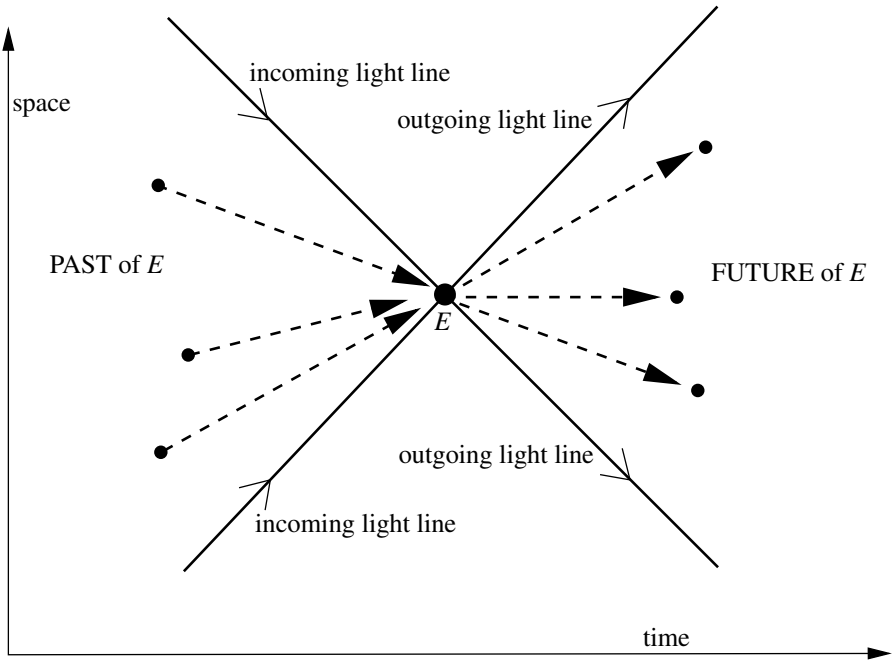
\begin{figure}[htb]
\labellist
\small\hair 2pt
 \pinlabel space [l] <0pt,0pt> at 4 379
 \pinlabel $E$ [t] <0pt,-5pt> at 305 251
 \pinlabel {outgoing light line} [tr] <0pt,0pt> at 411 380
 \pinlabel {incoming light line} [bl] <-5pt,0pt> at 178 391
 \pinlabel {incoming light line} [tl] <-5pt,0pt> at 182 103
 \pinlabel {outgoing light line} [br] <0pt,0pt> at 412 129
 \pinlabel {FUTURE of $E$} [B] <0pt,0pt> at 544 260
 \pinlabel {PAST of $E$} [B] <0pt,0pt> at 73 266
 \pinlabel time [b] <20pt,-37pt> at 423 74
\endlabellist
\centering
\includegraphics[width=.9\hsize]{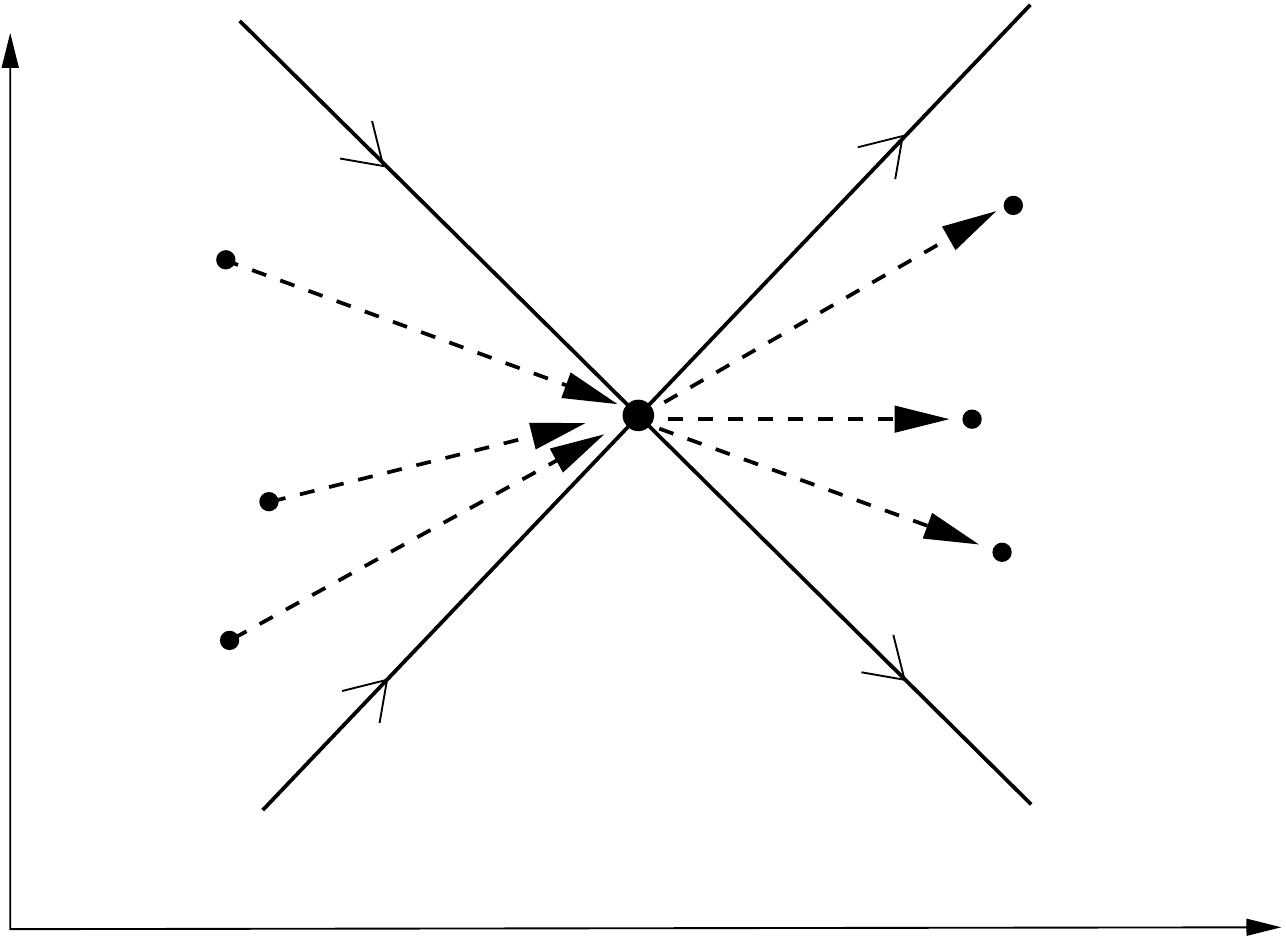}
\caption{Causality diagram with light lines.  The top and bottom
  regions can be regarded as ``simultaneous'' with $E$.}
\label{fig:cause2}
\end{figure}

\fullref{fig:cause} has been updated in \fullref{fig:cause2} with this
new information.  The set of events in the future which can be
affected by $E$ are bounded by the two outgoing light-lines from $E$
(one going up and the other going down).  The region comprising these
events is called the \emph{\ind{future} of $E$}.  Similarly the
\emph{\ind{past} of $E$} is bounded by the incoming light lines.  The
remaining events (the top and bottom regions in \fullref{fig:cause2})
can be regarded as simultaneous with $E$.\index{simultaneous events}
More precisely, they are simultaneous for particular choices
of world-line.  There will be more to be said about this in
\fullref{subsec:Lt} below.  Now light travels very fast indeed and, if
common units such as metres and seconds were used, then the
light-lines in the diagram would be very close to vertical.  To make
the diagram comprehensible, units have been used which make the speed
of light (usually denoted by the letter $c$) equal to $1$.  This puts
the light-lines at $45^\circ$.  For the most part, this book uses
these uncommon (aka ``natural'' or ``astronomical'') units, with time
expressed in terms of years and distance in light-years (the distance
travelled by light in one year).%
\index{natural units}\index{astronomical units}

\begin{figure}[htb]
\labellist
\small\hair 2pt
 \pinlabel {space 1} [l] <0pt,0pt> at 5 231
 \pinlabel {space 2} [lb] <10pt,-10pt> at 42 23
 \pinlabel {time} [b] <0pt,0pt> at 188 51
 \pinlabel $E$ [t] <0pt,-5pt> at 242 212
 \pinlabel {PAST} [B] <0pt,5pt> at 125 220
 \pinlabel {of $E$} [B] <0pt,-5pt> at 125 220
 \pinlabel FUTURE [B] <0pt,10pt> at 352 225
 \pinlabel {of $E$} [B] <0pt,0pt> at 352 225
 \pinlabel (inside) [B] <0pt,-10pt> at 352 225
\endlabellist
\centering
\includegraphics[width=.6\hsize]{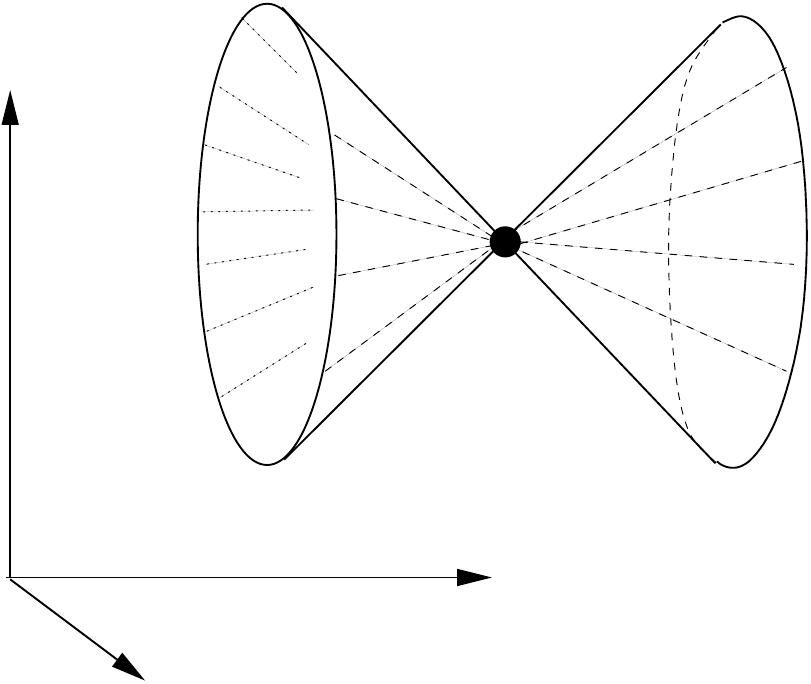}
\caption{$3$--dimensional causality diagram}
\label{fig:cones}
\end{figure}
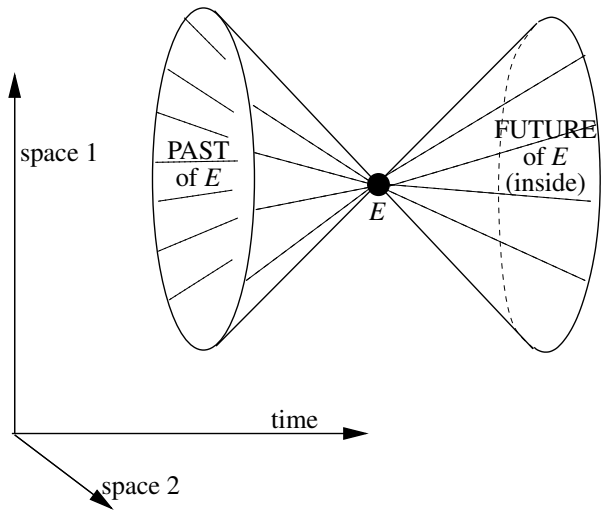

These diagrams both simplify ``space'' to be $1$--dimensional.  In
fact, of course, it is 3--dimensional and for an accurate diagram it
would be necessary to draw it in four dimensions.  This is difficult
to visualise but you can make a start with a $3$--dimensional diagram
where space is represented by two dimensions, \fullref{fig:cones}.  In
this diagram the outgoing light lines from $E$ fill out a cone called
the \emph{\ind{light-cone}} and the interior of this cone is the
\emph{future of $E$}.  Similarly the \emph{past of $E$} is bounded by
the incoming light-cone.

There is one other key fact about light that is needed.  The speed of
light in a vacuum as measured by any observer is always the same.  The
famous Michelson--Morley experiment was an attempt to find the
absolute velocity of the earth through the ether by comparing the
speed of light in two perpendicular directions.  Later experiments
also compared the speeds at two opposite points of the earth's orbit
around the sun.  In all cases, no difference was found.  This negative
result could have been explained by assuming that the earth drags the
ether with it.  The bold explanation which took some time to be
accepted was that there is no ether and all observers measure the
speed of light to be the same.  This bold hypothesis leads to
\ind{mathematical relativity} (aka special relativity) and has been
amply justified by the extensive applicability of the theory.

\section{Lorentz transformations}\label{subsec:Lt}
\index{Lorentz!transformation}

It is now necessary to explain how the mathematical theory of
relativity differs from the simple naive statement that all motion is
relative.  This is a philosophical truism neatly encapsulated in this
anecdote attributed to \ind{Wittgenstein}:\goodbreak

{\leftskip 25pt\small\sl
   Two philosophers meet in the hall. One says to the other, Why do
   you suppose people believed for such a long time that the sun goes
   around the earth, rather than that the earth rotates? The other
   philosopher replies, Obviously because it looks as though the sun
   is going around the earth. To which the first philosopher replies,
   But what would it look like if it looked as though the earth was
   rotating?\par}

Motion is always motion measured relative to something else.  There is
no difference in content between the statement that the sun goes round
the earth and the statement that the earth rotates.  Both describe the
same relative motion.  The former is more useful for earth-based
purposes whilst the latter is more useful for astronomical purposes.
It is not a case that one is \emph{true} and the other \emph{false}.
Both are valid.  This is the essence of a famous principle, known
as ``Mach's principle'', which is explained in \fullref{sec:Sciama}.

But this is not Mathematical Relativity.  Mathematical Relativity is a
theory which squares the naive principle that all motion is relative
with the apparently contradictory fact that the speed of light in a
vacuum as measured by any observer is always the same.%
\index{relativity!mathematical}

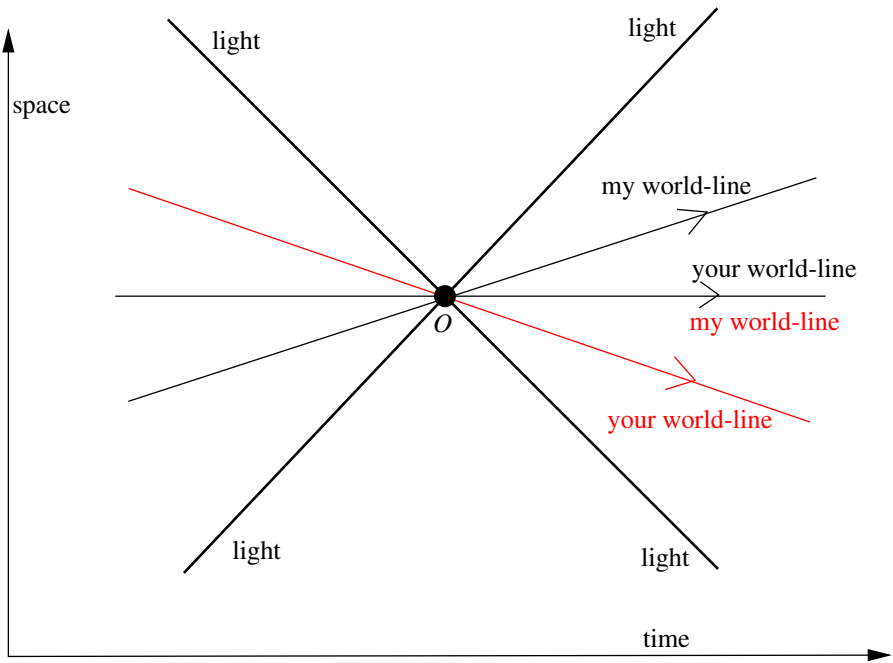
\begin{figure}[b!]
\labellist
\small\hair 2pt
  \pinlabel space [l] <0pt,0pt> at 4 379
 \pinlabel $O$ [t] <0pt,-5pt> at 305 251
 \pinlabel light [bl] <0pt,0pt> at 143 414
 \pinlabel light [br] <0pt,0pt> at 469 423
 \pinlabel light [tl] <5pt,0pt> at 148 86
 \pinlabel light [tr] <0pt,0pt> at 478 81
 \pinlabel time [b] <20pt,-37pt> at 423 74
 \pinlabel {my world-line} [rB] <0pt,0pt> at 522 321
 \pinlabel {your world-line} [b] <0pt,2pt> at 534 253
{\color{red}
 \pinlabel {your world-line} [tr] <10pt,-2pt> at 517 179
 \pinlabel {my world-line} [tl] <-5pt,0pt> at 482 243
}
\endlabellist
\centering
\includegraphics[width=.9\hsize]{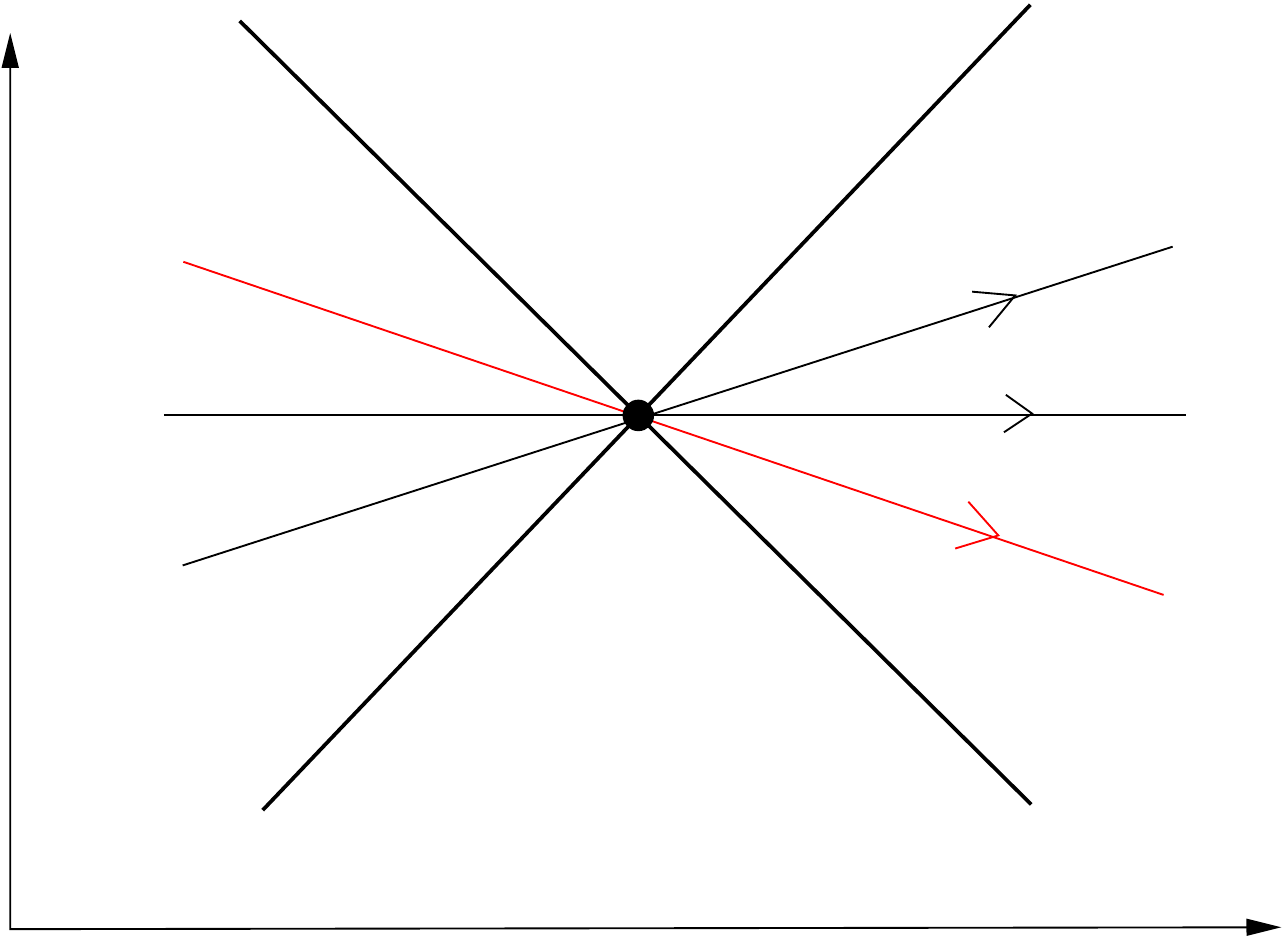}
\caption{Our world-lines. My point of view is red, yours is black}
\label{fig:mink1}
\end{figure}

The apparent contradiction is because if you measure the speed of a
beam of light coming from you to me and if I am moving towards you,
then I must measure the same beam travelling more quickly since my
speed must be added to the speed of light from you.  The resolution of
this contradiction is that either my time is different from yours OR
my measuring rods are shrinking with respect to yours because of my
motion.  In Mathematical Relativity BOTH these changes occur.  At this
point it is necessary to be a bit technical.

Make a simplifying assumption.  Suppose that we are in a universe
comprising $1$ dimension of space and $1$ of time and agree to use
natural units so that $c=1$.  In other words we are in the universe
illustrated in \fullref{fig:cause2}.  Suppose that you are at the
``origin'' at time zero (the point in the middle labelled $O$ in
\fullref{fig:mink1}) and not moving .  This means that your world-line
is the horizontal line to the right in the figure.  Suppose for
simplicity that I am also at the origin at time zero, but that I am
travelling upwards with constant velocity.  Then my world-line will be
a straight line inclined upwards as illustrated.  But as far as I am
concerned, it is I who am stationary and you who are travelling
(downwards).  My view of things is shown in red on the diagram.

\begin{figure}[b!]
\labellist
\small\hair 2pt
  \pinlabel space [l] <0pt,0pt> at 4 379
 \pinlabel $O$ [t] <0pt,-5pt> at 305 251
 \pinlabel light [bl] <0pt,0pt> at 143 414
 \pinlabel light [br] <0pt,0pt> at 469 423
 \pinlabel light [tl] <5pt,0pt> at 148 86
 \pinlabel light [tr] <0pt,0pt> at 478 81
 \pinlabel time [b] <20pt,-37pt> at 423 74
 \pinlabel {my time} [rB] <0pt,0pt> at 522 321
 \pinlabel {your time} [b] <0pt,2pt> at 534 253
 \pinlabel {my} [l] <0pt,0pt> at 368 447
 \pinlabel {space} [l] <-2pt,-8pt> at 368 447
 \pinlabel {your} [l] <1pt,0pt> at 305 453
 \pinlabel {space} [l] <1pt,-8pt> at 305 453
 {\color{red}
 \pinlabel {your time} [tr] <10pt,-2pt> at 517 179
 \pinlabel {my time} [tl] <-5pt,0pt> at 482 243
 \pinlabel {my} [r] <0pt,0pt> at 305 453
 \pinlabel {space} [r] <0pt,-8pt> at 305 453
 \pinlabel {your} [r] <0pt,0pt> at 241 444
 \pinlabel {space} [r] <2pt,-8pt> at 241 444
}
\endlabellist
\centering
\includegraphics[width=.9\hsize]{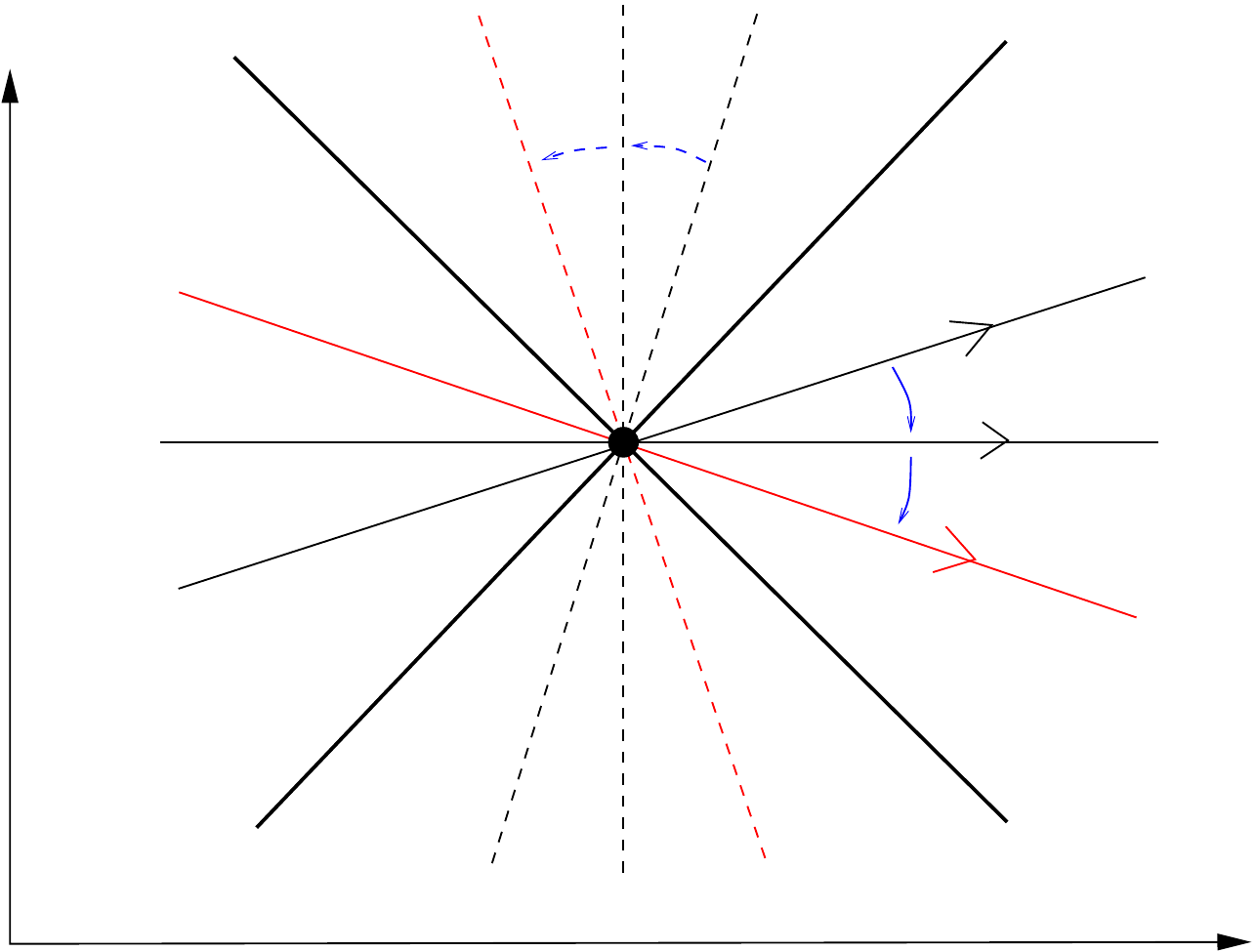}
\caption{Our space-times. My point of view is red, yours is black. The
  transformation taking my view to yours is blue.}
\label{fig:mink}
\end{figure}
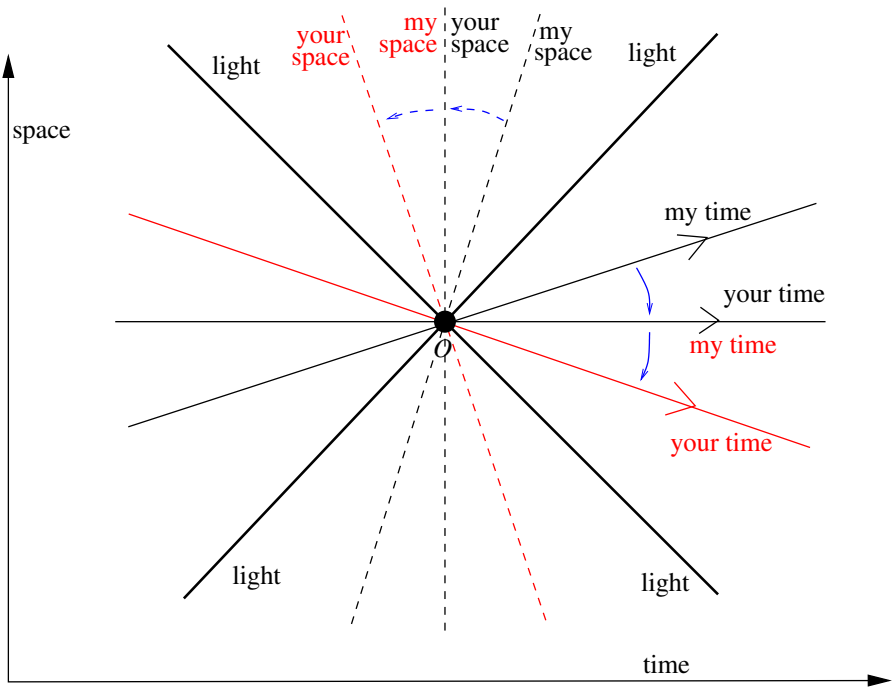
Now as far as we are concerned, our notion of ``time'' corresponds to
our motion along our world-lines, so you can think of the lines
labelled ``world-line'' as labelled ``time'' -- the time for the
observer moving along that world-line.  What does our corresponding
``space'' look like?  For you at rest, space is obviously the vertical
line through the origin.  The points of this line represent events
that are simultaneous with $O$ (from your point of view).  But from my
point of view space is represented by a line inclined to the right as
illustrated in \fullref{fig:mink}.  In other words the events that I
see as simultaneous with $O$ are not the same as the events that you
see.  In \fullref{fig:rel}, for simplicity, the observers' spaces are
drawn as perpendicular to their times.  This is true, but it is a
peculiar property of the space for Special Relativity, that
perpendicular does not always look perpendicular.  \fullref{fig:mink}
is the correct picture.

\begin{figure}[ht!]
\labellist
\small\hair 2pt
 \pinlabel {my world-line} [lt] <0pt,0pt> at 597 346
 \pinlabel {your world-line} [t] <0pt,0pt> at 590 220
\endlabellist
\centering
\includegraphics[width=.8\hsize]{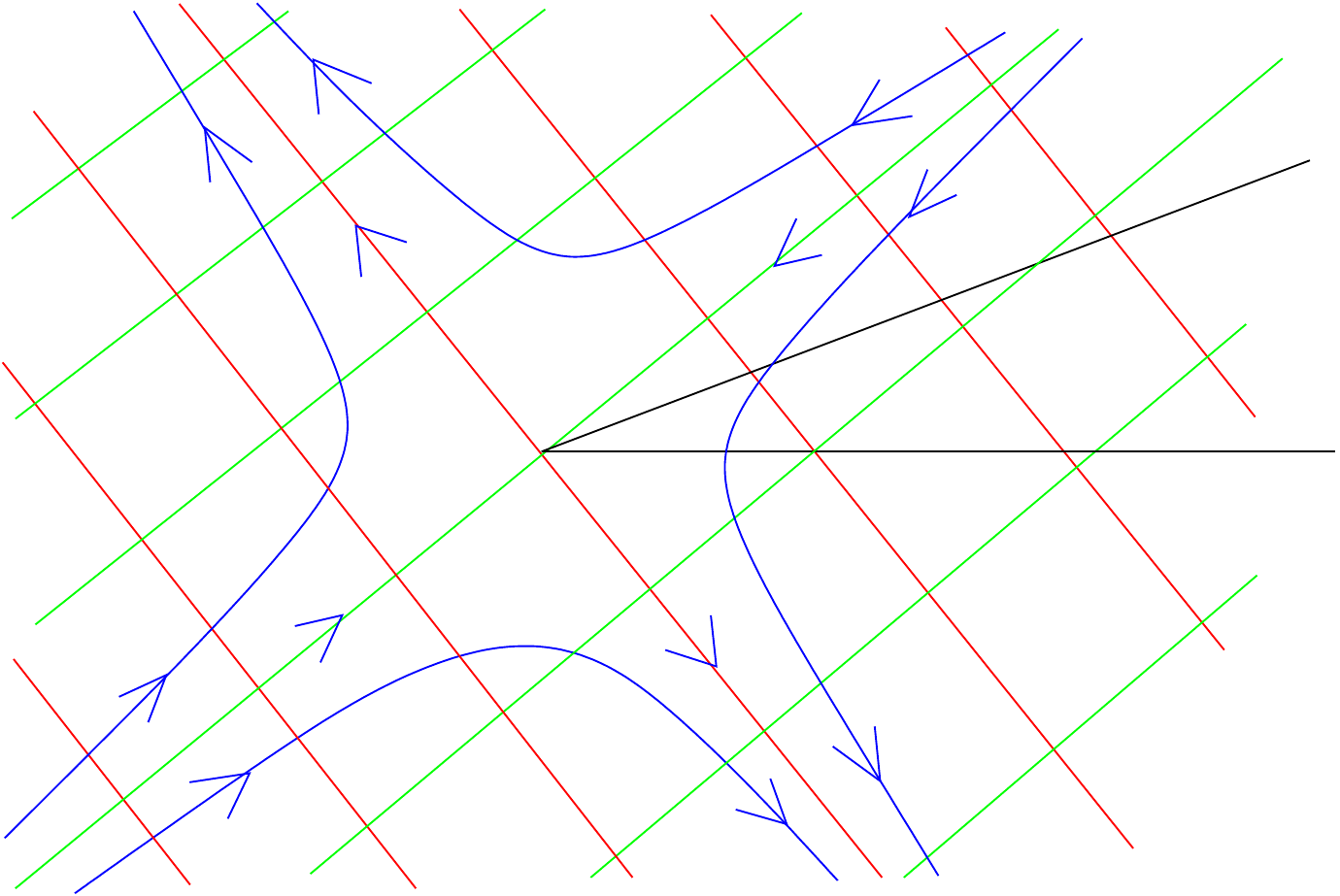}
\caption{The light grid and hyperbolic rotation}
\label{fig:grid}
\end{figure}

To justify this picture, it is necessary to describe how our two views
of the universe are related.  From my point of view, I see my space as
perpendicular to my time (as you do from your point of view).  We can
both make an accurate map of the world but a different map.  The key
to understanding special relativity is to understand how to compare
our two maps.  On your map a typical event has coordinates $(t,x)$ say
but on my map the same event has (usually) different coordinates
$(t',x')$.  The transformation that takes $(t,x)$ to $(t',x')$ is
indicated roughly in blue in \fullref{fig:mink}.  It takes my time and
space axes to yours.  The transformation has been drawn as if it was a
rotation.  Indeed it is, but it is a strange \ind{hyperbolic rotation}.  The
fundamental fact that needed is that we agree on light lines.  Look
now at \fullref{fig:grid}.

Upward light lines are drawn in green and downward ones in red.  They
form a grid which covers the whole map.  We agree on this grid, but we
do not need to agree on the spacing of lines in this grid.  We also
agree on the central point (the event where I meet you).  The
transformation that takes my view to yours (and takes my world-line to
yours) must shrink the green lines and stretch the red ones.  We
expect the transformation to be uniform over the whole plane (this is
justified by thinking of observers moving along parallel paths in
space-time) and therefore the stretch of the red lines is by a
constant factor $k$ say where $k>1$ and the shrinkage of the green
lines is by another constant $l<1$.  But if we reverse our roles, the
transformation that takes your view to mine stretches green lines by
$1/l$ and shrinks red ones by $1/k$.  But our roles are exactly
symmetric and it must be the case that $k=1/l$.

This shrinkage and stretching has been indicated by blue arrows in the
diagram.  The formula for this transformation is $(u,w)\mapsto
(ku,w/k)$ where $(u,w)$ are the grid coordinates.  Since $u$ and $w$
are transformed by reciprocal factors, the transformation preserves
the curves given by $uv=C$ for $C$ constant.  These are rectangular
hyperbolae.  In the figure, two of these have been drawn (in blue)
corresponding to $C=-1$ (the right and left curved lines) and $C=+1$
(the top and bottom curved lines).  In terms of usual coordinates ($x$
for space and $t$ for time) since $u=x+t$ and $v=x-t$ these curves are
given by $x^2-t^2 = C$ for varying $C$.

The transformation makes points flow along these hyperbolae, as
indicated by more blue arrows.  This is why this transformation is
called a hyperbolic rotation.  Now it is evident why my ``space''
which is transformed to yours by this rotation is inclined upwards to
the right (as drawn it in \fullref{fig:mink}) and therefore any point
in the upper or lower quadrants in \fullref{fig:cause2} are
simultaneous with $O$ for a suitable observer as also claimed earlier.

The hyperbolic rotation just arrived at is a simple example of a
Lorentz transformation.  A general Lorentz transformation is a
combination of hyperbolic rotations with ordinary translations and
rotations.  It is necessary to think of space as 3--dimensional
instead of 1--dimensional.  This 3--space can be moved around for
different points of view by translating and rotating (so called
Euclidean motions) and time can also be translated.  All these motions
together with hyperbolic rotations make up the set of Lorentz
transformations (the Lorentz group)\index{Lorentz!group}.
\index{Lorentz!transformation}
Incidentally the Lorentz group has been derived assuming that a
Lorentz transformation preserves all light lines and is also uniform.
There is a famous theorem of Christopher Zeeman which says that you
only need to consider the most basic fact that two observers must
agree on, namely causality, to obtain the Lorentz group \cite{Zeeman}.\index{Zeeman, Christopher}

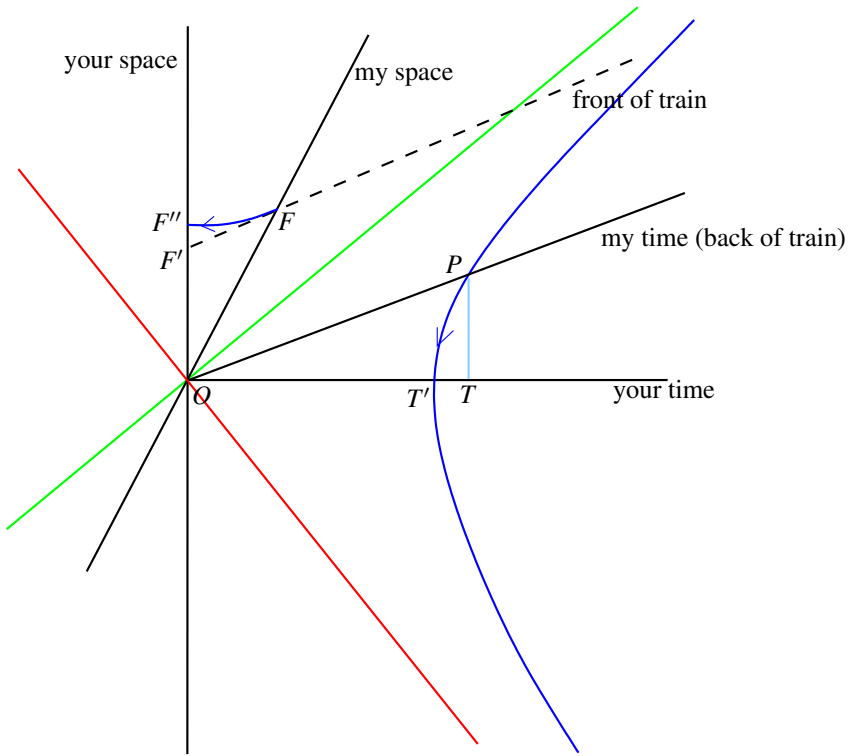
\begin{figure}[b!]
\labellist
\small\hair 2pt
 \pinlabel $O$ [tl] <2pt,-2pt> at 75 157
 \pinlabel $T'$ [tr] <0pt,-1pt> at 180 156
 \pinlabel $T$ [t] <0pt,-1pt> at 194 157
 \pinlabel $P$ [br] <0pt,0pt> at 193 199
 \pinlabel $F''$ [r] <0pt,0pt> at 77 222
 \pinlabel $F'$ [tr] <0pt,0pt> at 77 212
 \pinlabel $F$ [tl] <1pt,-1pt> at 112 228
 \pinlabel {my time (back of train)} [lt] <0pt,0pt> at 248 221
 \pinlabel {your time} [lt] <0pt,-2pt> at 253 159
 \pinlabel {front of train} [lt] <0pt,0pt> at 236 278
 \pinlabel {my space} [lt] <0pt,0pt> at 145 287
 \pinlabel {your space} [r] <0pt,0pt> at 75 287
\endlabellist
\centering
\includegraphics[width=.7\hsize]{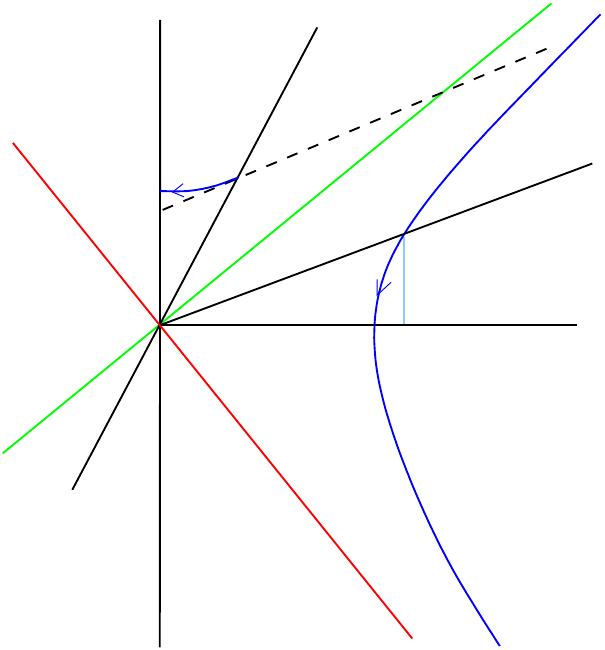}
\caption{Time dilation and length contraction}
\label{fig:scale}
\end{figure}

\section{Time dilation and \ind{length contraction}}\label{subsec:Tdlc}
\index{time dilation}

Now that there is a good picture of the relationship between our two
views of the world, it is possible to explain that other strange
property of motion in special relativity, namely that motion causes
time to appear to dilate and lengths to appear to contract.  Look now
at \fullref{fig:scale}.  Suppose I move along my world-line from $O$
to $P$.  From your point of view this takes a time equal to the length
$OT$.  But applying the transformation that takes my view to yours,
then $P$ moves to an point $T'$ \emph{closer} to $O$, as drawn.  So
the real elapsed time \emph{for me} is the length $OT'$ which is
smaller than $OT$.  \emph{You see my time dilated.} But of course the
situation is symmetric as always, so \emph{I see your time dilated by
  exactly the same factor.}  To see how length contraction works,
suppose that my motion happens because I am at the back of a train
moving upwards.  The front of the train moves on a parallel world
line, drawn dashed.  When the back of the train is at $O$ in my space
(which is the same as the train's space) the front is at $F$.  For you
the length of the train is $OF'$.  But applying the transformation
taking my space to yours the train really has length $OF''$ which is
larger.  \emph{You see my lengths contracted.}  Again by symmetry,
\emph{I see your lengths contracted.}

If you are good at calculation, then using the fact that the motion
that carries my space to yours is a hyperbolic rotation, you can
derive the following precise formulae for these dilation/contraction
effects from \fullref{fig:scale}.  Time is apparently dilated by a
factor $1/d$ and lengths are contracted by $d$ where $d=\sqrt{1-v^2}$
and $v$ is our relative velocity.  This is the formula in natural
units (with $c=1$).  In common units the formula is
$d=\sqrt{1-v^2/c^2}$.

\section{\ind{Minkowski space}}\label{subsec:Mink}

At this point a good description of a particular space-time which is
called Minkowski space has been achieved.  This space is the
fundamental space-time for relativity.  Special relativity takes place
in Minkowski space and general relativity is built upon it as will be
seen in the next section.

It is necessary to be very precise.  Minkowski space is
$4$--dimensional space with coordinates $(t,x,y,z)$ where the first
coordinate $t$ is called \emph{time} and the other three are
\emph{space}.  The transformations just called Lorentz transformations
act on Minkowski space and special relativity is the study of
properties which are unchanged by Lorentz transformations.  Home in on
one particular property, namely \emph{length}.  As usual, for
simplicity, assume there is just one dimension of space $x$.  When the
study of length in 2--dimensional Minkowski space is finished, it will
be easy to generalise back to 4 dimensions.

Consider two events (points of space-time) in Minkowski space with
coordinates $(t,x)$ and $(t',x')$.  The fundamental ``property'' of
the two events taken together is the ``number'' $s$ where
$s^2=-(t-t')^2+(x-x')^2$.  This has been put in inverted commas
because sometimes $s$ is the square root of a negative number, in
other words it may be imaginary.  To avoid having to think about
imaginary numbers use $s^2$ instead of $s$.  $s^2$ is the appropriate
number to be considered ``length'' (or rather the square of length) in
Minkowski space.\index{Minkowski space!length}

First notice that it doesn't change under translation -- ie replacing
$x$ by $x-a$ and $t$ by $t-b$ where $a$ and $b$ are constants.
Translate so that one of the points (say $(t',x')$) is at the origin
$(0,0)$.  Then the formula for $s^2$ is simpler $s^2=-t^2+x^2$.  Now
consider a hyperbolic rotation.  As found above, this preserves
rectangular hyperbolae $x^2-t^2=C$.  In other words it preserves $s^2$
(and hence $s$).  So $s^2$ is preserved by all Lorentz transformations
as ``length'' must.  There is an obvious analogy with Cartesian
(ordinary) length $s$ in Euclidean (normal) space, which by
Pythagoras' Theorem has the formula $s^2=x^2+y^2$.  But there are
obvious differences -- it can be zero, for example if $x=\pm t$, ie if
$(t,x)$ lies on a light line through the origin.  Lengths in ordinary
space are never zero!

\begin{figure}[ht!]
\labellist
\small\hair 2pt
 \pinlabel $O$ [t] <0pt,0pt> at 304 242
 \pinlabel $(t,x)$ [l] <0pt,0pt> at 457 289
 \pinlabel $(t,x)$ [tl] <0pt,0pt> at 448 397
 \pinlabel $(t,x)$ [b] <0pt,0pt> at 340 399
{\color{red}\large 
 \pinlabel $0$ [tr] <0pt,0pt> at 124 61

 \pinlabel $0$ [tl] <0pt,0pt> at 491 63
 \pinlabel $0$ [bl] <0pt,0pt> at 489 445
 \pinlabel $0$ [br] <0pt,0pt> at 112 438
 \huge
 \pinlabel $+$ <0pt,5pt> at 299 446
 \pinlabel $-$ <15pt,0pt> at 495 249
 \pinlabel $+$ <0pt,-5pt> at 307 62
 \pinlabel $-$ <-15pt,0pt> at 119 242
 }
\endlabellist
\centering
\includegraphics[width=.5\hsize]{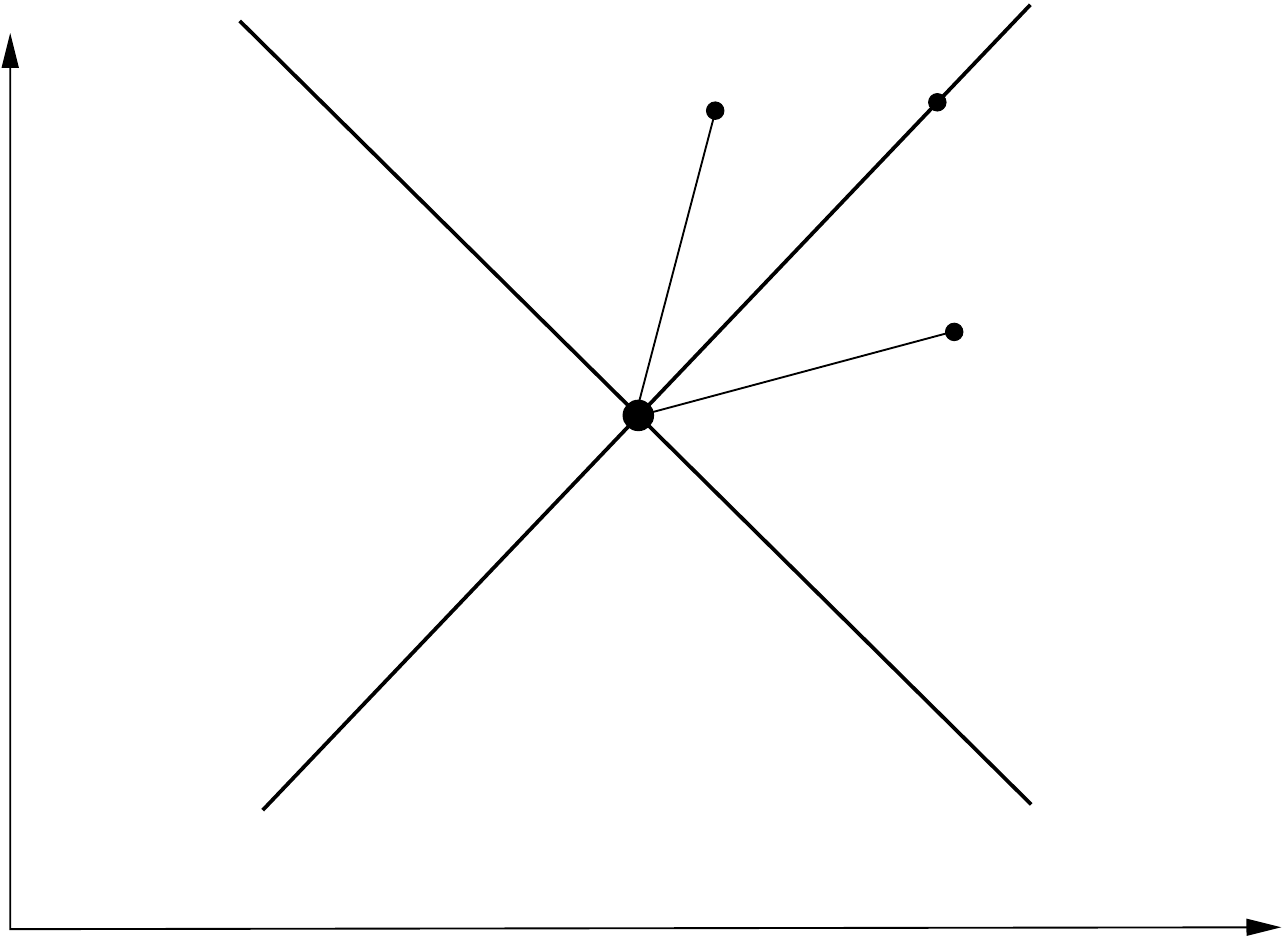}
\caption{The sign of $s^2$}
\label{fig:sign}
\end{figure}

Consider some other cases.  Suppose that $t$ is positive and
that $t>x$ or $t>-x$ in other words that $(t,x)$ lies in the right
hand quadrant.  Then $s^2=-t^2+x^2 $ is negative (length is imaginary
if you like).  A similar thing happens if $(t,x)$ lies in the left
hand quadrant.  If $t<x$ or $-t>x$ (top quadrant) $s^2$ is positive.
Finally if $x=t$ or $x=-t$ (the light lines through the origin) then
$s^2$ is zero.  These facts are illustrated in \fullref{fig:sign}.

Causality can be interpreted in terms of this new concept of
``length''.  Two events $P$ and $Q$ are causally related if the
(square) of the length of the interval $PQ$ is zero or negative and
this length can be thought of as being the time separating the two
events.  Similarly if the (square) of the length of the interval $PQ$
is positive then the events are not causally related and we can think
of the length as being the distance in space separating the two
events.\index{causality!in terms of length}

This length is the \emph{\ind{metric}} on Minkowski space.  More precisely,
moving back now to 4 dimensions, Minkowski space is $4$--dimensional
space with coordinates $(t,x,y,z)$ and metric (distance) $s$ given by
$$s^2=-t^2+x^2+y^2+z^2.$$
Metrics are often expressed in infinitesimal form using $ds$ (a tiny
step along $s$) etc:
$$ds^2=-dt^2+dx^2+dy^2+dz^2$$
\index{Minkowski space!metric}

\index{Einstein!special relativity|)}\index{relativity!special|)}

\section{General Relativity}\label{sec:genrel}
\index{Einstein!general relativity|(}\index{relativity!general|(}

Now move on to consider mathematical models for the universe.
Minkowski space is the simplest model but it is far too simple.
Einstein's deep insight was that the force of gravity -- the force
that keeps us anchored to the earth and which keeps the earth moving
around the sun -- should be thought of as encoded in the fabric of
space-time by means of curvature.  With this insight, Minkowski space
(which has no curvature) is a model for an empty universe: one with no
planets, stars or galaxies.  So more general models are needed.
Nevertheless Minkowski space remains fundamentally important because
it correctly describes the local geometry of the universe.  It is
accurate over small distances and for a small interval of time.  This
fact is taken as an axiom in general relativity.  What it says in
words is that the small scale geometry of space-time is the same
everywhere for all observers.  And of course, these local Minkowski
spaces are inertial frames.

\section{Manifolds and space-times}

\begin{figure}[ht!]
\cl{\includegraphics[width=1in]{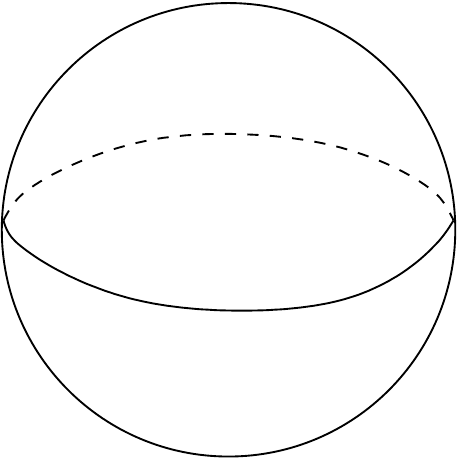}\qquad\qquad\includegraphics[width=2in]{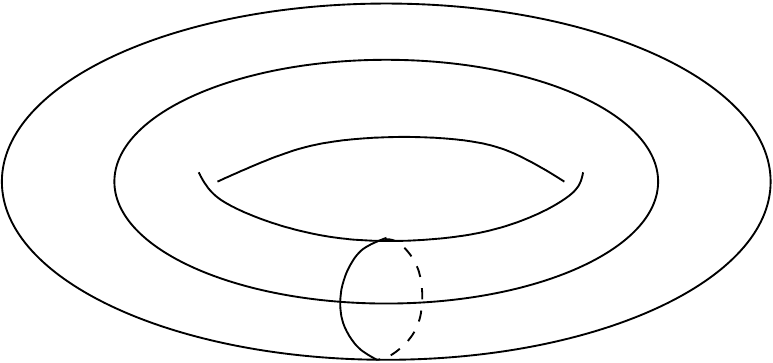}}
\caption{Sphere (left) and torus (right)}\label{fig:sphtor}
\end{figure}

A \emph{\ind{manifold}} is a space which is locally the same as ordinary
(Euclidean) space, but which might be quite different globally.  The
dimension is the dimension of the local Euclidean space.  A one
dimensional maniold is locally like a line but could be closed (as a
circle).  A 2--manifold is a surface of which the sphere and the torus
are examples (\fullref{fig:sphtor}). 

Here is a simple example of a 3--manifold which is not ordinary
3--space.  Think of a cube in $3$--space and make a conceptual leap by
assuming that the top of the cube is exactly the same as the bottom.
What this means is that if you move upwards through the top, you
immediately appear at the bottom.  Make the same leap for the other
two pairs of oposite faces, so that if you move out through the left
side you immediately appear at the right and similarly forwards and
backwards.  The space being described is the $3$--manifold known as
the $3$--\emph{torus}.\index{3--torus}

4--dimensionsal manifolds are needed for relativity.  But with the
essential difference that at any point some directions are time-like
and some space-like.  A \emph{space-time}, also called a
\emph{Lorentz} manifold, is a space locally like Minkowski space.  At
each point
\index{Lorentz!manifold}\index{space-time}\index{manifold!Lorentz}%
(event) there are two null cones representing incoming and outgoing
light lines as in figure \ref{fig:cones}.  The 3--torus can be made
into a space-time by adding one extra dimension for time and using
Minkowski space as a model for how this time dimension fits with the
three space dimensions.

There is one other important property of the manifolds used for
relativity. They are \emph{smooth} manifolds, which means they have a
smooth metric which is locally diffeomorphic (smoothly equivalent) to
the metric on Minkowski space.  Thus the light-cones at each point can
be defined, as in Minkowski space, as directions in which the metric
is null (points on the same light ray have zero separation in the
metric).  Further time-like directions are ones where $s^2$ (the
square of the metric) is negative and space-like ones where it is
positive.

\index{metric!general}
The notation used for a \ind{general metric} is
$$ds^2 = \sum_{i,\,j} g_{i,\,j}\, dx_i\, dx_j$$ and the array of
coefficients ${\bf g} = (g_{i,\,j})$ is also called the metric.  In
technical terms, {\bf g} is a bilinear form of index $(-1,3)$.  Here
$-1$ is for the time-like direction and $3$ for the three space-like
directions.  An important example is the \ind{Schwarzschild metric}
which is used repeatedly through the book\index{metric!Schwarzschild}
\begin{equation}\label{eq:Schw}
ds^2 = -(1-2M/r)\,dt^2 + (1/(1-2M/r))\,dr^2 + r^2\,(d\theta^2 + \sin^2\theta \, d\phi^2).
\end{equation}
Here $M$ is a constant interpreted as central mass and spherical
coordinates $(r,\theta,\phi)$ are used for space.  In this example
$g_{tt}=-(1-2M/r)$, $g_{rr}=1/(-g_{tt})$, $g_{\theta\theta}=r^2$,
$g_{\phi\phi}=r^2\sin^2\theta$ and the others are zero.  The manifold for this
metric is ordinary 3--space with the origin removed crossed with one
dimension (for time).

\section{Curvature}

\begin{figure}[ht!]
  \cl{\includegraphics[width = 2in]{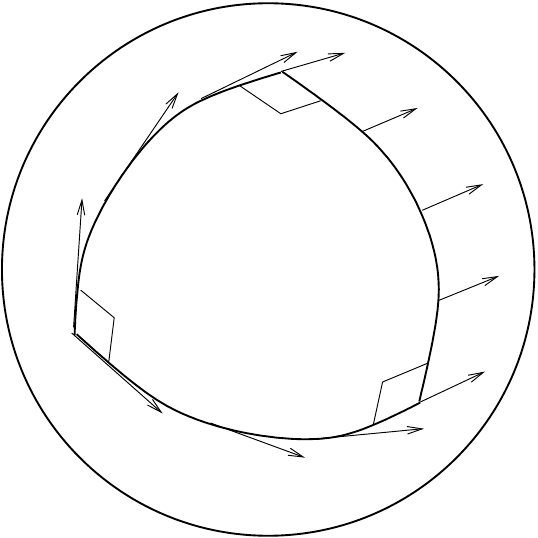}}
\caption{Transporting a vector around a trangle on the
  sphere (start at $O$)}\label{fig:transport}
\end{figure}

To proceed it is necessary to discuss the \emph{\ind{curvature}} of a
space-time.  This idea applies to any manifold and the simplest
example to think about is the curvature of a \emph{\ind{surface}} (or
$2$--manifold).  The most familiar curved surface is the \emph{sphere}
or the surface of a round ball.  It is obviously curved, but to
explain curvature in general it is necessary to encapsulate curvature in
mathematical terms.  Think about a triangle in the sphere and think
about carrying a vector around that triangle keeping it as parallel to
itself as possible (this is called \emph{\ind{parallel transport}}).
Whatever triangle you choose, the vector ends up pointing in a
different direction.  For example see \fullref{fig:transport}.

\ssh{\ind{Riemann curvature}} \index{curvature!Riemann} 

It is not necessary to use a triangle to detect the curvature.
Transporting a vector around almost any closed curve on the sphere
results in a non-parallel vector.  The same idea can be used in any
manifold with a metric.  By transporting vectors around small curves
it is possible to define curvature.  If the curve is chosen to lie in
a plane, this gives the notion of the curvature of that plane.  But
notice that this is a vector---the discrepancy after transport---not a
number.

The \emph{Riemann curvature} of a manifold is this idea used
exhaustively.  A space-time is a 4--manifold and, sticking to
coordinate planes gives a choice of 12 planes.  In each plane a
coordinate vector can be transported around a small curve and the
discrepancy, which is a vector, read.  The Riemann curvature tensor,
$R^{i}_{jkl}$, is an array of $4^4$ numbers obtained in this way.  The
definition of $R^{i}_{jkl}$ is: transport the $j\th$ coordinate vector
around a small curve in the $(k,l)$--plane and read the change in the
$i\th$ coordinate of the result.  It is a $(1,3)$--\emph{tensor}
because of the way it transforms under change of coordinates.  There
are many symmetries and identities amongst the components and there
are in fact only 20 independent components.  The Riemann curvature
gives all possible information about how the manifold curves.  However
for Einstein's General Relativity, only about half of this information
is needed, namely the Ricci curvature.

There are (fairly complicated) formulae for the Riemann curvature in
terms of the metric:
\begin{equation}\label{eq:Riem}
R^{i}_{jkl} = \d_k\Gamma^i_{lj} - \d_l\Gamma^i_{kl} + \sum_\lambda
(\Gamma^i_{k\lambda}\Gamma^{\lambda}_{lj}
-\Gamma^i_{l\lambda}\Gamma^{\lambda}_{kj})
\end{equation}
where the \emph{\ind{Christoffel symbols}} $\Gamma^m_{ij}$ are defined by
\begin{equation}\label{eq:Chris}
  \Gamma^m_{ij} = \sum_k\textstyle{\frac12}
  g^{km}(\d_ig_{kj}+\d_jg_{ik}-\d_kg_{ij}),
\end{equation}
$\d_i=\d/\d x_i$ means differentiation wrt to $x_j$ and $g^{ij}$ is the
inverse matrix to $g_{ij}$.  These formulae are useful in special cases
(eg for diagonal metrics where many terms vanish).
\index{curvature!Riemann!formula}

\ssh{\ind{Ricci curvature}}
\index{curvature!Ricci}

The \emph{Ricci curvature} is a contraction of the Riemann curvature.
It is another tensor (in fact a 2--tensor or bilinear form) and the
definition is $\Ric_{ij}= \sum_k R^{k}_{ikj}$.  There are again
symmetries and the number of independent components is 12 rather than
16.  It has a simple geometric interpretation.

A bilinear form is determined by values on single repeated vectors
(rather than general pairs of vectors) -- the associated quadratic
form -- and from the definition this has the following meaning.
Consider 2--planes containing the given vector $v$ and add the
curvature for 4 mutually perpendicular planes.  So this is the
``average'' curvature for planes containing $v$.  But there is a
simpler interpretation.  Consider a small cone of vectors near to $v$
and measure the 4--volume of this cone.  It differs from the result in
flat (Minkowski) space due to curvature.  This difference is the value
of (the associated quadratic form to) $\Ric$ on $v$.  So Ricci
curvature measures the way space-time expands (or contracts).

The diagonal components of the Ricci curvature are the sectional
curvatures which are directly analogous to the curvature of a surface;
these are the curvatures of four mutually perpendicular hyperplanes
measured in a perpendicular direction.  The contraction of $\Ric$ is
the \emph{scalar curvature} $S$ defined
by $$S=\sum_{i,j}g^{ij}\Ric_{ij}.$$

\section{Einstein's equations}\label{sec:Eequns}
\index{Einstein!equations}\index{Einstein!tensor}
\index{stress-energy tensor}

Einstein's idea of pure genius was to interpret the force of gravity
as due to curvature of space-time.  As has been seen, the Ricci
curvature determines the way volume grows.  If this is positive then
nearby parallel geodesics will tend to converge (as if under the
influence of a force).  The formulation that Einstein eventually found
after much effort was in terms of this, the Ricci curvature, rather
than the general Riemann curvature.  The \emph{Einstein tensor}
denoted $G_{ij}$ is not quite the Ricci curvature.  Einstein's
equations express the curvature in terms of the presence of matter.
There is a \emph{stess-energy} tensor $T$ which encodes the energy and
momentum of matter.  The idea was that the equations should say that
$G=kT$ for some suitable constant $k$.  Conservation of energy and
momentum implies that $\div T = 0$ where $\div$ is divergence.  But
$\div\Ric$ is non zero, in fact it is $\frac12 dS$, half the deriative
of the scalar curvature, so to achieve $\div G = 0$, define the
Einstein tensor $G$ to be $\Ric - \frac12 S{\bf g}$,
ie $G_{ij}=\Ric_{ij}-\frac12Sg_{ij}$ where $S$ is scalar curvature
and ${\bf g}$ is the metric.

Einstein's equations now read:
$$G=8\pi T$$ The constant $8\pi$ is found by considering simple
special cases and using natural units where Newton's gravitational
constant (also confusingly denoted $G$) is 1.  $T$ will not be
described explicitly here because, for the most part, this book is
concerned with vacuum solutions ($T = 0$) or modifications of these
due to inertial effects.  The interested reader can find many good
descriptions in the literature.  The vacuum equations are $G= \Ric -
\frac12S{\bf g}=0$.  But contracting this equation implies that $S=0$
and hence:
\medskip

\cl{\textbf{Einstein's vacuum equations are equivalent to $\Ric=0$.}}
\index{Einstein!equations!vacuum}

\ssh{Einstein's biggest blunder}
\index{Einstein!biggest blunder}

In order to have a static solution for the universe, Einstein modified
his basic equations by adding a \emph{\ind{cosmological constant}} $\kappa$
times ${\bf g}$ to his tensor:
$$G + \kappa{\bf g} = 8\pi T$$
or
\begin{equation}\label{eq:cosconst}
  \Ric + (\kappa- \textstyle{\frac12}S){\bf g} = 8\pi T.
\end{equation}
This happened before the observations of Hubble (preceded by Slipher
and Humason) suggested that the universe might not be static but
expanding.  Einstein then rescinded his cosmological constant $\kappa$
calling this his biggest blunder.  If he hadn't introduced it, he
could have predicted the observed expansion!  Since the 1998 WMAP
observations, most cosmologists are happy to keep the cosmological
constant since the universe seems now to approximate de Sitter space
which has a positive cosmological constant (as will be seen shortly).
From the author's point of view, Einstein's biggest blunder was the
reintroduction of a universal time in his (and consequently current
mainstream) models for the universe in the large.  There is no
universal time in either special or general relativity.  It is the
assumption of a universal time that leads to the (false) big bang
theory which dominates current cosmology.

\ssh{Vacuum equations with cosmological constant}
\index{Einstein!equations!vacuum cosmol const}

For the case of a vacuum ($T=0$) the ${\bf g}$ terms in equation
\ref{eq:cosconst} can be collected to give
\begin{equation}\label{eq:vaccosconst}
  \Ric = \Lambda{\bf g}
\end{equation}
where $\Lambda=\frac12S-\kappa$ is a scalar field.  This formulation
is slightly more general than Einstein's since it allows $\kappa$ to
vary over space-time.

\ssh{The Schwarzschild and de Sitter solutions}\label{sec:Birk}

Finding general solutions to the Einstein equations is not easy
because of their complication when expressed in terms of the metric,
but there is an important special case when it is fairly easy.  This
is the \ss\ case and is the appropriate case for studying the metric
near an isolated heavy body.  Spherical symmetry implies that the 
metric can be expressed the in the form:
\begin{equation}\label{eq:metric-gen-rev}
ds^2 = -Q\,dt^2 + P\,dr^2 + r^2\,d\Omega^2
\end{equation}
where $P$ and $Q$ are positive functions of $r$ and $t$ on a suitable
domain.  Here $t$ is time, $r$ is ``distance from the centre'' and
$d\Omega^2$, the standard metric on the unit 2--sphere $S^2$, is an
abbreviation for $d\theta^2 + \sin^2\theta \, d\phi^2$.  This metric
is diagonal which implies that many of the terms in equations
\ref{eq:Riem} and \ref{eq:Chris} are zero and it is not too hard to
compute the Ricci curvature, see for example Win \cite{Win}.  Then it
is fairly easy to prove that if equation \ref{eq:vaccosconst} holds
then $P$ and $Q$ are independent of $t$, $\Lambda$ is constant and
$$Q=\frac1P=1-\frac{\Lambda r^2}3 - \frac{2M}r$$ with $M$ constant.
For details here see \cite{uniqueness}.  This is mild generalisation
of \ind{Birkhoff's theorem}.

The special case $\Lambda=0$ is the \emph{\ind{Schwarzschild metric}} and
the case $M=0$ is the \emph{\ind{de Sitter metric}}.  The general case is
the Schwarzschild--de Sitter metric also called the \ind{Kottler metric}.

\ssh{Black holes}
\index{black hole}

The Schwarzschild metric is the unique \ss\ metric satisfying
Einstein's vacuum equations without a cosmological constant.  It is
given by \ref{eq:metric-gen-rev} with $Q=1/P = 1 - 2M/r$.  The
metric appears to go singular at $r=2M$ (the
\emph{\ind{Schwarzschild radius}}) where $P = 1/(1-2M/r)$ is infinite.
The solution was discovered in 1915 just a few months after Einstein
published his theory and for nine years it was believed that this
singularity was a real property of the space and the boundary $r=2M$
separated real space (outside the Schwarzschild radius) from the
virtual space inside.  This belief continued until in 1924, when Arthur
Eddington showed that the singularity disappeared after a suitable
change of coordinates.  Nevertheless the Schwarzschild boundary has a
real significance for a distant observer.  A photon starting at or
inside the Schwarzschild boundary cannot cross this boundary.  The
whole of the future of an event on the boundary lies inside the
Schwarzschild radius.  To a stationary outside observer the
boundary appears completely black -- a \emph{black hole} in fact.

Black holes have captured the imagination of the general scientific
public and many good treatments of them can be found in the literature
to follow up the bare bones given here.

\ssh{De Sitter space}
\index{de Sitter space}

The de Sitter metric defines a space called de Sitter space.  It is of
fundamental importance for the new paradigm presented in this book
because the new model for the universe with observed redshift is based
on it (see \fullref{sec:red}) and also the new explanations for the
CMB and gamma ray bursts (see \fullref{sec:CMB} and
\fullref{sec:revis}).

This space is explored in some detail in \fullref{sec:deS} as part of
the explanation of redshift.  A fuller treatment can be found in
\fullref{app:deS}.  \index{Einstein!general relativity|)}
\index{relativity!general|)}

\chapter{De Sitter space}\label{app:deS}
\index{de Sitter space|(}

\sh{Minkowski space}

{\em Minkowski $n$--space} $M^n$ is $\R^n=\R\times\R^{n-1}$ (time
times space) equipped with the standard (pseudo)-metric of signature
$(-,+,\ldots,+)$:
$$ds^2 = - dx_0^2 + dx_1^2 + \ldots + dx_{n-1}^2$$ The time coordinate
is $x_0$ and the space coordinates are $x_1,\ldots,x_{n-1}$.  For $x,
y\in M^n$, the (pseudo-)inner product $\langle x, y\rangle$ is defined
to be $-x_0y_0+x_1y_1+\ldots+x_{n-1}y_{n-1}$.

\iffalse  We write $\| x\|$ for
$\pm\sqrt{\pm \xx}$ where the signs agree and the inner sign is chosen
to make the argument of the square root positive and we think of
$\|x\|$ as the ``length'' of $x$.
\fi

The {\em Lorentz $n$--group} is the group of ``isometries''
(transformations preserving the inner product) of Minkowski space,
fixing ${\bf 0}$, and preserving the time direction.  This implies
that a Lorentz transformation is an linear isomorphism of $\R^n$ as a
vector space.  If, in addition to preserving the time direction, it
also preserves space orientation then the resulting group can be
denoted $\SO(1,n-1)$.  Notice that a Lorentz transformation which
preserves the $x_0$--axis is an othogonal transformation of the
perpendicular $(n-1)$--space, thus $\SO(n-1)$ is a subgroup of
$\SO(1,n-1)$ and elements of this subgroup are (Euclidean)
rotations about the $x_0$--axis.

Minkowski $4$--space is simply called {\em Minkowski space} and is the
simplest example of a space-time.  The Lorentz $4$--group is called the
{\em Lorentz group}.

\sh{Space-times}

A pseudo-Riemannian manifold $L$ is a manifold equipped with
non-degenerate quadratic form $g$ on its tangent bundle called the
{\em metric}.  A {\em space-time} is a pseudo-Riemannian 4--manifold
equipped with a metric of signature $(-,+,+,+)$.  Minkowski space is
the simplest example of a space-time and in general the Lorentz gropup
acts as structure group for the tangent bundle of a space-time.  The
metric is often written as $ds^2$, a symmetric quadratic expression in
differential 1--forms as above.  A tangent vector $v$ is {\em
  time-like\/} if $g(v)<0$, {\em space-like\/} if $g(v)>0$ and {\em
  null\/} if $g(v)=0$.  The set of null vectors at a point form the
{\em light-cone\/} at that point and this is a cone on two copies of
$S^2$.  The set of time-like vectors at a point breaks into two
components bounded by the two components of the light cone.  A choice
of one of these components determines the {\em future\/} at that point
and {\em time orientability\/}, ie a global choice of future pointing
light cones is always assumed.  An {\em observer field\/} on a
space-time $L$ is a smooth future-oriented time-like unit vector field
on $L$.

\sh{de Sitter and hyperbolic spaces}

Now go up one dimension.  {\em Hyperbolic $4$--space} is the subset
$$\H^4= \{\xx = -a^2,\ x_0>0 \mid x\in M^5\}.$$
{\em de Sitter space} is the subset
$$\deS=\{\xx = a^2\mid x\in M^5\}.$$ There is an isometric copy
$\H^4_q$ of hyperbolic space with $x_0<0$.  The induced metric on
hyperbolic space is Riemannian and on de Sitter space is Lorentzian.
Thus de Sitter space is a space-time.  It is a solution of Einstein's
equations with positive cosmological constant $\Lambda = 3/a^2$ and no
matter.

The {\em light cone} is the subset $$L=\{\xx = 0\mid x\in M^5\}.$$ and
is the cone on two $3$--spheres with natural conformal geometries (see
hyperbolic geometry below).  These are $S^3$ and $S^3_q$ where $S^3$
is in the positive time direction and $S^3_q$ negative.

\sh {Projective geometry}

Points of 
\begin{equation}
S^3\cup  S^3_q\cup\deS\cup\,\H^4\cup\H^4_q\label{type}
\end{equation}
are in natural bijection with half-rays from the origin and this is
called {\em half-ray space}.  Considering full rays (lines) through
the origin gives a copy of projective $4$--space $\P^4$ which has
half-ray space as its unique double cover.  $\SO(1,4)$ acts faithfully
on $\P^4$ by projective transformations.  Each of $S^3, S^3_q,\H^4$
and $\H^4_q$ is faithfully represented as a subset of $\P^4$ with
copies identified.  $\SO(1,4)$ acts on $\H^4$ by hyperbolic isometries
(see below).  The natural linear structure on $\P^4$ (given by
subspaces of $\R^5$) induces a natural linear structure on each of
$\H^4$ and $\deS$ and in particular planes through the origin cut
$\H^4$ and $\deS$ in geodesics and all geodesics are of this form
(this is proved below).

\begin{figure}[ht!]
\labellist\small
\pinlabel time [r] at 220 351
\pinlabel light [tl] at 316 312
\pinlabel space [l] at 425 211
\endlabellist
\cl{\includegraphics[width=2.2in]{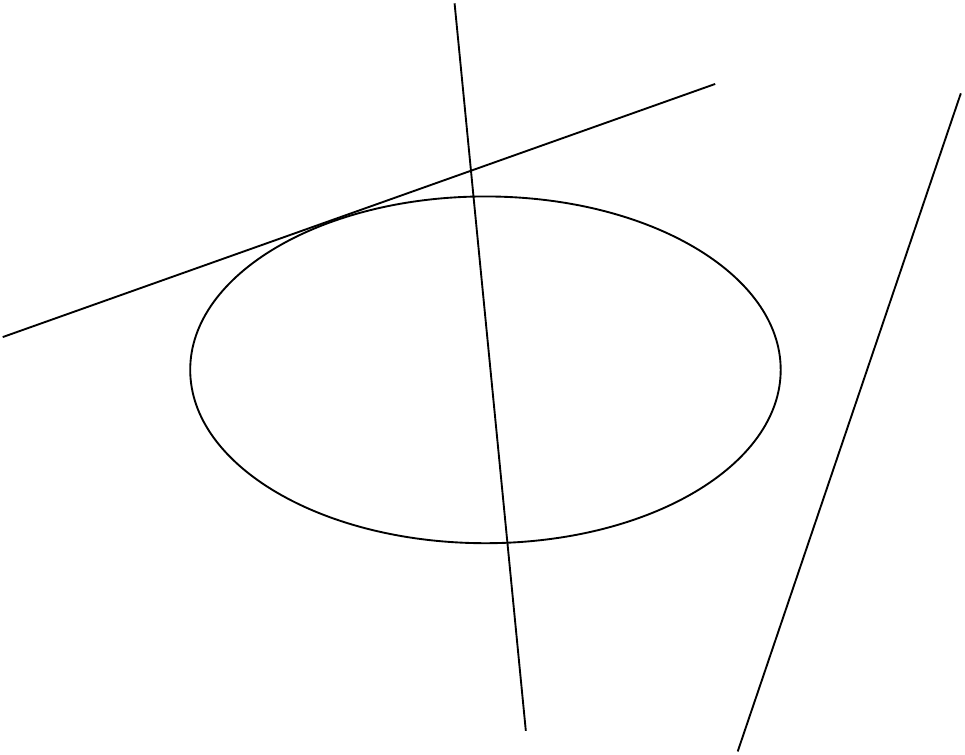}}
\caption{three types of lines}\label{fig:lines}
\end{figure}

Geodesics come in three types: {\em light-like} corresponding to
planes tangent to the light cone; {\em time-like} corresponding to
planes which intersect $\H^4$ and {\em space-like} corresponding to
planes which meet $L$ only at the origin (and hence miss $\H^4$).
\fullref{fig:lines} is a projective picture illustrating these types.

\sh{Hyperbolic and de Sitter geometry}

The subset $\H^4$ of $\P^4$ with the action of $\SO(1,4)$ is the {\em
  Klein model} of hyperbolic $4$--space.  $S^3$ is then the sphere at
infinity and $\SO(1,4)$ acts by conformal transformations of $S^3$ and
indeed is isomorphic to the group of such transformations.  $\SO(1,4)$
acts as the group of time and space orientation preserving isometries
of $\deS$ and is also known as the {\em de Sitter group} as a result.

\sh{Transitivity of points and geodesics}

There is an element of $\SO(1,4)$ carrying any half-ray to any other
of the same type (corresponding to the decomposition (\ref{type})).
Here is an explicit way to see this which also proves transitivity on
geodesics and checks the characterisation of geodesics mentioned
earlier.  A Lorentz transformation of $M^2$, called a {\em shear},
namely $x\mapsto {\tiny\bigl(\!
\begin{array}{rr} \cosh q&\sinh q\\\sinh q&\cosh q\end{array}\!
  \bigr)}x$,
for suitable choice of $q\in\R$ acting in a vertical plane (one
containing the $x_0$--axis) and crossed with the identity on the
perpendicular $3$--space, will move any point of $\H^4$ to the centre
(intersection with the $x_0$--axis) and any point of $\deS$ to a point
on the equator (intersection with the $(x_1,x_2,x_3,x_4)$--space).
Then a rotation about the $x_0$--axis carries it to any other point.
This proves transitivity for points in $\H^4$ (and similarly $\H^4_q$)
and $\deS$.  For $S^3$ (and similarly $S^3_q$) a rotation about the
$x_0$--axis carries one point to any other.

This argument is now extended to prove transitivity on geodesics.  Since
it has not yet proved that geodesics are intersections with planes
through the origin, such intersections are called {\em lines} and it
will be seen that lines are in fact geodesics shortly.  Notice that a
time-like plane (one meeting $\H^4$, see terminology introduced above)
meets $\deS$ in {\em two} antipodally opposite lines which get
identified in $\P^4$ and which we call a line-pair.

Transitivity will be proved for lines of the same type.  In $\H^4$ all
lines are time-like and any point of each line can be moved to the
centre and then a rotation car ries one line into the other.  Turning now to de
Sitter space, the same sequence of transformations applied to $\deS$
carries any time-like line-pair to any other and a rotation can be
used to swap the lines if necessary.  A space-like line can be carried
to the equator by two perpendicular shears.

\begin{figure}[ht!]\small
\cl{\SetLabels
\L(.6*.5) $O$\\
\L\E(.99*.99) future of $O$\\
\L\E(.81*.35) past of $O$\\
\endSetLabels
%\ShowGrid
\AffixLabels{\includegraphics[width = 2.5in]{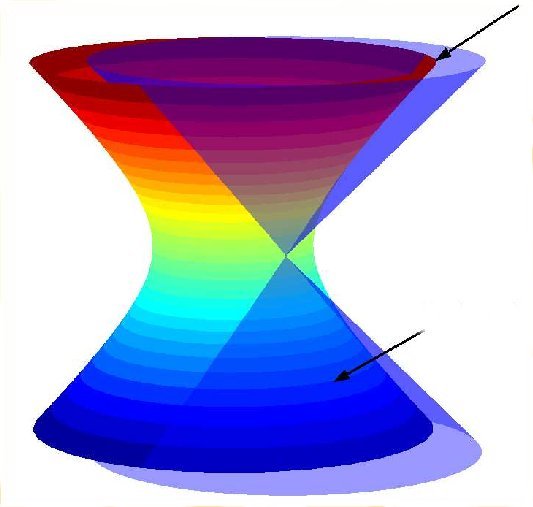}}}
\caption{Light-cones: the light-cone in $\deS$ is the cone on a
  $0$--sphere (two points) in the dimension illustrated, in fact it is
  the cone on a $2$--sphere.  The figure is reproduced from
  \cite{Mosch}.}\label{light-cones}
\end{figure}

To prove transitivity for light-like lines, observe that in
Euclidean terms $\deS$ is a hyperboloid of one sheet ruled by lines
and that each tangent plane to the light cone meets $\deS$ in two
ruling lines.  These lines are light-lines in $M^5$ and hence in
$\deS$.  (See \fullref{light-cones}, which is taken from Moschella
\cite{Mosch}.)
Now a point on a given light-line can be carried to a point of
the equator by a shear followed by a rotation.  But the rotations
fixing this point now carry the light-line around the light cone in
$\deS$ and hence any two are equivalent under an isometry.

The proof of transitivity shows that there are isometries which move
any line along itself and further it can now be seen that the isometries
which fix a given line point-wise form a $\SO(3)$ subgroup of
rotations.  Thus by symmetry, parallel transport along a line carries
a line into itself and they are, as claimed, geodesics.  Since there
are lines through each point in each direction, all geodesics are of
this form.  Thus:

\begin{prop}
Let $l,m$ be geodesics in $\deS$ of the same type and let $P\in l$ and
$Q\in m$.  Then there is an element of $\SO(1,4)$ carrying $l$ to $m$
and $P$ to $Q$.\label{prop:PCP}
\end{prop}

\sh{The expansive metric}

Let $\Pi$ be the $4$--dimensional hyperplane $x_0 + x_4 =0$.  This
cuts $\deS$ into two identical regions.  Concentrate on the upper
complemetary region $\Exp$ defined by $x_0 +x_4 > 0$.  $\Pi$ is
tangent to both spheres at infinity $S^3$ and $S^3_q$.  Name the
points of tangency as $P$ on $S^3$ and $P'$ on $S^3_q$.  The
hyperplanes parallel to $\Pi$, given by $x_0 + x_4 = k$ for $k>0$, are
also all tangent to $S^3$ and $S^3_q$ at $P,P'$ and foliate $\Exp$ by
paraboloids.  Denote this foliation by $\mathcal{F}$.  It will be seen
that each leaf of $\mathcal{F}$ is in fact isometric to $\R^3$.  There
is a transverse foliation by the time-like geodesics passing through
$P$ and $P'$.

\begin{figure}[ht!]
\labellist\small
\pinlabel $\Exp$ [l] at 500 376
\pinlabel $\Pi$ [tr] at 340 30
\pinlabel {leaves of $\mathcal{F}$} [bl] at 505 301
\pinlabel {hyperplanes parallel to $\Pi$} <-2pt, 0pt> [tl] at 566 21
\endlabellist
\cl{\includegraphics[width = 2.7in]{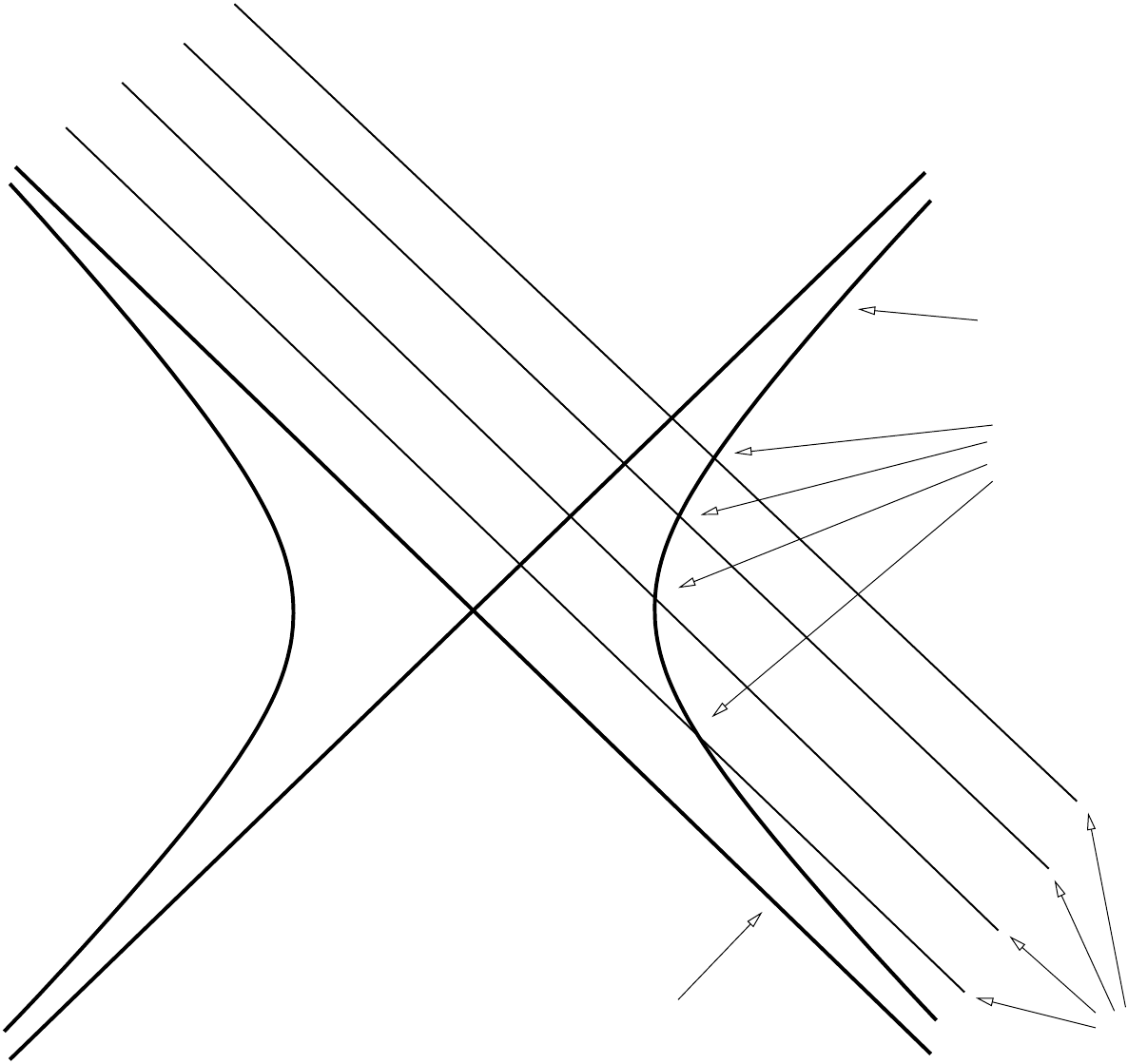}}
\caption{The foliation $\mathcal{F}$ in the $(x_0,x_4)$--plane}\label{foli}
\end{figure}
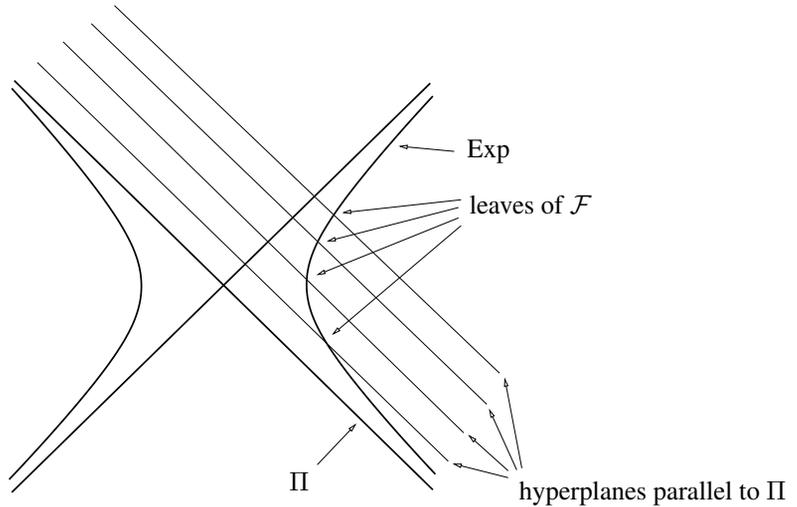

\begin{figure}[ht!]\small
\SetLabels
\R\E(.06*.49) horizons\\
\endSetLabels
%\ShowGrid
\cl{\includegraphics[height=2.05in]{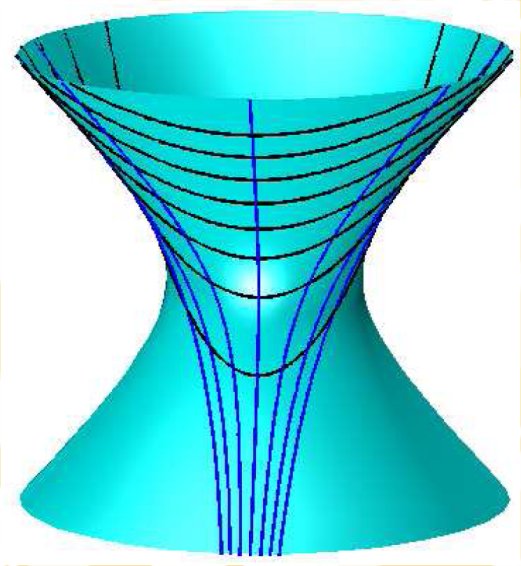}\qquad\qquad\qquad
\AffixLabels{\includegraphics[height=2in]{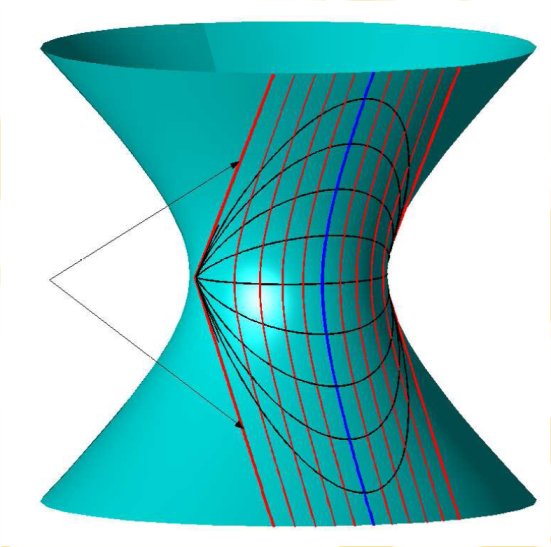}}}
\caption{Two figures reproduced from \cite{Mosch}.  The left-hand
  figure shows the foliation $\mathcal{F}$ (black lines) and the
  transverse foliation by geodesics (blue lines).  The righthand
  figure shows the de Sitter metric as a subset of
  $\deS$.}\label{Mosch_figs}
\end{figure}

 These foliations are illustrated in Figures \ref{foli} and
 \ref{Mosch_figs}.  \fullref{foli} is the slice by the
 $(x_0,x_4)$--coordinate plane and \fullref{Mosch_figs} (the left-hand
 figure) shows the view from the $x_4$--axis in $3$--dimensional
 Minkowski space ($2$--dimensional de Sitter space).  This figure and
 its companion are again taken from Moschella \cite{Mosch}.

Let $G$ be the subgroup of the Lorentz group which fixes $P$ (and
hence $P'$ and $\Pi$).  $G$ acts on $\Exp$.  It preserves both
foliations: for the second foliation this is obvious, but all Lorentz
transformations are affine and hance carry parallel hyperplanes to
parallel hyperplanes; this proves that it preserves the first
foliation. Furthermore affine considerations also imply that it acts
on the leaves of $\mathcal{F}$ by scaling from the origin: in other
words there is a scale factor $\lambda(g)$ for each $g\in G$ which
maps the leaf at distance $\mu$ from $\Pi$ to the one at distance
$\lambda(g)\mu$ (here distance means Euclidean distance in $\R^5$).
The map $\lambda\co G\to \R_*$ is a homomorphism from $G$ to the
positive reals under multiplication.

Now consider the action of $G$ on $S^3$ (the light sphere at
infinity).  As remarked earlier $\SO(1,4)$ acts on $S^3$ as the group
of conformal isomorphisms.  Thus $G$ is the subgroup of conformal
isomorphisms of $S^3$ fixing the point $P$.  But this is a very
familiar group.  Use stereographic projection to identify $S^3 - P$
with $\R^3$ then $G$ acts as the group of conformal isomorphisms of
$\R^3$, which is precisely the group of similarities of $\R^3$ ie the
group generated by isometries and dilations.  There is then another
homomorphism $\sigma\co G\to \R_*$ which maps each element to its
scale factor.  It can be checked that $\lambda=\sigma$.  One way to
see this is to list the non-trivial normal subgroups of $G$ (this is
not hard) and check that there is only one homomorphism to $\R_*$.
Another way is to check directly that a standard shear $j$, say, in the
$(x_0,x_4)$--plane is a pure dilation at infinity by exactly
$\lambda(j)$.  (It is easy to see that $j$ acts on leaves by dilation
to first order and at infinity.  The scaling is given by affine
considerations.) 

Now define $H$ to be the kernel of $\lambda=\sigma$.  This comprises
elements of $G$ which maps the leaves of $\mathcal{F}$ into themselves,
and in terms of the action on $S^3 - P = \R^3$, it is the
subgroup of Euclidean isometries.  Thus the Euclidean group acts on each
leaf of $\mathcal{F}$.  By dimension considerations it acts an the
full group of isometries of each leaf.  It follows that each leaf has
a flat Euclidean metric.  

It can now be seen that the metric on $\Exp$ is the same as the
Robertson--Walker metric for a uniformly expanding infinite universe.
The transverse foliation by time-like geodesics determines the
standard observer field and the distance beetween hyperplanes defining
$\mathcal{F}$ gives a logarithmic measure of time.  Coordinates are
given below.  Notice that it has been proved that every isometry of
$\Exp$ is induced by an isometry of $\deS$.

It is worth remarking that exactly the same analysis can be carried
out for $\H^4$ where the leaves of the foliation given by the same set
of hyperplanes are again Euclidean.  This gives the usual
``half-space'' model for hyperbolic geometry with Euclidean horizontal
sections and vertical dilation.

\sh{Time-like geodesics in $\Exp$}

There is a family of time-like geodesics built in to $\Exp$ namely the
observer field mentioned above.  These geodesics are all {\em static}.
They are all equivalent by a symmetry of $\Exp$ because we can use a
Euclidean motion to move any point of one leaf into any other point.
Other time-like geodesics are {\em non-static}.  Here is a perhaps
surprising fact:

\begin{prop}\label{prop2}
Let $l,m$ be any two non-static geodesics in $\Exp$.  Then there is an
element of $G = \Isom(\Exp)$ carrying $l$ to $m$.
\end{prop}

Thus there in no concept of conserved velocity of a geodesic with
respect to the standard observer field in the expansive metric.  This
fact is important for the analysis of black holes in a universe with a
fixed cosmological constant, cf \cite{redshift}.

The proof is easy if one thinks in terms of hyperbolic geometry.
Time-like geodesics in $\Exp$ are in bijection with geodesics in
$\H^4$ since both correspond to 2--planes through the origin which
meet $\H^4$.  But if using the upper half-space picture for $\H^4$
static means vertical and two non-static geodesics are represented by
semi-circles perpendicular to the boundary.  Then there is a conformal
map of this boundary (ie a similarity transformation) carrying any two
points to any two other: translate to make one point coincide and then
dilate and rotate to get the other ones to coincide.

\sh{The de Sitter metric}

There is another standard metric inside de Sitter space which is the
metric which de Sitter himself used.  This is illustrated in
\fullref{Mosch_figs} on the right.  This metric is static, in other
words there is a time-like Killing vector field (one whose associated
flow is an isometry).  The region where it is defined is the
intersection of $x_0+x_4>0$ defining $\Exp$ with $x_0-x_4<0$ (defining
the reflection of $\Exp$ in the $(x_1,x_2,x_3,x_4)$--coordinate
hyperplane).  The observer field, given by the Killing vector field,
has exactly one geodesic leaf, namely the central (blue) geodesic.
The other leaves (red) are intersections with parallel planes not
passing through the origin.  There are two families of symmetries of
this subset: an $\SO(3)$--family of rotations about the central
geodesic and shear along this geodesic (in the $(x_0,x_1)$--plane).
Both are induced by isometries of $\deS$.

This metric accurately describes the middle distance neighbourhood of
a black hole in empty space with a nonzero cosmological constant.  The
embedding in $\deS$ is determined by the choice of central time-like
geodesic.  \fullref{prop2} then implies that there are precisely two
types of black hole in a standard uniformly expanding universe.  

\sh{Explicit formulae}

Here are explicit formulae for these two metrics in terms of the $x_i$
adapted from wikipedia.

\textbf{Expansive metric}\qua Let
\begin{align*}x_0 &= a \sinh(t/a) + r^2\exp(t/a) / 2a\\
x_4 &= a \cosh(t/a) - r^2\exp(t/a) / 2a\\
x_i &= \exp(t/a)\, y_i  \text{\ \ for\ \ }  1 \le i \le 3
\end{align*}
where $r^2 = \sum_i y_i^2$.
Then in the $(t,y_i)$ coordinates the metric reads
$$
ds^2 = - dt^2 + \exp(2t/a)\, du^2
$$
where $du^2=\sum dy_i^2$ is the flat metric on $y_i$'s.

\textbf{de Sitter coordinates (static coordinates)}\qua
Let
\begin{align*}
x_0 &= \sqrt{a^2 - r^2} \sinh(t/a)\\
x_4 &= \sqrt{a^2 - r^2} \cosh(t/a)\\
x_i &= r\, z_i \text{\ \ for\ \ }  1 \le i \le 3
\end{align*}
where $z_i$ gives the standard embedding of $S^2$ in $\R^3$ then
in $r, z_i$ coordinates the metric reads
\begin{equation}\label{eq:deS_metric}
ds^2 = -Q dt^2 + 1/Q dr^2 + r^2 d\Omega^2  
\end{equation}
where $Q = 1 - (r/a)^2$ and
$d\Omega^2$ is the standard metric on $S^2$.
\index{de Sitter space|)}

\np\thispagestyle{empty}

\chapter{Quasars: technical material}\label{app:3author}
\index{quasar|(}

This appendix contains the technical material from the three author
paper \cite{BHQR} deferred from \fullref{sec:quasars}.  This paper,
which is joint work with Robert MacKay and Rosemberg Toala Enriques,
is still in draft form: the next version will deal accurately with
ionisation of the incoming gas$/$plasma stream and model the settling
region.  Note that, throughout this appendix, scientific (MKS) units
are used rather than the natural units used elsewhere and that $G$ is
Newton's gravitational constant and not the Einstein tensor.  The appendix
starts with the omitted details for the Bondi sphere radius and the
accretion rate.

\index{MacKay, Robert}\index{Toala Enriques, Rosemberg}

\sh{Bondi sphere radius and accretion rate}\label{sec:Bondi}

The \ind{Bondi sphere} of radius $B$ is defined by equating the root mean
square velocity $\sqrt{3kT/m_H}$ of Hydrogen atoms in the medium with the
escape velocity $\sqrt{2GM/B}$.  Here $T$ is the temperature of the
medium at the Bondi radius, $M$ is the mass of the BH, $G$ is the
gravitational constant, $k$ is \ind{Boltzmann's constant} and $m_H$ is the
mass of a proton.  Thus:
\begin{equation}B = \frac{2GM m_H}{3kT} \label{eq:Bondi}
\end{equation}
Note that the Newtonian formula for escape velocity has been used,
which, as will be seen later, is also correct in Schwarzschild
geometry.

The significance of the Bondi sphere is that protons in the medium are
trapped (on average) inside this sphere because they have KE too small
to escape the gravitational field of the BH.  The mass of matter per
unit time trapped in this way is called the \emph{accretion rate} $A$
and can be calculated as\index{accretion!rate}
\begin{equation}
A = 2 B^2 n \sqrt{2\pi kTm_H} 
\label{eq:A}
\end{equation}
where $n$ is the density of the medium (number of protons per unit
volume). 

Here are the details for this calculation.  Maxwell's distribution for
the radial velocity $v_r$ has density $\sqrt{m_H/2\pi kT}
e^{-m_Hv^2/2kT}$, so the mean $\bar v_r$ over inward velocities is 
$$
\int_0^\infty 2\sqrt{\frac {m_H}{2\pi kT}} e^{-m_Hv^2/2kT} v\, dv\,.
$$
Put $u=m_Hv^2/2kT$ to obtain
$$
\int_0^\infty 2\sqrt{\frac {kT}{2\pi m_H}} e^{-u}\, du = 2\sqrt{\frac
  {kT}{2\pi m_H}}. 
$$
Then $A = 4\pi B^2 nm_H\bar v_r/2 = 2B^2 n \sqrt{2\pi kT m_H}$.

\sh{Kinetic energy, escape velocity and redshift}\label{sec:KE}

This is the start of the detailed calculations of the energy production.  

Throughout the appendix the standard Schwarzschild metric is used
\begin{equation}\label{eq:metric-sch}
c^2\,ds^2 = -Q\,c^2\,dt^2 + \frac1Q\,dr^2 + r^2\,d\Omega^2 ,
\end{equation}
where $Q=1-S/r=1-2GM/c^2r$.  Here $t$ is thought of as time, $r$ as
radius and $d\Omega^2$, the standard metric on the 2--sphere, is an
abbreviation for $d\theta^2 + \sin^2\theta \, d\phi^2$ (or more
symmetrically, for $\sum_{j=1}^3dz_j^2$ restricted to
$\sum_{j=1}^3z_j^2=1$ and $S=2GM/c^2$ is the Schwarzschild radius.
Note that $\sqrt{-ds^2}$ can be regarded as proper time.

It is necessary to discuss KE.  As remarked earlier, this is not a
relativistic concept.  It makes sense in Minkowski space where there
is the Einstein formula for the KE of a particle of mass $m$ moving
with velocity $v$
\begin{equation}
mc^2 \left(\frac1{\sqrt{1-v^2/c^2}} - 1\right) 
\label{eq:KE}
\end{equation}
and therefore it makes sense in an inertial frame of reference.  

Consider a particle falling freely and radially into a Schwarzschild
\BH\ (and hence following a geodesic).  Use $\tau$ for proper time along
this geodesic.  Let $\rdot$ denote $dr/d\tau$.  The MacKay--Rourke
paper \cite{nat-flat} describes two natural flat observer fields, the
escape field and the dual capture field.  Use the latter.  This
gives a foliation by geodesics following inward freefall paths with
orthogonal flat space slices (ie isometric to Euclidean 3--spaces).
Thus there are local coordinates with time being proper time along the
geodesics and space defined by flat Euclidean coordinates in the
orthogonal space slices.  These local coordinates provide convenient
inertial frames in which to measure KE.
\index{inertial frame!Schwarzschild space}

Now the flat slices are derived by making the distance between spheres
of area $4\pi r_1^2$ and $4\pi r_2^2$ be $|r_2-r_1|$ and hence $r$ is
a Euclidean coordinate and it follows that $\rdot$ is the correct
definition of radial velocity for calculating KE.  For tangential
velocity, $\theta,\phi$ provide standard spherical coordinates in this
inertial frame and the usual Euclidean formula for velocity in
$(r,\theta,\phi)$ (again measured wrt $\tau$) provides the correct
velocity $v$ to measure KE in equation (\ref{eq:KE}).

A formula for escape velocity is also needed.  MacKay and Rourke
provide this in \cite[Equation (10)]{nat-flat} namely
$\rdot=c\sqrt{1-Q} = \sqrt{2GM/r}$.  [MacKay and Rourke use natural
  units with $G=c=1$, a factor $c$ has been added to convert to MKS
  units.]

In the next section these formulae are derived by a simple direct
analysis but first here is the promised formula from which the
\ind{Eddington radius} $R$ can be read.

Recall from \fullref{sec:3spheres} that the \emph{\ind{Eddington
    sphere}} of radius $R$ is defined by equating outward radiation
pressure on the protons in the medium with inward gravitational
attraction from the \BH.  Also recall the standard equation for the
luminosity at the Eddington limit, \cite[page 5]{meier}
\begin{equation}L_E = \frac{4\pi}\kappa GMc  
\label{eq:Edd}
\end{equation}
where $\kappa$ is the radiative opacity for electron scattering which
is usually taken to be $0.4 {\rm cm}^2 / {\rm g}$ or $4\times 10^{-2}$
in MKS units \cite[page 5]{meier}.  The Eddington sphere is defined by
the same considerations and hence this gives the radiation emitted
from this sphere.  Note that this formula does not depend on the
radius of the radiating sphere.  Since it corresponds to local balance
of forces, it is true in a relativistic setting provided it is stated
exactly where it is applied.  It is applied near the \ind{Eddington
  sphere}.

Now assume that the luminosity is, within a factor $X$, the same as
the KE of accreted matter falling onto the Eddington sphere.  The
intuitive description that given in \fullref{sec:3spheres} of the
nature of the Eddington sphere suggests that about $1/2$ of the KE
released on ``impact'' should be radiated outwards and about $1/2$
absorbed into the medium below so that $X$ is roughly $1/2$.  But, as
will be seen later, there is also energy arriving upwards from inside
the sphere, and this suggests a larger figure for $X$.  This estimate
will be revisited later, but for now keep $X$ as a parameter to be
determined.

Equating $X$ times the KE released on impact with the Eddington
luminosity gives
\begin{equation}
X\,A\, c^2 \left(\frac1{\sqrt{1-v^2/c^2}} - 1\right) = \frac{4\pi}\kappa GMc
\label{eq:PreTrap}
\end{equation}
where $v=\sqrt{2GM/R}$ is the escape velocity at $R$, the velocity of freely
infalling matter.  Matter does not in fact arrive radially because of
tangential motion, which is amplified by conservation of angular
momentum as described earlier.  However the energy of motion available
to be absorbed and re-radiated is unaffected by the transfer of energy
from radial to partially tangential and therefore there is no error in
assuming that motion is radial here.

It is worth digressing a little here.  A particle in the outer region
with significant tangential velocity may not reach the Eddington
sphere.  This happens if the tangential velocity, amplified by
conservation of angular momentum, absorbs all the KE and the radial
velocity slows to zero.  But, because of the mechanics near the Bondi
sphere described earlier, particles cannot escape the outer region in
significant numbers.  It is implicitly assumed that there is a steady
state on timescales short compared with that given by the accretion
rate.  It follows that excess tangential velocity in the outer region
must be transmuted into radial velocity by non-thermal particle
interaction as suggested earlier.  Thus in this region particle
interaction allows the plasma to ``settle'' inwards towards the
Eddington sphere, without significant loss of KE.  This settling
process will need to be modelled in detail in the next version of this
work.  At this stage just assume that it takes place.  There are some
features of the process that can be deduced from observations
discusssed in \fullref{sec:conc}.  \index{outer region!settling}

It is not hard to solve equation (\ref{eq:PreTrap}) to find an
explicit formula for the Eddington radius $R$ in terms of the other
parameters.  For calculation purposes however, it is far more
convenient to use redshift which has a simple relationship to $R$.
For a \BH\ with Schwarzschild radius $S= 2GM/c^2$, redshift $1+z$ at a
radius with escape velocity $v$ is
$1/\sqrt{1-v^2/c^2}=1/\sqrt{1-S/R}$, since $v=2GM/R$, and hence $1-S/R
= (1+z)^{-2}$ or
\begin{equation}
S=R(1-(1+z)^{-2}).
\label{eq:Rz}
\end{equation}
But in terms of $z$, equation (\ref{eq:PreTrap}) gives the following
simple formula for the observed redshift for a \BH\ radiating from the
Eddington sphere:
\begin{equation}
z = \frac{4\pi MG}{A c\kappa X}
\label{eq:z}
\end{equation}
and then  substituting for $A$ and $B$ gives:
$$
z = \frac{4\pi MG}{2 (\frac{2GM
      m_H}{3kT})^2 n\, \sqrt{2\pi kT m_H}\, c\kappa X}
$$
and collecting terms:
\begin{equation}
z = 2^{-1}\,9\sqrt{\pi/2}\,\kappa^{-1}\, M^{-1}\, n^{-1}\, (kT)^{1.5}\, m_{\rm
  H}^{-2.5}\, G^{-1}\, c^{-1}\, X^{-1}
\label{eq:zsubs}
\end{equation}

\sh{Potential and kinetic energy in Schwarzschild space-time}\label{sec:Ros}

This section gives the promised direct calculation using
Schwarzschild geometry for the formulae used in \fullref{sec:KE} for
KE and escape velocity.

Take the approach that a particle is fundamentally described
by its 4--momentum, that is, by $P = m U$, where $m = \sqrt{-\langle
  P,P \rangle}$ is the rest mass of the particle and $U = (\dot{t},
\dot{r}, \dot{\theta}, \dot{\phi})$ is its 4--velocity and dot
represents differentiation with respect to proper time.

Consider a particle falling freely in Schwarzschild spacetime, that is
following a geodesic path. There are conserved quantities associated
to the symmetries of the Schwarzschild spacetime, for example
	\[
	 E_0 = - \langle P, \partial_t \rangle .
	\]
It is tempting to interpret $E_0$ as the energy measured by a static
observer, however this is misleading since $\partial_t$ does not have
unit-length and hence does not correspond to a physical
observer. There is one exception though, at infinity $\partial_t$
corresponds to an observer comoving with the gravitational source, so
it makes sense to interpret $E_0$ as the energy of the particle
measured at infinity by a static observer.

Correspondingly, 
	\begin{eqnarray*}
	E := - \langle P, \frac{1}{\sqrt{Q}} \partial_t \rangle =
        \frac{E_0}{\sqrt{Q}},
	\end{eqnarray*}		
is regarded as the energy measured by an \emph{interior} static
observer, where $Q=1-\frac{2GM}{c^2r}$. Explicitly, $ E =
\dot{t} E_0 \sqrt{Q}$.

As the particle falls inwards it gains potential energy
	\begin{eqnarray*}
	{\rm PE} := E_0 - E = E_0 \left(1-\frac{1}{\sqrt{Q}} \right)
	\end{eqnarray*}		
and the relativistic expression for the Kinetic energy can be written
as the difference between the observed energy and the rest energy of
the particle,
	\begin{eqnarray*}
	{\rm KE} := E - m c^2 
	\end{eqnarray*}		
and this gives a conservation law of the form
	\begin{eqnarray*}
	{\rm KE} + {\rm PE}  = E_0 - m c^2 
	\end{eqnarray*}		
where the RHS can be interpreted as the kinetic energy available at
infinity. For example, it vanishes when the particle is falling at
escape velocity, cf equation $(\ref{3})$.

Now elaborate the formula for KE. The proper time parametrisation
condition translates to
	\begin{eqnarray*}
	\langle P, P \rangle =-m^2
	\end{eqnarray*}			
which, for a particle falling radially, reduces to 
	\begin{eqnarray} \label{1}
	- Q c^2 \dot{t}^2 +  Q^{-1} \dot{r}^2 = -c^2
	\end{eqnarray}
This in turn can be written as a single ODE for $r$, using the conservation of ``energy'',
	\begin{eqnarray} \label{2}
	 \dot{r}^2 = c^2\left(\frac{E_0^2}{m^2c^4} - Q\right).
	\end{eqnarray}
From this it is possible to deduce the escape velocity as measured by
proper time.  Note that for the particle to get asymptotically to
infinity ($\dot{r}= 0$ at $r=\infty$) it is necessary that
$mc^2=E_0$. Hence the velocity necessary to achieve this is
	\begin{eqnarray} \label{3}
	\dot{r}_{\rm escape} = \pm c \sqrt{1-Q}= \pm \sqrt{\frac{2GM}{r}},
	\end{eqnarray}
which recovers the classical value. 

\textbf{Remark}\qua These geodesics, namely the ones that follow
$(\dot{t}, \dot{r}) = (\frac{1}{Q}, \pm c \sqrt{1-Q})$, are precisely
the natural observer fields found by MacKay and Rourke and they
correspond to a stream of test particles falling at precisely
escape velocity.

Returning to kinetic energy, note that
	\begin{eqnarray*}
		{\rm KE}  &=&  mc^2 \left( \frac{E}{m c^2} - 1  \right) \\
			&=& mc^2 \left( \frac{\dot{t} E_0 \sqrt{Q}}{m c^2} - 1  \right).
	\end{eqnarray*}	
Dividing $(\ref{1})$ by $\dot{t}^2$ gives
	\begin{eqnarray*}
		\dot{t}  = \sqrt{\frac{Q}{Q^2-u^2/c^2}},
	\end{eqnarray*}	
where $u=\frac{\dot{r}}{\dot{t}}$ is the velocity measured by the static coordinates. However, it will be convenient to use the velocity measured by the MacKay--Rourke natural flat observers, that is
	\begin{eqnarray*}
		v= \frac{dr}{d\tau}=\frac{dr}{dt} \frac{dt}{d\tau}=\frac{u}{Q}	
	\end{eqnarray*}
Therefore the kinetic energy can be written as:
	\begin{eqnarray*}
		{\rm KE}  = mc^2 \left( \frac{E_0 }{m c^2\sqrt{1-v^2/c^2}} - 1  \right)
	\end{eqnarray*}	
Note that for the case of a particle falling at escape velocity this reduces to: 
	\begin{eqnarray*}
		{\rm KE}  = mc^2 \left( \frac{1 }{\sqrt{1-v^2/c^2}} - 1  \right)
	\end{eqnarray*}

\sh{The critical radius and high redshift \BH s}
\index{critical radius}\index{black hole!high redshift}

Before inserting numbers to compare with observations, there are a
couple more pieces of theory.  Consider a particle infalling from
outside the \BH\ and suppose that at radius $r$ it releases all its KE,
which radiates outwards.  The KE is $\mathrm{KE}(r) = m c^2 (
1/\sqrt{Q} -1)$ where $Q = 1 - 2GM/rc^2 = 1 - v^2/c^2$ and $v =
\sqrt{2GM/r}$ the escape velocity at $r$.  The energy $E(r)$ received
outside the \BH\ is $Q = 1/(1+z)^2$ times this, in other words
\begin{equation}
E(r) = mc^2 ( \sqrt{Q} - Q )
\label{eq:energy_rec}
\end{equation}
which is $\ge 0$ and zero when $v = 0$ and when $v=c$.  The first is
natural and obvious but the second is counterintuitive.  KE $\to
\infty$ as the particle approaches the speed of light at the
Schwarzschild radius and you expect the released energy to $\to
\infty$ as well.  It doesn't.

This mistake occurs in the literature in several places.  See for
example the discussion in the introduction to \cite{BLN}.  There is no
observational difference between a \BH\ and a super-dense neutron star
whose surface is just a little bit above the event horizon.  The error
is to ignore the redshift reduction in radiated energy.
\index{Schwarzschild radius!received energy}

$E(r)$ has a simple maximum when $Q=1/4$ so there is a maximum energy
released.  This depends \emph{only on} $m$ \emph{and not on} $M$.
Again highly counterintuitive.  What does depend on $M$ is the
\emph{critical radius} $r= 4S/3$ at which this maximum is achieved.
Here $1-v^2/c^2 = 1/4$ or $v = c \sqrt3/2$ and $E(r) = mc^2/4$.

Inside the critical radius the received energy drops off sharply and
this allows us to obtain a bound on the radiated energy for \BH s whose
Eddington radius is $\le 4S/3$ or equivalently with redshift
(calculated at the Eddington sphere) $1+z\ge2$ or $z\ge1$.  Let's call
these \BH s \emph{high redshift \BH s}.

The KE for an infalling particle $P(r) = \mathrm{KE}(r) = m c^2 (
1/\sqrt{Q} -1)$ represents the maximum energy available to be
converted into radiation at that radius, see \fullref{sec:Ros}.  This
conversion as analogous to friction.  The medium inside the Eddington
radius is ``sticky'' and slows the particle down, releasing energy.
Now normalise so that all radiated energy is measured as received
outside the \BH.  To do this multiply by $1/(1+z)^2 = Q$.  Assume that
the emissions come from inside the critical radius so that the
received energy per unit $r$--distance is decreasing monotonically.
Once a portion of $P(r)$ is converted to radiation, it is not
replaced, so for maximum effect it needs to be radiated outwards as
soon as possible.  In other words the maximum possible radiation
outwards is obtained by keeping the inward velocity as low as possible
(very small KE).  So for a bound assume all the KE available at the
Eddington radius is radiated outwards and within the Eddington radius
set $\rdot =0$ and this gives an upper bound for the extra energy
received outside the \BH\ from below the Eddington radius $R$:
\begin{align*}
- mc^2 &\int_S ^R Q\, \frac {dQ^{-\frac12}}{dr} dr\\
      	& = mc^2  \int_S ^R Q\, \frac{Q^{-\frac32}}2 \frac{dQ}{dr} dr\\
	& = mc^2  [\sqrt{Q}] \text{ evaluated at }R
\end{align*}
Since $Q\le (1/2) \sqrt{Q}$ in this range, this is within a factor 2
of the KE arriving at the Eddington radius from above, and hence the
total possible energy radiated outwards is 3 times this KE.  In other
words, in terms of the notation of \fullref{sec:KE}, it has been
proved that $X\le3$.  However, the assumption that all this energy
radiates outwards is unrealistic and the earlier estimate of $X=1/2$
is much more reasonable.

\textbf{Note}\qua The same analysis gives a rough upper bound for \BH s
with small redshift but the result $\sqrt{Q}$ evaluated at the
Eddington radius may be far larger than the Eddington luminosity and
not provide a useful upper bound.  Indeed as $r\to\infty$ it tends to
$mc^2$.

\sh{Calculations}\label{sec:calc}
\index{quasar!calculations}

The model will now be compared numerically with observations.  In this
section various parameters are calculated and, in the next section,
their fit with data is tested.  MKS units are used throughout, work to
3 significant figures, and use the following constant values:

$\kappa=4\times 10^{-2}$, $k = 1.38 \times 10^{-23}$, $m_{\rm H} = 1.67
\times 10^{-27}$, $G = 6.67 \times 10^{-11}$, $c= 3\times 10^{8}$.

\ssh{Redshift in terms of medium factor and mass}

The key equation is the redshift equation (\ref{eq:zsubs}):
\begin{equation*}
z = 2^{-1}\,9\sqrt{\pi/2}\,\kappa^{-1}\, M^{-1}\, n^{-1}\, (kT)^{1.5}\, m_{\rm
  H}^{-2.5}\, G^{-1}\, c^{-1}\, X^{-1}
\end{equation*}
For convenience (and familiarity) express $M$ in solar masses; in
other words write $M=\K\,M_{\rm sun}= 2\times 10^{30}\, \K$, where
$\K$ is the \BH\ mass in solar masses.  Substituting for $\kappa, k,
m_H, G, c$ gives the numerical version which was previewed as equation
(\ref{eq:red-prev}):
\begin{equation}
z = 1.27\times 10^7\, \K^{-1}\, n^{-1}\, T^{1.5}\, [1/(2X)]
\label{eq:znum}
\end{equation}
For simplicity use the default value ($\frac12$) for $X$ which is the
same as ignoring the expression in square brackets.  If further
information on $X$ comes to light, it can be reinstated.

The factor $n^{-1}\, T^{1.5}$ depends only on the ambient medium; and is
called the \emph{ambient coefficient}, with the notation $\AC$.  Recall
that $n$ is the density in particles (protons) per cubic metre and $T$
is the ambient temperature in degrees Kelvin.  

The equation now takes the simple form:
\begin{equation}
z = 1.27\times 10^7\, \frac{\AC}\K
\label{eq:znumsimp}
\end{equation}
To get an idea of the range of possible values for $\AC$, interstellar
density is estimated at between $10^2$ and $10^{12}$ where the thinner
regions are associated with higher temperatures, which vary inversely
with the density from about $10^5$ to $10$ \cite{wiki:ISM}.  Thus
$\AC$ varies from about $10^{5.5}$ at the high end (hot thin plasma)
to $10^{-10.5}$ at the low end (cold dense gas).  [An aside here:
  ``dense'' is a relative term.  The density of the atmosphere is
  $10^{25}$, and the interstellar density is always far smaller then a
  laboratory ``high vacuum'' of about $10^{16}$.]

As you can see immediately, the redshift depends critically on the
nature of the ambient medium, which can cause it to vary by 16 orders
of magnitude.  By contrast, the variation with mass, which might be in
the range $10^4$ to $10^8$ solar masses, is far smaller, a further 4
orders of magnitude.  For example, given a \BH\ of mass $10^7\,M_\sun$
(a little bigger than Sgr\astar), so that $10^7\, \K^{-1}=1$, then
avoiding the extremes for the ambient coefficient, the redshift might
vary from $10^{-7}$, in other words so small that there is no
measurable redshift, up to $10^3$ which is so big that the redshift
reduction factor in received luminosity, $(1+z)^{-2}$ or about
$10^{-6}$, makes it extremely unlikely that the quasar could be
detected, unless, like Sgr\astar, it is very close to us.

Two remarks at this point:
(1)\qua In \fullref{sec:quasars} it was promised to comment on the
maximum density that supports the observed forbidden lines.  This is
estimated by Greenstein and Schmidt to be about $3\times 10^{10}$
\cite[third paragraph of abstract]{G-S} which fits nearly all the
densities that have been considered, missing just the extreme cold,
dense media.

(2)\qua It is worth looking at the data for Sgr\astar\ since it has
just been mentioned.  This has mass $4.6\times 10^6\, M_\sun$ and
according to the model should have redshift varying from about
$10^{-10}$ to $10^{6}$.  A redshift of $10^4$ would imply that the
received luminosity was $10^{-8}$ of the Eddington limit, which is
exactly what is observed \cite[page 1357 top right]{BLN}.  Thus the
model suggests that the lack of luminosity for Sgr\astar\ is due to a
rather hot, thin medium near this \BH.\index{Sgr A*}

The data for Sgr\astar\ will be examined in detail, at the end of
\fullref{sec:data}.

\ssh{Three types of redshift and the Hubble formula}
\index{redshift!three types and Hubble formula}

The redshift $z=z_{\rm grav}$ used by the model (and quantified above)
is the \emph{gravitational} aka \emph{intrinsic} redshift.  But when
you observe a quasar, you see the \emph{observed} redshift $z_{\rm
  obs}$ which depends on both the gravitational redshift $z_{\rm
  grav}$ and the cosmological redshift $z_{\rm cos}$ which is a
function of distance.

The relationship between the three is
$$
1+z_{\rm obs} = (1+z_{\rm grav})(1+z_{\rm cos})
$$
which, provided at least one of $z_{\rm grav}$ or $z_{\rm cos}$ is fairly small,
can conveniently be approximated as:
$$
z_{\rm obs} \approx z_{\rm grav}+z_{\rm cos}
$$
From the cosmological redshift you can read the distance $d$ by the
Hubble formula $d=cz_{\rm cos}/H$ where $H$ is the Hubble constant
$2.2 \times 10^{-18} {\rm sec}^{-1}$.  Substituting for $c$ gives:
\begin{equation}
d= 1.35\times 10^{26}\, z_{\rm cos}
\label{eq:dH}
\end{equation}
The other observed datum is magnitude which is discussed below.  From
the magnitude and the distance you can calculate the mass.  But you
need the cosmological redshift, which is not observed, to find the
distance.  Deciding how to split the observed redshift into intrinsic
and cosmological is not simple.  The best that can be done is to try
various splits and see how they fit.  There are however examples
(which are referred to as \emph{Arp} quasars) where the observations
suggest a galaxy at the same distance as the quasar so that 
the redshift for this galaxy for $z_{\rm cos}$ can be used.

Specific examples of both these will be looked at in the next section.

\ssh{Luminosity and magnitude}
\index{luminosity and magnitude}

The main observed data for a quasar are redshift and
luminosity, which has a simple relationship to magnitude:
$$L_{\rm obs} = 2.87 \times 10^{-8} \times 10^{-\frac25{\rm mag}}$$
This is the received luminosity in $W/m^2$ and the calculation is
based on comparison with the solar luminosity ($1.3 kw/m^2$) and
magnitude ($-26.7$).  In the model, the emitted luminosity is always
the Eddington luminosity which depends purely on the \BH\ mass:
\begin{equation}
L_E = \frac{4\pi}\kappa GMc  = 1.26 \times 10^{31} \K
\label{eq:Edd-num}
\end{equation}
From this you can calculate the received luminosity by applying three
correction factors.  The first two are straightforward.  Use the
inverse square law and divide by $1/4\pi$ to convert from total
emitted luminosity to received luminosity per unit area and secondly
apply redshift correction $(1+z_{\rm obs})^{-2}$.  (If redshifts are
small, this second factor can be ignored.)

The third factor is more problematic.  Magnitude is usually measured
using visible wavelengths, but \BH\ radiation covers a far wider
spectrum.  This implies that the observed magnitude underestimates the
luminosity by a factor of perhaps 10 or larger.  Further the radiation
from the \BH\ is attenuated by intervening clouds for which there is
strong evidence (see the discussion in \fullref{sec:conc}) and this
gives a further underestimate, which is again difficult to quantify
but which might also be up to a factor of 10.  Let's call the result
of these two the \emph{magnitude correction factor}, denoted $\MCF$,
and note that it might vary between 1 and 100 or more.

Thus
$$
L_{\rm obs} = \frac{L_E}{4\pi\,\MCF\, d^2(1+z_{\rm obs})^2}
$$ 
and substituting for the luminosities and distance (using equation
(\ref{eq:dH})), gives the following formula for mass in terms of
magnitude and redshifts:
$$
\K = \frac {2.87}{1.26} 10^{-31}\times 4\pi\,\MCF\times(1.35)^2\times
10^{52}\times z_{\rm cos}^2 (1+z_{\rm obs})^2\times 10^{-8} \times 10^{-\frac25 {\rm mag}}
$$
which simplifies to:
$$
\K=\MCF\times 5.22\times 10^{(14-\frac25 {\rm mag})}\times z_{\rm cos}^2 (1+z_{\rm obs})^2
$$
To get a feeling for this formula, anticipate the first example in the
next section where the data are treated more accurately.  Objects 2
and 3 in NGC7603 (see \fullref{fig:NGC7603}) both have $\rm mag
\approx 20$ and $z_{\rm cos}\approx .03$ (taken from the main galaxy)
so the formula gives approximately:
$$\K = \MCF\times 5 \times 10^3$$ 
The gravitational redshift is approx $0.3$ and substituting for $\K$ in
the redshift formula (\ref{eq:znumsimp}) gives:
$$\MCF \approx 10^4 \AC$$
Thus $\MCF=1$ (no magnitude correction) corresponds to a black hole of
mass about $5\times 10^3$ solar masses floating in a medium of ambient
coefficient $10^{-4}$ which is pretty cold and dense medium.  Perhaps
the visible filament in which these objects appear to be immersed is a
cold dense cloud.  Or perhaps, the magnitude correction should be
about 100 and the mass $5\times 10^5$, which seems a more likely mass
for a quasar, with the medium having a less extreme ambient
coefficient of about $10^{-2}$.

This section finishes with formulae for the Eddington radius and the
temperature of the Eddington sphere (assuming the radiation is black
body).

\ssh{\ind{Eddington radius}}

Recall $1-S/R = (1+z)^{-2}$ where $S$ is Schwarzschild radius and $R$
is Eddington radius.  Write $\zeta =S/R= 1-(1+z)^{-2}$ and notice that for
small $z$, $\zeta = 2z + O(z^2)$.  Since the Schwarzschild radius of the sun
is $3\times 10^3$m this gives:
\begin{equation}
R = 3\times 10^3\, \K/\zeta
\label{eq:ELum}
\end{equation}

\ssh{Radiant temperature}
\index{quasar!radiant temperature}

Suppose the radiation is effectively black body with temperature $T_B$
(notation intended to keep distinct from $T$ which is ambient
temperature used earlier).  Stefan-Boltzmann gives total luminosity
$4\pi R^2 \sigma T_B^4$, where $\sigma = 5.67 \times 10^{-8}$ and
equating this with Eddington luminosity gives:
$$ 4\pi \times 9 \times 10^6\, \K^2 \times 5.67 \times 10^{-8}\,
T^4_B/\zeta^2 = 1.26 \times 10^{31}\, \K$$
which gives:
\begin{equation}
T^4_B = 1.96\times10^{30}\, \K^{-1}\, \zeta^2
\label{eq:RadT}
\end{equation}
\emph{Example}\qua $\K=10^6$, $z=0.1$ so that $\zeta^2\approx 0.04$ then 
$T_B \approx 1.67 \times 10^5$.

\sh{Data}\label{sec:data}
\index{quasar!data}

Now proceed to examples, that is, given the data $z_{\mathrm{cos}}$,
$z_{\mathrm{grav}}$ and magnitude use the model to deduce luminosity,
mass, ratio ${R}/{S}$, distance to Earth and temperature of the source
as if it were a black body.

Continue to use the default value $\frac12$ for $X$ and ignore the
correction factor $\MCF$ (ie assume that it is $1$). To take these
into account, use the following rules.  Multiply $\AC$ by $X/2$ and
further multiply both $\K$ and $\AC$ by $\MCF$.

First consider the system around NGC 7603, previewed in the last
section, which appears to contain two Arp quasars (objects 2 and 3 in
\fullref{fig:NGC7603}).  Lopez Corredoira and Gutierrez \cite{LCG}
report $z = 0.0295$ and $B = 14.04$ mag for the main galaxy, NGC
7603. A fact that attracted attention is its proximity to NGC 7603B
(Object 1 hereafter), a spiral galaxy with higher redshift $z
=0.0569$, moreover a filament can be observed connecting both
galaxies. They also found two objects superimposed on the filament
with redshifts $0.394\pm0.002$ and $0.245\pm 0.002$ for the objects
closest to and farthest from NGC 7603, Objects 3 and 2, respectively.
$B$--magnitudes corrected for extinction (due to the filament) are
respectively $21.1\pm1.1$ and $22.1\pm 1.1$ \cite{LCG}.

They go on to say ``If we consider the redshifts as indicators of
distance, the respective absolute magnitudes would be : $M_V = -21.5
\pm 0.8$ and $-18.9 \pm 0.8$. However, if we consider an anomalous
intrinsic redshift case (in such a case, in order to derive the
distance, we set $z = 0.03$), the results are: $M_V = -15.2 \pm 0.8$
and $-13.9 \pm 0.8$ resp. In this second case, they would be on the
faint tail of the HII-galaxies, type II; they would be dwarf galaxies,
`tidal dwarfs', and this would explain the observed strong star
formation ratio: objects with low luminosity have higher
EW(H$_{\alpha}$). Of course, this would imply that we have
non-cosmological redshifts. \dots From several absorption lines we
estimated the redshift of the filament apparently connecting NGC 7603
and NGC 7603B as $z = 0.030$, very similar to the redshift of NGC 7603
and probably associated with this galaxy.''

This analysis suggests setting $z_{\mathrm{cos}} = 0.03$ for the
group and $z_{\mathrm{grav}} = z - z_{\mathrm{cos}}$.  Hence the
Hubble distance, $d = c \times z_{\mathrm{cos}}/H = 13.5 \times
10^{25} \times z_{\mathrm{cos}} $, is $4.05\times 10^{24}$ metres in
this case.

Next, the ratio between the Eddington radius and the Schwarzschild
radius is $R/S= 1/{1-(1+z_{\mathrm{grav}})^{-2}}$, this gives 18.6,
3.12 and 2.17 for Objects 1, 2 and 3, where $z_{\mathrm{grav}}$ has
been taken to be equal to 0.028, 0.213 and 0.361, respectively.

The luminosity (in $W/m^2$ received at Earth) is given in terms of the
magnitude by $ L_\mathrm{mag} = 2.87 \times 10^{-8} \times
10^{-\frac25 mag} $.  This gives $5.468\times 10^{-15}$, $5.468\times
10^{-17}$ and $7.904\times 10^{-17}$ for Objects 1, 2 and 3,
respectively.

Obtain the mass by comparing the formulae for the Eddington
luminosity and the magnitude luminosity, $\K= M/M_{\mathrm{sun}} = 4
\pi d^2 L_\mathrm{mag} \times (1+z)^2\times 1.26^{-1} \times 10^{-31}
$.  Thus $\K= 9.45\times 10^{4}$ , $1.32\times 10^{3}$ and $
2.39\times 10^{3} $ for objects 1, 2 and 3, respectively.

The temperature of the quasar as if it were a black body is given by
Stefan's law $ T_B = \left( \lfrac{L (1+z)^2}{ \sigma 4\pi R^{2}}
\right)^{\frac14}$ and in terms of previous data it is
	\begin{eqnarray*}
			T_B = \left( {L_\mathrm{mag}} (1+z)^2 \times
                        \lfrac{1}{\sigma} \times d^2 \times
                        \left(\lfrac{S}{R}\right)^2 \times
                        \left(\lfrac{1}{\K}\right)^2 \times \left(
                        \lfrac{1}{S_{\mathrm{sun}} }\right)^2
                        \right)^{\frac14} .
	\end{eqnarray*}
For Objects 1, 2 and 3 this gives $4.95\times 10^{5}$, $3.52\times 10^{6}$
and $3.63\times 10^{6}$, respectively.
	
Finally, the ambient coefficient is defined by $\AC = 10^{-7}\,z\, \K
$, which helps to constrain the possible values of the ambient density
and temperature.  For the case at hand this gives $2.07\times
10^{-4}$, $2.19\times 10^{-5}$ and $6.75\times 10^{-5}$ for objects 1,
2 and 3, resp.

A spreadsheet has been used for these calculations, and the results
for these and several more examples, are in the tables which follow.
Included are two quasars (3C273 and 3C48) for which the redshift split
is unknown and for which various splits have been tried.  The examples
come from Galianni, Arp, Burbidge, etal \cite{Getal}, Lopez Corredoira
and Gutierrez \cite{LCG2,LCG}, Greenstein and Schmidt \cite{G-S}, and
Hoyle and Burbidge \cite{H-B}. 

\begin{center}\small
  Lopez Corredoira-Gutierrez
\index{quasar!data!Lopez Corredoira-Gutierrez}
\end{center}\nobreak
\begin{adjustbox}{max width=\textwidth}
\begin{tabular}{|ccccc | ccccc ccc|}
\hline
	&	&	INPUTS&	&	&	&	&	& OUTPUTS	&	&	&	&\\
	&	&z	&	&Magnitude	&R/S	&$L_{\textrm{mag}}$	&Solar masses	&Distance	&$T_B$	&$T_B$ * 1/1+z	&	&Ambient coefficient\\
	&Obs	&Cos	&Grav	&	&	&$W/m^2$	&	&	&	&	&X	&$n^{-1} * T^{1.5}$\\
\hline
NGC 7603	&0.029	&0.03	&0	&14.04	&-	&6.948E-14	&1.136E6	&4.050E24	&-	&-	&-	&-\\
Object 1	&0.058	&0.03	&0.028	&16.8	&1.861E1	&5.469E-15	&9.449E4	&4.050E24	&4.951E5	&4.816E5	&0.5	&2.067E-4\\
Object 2	&0.243	&0.03	&0.213	&21.8	&3.121E0	&5.469E-17	&1.316E3	&4.050E24	&3.519E6	&2.901E6	&0.5	&2.189E-5\\
Object 3	&0.391	&0.03	&0.361	&21.4	&2.173E0	&7.905E-17	&2.394E3	&4.050E24	&3.631E6	&2.668E6	&0.5	&6.752E-5\\
NEQ 3	&	&	&	&	&	&	&	&	&	&	&	&\\
Object 1	&0.1935	&0.12	&0.0735	&19.8	&7.562E0	&3.450E-16	&1.040E5	&1.620E25	&7.582E5	&7.063E5	&0.5	&5.973E-4\\
Object 2	&0.1939	&0.12	&0.0739	&19.6	&7.525E0	&4.148E-16	&1.252E5	&1.620E25	&7.257E5	&6.758E5	&0.5	&7.226E-4\\
Object 3	&0.2229	&0.12	&0.1029	&20.2	&5.621E0	&2.387E-16	&7.596E4	&1.620E25	&9.513E5	&8.625E5	&0.5	&6.107E-4\\
Object 4	&0.1239	&0.12	&0.0039	&17.3	&1.290E2	&3.450E-15	&9.097E5	&1.620E25	&1.068E5	&1.064E5	&0.5	&2.772E-4\\
	&	&	&	&	&	&	&	&	&	&	&	&\\
GC 0248+430	&0.051	&-	&-	&-	&-	&-	&-	&-	&-	&-	&	&-\\
QSO 1	&1.311	&0.051	&1.26	&17.45	&1.243E0	&3.005E-15	&7.253E5	&6.885E24	&1.151E6	&5.091E5	&0.5	&7.140E-2\\
QSO 2	&1.531	&0.051	&1.48	&21.55	&1.194E0	&6.885E-17	&2.001E4	&6.885E24	&2.881E6	&1.162E6	&1.5	&6.940E-3\\
	&	&	&	&	&	&	&	&	&	&	&	&\\
B2 1637+29	&0.086	&-	&-	&-	&-	&-	&-	&-	&-	&-	&	&-\\
Partner	&0.104	&0.086	&0.018	&-	&-	&-	&-	&-	&-	&-	&	&-\\
Aligned QSO	&0.568	&0.086	&0.482	&20	&1.836E0	&2.870E-16	&8.470E4	&1.161E25	&1.620E6	&1.093E6	&1.5	&9.568E-3\\
\hline
\end{tabular}
\end{adjustbox}
\eject

\begin{center}\small
\index{quasar!data!Hoyle-Burbidge}
\index{quasar!data!Arp-Burbidge-et al}
Hoyle-Burbidge, Arp-Burbidge-et al
\end{center}
\begin{adjustbox}{max width=\textwidth}
\begin{tabular}{|ccccc | ccccc ccc|}
\hline
	&	&	INPUTS&	&	&	&	&	& OUTPUTS	&	&	&	&\\
	&	&z	&	&Magnitude	&R/S	&$L_{\textrm{mag}}$	&Solar masses	&Distance	&$T_B$	&$T_B$ * 1/1+z	&	&Ambient coefficient\\
	&Obs	&Cos	&Grav	&	&	&$W/m^2$	&	&	&	&	&X	&$n^{-1} * T^{1.5}$\\
\hline
NGC 4319	&0.0057	&0.0057	&0	&-	&-	&-	&-	&7.695E23	&-	&-	&	&-	\\
MK 205	&0.07	&0.0057	&0.0643	&14.5	&8.534E0	&4.549E-14	&3.041E4	&7.695E23	&9.706E5	&9.120E5	&0.5	&1.528E-4	\\
	&	&	&	&	&	&	&	&	&	&	&	&	\\
NGC 3067	&0.0047	&0.0047	&0	&-	&-	&-	&-	&6.345E23	&-	&-	&	&-	\\
3C 232	&0.533	&0.0047	&0.5283	&15.8	&1.749E0	&1.374E-14	&1.288E4	&6.345E23	&2.658E6	&1.739E6	&0.5	&5.314E-4	\\
	&	&	&	&	&	&	&	&	&	&	&	&	\\
ESO 1327-2041	&0.018	&0.018	&0	&-	&-	&-	&-	&2.430E24	&-	&-	&	&-	\\
QSO 1327-206	&1.17	&0.018	&1.152	&16.5	&1.275E0	&7.209E-15	&1.965E5	&2.430E24	&1.575E6	&7.318E5	&0.5	&1.769E-2	\\
	&	&	&	&	&	&	&	&	&	&	&	&	\\
Gal 0248+430	&0.051	&0.051	&0	&-	&-	&-	&-	&6.885E24	&-	&-	&	&-	\\
Q 0248 +430	&1.1311	&0.051	&1.0801	&17.45	&1.301E0	&3.005E-15	&6.144E5	&6.885E24	&1.173E6	&5.638E5	&0.5	&5.185E-2	\\
	&	&	&	&	&	&	&	&	&	&	&	&	\\
Gal Abell 2854	&0.12	&0.12	&0	&-	&-	&-	&-	&1.620E25	&-	&-	&	&-	\\
2319+272 (4C 27.50)	&1.253	&0.12	&1.133	&18.6	&1.282E0	&1.042E-15	&1.240E6	&1.620E25	&9.911E5	&4.646E5	&0.5	&1.098E-1\\
	&	&	&	&	&	&	&	&	&	&	&	&	\\
NGC 3079	&0.00375	&0.00375	&0	&-	&-	&-	&-	&5.063E23	&-	&-	&	&-	\\
0958+559	&1.17	&0.00375	&1.16625	&18.4	&1.271E0	&1.253E-15	&1.502E3	&5.063E23	&5.336E6	&2.463E6	&0.5	&1.368E-4 \\
	&	&	&	&	&	&	&	&	&	&	&	&	\\
Arp, Burbidge, et al.	&	&	&	&	&	&	&	&	&	&	&	&	\\
NGC 7319	&0.022	&0.022	&0	&-	&-	&-	&-	&2.970E24	&-	&-	&	&-	\\
QSO	&2.114	&0.022	&2.092	&21.79	&1.117E0	&5.519E-17	&4.640E3	&2.970E24	&4.293E6	&1.388E6	&0.5	&7.583E-4	\\
\hline
 \end{tabular}
\end{adjustbox}

\begin{center}\small
\index{quasar!data!Greenstein-Schmidt}
  Greenstein-Schmidt
\end{center}
\begin{adjustbox}{max width=\textwidth}
\begin{tabular}{|ccccc | ccccc cccc|}
\hline
	&	& INPUTS	&	&	&	&	& OUTPUTS	&	&	&	&	&	&\\
	&	&z	&	&Magnitude	&R/S	&$L_{\textrm{mag}}$	&Solar masses	&Distance	&$T_B$	&$T_B$ * 1/1+z	&	&Ambient coefficient	&Spectral index\\
	&Obs	&Cos	&Grav	&	&	&$W/m^2$	&	&	&	&	&X	&$n^{-1} * T^{1.5}$	&\\
\hline
3C 273	&0.1581	&0.001	&0.1571	&12.6	&3.951E0	&2.617E-13	&4.768E3	&1.350E23	&2.437E6	&2.106E6	&0.5	&5.852E-5	&0.9\\
	&0.1581	&0.01	&0.1481	&12.6	&4.143E0	&2.617E-13	&4.768E5	&1.350E24	&7.496E5	&6.529E5	&0.5	&5.517E-3	&0.9\\
	&0.1581	&0.05	&0.1081	&12.6	&5.388E0	&2.617E-13	&1.192E7	&6.750E24	&2.888E5	&2.606E5	&0.5	&1.007E-1	&0.9\\
	&0.1581	&0.1	&0.0581	&12.6	&9.363E0	&2.617E-13	&4.768E7	&1.350E25	&1.514E5	&1.431E5	&0.5	&2.164E-1	&0.9\\
	&0.1581	&0.158	&1E-04	&12.6	&5.001E3	&2.617E-13	&1.190E8	&2.133E25	&5.066E3	&5.066E3	&0.5	&9.299E-4	&0.9\\
	&	&	&	&	&	&	&	&	&	&	&	&	&\\
3C 48	&0.3675	&0.001	&0.3665	&16.2	&2.153E0	&9.503E-15	&3.224E2	&1.350E23	&6.022E6	&4.407E6	&0.5	&9.231E-6	&1.25\\
	&0.3675	&0.01	&0.3575	&16.2	&2.187E0	&9.503E-15	&3.182E4	&1.350E24	&1.896E6	&1.397E6	&0.5	&8.886E-4	&0.95\\
	&0.3675	&0.05	&0.3175	&16.2	&2.359E0	&9.503E-15	&7.492E5	&6.750E24	&8.286E5	&6.290E5	&0.5	&1.858E-2	&0.95\\
	&0.3675	&0.1	&0.2675	&16.2	&2.649E0	&9.503E-15	&2.774E6	&1.350E25	&5.638E5	&4.448E5	&0.5	&5.797E-2	&0.95\\
	&0.3675	&0.2	&0.1675	&16.2	&3.754E0	&9.503E-15	&9.413E6	&2.700E25	&3.489E5	&2.988E5	&0.5	&1.232E-1	&0.95\\
	&0.3675	&0.367	&0.0005	&16.2	&1.001E3	&9.503E-15	&2.328E7	&4.955E25	&1.704E4	&1.703E4	&0.5	&9.093E-4	&0.95\\
\hline
 \end{tabular}
\end{adjustbox}
\medskip

Finally consider data for Sgr\astar.  According to \cite{BLN}, the
received luminosity is $1.85\times 10^{-13}$ $W/m^2$ which is
approximately $10^{-8}$ of the Eddington limit.  Accordingly set
$z_{\mathrm{grav}} = 10^{-4}$, which gives the following data in
the same format as above.\index{Sgr A*}

\begin{center}\small
\index{quasar!data!Sgr A*}Sgr A* data
\end{center}
\begin{adjustbox}{max width=\textwidth}
\begin{tabular}{|ccc|cc ccccc c|}
\hline
	&z	&	&R/S	&$L_{\textrm{mag}}$	&Solar masses	&Distance	&$T_B$	&$T_B$ * 1/1+z	&	&Ambient coefficient\\
Obs	&Cos	&Grav	&	&$W/m^2$	&$\K$	&	&	&	&X	&$n^{-1}* T^{1.5}$\\
\hline
$10^{4}$	&0	&$10^{4}$	&100	&1.185E-13	&4.300E6	&2.592E20	&5.269E5	&5.268E1	&0.5	&3.516E3\\
\hline
 \end{tabular}
\end{adjustbox}
\medskip

\begin{figure}[ht!]
\cl{\includegraphics[width=.7\hsize]{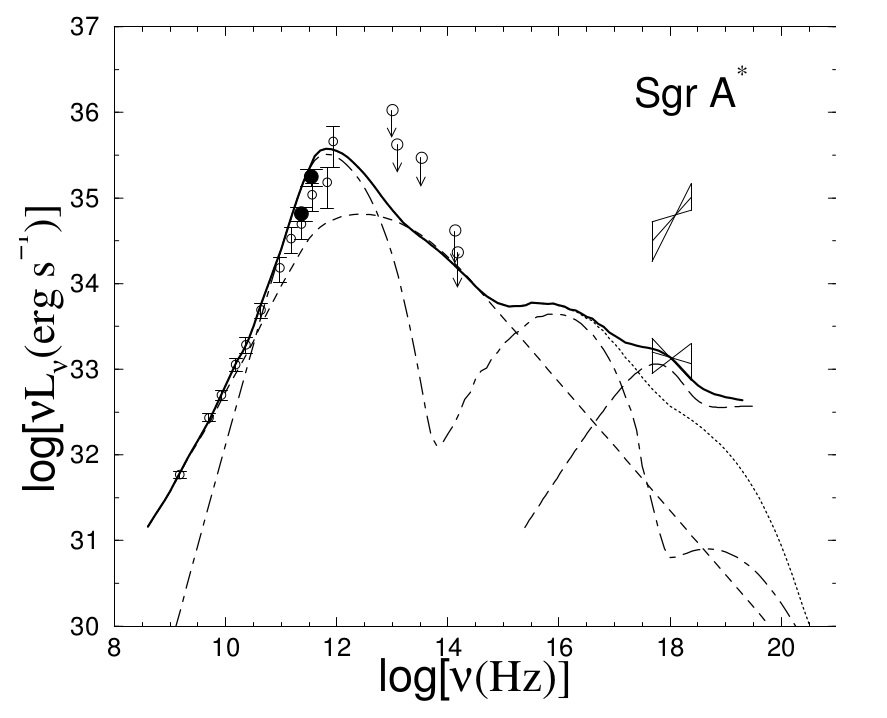}}

\caption{Figure 3 from \cite{ADA} where the following references can
  be found.  The radio data are from Falcke et al (1998; open
  circles) and Zhao et al (2003; filled circles), the IR data are
  from Serabyn et al (1997) and Hornstein et al. (2002), and the two
  ``bow-ties'' in the X-ray band correspond to the quiescent (lower)
  and flaring (higher) data from Baganoff et al (2001, 2003). }
\label{fig:Sgr-spec}
\end{figure}

This table predicts the observed temperature for Sgr\astar\ of about
50 K, which fits well with observations in the radio frequency range.
The spectrum of Sgr\astar\ from Narayan--McClintock \cite[page
  6]{ADA} is reproduced in \fullref{fig:Sgr-spec}.\index{Sgr A*!temperature}

Ignoring the solid and dotted lines (which are attempts to fit the
data with current models), the radio frequency observations and
infra-red observations (up to about $10^{14}$ Hz) are a pretty good
fit for a black body radiator with peak output at about
$5\times10^{12}$ Hz which corresponds to a temperature of about
50 K (see the frequency-dependent formulation of
Wien's law in \cite{wiki:Wien}) and fits the data well.  Note that the
actual temperature of the Eddington sphere is $5\times 10^5\,
\mathrm{K}$; it is the apparent temperature, after redshift
adjustment, which is 50 K.  The extreme redshift of
Sgr\astar\ explains why the principal radiation falls in the radio
frequency range.  The two ``bow-ties'' are probably due to activity
remote from the actual \BH, perhaps associated with orbiting clouds in
the outer region.  This illustrates clearly that the model is merely a
first approximation to reality, applying only to the main \BH\ radiator,
and omits other important features.

\sh{Conclusions}\label{sec:conc}
\index{quasar!data!conclusions}

\fullref{sec:quasars} and this appendix has investigated a very simple
model for \BH\ radiation which appears to explain the observations of
Arp and the paper of Hawkins \cite{H}, both of which suggest that
quasars typically exhibit redshift that is not cosmological.

\begin{figure}[ht!]
\cl{\includegraphics[width=.4\hsize]{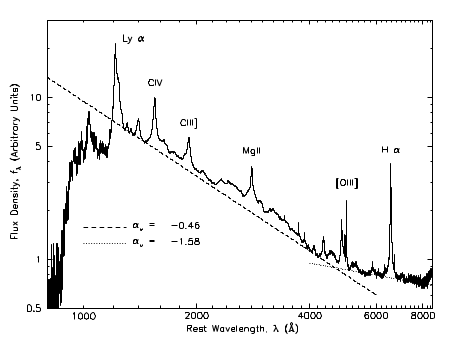}\qquad \includegraphics[width=.4\hsize]{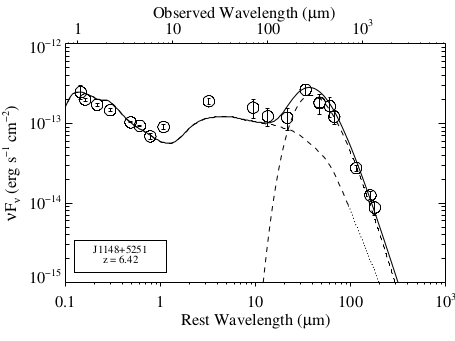}}
\caption{Left: composite spectrum (figure 3 from \cite{comp})\qquad
  Right: spectrum of the $z = 6.42$ quasar SDSS J1148+5251 (figure 1
  from \cite{dust})}
\label{fig:spectra}
\end{figure}

It is not suggested that the model is a perfect fit for all the facts.
One obvious set of data that need a more complicated model are the
Spectral Energy Distributions (SEDs) for quasars which are typically
quite complicated and far from simple black body graphs; for a fairly
simple example see \fullref{fig:spectra} right.  By contrast, the
composite spectrum on the left does have the rough outline of a black
body, suggesting that
the basic mechanism for radiation is by thermal excitation, as in the
model.  One obvious suggestion for correcting SEDs is to take into
account the orbiting clouds, responsible for the observed variation
in radiation and which absorb radiation.  The spectrum on the right
could plausibly result from a black body spectrum which is partially
obscured causing the two dips at the top.  Or perhaps, like Sgr
\astar\ there is a black body radiator in the longer wavelengths with
some short wavelength activity from the outer region superimposed.

\begin{figure}[ht!]
\cl{\includegraphics[width=.7\hsize]{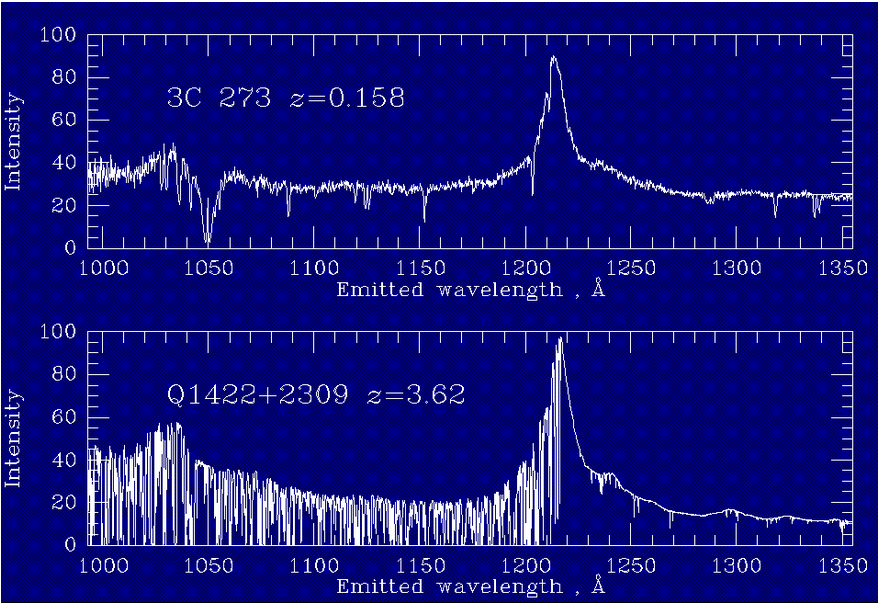}}
\caption{The Lyman Alpha Forest at low and high redshift, taken from \cite{LAF}}
\label{fig:forest}
\end{figure}

Another strong piece of evidence (apart from variability) for the
existence of orbiting clouds is the so-called ``\ind{Lyman-alpha-forest}''.
The clouds on the path to us cause absorption lines and the principal
line is the $L_\alpha$--line.  The clouds are all at different
redshifts and these lines form a forest, see \fullref{fig:forest}.
The existence of the $L_\alpha$--forest is used by Wright \cite{LAF}
to prove (fallaciously) that Arp is wrong about intrinsic redshift.
He assumes that if the redshift is intrinsic then it jumps down
suddenly away from the quasar and therefore there should be a gap to
the left of the main $L_\alpha$ emission line before the forest
starts.  But the absorption clouds can orbit as close as they like to
the Eddington sphere, and there is no reason for there to be a gap.  

The $L_\alpha$--forest suggests strongly that the settling process,
that is hypothesised to take place in the outer region, tends to form
strata.  This is plausible because once a stratum of greater density
starts to build up, then interaction with other particles becomes more
likely, and this will often result in material added to the stratum.
This is analogous to the instability observed in many queuing or
draining situations (for example traffic congestion with most of the
traffic locked up in stationary bands at any one time).  These strata
are responsible both for the observed $L_\alpha$--forest and the
quasar variability.  As remarked earlier, the outer region needs
proper modelling, and the authors intend to return to this in a later
paper.

However, there are complicated features for many quasars which are not
adequately explained by the simple model exposited in this appendix, even
with added absorption clouds and strata.  For heavier quasars, whose
redshift is largely cosmological, the current theory is probably much
more appropriate, especially when there are features such as jets
which can be observed.  It is only suggested that the theory fits smaller
\BH s with high intrinsic redshifts, which are probably much smaller and
closer than current theory suggests.  Note that very high redshift
examples are very dim because of the redshift reduction in energy
received and therefore unlikely to be observed.

\index{quasar|)}

\chapter{Local stellar velocities}\label{app:lsv}
\index{local stellar velocities|(}

As remarked in \fullref{sec:obs}, there has been a huge effort
expended mapping the velocities of stars in our neighbourhood.  There
are some (apparently) paradoxical properties of these excellent
observations which all have natural explanations in the new model
exposited in this book.  These (fairly technical) explanations are
given in this appendix.

\sh{The observations}

The discusson is based on the excellent treatment in Binney and
Merrifield \cite[Section 10.3]{BM}.  The first and most important
point that must be understood is that the observations are all {\em
  relative to the Sun}.  There is no way of determining absolute
motion (eg with respect to the centre of the galaxy) from these
observations.  If a model for galactic motion is chosen (eg the
current conventional model of roughly circular motion in the plane of
the galaxy) then absolute motion can be deduced, but other models give
other results.

The coordinate system used to express observations is $(x,y,z)$ where
$x$ points from the Sun to the centre of the galaxy, $y$ is
perpendicular to $x$ in the plane of the galaxy and points roughly in
the direction that the Sun is moving and $z$ is perpendicular to both
and points to the galactic north pole.  By convention, velocities in
these three directions are denoted $U,V,W$ respectively.

The salient features of the observations are:

\begin{itemize}
\item[(1)] The Sun is moving with velocity $(U,V,W)\approx
  (10,5,7)$km/sec with respect to the average velocity of nearby
  stars. 
\end{itemize}
\index{velocity of sun}

This velocity is well within the observed variations for stellar
velocities for all types of stars in our neighbourhood and therefore
this observation is completely unremarkable, unlike the remaining
ones.  Note that this does {\em not\/} imply that the Sun is moving
towards the centre ($U>0$) but merely that its velocity measured with
respect to the average velocity for nearby stars has a component
towards the centre.  In the model presented in this book, stars are
moving around the galaxy at the usual tangential velocity of about
200km/sec and also outwards at perhaps 20km/sec, so the Sun is also
moving outwards at perhaps 10km/sec.

The remaining observations concern the statistics of the observed
velocities for subsets of stars of a given stellar type.  The main
variable considered is colour ``B--V'' which for Main Sequence stars
is largely determined by age (or rather by metalicity, which for stars
in our neighbourhood is inversely correlated with age, see
\fullref{sec:pop}).  The reddest observations are ignored to improve
the correlation with age, see the comments at the top of page 630 of
\cite{BM}.

\begin{itemize}
\item[(2)] The average velocity of Main Sequence stars in our
  neighbourhood {\em decreases monotonically} with respect to age.
\index{stellar velocity!decreases with age}
  
\item[(3)] The variation in velocities (measured for example as the
  square of the standard deviation of the velocities from the mean
  velocity) {\em increases monotonically} with respect to age.
  \index{stellar velocity!variation increases with age}
\end{itemize}

For details here see \cite[Figures 10.10, 10.12]{BM}.

These observations are very remarkable.  At first sight there is no
reason at all to expect any dynamic properties of stars in the galaxy
to depend systematically on age.  The two observations can be combined
to give a linear relation between velocity and variation, which is
called {\em asymmetric drift\/}: for all types of stars, velocity
decreases linearly with respect to squared variation in velocity
\cite[Figure 10.11, page 628]{BM}.

Now consider the variation in velocity as a function of direction.  To
first approximation, squared standard deviation can be modelled as a
quadratic form.  This is the so-called {\em velocity ellipsoid}
\index{velocity!ellipsoid}
\cite[Box 10.2]{BM}. The distance of the ellipsoid surface from the
origin in a given direction gives the standard deviation for
velocities in that direction.  The principal axes of this ellipsoid
give intrinsic directions related to the velocity variation.  As
expected from symmetry considerations, for all types of stars one of
the principal axes is parallel to the $z$--axis (ie towards the
galactic north pole) and this is the shortest principal axis.  The
other two lie in the galactic plane.  For the current model of
galactic motion in which stars are supposed to move in roughly
circular orbits, the $x$--axis should be a line of symmetry.  The
final, and most remarkable of these observations is that this is not
the case.  The major axis of the velocity ellipsoid lies in the
galactic plane and points, not towards the galactic centre, but makes
an non-zero angle with the $x$--axis on the side of the positive
$y$--axis of between 10 and 30 degrees approximately.  This non zero
angle is called {\em \ind{vertex deviation}}. The final and most remarkable
observation is the following.

\begin{itemize}
\item[(4)] Vertex deviation {\em decreases} with stellar age.
\end{itemize}
\index{vertex deviation!decreases with age}

\sh{The explanations:\qua Velocity variation increases with age}
\index{stellar velocity!variation increases with age}

Recall that stellar systems form in the spiral arms by condensation of
the background gas stream, together with dust and contaminants from
supernova explosions etc, see \fullref{sec:pop}.

At birth, a star's velocity will be much the same as the average
velocity of the gas stream, but once born it is subject to various
gravitational forces of a random character from nearby stars and
groups of stars and its velocity tends to vary from average in a
statistical sense.  Thus the older a star is, the longer time it has
to acquire random variations and the more variation you would expect.
This is observation (3).

\sh{Asymmetric drift}
\index{asymmetric drift}

Variations in velocity are mostly due to interactions between nearby
stars and groups of stars.  Therefore they conserve kinetic energy.
When uniform velocities vary randomly from a common average preserving
kinetic energy then average velocity {\em decreases} with the average
decrease proportional to the average squared deviation.  This explains
asymmetric drift and observation (2) follows from observation (3).
This is a well-known phenomenon and proved in for example
\cite[Section 4.2.1]{BT}.  Here is an elementary proof which gives the
dependence on average velocity explicitly.

\begin{figure}[t!]
\labellist\small 
\pinlabel {spiral arm} [rb] at 116 449
\pinlabel {spiral arm} [rt] at 39 11
\pinlabel {centre of galaxy} [l] at 136 255
\pinlabel {direction of rotation of galaxy} [lb] at 164 387
\pinlabel $x$ [b] at 107 144
\pinlabel STAR [r] at 86 65
\pinlabel $y$ [l] <0pt,0pt> at 191 58
\pinlabel {$\phantom{y}$\quad direction of tangential} [l] <0pt,5pt> at 191 58
\pinlabel {$\phantom{y}$\quad rotation and $\mathbf{v}_{\rm rot}$} [l] <0pt,-5pt> at 191 58
\pinlabel {direction of major} [lb] at 182 145
\pinlabel {variation in velocity} [lb] <0pt,-10pt> at 182 145
\pinlabel $\mathbf{v}$ [tr] at 155 27
\pinlabel {resulting actual motion} [t] <0pt,-2pt> at 195 9
\pinlabel {motion along arm} [rB] at 71 38
\pinlabel {$\mathbf{v}_{\rm inert}$} <-5pt, 0pt> [t] at 84 1
\pinlabel {$= \mathbf{v}{-}\mathbf{v}_{\rm rot}$} <-5pt,-8pt> [t] at 84 1
\endlabellist
\cl{\includegraphics[width=2in]{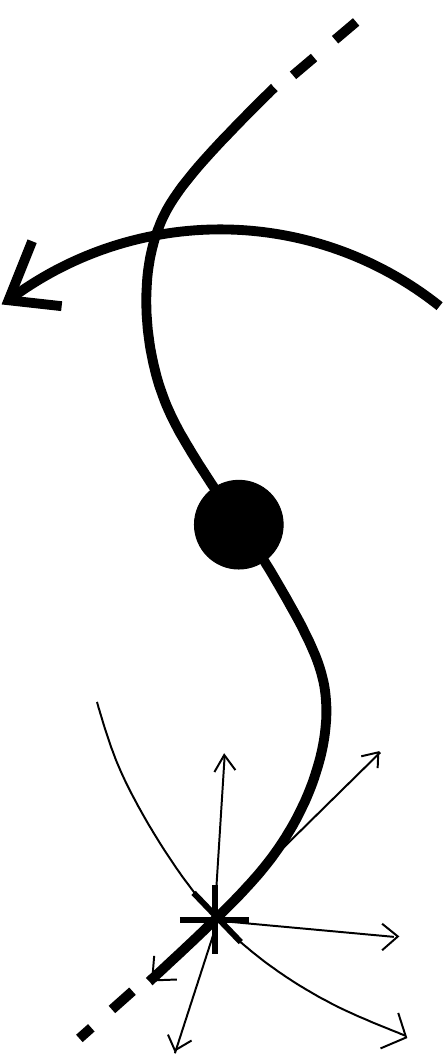}}
\vspace{12pt}
\caption{The velocities near a star}
\label{fig:vel}
\end{figure}
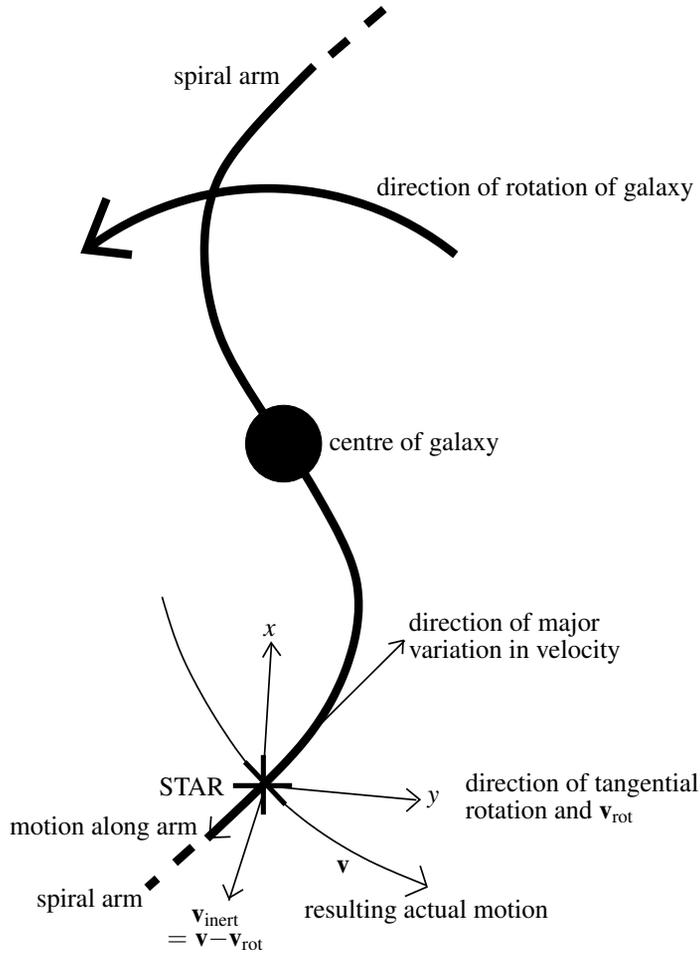

Assume for simplicity that there is a group of $N$ stars of equal mass
all travelling with the same velocity vector $\mathbf{v}$ subject to
small random changes preserving kinetic energy.  Let the new velocity
of the $i$--th star be $\mathbf{v}+\mathbf{e_i}$ then conservation of
kinetic energy gives:
$$
\sum_i ||\mathbf{v}+\mathbf{e_i}||^2 = \sum_i ||\mathbf{v}||^2
$$ 
which implies 
$$
\sum_i 2\mathbf{v}.\mathbf{e_i} = - \sum_i ||\mathbf{e_i}||^2
$$ divide both sides by $2Nv$ where $v=||\mathbf{v}||$ and the left
hand side becomes the average increase in velocity in the $\mathbf{v}$
direction (negative and therefore a decrease) and the right hand side
is $-1/(2v)$ times the average squared variation.  It will be seen
shortly that, after correcting for the effect of inertial
drag (replacing $\mathbf{v}$ by $\mathbf{v}_{\rm inert}$ see
\fullref{fig:vel}), the principal source of velocity variation is
roughly in the $\mathbf{v}$ direction and therefore the major change
of velocity is roughly parallel to $\mathbf{v}$ and hence the
average velocity decrease is proportional to average squared variation
with the constant of proportionality being $1/(2v)$.

Comparing this with \cite[Equation 10.12]{BM} gives $2v = 80$km/sec
and hence $v=40$km/sec approximately.  This is not the observed
velocity for the Sun against distant objects which, in common with all
observed rotation curves, is approximately 200km/sec.  To explain this
discrepancy, which is caused by \id, it is necessary to recall the
analysis of \fullref{sec:rot_curve}.  Write the velocity vector
of a particle moving in the plane of the galaxy as the sum:
$$
\mathbf{v}=\mathbf{v}_{\rm rot}+\mathbf{v}_{\rm inert}
$$ where $\mathbf{v}_{\rm rot}$ is the velocity due to rotation of the
local inertial frame and $\mathbf{v}_{\rm inert}$ is the velocity
measured in the local inertial frame.  Note that the notation here is
not the same as used in \fullref{sec:rot_curve}, where $v$ was tangential
velocity and not total velocity.  The use of bold face is intended to
make this distinction clear.  Now the conservation of energy applies
only to $\mathbf{v}_{\rm inert}$ and it is twice the size of this
velocity which is the inverse of the constant of proportionality.
Tangential velocity is mostly rotational near the centre and moving
outwards, the inertial part of tangential velocity grows
asymptotically to a maximum of half the asymptotic limit of 200km/sec.
Radial velocity is all inertial but decreases outwards as you would
expect.  The nett effect is that an inertial volocity on average of
size roughly 40km/sec is consistent both with the model and with
observations.

\sh{Vertex deviation}
\index{vertex deviation}

To understand vertex deviation it is necessary need to think carefully
about the geometry that was analysed in \fullref{sec:spiral_struc},
which produces the classic spiral structure.  Consider
\fullref{fig:vel}; a star moves tangentially with the rotating galaxy
and also outwards along the arm in which it lies.  The nett effect is
a spiral in the {\em opposite} direction as illustrated.

Some preliminary remarks are needed.  The variation in velocity is a
relative effect and depends only on the interaction of stars in the
frame moving with a star.  Therefore it is the apparent arrangement of
stars which is important, in other words the visible spiral structure,
called the \emph{spiral frame}, in which stars move along the
arms.  Further the direction of variations is preserved as the stars
move in the spiral frame.

Now the main source of random variations in stellar velocities is from
the fact that the arms, created as they are by a series of explosions
in the belt, are not uniform.  Hence the component of velocity {\em
  along} the arm is subject to the major variation.  But this is in a
direction towards the centre near the root of the arm and then turns
away in a direction towards the direction of rotation as illustrated
in \fullref{fig:vel}.  Thus the major variation is not towards the
centre (along the $x$--axis) but has a component along the $y$--axis
which is precisely the observed vertex deviation.  But by inspecting
the shape of the arm, it can be seen that the younger a star is, the
greater will be the proportion of its life spent in the outer region
of the arm, where the direction of variation is further from the
centre and hence the greater will be the vertex deviation.  This is
observation (4).
\index{local stellar velocities|)}
%\newpage

\np\thispagestyle{empty}

\chapter{Optical distortion in the Hubble Ultra-Deep Field}\label{app:HUDF}
\index{Hubble ultra-deep field|(}

\sh{Introduction}

The Hubble Ultra-Deep Field (HUDF) \cite{HUDF} provides a unique
snapshot of the universe at a great distance (and hence time) removed
from our immediate neighbourhood.  There are many strange looking
galaxies in the field and the purpose of this appendix is to examine a
selection of these galaxies and to suggest that their strange
appearance is not instrinsic but rather due to optical distortion
caused by non-uniformity in the intervening space-time, and that the
galaxies being viewed are in fact similar to a field of comparable
size in a closer neighbourhood.

Patterns of non-uniformity in space-time are usually called
``\ind{gravitational waves}'', which expresses graphically the way that they
propagate with respect to a particular time parameter and this
terminology will be used frequently.  Now one of the main hypotheses
of this book is that big spiral galaxies are rotating and in so doing
they create \id\ fields which propagate at the speed of light.  This
implies that the universe is filled with low level gravitational
disturbance, and therefore the effects of this are expected to be seen.
There are also gravitational disturbances coming from movements of
heavy objects other than rotation and indeed natural observer fields
which are also associated with heavy objects also have a distorting
effect on space-time.  (This is used in the explanation for redshift
in \fullref{sec:red}.)

\sh{The face galaxy}
\index{Hubble ultra-deep field!face galaxy}

\begin{figure}[ht!]\centering
\includegraphics[scale=4]{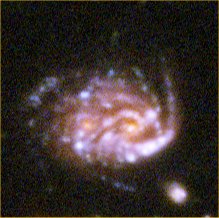}
\caption{The face galaxy}\label{face}
\end{figure}

The discussion starts by examining the clearest example and one where
it is possible to describe a simple gravitational field which produces
the visible distortion.  This is the ``face galaxy'' copied in
\fullref{face}.

\medskip
{\bf Note}\qua You are recommended to download a copy of the highest
resolution jpeg of the HUDF as instructed in the bibliography at
\cite{HUDF}.  To help you find a particular galaxy or image instrinsic
coordinates are given from the bottom left, where the height and width
are 1 unit and coordinates are taken mod 1 (so that a negative number
is a coordinate from the right or top).  The face galaxy is at $(.42,
-.09)$.

\begin{figure}[ht!]\centering
\includegraphics[scale=4]{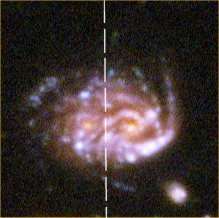}
\caption{The rough line of symmetry}\label{face2}
\end{figure}

\medskip
If the face galaxy is an accurate representation of a real galaxy,
then it is one of the weirdest galaxies you can imagine.  It has two
centres.  They must be in the process of merging.  A far more chaotic
structure would be expected from such a merger and moreover there is
no reason at all to expect the colours to match so accurately.  Far
more plausible is that the two centres are the same and that the
appearance is due to some kind of optical reflection process.  Looking
more closely, there is a rough line of symmetry in the centre (marked
with dashes in \fullref{face2}).

\index{Hubble ultra-deep field!face galaxy!line of symmetry}

\begin{figure}[ht!]\centering
\includegraphics[scale=4]{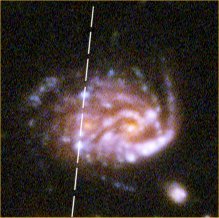}
\caption{The second mirror line}\label{face3}
\end{figure}

The symmetry is near perfect in the top half (near the line of
symmetry) and not so accurate in the bottom.  So apart from this
reflection, there is some other distortion going on.  Looking
carefully at the line of symmetry, there are some white dots as it
crosses some of the denser parts of the galaxy.  If the reflection is
due to a lensing effect then there will be an element of focussing at
the line of reflection and this will produce a bunching of light paths
and explain these white dots (more detail on this will be given
below).  The symmetry breaks down at the outside where there are clear
spirals going the same way and not mirror images, but now that it is
known how to recognise a mirror line then another slightly slanting to
the left (dashed in \fullref{face3}) can be seen.

\begin{figure}[ht!]\centering
\includegraphics[scale=.45]{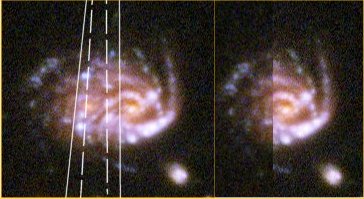}
\caption{Cutting and gluing}\label{face4}
\end{figure}

Finally, fold along these mirror lines and cut out the middle
(and the spurious white dots) and paste the outsides together.  This
has been done on the right in \fullref{face4}.  On the left in
\fullref{face4} is the original galaxy with the two mirror lines
dashed and the two cut lines (which coincide after both reflections)
shown solid.  The final picture on the right in \fullref{face4} is
obtained by cutting along the cut lines, discarding the middle and
pasting the two outside pieces together.  It is close to a standard
spiral galaxy (with just a little residual distortion).

\sh{Gravitational solitons}

Before examining other funny objects in the HUDF it is worth pointing
out that there is a simple gravitational field which produces exactly
the distortion seen in the face galaxy (reflection in two roughly
parallel mirrors) namely a \emph{\ind{gravitational soliton}}. 

The description of this field is in terms of distortion of the metric
on space and ignores the accompanying distortion in time.  This is
justified since the spatial distortion is small and relativistic
effects minimal.  A proper treatment would treat both space and time.

Suppose given two concentric spheres of fairly large radius with a
relatively small gap between them.  Suppose that the metric on the gap
is altered so that radial distance is changed by a fixed scale factor
close to 1, tangential distance being unaltered.  If the factor is
greater than 1, this is called a \emph{positive} soliton and if the
factor is less than 1, a \emph{negative} soliton.  It is not hard to
describe the geodesics in this metric.  Outside the gap, they are of
course straight lines.  In the gap they are circles.  This is easy to
see for the positive case where a plane section through the centre is
isometric to a portion of a cone, which can be flattened and the
geodesics drawn.  In this case the circles are concave towards the
centre of the spheres.  In the flat case the geodesics in the gap are
straight lines and, by extrapolation, in the negative case, they are
circles concave outwards.  When a geodesic crosses one of the spheres
it makes an apparent bend, namely the tangent of the angle to the
tangent plane is scaled by the same factor as the metric scale.  The
bend is describes as ``apparent'' because the geodesic is straight as
it crosses the sphere if the local metric scaling is performed.

%The general supposition I am making here is that this field is part of
%a field of gravity waves emanating from some source and that a soliton
%wave of this type is likely to appear and to be persistent for a while
%through propagation.

\begin{figure}[ht!]\centering
\labellist\small\hair 6pt
\pinlabel EYE [t] at 224 0
\pinlabel GALAXY [b] at 235 534
\hair 2pt
\pinlabel soliton [tr] at 126 456
\pinlabel 1 [l] at 235 178
\pinlabel 2 at 220 152
\pinlabel 3 [r] at 203 138
\endlabellist
\vspace{10pt}\includegraphics[width=1.5in]{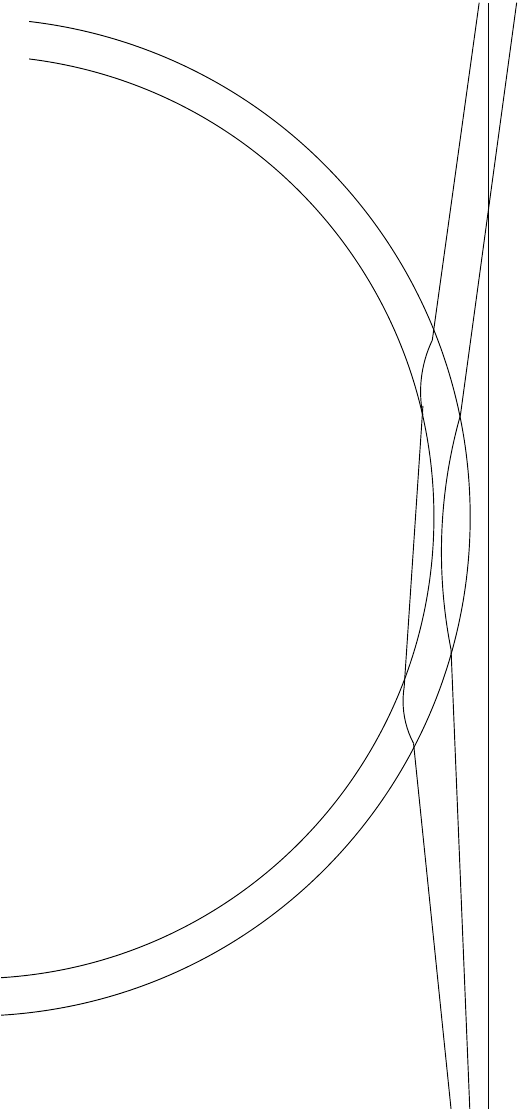}\vspace{10pt}
\caption{The soliton in action}\label{soliton}
\end{figure}

Now suppose that there is a negative gravitational soliton between us
and the face galaxy with a tangent plane passing through our eye and
the galaxy.  It can be seen that the image of the face galaxy has
two roughly parallel mirrors.  Look at \fullref{soliton}.

Three typical light paths from our eye to the galaxy have been drawn.
Path 1 is straight and panning left, moving in from the right, paths
stay straight until they reach tangency to the outer sphere.  At this
point they start to contain a portion of a circle which is concave to
the right and causes the paths to bend to the right as typified by
path 2.  This bending increases (and the far end of the path pans to
right) until tangency to the inner sphere is reached, when the path
becomes three straight lines with two smaller circular portions as
typified by path 3.  The paths now continue to pan to the left.  Thus
there are two places where movement of the far ends of the paths
reverses and this gives the double mirror effect.

Robert MacKay points out that a mirage has a similar mechanism and may
be more familiar than a gravitational soliton!

Finally notice that at the points of reflection there will be a
focussing effect.  The metric described is not $C^\infty$ but merely
$C^1$.  If a $C^\infty$ approximation is used then there is a non-zero
angle of paths all roughly converging to the same point at the
reversal times and this gives rise to the white blobs seen on the
miror lines in the face galaxy (assuming that the mechanism at work in
the face galaxy is similar to the one described here).

\sh{The companion face}
\index{Hubble ultra-deep field!companion face}

\begin{figure}[ht!]\centering
\includegraphics[scale=4]{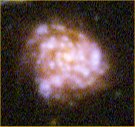}\qquad\includegraphics[scale=4]{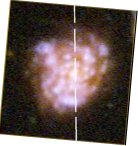}
\caption{The companion face (left) and rotated (right)}\label{cface}
\end{figure}

At $(.40, .50)$ there is a very similar object, \fullref{cface} (left)
The similarity is more apparent if it is rotated, \fullref{cface}
(right).

Now there is a clear (and very rough) vertical line of symmetry
(marked dotted) but there the analogy with the face galaxy stops.  It
is difficult to finish the description of the precise distortion that
must have happened to make a standard spiral galaxy look like
this.  But it is clear that this is again a distorted spiral galaxy.

The colouring is very similar to the original face (\fullref{face}) and
it is just possible that both these two galaxies are two distorted
images of the same galaxy.

\sh{The group of four}
\index{Hubble ultra-deep field!group of four}

\begin{figure}[ht!]\centering
\includegraphics[scale=4]{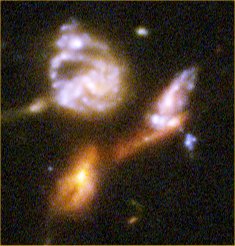}
\caption{The group of four}\label{group}
\end{figure}

At $(.39, -.16)$ is a group of four galaxies: two ``white'' and two
``orange'', \fullref{group}.

The left-hand white galaxy is clearly an ordinary spiral galaxy
showing optical distortion: the centre has been elongated (top-left to
bottom-right) and, to the left and top, there is a pair of spiral arm
sections which have been dragged out; they look as though they are on
a sheet which has been bent up.  The other white galaxy is severely
distorted with a clear sloping ``cut-off'' plane to the left.  This
would be due to a planar gravitational wave front in the intervening
space.  Moreover these two galaxies have very similar colour and light
distribution and most probably they are in fact two images of the same
galaxy.  The ``reflection'' plane would be associated with the same
wave front that is causing the cut-off in the right-hand image.  

The two orange galaxies are both severely distorted and again are
quite likely to be different images of the same galaxy.

\sh{Four distorted spirals}
\index{Hubble ultra-deep field!four distorted spirals}

\begin{figure}[ht!]\centering
\includegraphics[scale=.4]{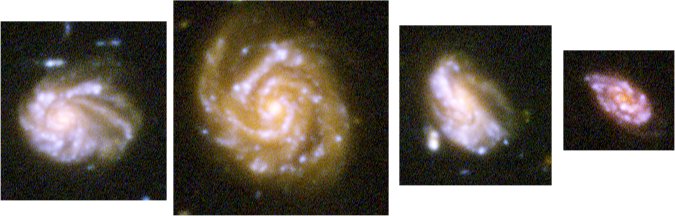}
\caption{Four distorted spirals}\label{four}
\end{figure}

In \fullref{four} are four galaxies from different parts of the field.  Their
coordinates are $(.31,-.16)$, $(.13,-.33)$, $(-.09,-.14)$, $(.12,-.24)$ respectively.

Each is a spiral galaxy with optical distortion.  On the left is a
galaxy having a ``bad hair day'' caused by image distortion on the
right-hand side.  Middle-left is a spiral galaxy with anomalous
straight section in one arm (top left).  Although this could plausibly
be an undistorted image, it seems more likely, given the distortion
that seen elsewhere, that this straight section is caused by
focussing at a wave front in the intervening space.  Middle-right is a
distorted spiral with several different kinds of distortion and to the
right is a spiral with quite simple distortion causing a ``toothpick''
appearance.  

\sh{Miscellanea}
\index{Hubble ultra-deep field!miscellanea}

\begin{figure}[ht!]\centering
\includegraphics[scale=.4]{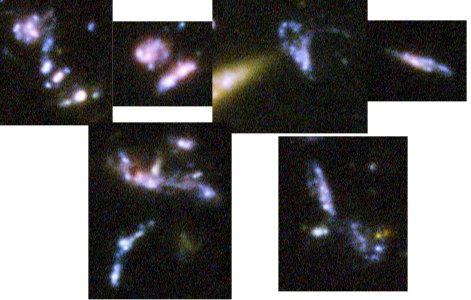}
\caption{Miscellanea}\label{misc}
\end{figure}

Finally in \fullref{misc} is a collection of miscellaneous objects
from various parts of the field.  The coordinates are (top row) $(.04,
.44)$, $(-.04,-.12)$, $(-.25,.09)$, $(.24,.32)$ and (bottom row)
$(.14, .37)$, $(.19,.29)$.

Top row left and centre-left are two sets of possibly repeated images
of the same object.  Top row centre-right (the blue ring galaxy) is
probably a highly distorted image of a regular spiral with the ring
being a distorted arm with a similar distortion to the left-hand white
galaxy in \fullref{four}.  This galaxy is probably a long way behind
the regular edge-on spiral to the left and not interacting with it.
Top-row right is a toothpick galaxy, a more extremely distorted (and
distant) version of the right-hand galaxy in \fullref{four}.  The
bottom row shows two collections of distorted fragments, which could
both be images of the same galaxy or pair of galaxies.

The images in \fullref{misc} are typical of many other images in the
field.  There is a collection of ``tadpole'' galaxies from the field
on the Hubble site (search tadpole) similar to the toothpick galaxies
given above, and there are collections of fragments like bottom images
all over the field.

One final remark.  Most of the distant objects in the field show
repeated white dots similar to those found on the mirror line in
\fullref{face2}.  These probably have a similar origin in local
focussing effects in the distorting gravitational fields between us
and these distant objects.  For example the ring in \fullref{misc}
(top centre-right) is probably the image of a fairly smooth arm of a
regular spiral with focussing effects causing the grainy appearance.

\sh{Conclusion}
\index{Hubble ultra-deep field!conclusion}

All the strange shapes and unfamiliar objects in the HUDF can be
explained as optically distorted images of familiar galaxies.  Given
the clear evidence of such distortion in the field, there are no
grounds for concluding that an undistorted view of the universe in the
region covered by the field would be qualitatively different from a
more local region.
\index{Hubble ultra-deep field|)}

\np\thispagestyle{empty}
\chapter{Gamma Ray Bursts}\label{app:GRB}
\index{gamma ray bursts|(}

This appendix reproduces a short version of the joint paper with
Robert MacKay \cite{GRB}.

\bigskip\cl{\textbf{A kinematic explanation for gamma-ray bursts}}

\bigskip\cl{\textsc{Robert S MacKay and Colin Rourke}}
\bigskip

{\small\textbf{Abstract}\qua Gamma-ray bursts are flashes of gamma-rays
lasting from milliseconds to a few minutes, which then soften
progressively to X-rays and ultimately to radio waves.  They are
observed from all directions in space, roughly uniformly.  They have
been attributed to cataclysmic events.  We propose, however, that many
of them may be optical illusions, simply the result of our entry into
the region illuminated by a continuously emitting object.  At such an
entry, the emitter appears infinitely blue-shifted and infinitely
bright.  We demonstrate the phenomenon in de Sitter space, where much
can be calculated explicitly, and then extend the idea to more general
space-times.

\textbf{Keywords}\qua {Gamma-ray bursts; kinematic effect; de Sitter space}

\textbf{PACS codes}\qua {98.70.Rz: Gamma ray bursts,
98.62.Py: Distances, redshifts, radial velocities,
04.20.Jb: Classical general relativity - exact solutions}}

\section{Introduction}

Gamma-ray bursts were first observed in 1967 during monitoring of the
\ind{nuclear test ban treaty}, but were subsequently realised to come from
outside our solar system, indeed outside our galaxy.  Dedicated
instruments have now detected and continue to detect many of them.
There is a highly developed theory of their origins in various types
of cataclysmic event, such as collapse of a high-mass star to a
neutron star, or capture of a star by a black hole.  For a review, see
\cite{Mes}.

We propose, however, that many gamma-ray bursts may be optical
illusions.  If space-time is geodesically complete but an emitting
object does not illuminate the whole of space-time, then on our entry
into the illuminated region we see the emitter infinitely blue-shifted
and infinitely intense.  Both the blue-shift and intensity fall off
with receiver time.  This produces an effect qualitatively similar to
the observations of gamma-ray bursts.

We believe the effect has been ignored so far because of Weyl's
coherency postulate \cite{We} and the subsequent standard assumption
that all matter moves along the Hubble flow in a big-bang Friedmann
universe.  It can occur, however, in Friedmann universes if they have
infinite past and emitter and receiver are not both on the Hubble
flow.\index{Weyl!coherency postulate}

We first demonstrate the phenomenon in de Sitter space, where much can
be calculated explicitly.  Then we extend the idea to more general
space-times.  Details are given in \cite{GRB}.

\section{Geodesics in de Sitter space}\label{sec:goeddeS}
\index{de Sitter space!geodesics}

De Sitter space ${\cal DS}$ is the Lorentzian manifold given by
restricting 5-dimensional Minkowski space ${\cal M}^5$ with
metric $$ds^2 = -dx_0^2 + \sum_{i=1}^4 dx_i^2$$ to the
hyperboloid $$-x_0^2 + \sum_{i=1}^4 x_i^2 = R_{DS}^2.$$ The constant
$R_{DS}$ is called the de Sitter radius.  The metric $g$ on $\cal{DS}$
satisfies Einstein's equation in vacuum $\mathrm{Ric} = \Lambda g$
with cosmological constant $\Lambda = 3/R_{DS}^2$.  Our universe is
believed to be entering a de Sitter phase with $R_{DS}$ around 12
billion light-years.  We choose units in which $R_{DS}=1$.\fnote{The
  notation used in this appendix differs slightly from that used in
  \fullref{sec:deS} and \fullref{app:deS} where $a$ is used for the
  de Sitter radius -- a symbol used here for the top-left matrix
  entry, see below.}

The time-like geodesics in ${\cal DS}$ are the components of its
intersections with hyperplanes through the origin of ${\cal M}^5$ of
slope steeper than $45^\circ$.  The null geodesics of ${\cal DS}$ are
the components of the intersections with hyperplanes through the
origin of slope $45^\circ$; note that they are null geodesics of
${\cal M}^5$.

Typical pairs of time-like geodesics in ${\cal DS}$ separate
exponentially in both forwards and backwards time.  Indeed the
time-like geodesic flow is Anosov \cite{nat-flat}.  Exceptionally,
pairs of time-like geodesics may converge together in backward time or
in forward time.

We consider the null geodesics from a time-like emitter geodesic $e$ to a time-like receiver geodesic $r$.
By an isometry of ${\cal DS}$ we can bring the receiver geodesic to
the form $x_0 = \sinh t, x_1 = \cosh t, x_j = 0$ for $j = 2,3,4$, with
proper time $t$.  The emitter geodesic can be expressed as $e =Mr$ for
some future-preserving isometry $M$ of ${\cal DS}$, equivalently, a
linear isometry of ${\cal M}^5$.  We parametrise the emitter geodesic
by its proper time $u$, the image of $t$ under $M$.
\index{de Sitter space!null geodesics}

Since the null geodesics in ${\cal DS}$ are null in ${\cal M}^5$, the set of pairs $(t,u)$ for which there is a future-pointing null geodesic from $u$ on $e$ to $t$ on $r$ is given by
\begin{equation}
-(a \sinh u + b \cosh u) \sinh t + (c \sinh u + d \cosh u) \cosh t = 1,
\label{eq:null}
\end{equation}
with $ a \sinh u + b \cosh u < \sinh t$, where
$\left[\begin{array}{cc}
a & b \\ c & d 
\end{array} \right]$
is the top $2\times 2$ block of the matrix representing $M$.
There are constraints on the values of $a,b,c,d$ for them to come from an isometry matrix, namely
$$ (ab-cd)^2 \le (a^2-c^2-1)(b^2-d^2+1),$$
both factors on the right are non-negative, and $a \ge 1$.

Condition (\ref{eq:null}) can be written conveniently in terms of $T=e^t$ and $U=e^u$ as
$$-ATU+BT/U+CU/T-D/TU = 2,$$
with 
\vspace*{-12pt}
\begin{eqnarray}
2A &=& a+b-c-d \\
2B &=& a-b-c+d \\
2C &=& a+b+c+d \\
2D &=& a-b+c-d ,
\end{eqnarray}
which are all non-negative.
This has the causal solution
$$T = \frac{U+\sqrt{BD+(1-BC-AD)U^2+ACU^4}}{B-AU^2}$$
for $U < \sqrt{B/A}$.
Equivalently we can write the emitter time as a function of receiver time.  Each is monotone increasing in the other.

\begin{figure}[htbp] %  figure placement: here, top, bottom, or page
   \centering
   \includegraphics[width=3.5in]{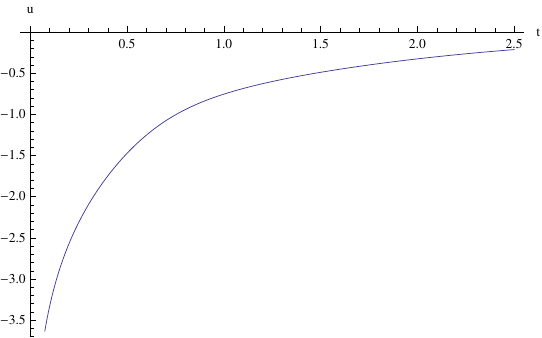} 
   \caption{Emitter time $u$ as a function of receiver time $t$ for a typical pair of time-like geodesics in ${\cal DS}$; the origins of receiver and emitter time have been shifted to $t^*,u^*$ respectively.}
   \label{fig:uvt}
\end{figure}

If $D,B,A \ne 0$ there is a first time $t^*$ at which the emitter becomes visible, given by $T = \sqrt{D/B}$.  Thus there is a sudden start to seeing the emitter, just as for gamma-ray bursts. We see its infinite past in a short interval of receiver time $t$.  As $t \to +\infty$, we see the emitter up to a last emitter time $u^*$ given by $U = \sqrt{B/A}$, but not beyond.  See Figure~\ref{fig:uvt}.

In the exceptional case $D=0$ ($e$ and $r$ backward asymptotic) then $t^*=-\infty$; similarly if $A=0$ (forward asymptotic) then $u^*=+\infty$.  Finally, if $B=0$ ($e$ past asymptotic to the antipodal geodesic to $r$) then both $t^*=-\infty$ and $u^*=+\infty$.  

Weyl \cite{We} abhorred the idea that an object might suddenly become visible, so hypothesised that all emitter geodesics are backward asymptotic to ours.  This eliminates, however, precisely the case we believe to be important for gamma-ray bursts.\index{Weyl} 

We now study the redshift and intensity of the received light.

The redshift $z$ of an emitter relative to a receiver is defined by
$$1+z = \frac{dt}{du} = \frac{U}{T} \frac{dT}{dU} .$$
An emitter frequency $\omega_e$ is transformed to a received frequency $\omega_r = \omega_e/(1+z)$.
In the generic case $D,B,A > 0$, the redshift goes monotonically from $-1$ at $t^*$ to $+\infty$ as $t \to +\infty$.  Thus at its first appearance, the emitter is seen infinitely blue-shifted.  Whatever it emits is seen as even higher frequency electromagnetic waves than gamma rays.  If we assume the emitter spectrum is roughly constant in emitter time, then as receiver time advances, the received light descends through gamma rays to X-rays, visible and microwaves to radio waves, just as for gamma-ray bursts.  For short time after the first appearance we have the asymptotic relation 
\begin{equation} 
1+z \sim t-t^* .
\label{eq:zasympt}
\end{equation}  

The emitter remains blue-shifted up to the time defined by $UT = \sqrt{D/A}$.  The duration $t_B$ of receiver time for which the emitter is seen blue-shifted comes out to
$$ t_B = \frac12 \log{\frac{1+\sqrt{AD}+\sqrt{1+2\sqrt{AD}+AD-BC}}{\sqrt{AD}}},$$
which provides a natural measure of the duration of the burst.
In exceptional cases, $z$ goes from 0 to $+\infty$ (backward asymptotic), or $-1$ to 0 (forward asymptotic) or jumps across 0 (intersecting geodesics).
\index{gamma ray bursts!blue-shift time}

\begin{figure}[htb!] %  figure placement: here, top, bottom, or page
   \centering
   \includegraphics[width=3.5in]{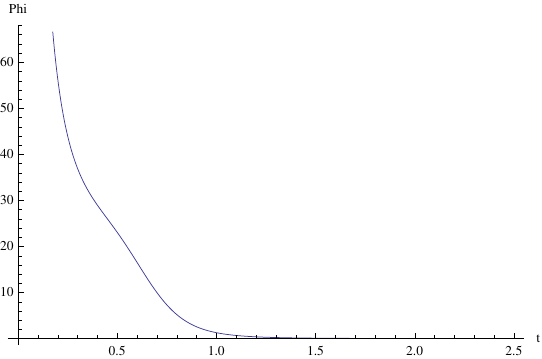} 
   \caption{An example of received flux $\mathrm{Phi}$ as a function of receiver time $t$ since $t^*$, assuming constant emitter power $P$.}
   \label{fig:Phit}
\end{figure}

The received flux $\Phi$ is related to the emitted power $P$ per unit solid angle by \cite{P}
$$\Phi = \frac{P}{((1+z)\rho)^2},$$
where $\rho$ is called the ``corrected luminosity distance'', which accounts for the geometric expansion of the bundle of rays leaving a point on the emitter.  In de Sitter space, $\rho$ is given by the change in affine parameter along the null geodesic, scaled to correspond to elapsed time in the emitter frame initially.  This yields
\begin{equation}
\rho = 1 - (\frac{C}{T}-AT)U .
\label{eq:rho2}
\end{equation}
In the generic case $A,B,D > 0$, $\rho$ starts from $1$ (the de Sitter radius) at $t^*$ and goes to $+\infty$ as $t \to  +\infty$.  
%It might seem strange that $\rho$ starts finite, but the apparently infinite distance is offset by Lorenz transformation of isotropic emission into a narrow forward beam of angle $2(1+z)$.
Thus if the emitter power $P > 0$ at $u=-\infty$, the factor $(1+z)^2$ makes the received flux infinite initially.   Even more, the received energy per unit area diverges for any receiver time interval including $t^*$, because of (\ref{eq:zasympt}).  Fig.~\ref{fig:Phit} shows an example for constant $P$. 
In reality, we should expect $P$ to be integrable as a function of emitter time $u$, thus the received flux is not infinite initially nor is the received energy infinite.  Yet both may be extremely large, just as for gamma-ray bursts.

\begin{figure}[htb!] %  figure placement: here, top, bottom, or page
   \centering
   \includegraphics[width=2.5in]{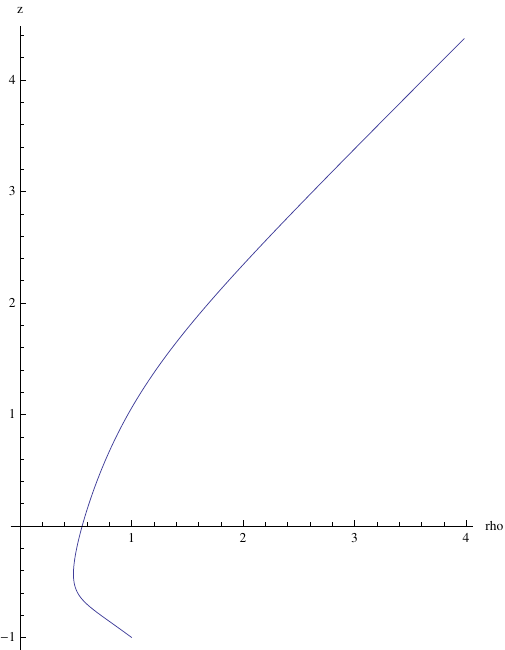} 
   \caption{Hubble diagram of redshift $z$ against corrected luminosity distance $\mathrm{rho}$ for one emitter throughout its visible life.}
   \label{fig:zrho}
\end{figure}

Note that $\rho$ decreases initially if $BC > AD$ (put $T=\sqrt{D/B}$ in (\ref{eq:rho2})).  An example is shown in Fig.~\ref{fig:zrho}.  
\index{gamma ray bursts!Hubble diagram}
This leads to an enhancement of the received flux.  Indeed the received energy in time interval $(t^*,t)$ can be written as
$$\int_{-\infty}^{u(t)} \frac{1}{\rho^3} \left(1+\frac{D}{T}e^{-u} - \frac{C}{T} e^u\right) P(u)\ du.$$
The regime $BC \gg AD$ corresponds to that of short blue-shift period $t_B$.  Thus we see that the brightest emitters are those with the shortest blue-shift period.  This fits another feature of gamma-ray bursts, namely that those observed are very short compared to the de Sitter timescale.

\begin{figure}[htbp] %  figure placement: here, top, bottom, or page
   \centering
   \includegraphics[width=4in]{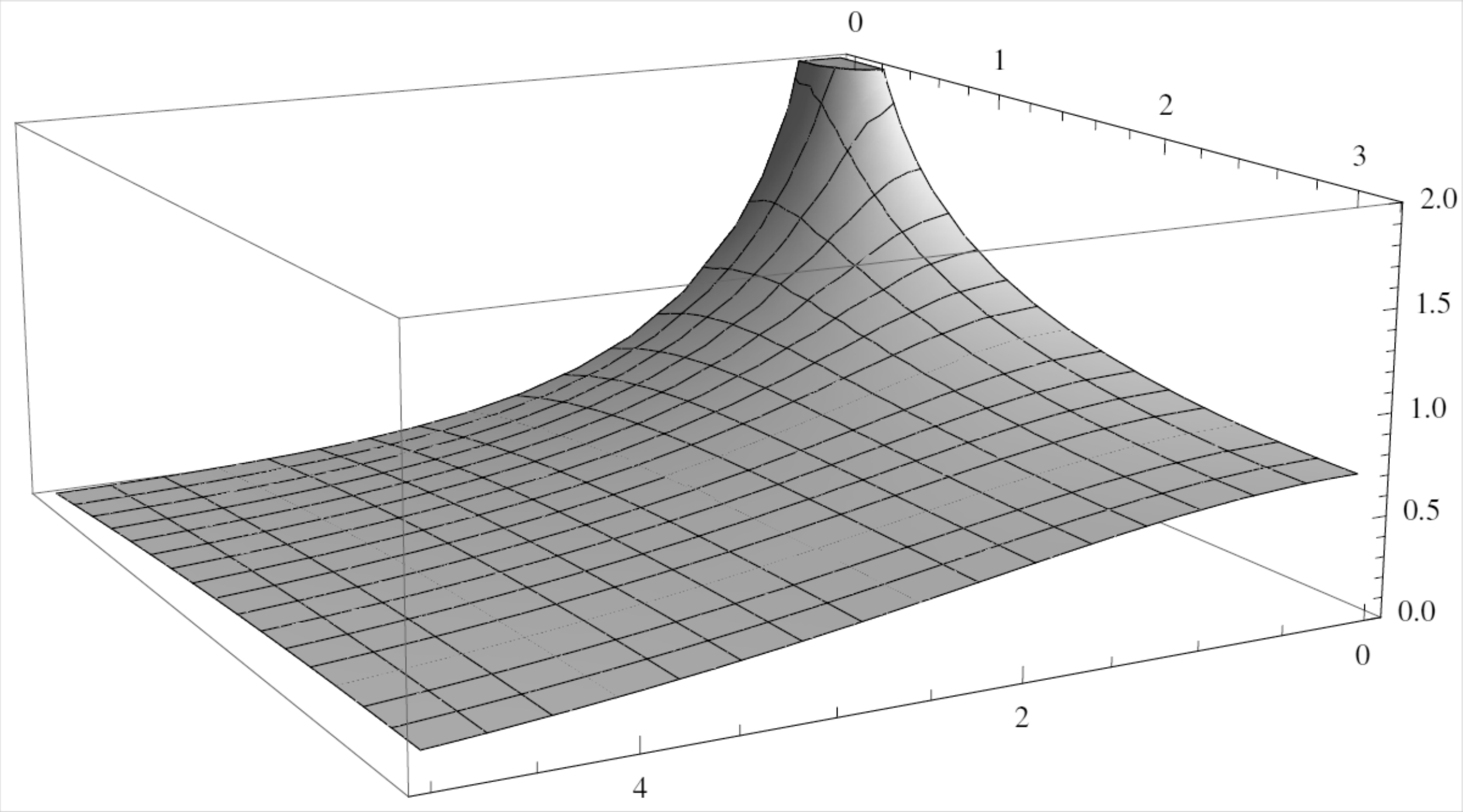}
   %BlueshiftPeriodN_gr1.eps} 
   \caption{Blue-shift period $t_B$ as a function of $\phi$ (on the bottom axis) and $\theta$ (on the top axis) for the reduced family.}
   \label{fig:tB}
\end{figure}

To study the received flux further, it is convenient to apply isometries to reduce the generic case to $a=\cosh \phi$, $b=c=0$, $d=\cos{\theta}$, for $\phi \ge 0$ and $\theta \in [0,\pi]$.  Then $A=D=(a-d)/2$ and $B=C=(a+d)/2$.  The null geodesic condition reduces to
$$ -a \sinh u \sinh t + d \cosh u \cosh t = 1,$$
and the redshift is given by
$$ 1+z = \frac{d \tanh u - a \tanh t}{a \tanh u - d \tanh t}.$$
The blue-shift period for this reduced case can be written 
\begin{equation}
t_B = \log{\frac{\sqrt{a+1}+\sqrt{1-d}}{\sqrt{a-d}}},
\label{eq:tB}
\end{equation} 
and is plotted in Fig~\ref{fig:tB}.
\index{mathematica!blue-shift plot}

\begin{figure}[ht!]
\cl{\includegraphics[width=.48\hsize]{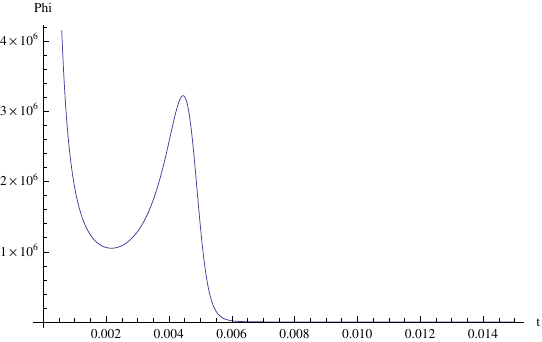}\quad
\includegraphics[width=.48\hsize]{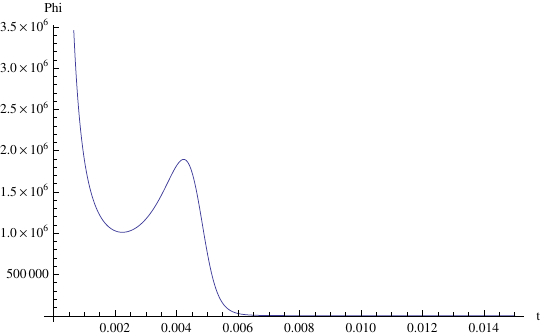}}
\cl{\includegraphics[width=.48\hsize]{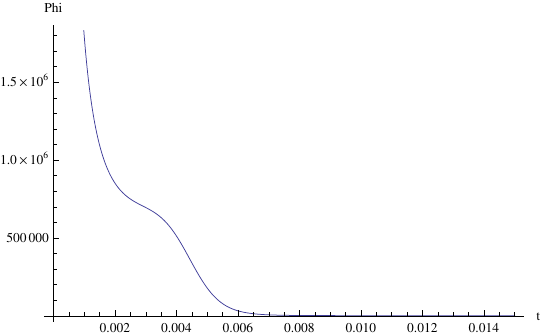}\quad
\includegraphics[width=.48\hsize]{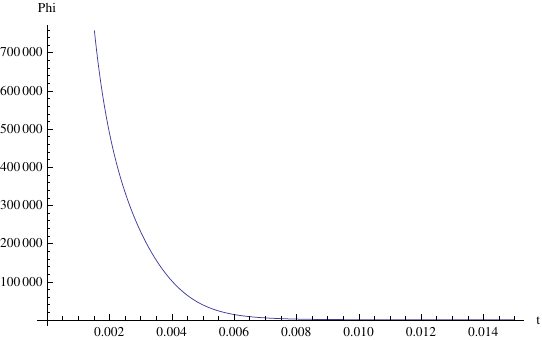}}
\caption{Some light curves for constant emitters in ${\cal DS}$ with short blue-shift period}\label{fig:lightcurves}
\end{figure}

We see that the shortest blue-shift periods are for $\phi$ large and $\theta$ near 0.  Fig.~\ref{fig:lightcurves} shows some light curves for short blue-shift periods.  
For the plots, we shifted the origin of $t$ to $t^* = - \mathrm{arctanh}{\frac{d}{a}}$.  Notice that a second hump occurs in some cases; this is due to $(1+z)\rho$ coming to a local minimum.  Such a second hump is a feature of many observed gamma-ray bursts (e.g.~figures in \cite{Mes}).  For the reduced case, $\rho$ can be written as
$$ \rho = a \sinh t \cosh u - d \cosh t \sinh u ,$$
and we calculate there is a second hump iff $d > \sqrt{8}/3$.

Observed light curves are more complicated than ours, but one possible explanation is that the emitter power varies with emitter time, and any variations are compressed into a short interval of receiver time.  We describe another possible contribution to the variability near the end of the paper.

\index{gamma ray bursts!distribution of duration}
We can also predict the distribution of durations.  For definiteness, we use the blue-shift period as our measure of duration.  We propose that the natural distribution for emitter geodesics in de Sitter space is invariant under isometries.    This implies that the distribution on our two-parameter space of $(\phi, \theta)$ is proportional to $\sinh^2{\phi}\ \sin{\theta}\ d\phi \ d\theta$ (the distribution is non-normalisable).  This can be written as $\sqrt{a^2-1}\ da \ dd$.  Using (\ref{eq:tB}) we obtain that the natural distribution for $t_B$ is asymptotically $\frac{16}{3}t_B^{-5}\ dt_B$ for $t_B$ small.  So the natural distribution is heavily skewed to short blueshift period, which again fits well with observations of gamma-ray bursts.  The difficulty is to explain why the observed density of durations (e.g.~$T_{90}$ in \cite{K+}) decreases for durations less than 20 seconds, but this could be because the relevant part of space-time deviates a lot from ${\cal DS}$ or there are few emitters in the region corresponding to $t_B < 20$ seconds.

\section{Critique}
\index{gamma ray bursts!critique}

\begin{figure}[htbp] %  figure placement: here, top, bottom, or page
   \centering
  \includegraphics[width=4in]{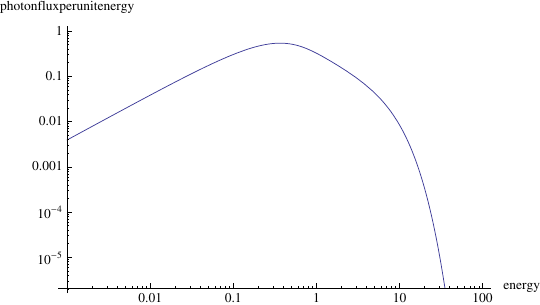} 
   \caption{Time-averaged spectrum of photons per unit time, area and energy, for the received flux from a constant emitter in ${\cal DS}$.}
   \label{fig:spec}
\end{figure}

A common criticism of our proposal is that the observed bursts have non-thermal spectrum.  There is no great reason to suppose that the emitter spectrum is thermal, but even if it is, we believe we can explain the non-thermal observations as an effect of averaging over time.  Photon count rates are often very low, of the order of at most 10 per second, so to estimate a spectrum observations are averaged over a significant interval of time.  If the emitter has temperature $\Theta$ then the received spectrum is thermal with temperature $\Theta/(1+z)$.  In our model, $1+z$ varies rapidly initially.  Averaging the received flux over a time interval produces spectra like that of Fig.~\ref{fig:spec}, which agree well with observations like those of \cite{G+}.

Also it is reported that the spectra at different stages of a burst are not simply Lorentz-boosted versions of each other.  We have no problem with that, however, because it is perfectly natural that the emission spectrum (as well as the power) vary non-trivially during the emitter's life-time.  We see the early history of the emitter compressed into a short interval of receiver time, so any such variations are accentuated.

Another criticism is that it is claimed that many gamma-ray bursts are associated with a distant galaxy which is in fact receding from us.  This is done on the basis of searching for potential host galaxies immediately after a gamma-ray burst is observed, or detecting red-shifted absorption bands in the afterglow.  We think this association may often be spurious.  The gamma-ray burst could be from an emitter way beyond the purported host galaxy and which is approaching us rapidly.  The absorption could indeed be by gas in an intermediate galaxy, but that does not imply the emitter is in that galaxy.

The most serious criticism is that our universe is believed to be
nothing like ${\cal DS}$ in the past.  It is said to contain
sufficient matter and radiation to have made it collapse to a
finite-time singularity in the past.
%, but the Penrose-Hawking trapping surface theorems give only that
%some geodesics can not be infinitely extended, not that all geodesics
%can not.
We will not address the case for the big bang in this paper, but first
we note that ${\cal DS}$ is not so far from the standard $\Lambda CDM$
model.  ${\cal DS}$ contains a flat Friedmann space-time with scale
factor $S(t) = e^{t/R_{\cal DS}}$, which was in fact de Sitter's
original space \cite{deS0}, namely the projection to space-time of the
unstable manifold of a given time-like geodesic, so this part looks
like an expanding universe, albeit going back to time $-\infty$ rather
than a finite-time singularity.  The key feature that seems to have
been ignored since Weyl is that flat Friedmann space-times may be
geodesically incomplete in other ways than a big bang.  Weyl completed
de Sitter's space in the way we presented it in this paper.  In
contrast to Weyl, however, we see no reason why objects should not
suddenly become visible to us.  Indeed, as we recall shortly, objects
suddenly become visible in conventional Friedmann models.
Nevertheless, we must examine how much of our mechanism survives
deviations from ${\cal DS}$.\index{Weyl}\index{Friedmann}

\begin{figure}[htbp] %  figure placement: here, top, bottom, or page
   \centering
   \includegraphics[width=2.0in]{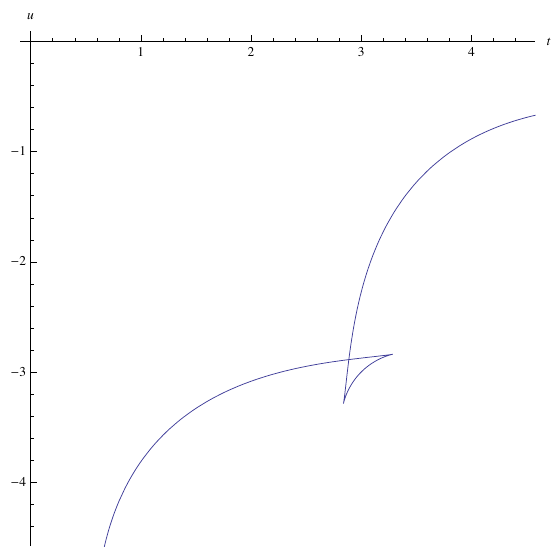} \quad
   \includegraphics[width=2.0in]{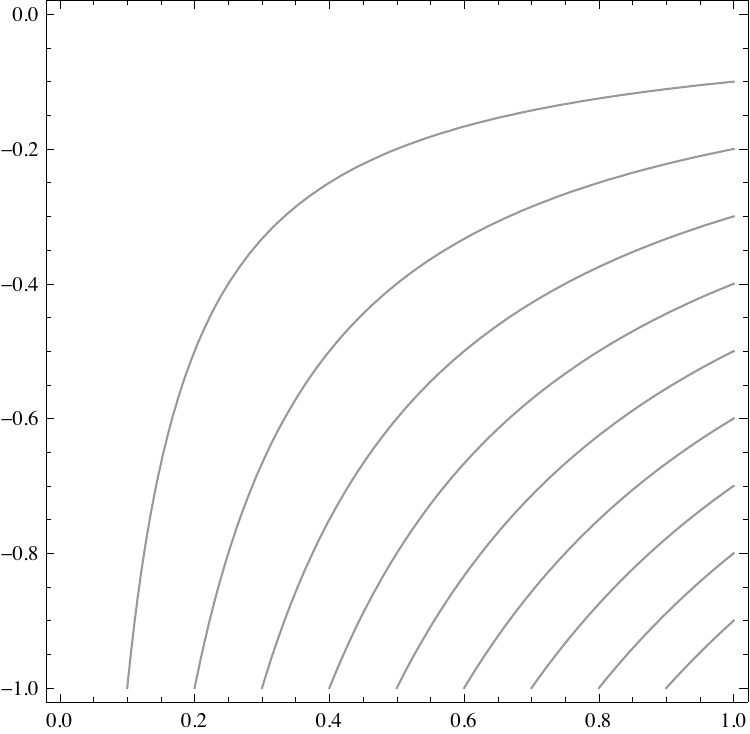}
   \caption{Perturbed $(u,t)$ relations with (a) two cusps, (b) an infinite sequence of branches.}
   \label{fig:perturbed}
\end{figure}

Small deviations of the metric from that of ${\cal DS}$ produce qualitatively the same time-like geodesic flow, because of the structural stability of Anosov systems.  This means there is a near-identity homeomorphism taking time-like geodesics of any $C^2$-small perturbation of the metric to those of ${\cal DS}$.  The proof does not extend to null geodesics, however, so there could be qualitative changes in the set of null geodesics connecting an emitter to a receiver.  
A suggested example is sketched in Figure~\ref{fig:perturbed}(a), corresponding to a swallowtail pleat in the forward light-cone of the emitter passing over the receiver.  
The cusps in the $(u,t)$-relation produce infinite intensity like $|t-t_0|^{-1/2}$ but the singularity is integrable, and the tangent to the cusp has slope in $(0,\infty)$ so there is no exceptional red or blue shifting.  This is a relativistic version of the twinkling of stars.  It is estimated that there are around $10^{22}$ caustics in our backward light-cone \cite{EBD}.  Each caustic of the emitter light-cone crossed by the receiver creates a cusp in the $(u,t)$ diagram.  So we should expect many cusps.  

An alternative deformation of the $(u,t)$ curve is to make a fold at an earlier time than $t^*$; then a precursor will be observed before the main burst, as in some observed cases.

If large perturbation from de Sitter space is considered then larger effects can be expected.  For example, on introducing a Schwarzschild black hole, as in Kottler space, the $(u,t)$-diagram gains an infinite series of curves, corresponding to light paths making different numbers of turns around the black hole, as sketched in Figure~\ref{fig:perturbed}(b).  The added travel time per turn in the black hole's frame is asymptotically $6\pi\sqrt{3}M$ for black hole mass $M$.  Successive curves are presumably fainter.

\index{gamma ray bursts!Friedmann universes}
Let us turn to Friedmann universes, those with metric $ds^2 = -dt^2 + S(t)^2 |dx|^2$ for some scale factor $S(t)>0$, and suppose there is a big bang, i.e.~$S(t)$ is defined for $t>0$ only and $S(t) \to 0$ as $t \to 0$.  If we are on a Hubble flow line $x=$ constant then we see every time-like geodesic redshifted initially.  There is a first time $t^*>0$ we begin to see it, corresponding to emitter time 0.  We see it infinitely redshifted, unless its velocity is directly towards us when it is just finitely redshifted.  The calculations are in \cite{nat-flat}.  If one allows cases with infinite past, however, like the case $S(t)=e^t$ for which the Friedmann universe is half of de Sitter space, and if emitter or receiver is not on the Hubble flow then our scenario for gamma-ray bursts occurs.  

We believe there is room in between de Sitter space and big-bang universes for our mechanism for gamma-ray bursts to apply.

\section{Final remark}\label{sec:GRBFR}
We conclude by remarking that all emitters in de Sitter space except
those converging to us in forwards time exhibit an asymptotic
Lemaitre-Hubble law $z \sim \rho$ for large positive time.  Those on
our unstable manifold do so exactly, but to obtain a good fit there is
no need to require all visible matter to be converging together in
backwards time.
\index{gamma ray bursts|)}

\np\thispagestyle{empty}

\newpage

\lhead[\fancyplain{}{\small\bfseries\thepage}]%
      {\fancyplain{}{\small\bfseries Index}}
\rhead[\fancyplain{}{\small\bfseries Index}]%
      {\fancyplain{}{\small\bfseries\thepage}}

\addcontentsline{toc}{chapter}{Index}
\printindex

\end{document}